\documentclass[a4paper,12pt,openright]{DLG_PhDThesis2017}
\usepackage{cleveref} 
\usepackage{tikz,tkz-fct,tikz-3dplot} 
\usepackage{pgfplots,pgfplotstable}
\usetikzlibrary{arrows}
\usetikzlibrary{shapes}
\usetikzlibrary{positioning}
\usetikzlibrary{matrix}
\usetikzlibrary{backgrounds}
\usetikzlibrary{arrows.meta}

\usepgfplotslibrary{external}
\usepgfplotslibrary{units}
\usepgfplotslibrary{groupplots}
\usepgfplotslibrary{colormaps}
\usepgfplotslibrary{fillbetween}

\titleformat{\chapter}[display]
  {\vspace*{6ex}\bfseries\Huge}
  {\thispagestyle{empty}\titleline[l]{\S\!\chaptertitlename\ \thechapter}}    {\dimexpr-\baselineskip\relax}
  {\filleft}

\titleformat{\section}[hang]{\raggedright\LARGE\bfseries}{\S\,\thesection}{1em}{}
\titleformat{\subsection}[hang]{\raggedright\Large\bfseries}{\S\,\thesubsection}{1em}{}
\titleformat{\subsubsection}[hang]{\raggedright\normalsize\bfseries\itshape}{}{0em}{}

\crefname{figure}{\textsc{Fig.}\!}{\textsc{Figs.}\!}
\crefname{table}{\textsc{Tab.}\!}{\textsc{Tabs.}\!}
\crefname{equation}{\textsc{Eq.}\!}{\textsc{Eqs.}\!}
\crefname{enumi}{\textsc{step}\!}{\textsc{steps}\!}
\crefname{chapter}{\S\!\!}{\S\S\!\!}
\crefname{appendix}{\S\!\!}{\S\S\!\!}
\crefname{section}{\S\!\!}{\S\S\!\!}
\crefname{subsection}{\S\!\!}{\S\S\!\!}
\crefname{subsubsection}{\S\!\!}{\S\S\!\!}
\usepackage[title]{appendix}

\newlength{\dhatheight}
\newcommand{\hhatTS}[1]{%
    \settoheight{\dhatheight}{\ensuremath{\hat{#1}}}%
    \addtolength{\dhatheight}{-0.25ex}%
    \hat{\vphantom{\rule{1pt}{\dhatheight}}%
    \smash{\hat{#1}}}}    
\newcommand{\hhatS}[1]{%
    \settoheight{\dhatheight}{\ensuremath{\scriptstyle{\hat{#1}}}}%
    \addtolength{\dhatheight}{-0.175ex}%
    \hat{\vphantom{\rule{1pt}{\dhatheight}}%
    \smash{\hat{#1}}}}
\newcommand{\hhatSS}[1]{%
    \settoheight{\dhatheight}{\ensuremath{\scriptscriptstyle{\hat{#1}}}}%
    \addtolength{\dhatheight}{-0.07ex}%
    \hat{\vphantom{\rule{1pt}{\dhatheight}}%
    \smash{\hat{#1}}}}
\newcommand{\hhat}[1]{\mathchoice{\hhatTS{#1}}{\hhatTS{#1}}{\hhatS{#1}}{\hhatSS{#1}}}

\newcommand{\vhatTS}[1]{%
    \settoheight{\dhatheight}{\ensuremath{\vec{#1}}}%
    \addtolength{\dhatheight}{-0.25ex}%
    \hat{\vphantom{\rule{1pt}{\dhatheight}}%
    \smash{\vec{#1}}}}    
\newcommand{\vhatS}[1]{%
    \settoheight{\dhatheight}{\ensuremath{\scriptstyle{\vec{#1}}}}%
    \addtolength{\dhatheight}{-0.175ex}%
    \hat{\vphantom{\rule{1pt}{\dhatheight}}%
    \smash{\vec{#1}}}}
\newcommand{\vhatSS}[1]{%
    \settoheight{\dhatheight}{\ensuremath{\scriptscriptstyle{\vec{#1}}}}%
    \addtolength{\dhatheight}{-0.07ex}%
    \hat{\vphantom{\rule{1pt}{\dhatheight}}%
    \smash{\vec{#1}}}}
\newcommand{\vhat}[1]{\mathchoice{\vhatTS{#1}}{\vhatTS{#1}}{\vhatS{#1}}{\vhatSS{#1}}}
\newcommand{\ccdot}{\!\cdot\!}

\makeatletter
\let\oldr@@t\r@@t
\def\r@@t#1#2{%
\setbox0=\hbox{$\oldr@@t#1{#2\,}$}\dimen0=\ht0
\advance\dimen0-0.2\ht0
\setbox2=\hbox{\vrule height\ht0 depth -\dimen0}%
{\box0\lower0.4pt\box2}}
\LetLtxMacro{\oldsqrt}{\sqrt}
\renewcommand*{\sqrt}[2][\ ]{\oldsqrt[#1]{#2}}
\makeatother

\DeclareMathOperator{\Tr}{Tr\!}
\DeclareMathOperator{\dd}{d\!}
\DeclareMathOperator{\e}{e}
\renewcommand{\Re}{\mathrm{Re}}
\renewcommand{\Im}{\mathrm{Im}}
\newcommand{\Unit}{\bm{\mathbbm{1}}}
\newcommand{\Real}{\mathbb{R}}

\newcommand{\Complex}{\mathbb{C}}
\newcommand{\Gradop}{\nabla_{}^{}}
\newcommand{\Hessop}{\nabla_{}^2}
\DeclareMathOperator{\atan2}{atan2}
\DeclareMathOperator{\sign}{sign}
\newcommand{\mA}{\mathbf{A}}
\newcommand{\mB}{\mathbf{B}}
\newcommand{\mC}{\mathbf{C}}
\newcommand{\mD}{\mathbf{D}}
\newcommand{\mLbd}{\mathbf{\Lambda}}
\newcommand{\mM}{\mathbf{M}}
\newcommand{\mQ}{\mathbf{Q}}
\newcommand{\mL}{\mathbf{L}}
\newcommand{\mS}{\mathbf{S}}

\newcommand{\mV}{\mathbf{V}}
\newcommand{\Zero}{\mathbf{0}}
\newcommand{\BigO}{\mathcal{O}}
\newcommand{\eps}{\textsc{eps}}
\newcommand*\scptspc{\vphantom{+}^{\vphantom{|}}_{\vphantom{|}}}
\newcommand{\mX}{\mathbf{X}}
\newcommand{\mR}{\mathbf{R}}
\newcommand{\mE}{\mathbf{E}}
\newcommand{\mI}{\mathbf{I}}

\newcommand{\trm}{\delta}
\newcommand{\rfo}{\alpha}
\newcommand{\eshft}{\sigma}
\newcommand{\condno}{\zeta}
\newcommand{\mineig}{\lambda_{\min}}
\newcommand{\maxeig}{\lambda_{\max}}

\newcommand{\Fid}{\mathcal{J}}
\newcommand{\Pen}{\mathcal{K}}
\newcommand{\targ}{\sigma}
\newcommand{\init}{\rho_0^{}}
\newcommand{\state}{\rho}
\newcommand{\Astate}{\chi}

\newcommand{\Ham}{\mathcal{H}}
\newcommand{\HamH}{\hat{\mathcal{H}}}
\newcommand{\HamZ}{\HamH_{\!\mathcal{Z}}^{}}
\newcommand{\HamNN}{\HamH_{\!\mathcal{N\!N}}^{}}
\newcommand{\HamNE}{\HamH_{\!\mathcal{N\!E}}^{}}
\newcommand{\HamEE}{\HamH_{\mathcal{EE}}^{}}
\newcommand{\HamC}{\HamH_{k}^{}}
\newcommand{\HamHH}{\hhat{\mathcal{H}}}
\newcommand{\Prop}{\hhat{\mathcal{P}}\!}
\newcommand*\pwr{p_c^{}}

\newcommand{\HamB}{\mathbf{H}}
\newcommand{\HamKB}{\mathbf{H}_k}
\newcommand{\HamJB}{\mathbf{H}_j}
\newcommand{\PropB}{\mathbf{P}\!}

\newcommand*\sfrac[2]{\ensuremath{{}^{#1}\!/_{\!#2}}}
\newcommand*\B[1]{\ensuremath{\mathrm{B_{#1}}}}
\newcommand*\spinfrac[2]{\ensuremath{\mathrm{spin\!-\!\sfrac{#1}{#2}}}}
\newcommand*\spinint[1]{\ensuremath{\mathrm{spin-{#1}}}}
\newcommand*\slen[1]{\ensuremath{\varepsilon_{#1}^{}}}
\newcommand*\snewt{\ensuremath{\scptspc\bm{\delta}}}
\newcommand*\itr{\ensuremath{\jmath}}
\newcommand*\lsconst[1]{\ensuremath{\kappa_{#1}^{}}}
\newcommand*\Obj[1]{\ensuremath{\scptspc J(#1)}}
\newcommand*\objv[1]{
\ifthenelse{\equal{#1}{}}{\ensuremath{\scptspc\bm{c}}}{\ensuremath{\bm{c}^{\vphantom{|}}_{#1}}}
}
\newcommand*\OBJV[1]{
\ifthenelse{\equal{#1}{}}{\ensuremath{\scptspc\bm{C}}}{\ensuremath{\bm{C}^{\vphantom{|}}_{#1}}}
}
\newcommand*\sdir[1]{
\ifthenelse{\equal{#1}{}}{\ensuremath{\scptspc\bm{d}}}{\ensuremath{\bm{d}^{\vphantom{|}}_{#1}}}
}
\newcommand*\Grad[1]{
\ifthenelse{\equal{#1}{}}{\ensuremath{\scptspc\Gradop\!\Obj{\objv{}}}}{\ensuremath{\scptspc\Gradop\!\Obj{\objv{#1}}}}
}
\newcommand*\Hess[1]{
\ifthenelse{\equal{#1}{}}{\ensuremath{\scptspc\Hessop\!\Obj{\objv{}}}}{\ensuremath{\scptspc\Hessop\!\Obj{\objv{#1}}}}
}
\newcommand*\HessA[2]{
\ifthenelse{\equal{#1}{}}{\ensuremath{\mathbf{H}^{#2}_{\vphantom{|}}}}{\ensuremath{\mathbf{H}^{#2}_{#1}}}
}
\newcommand*\HessAI[2]{
\ifthenelse{\equal{#1}{}}{\ensuremath{\mathbf{I}^{#2}_{\vphantom{|}}}}{\ensuremath{\mathbf{I}^{#2}_{#1}}}
}
\newcommand*\dobjvar[1]{\ensuremath{\scptspc\Delta\objv{}}}
\newcommand*\dGrad[1]{\ensuremath{\scptspc\Delta\mathbf{g}}}
\newcommand*\dOBJVAR[1]{\ensuremath{\scptspc\Delta\OBJV{}}}
\newcommand*\dGRAD[1]{\ensuremath{\scptspc\Delta\mathbf{G}}}

\newcommand*\ctrl[1]{
\ifthenelse{\equal{#1}{}}{\ensuremath{{c}}}{\ensuremath{{c}_{\scriptscriptstyle #1}^{\vphantom{|}}}}
}
\newcommand*\ctrlv[1]{
\ifthenelse{\equal{#1}{}}{\ensuremath{\bm{c}}}{\ensuremath{\bm{c}_{#1}^{\vphantom{|}}}}
}
\newcommand*\ctrlvn[1]{
\ifthenelse{\equal{#1}{}}{\ensuremath{\tilde{\bm{c}}}}{\ensuremath{\tilde{\bm{c}}_{#1}^{\vphantom{|}}}}
}
\newcommand{\DFid}{\scptspc\Gradop\!\Fid}
\newcommand{\DDFid}{\scptspc\Hessop\!\Fid}
\newcommand*{\DDFidA}[2]{
\ifthenelse{\equal{#1}{}}{\ensuremath{\mathbf{H}^{#2}_{\vphantom{|}}}}{\ensuremath{\mathbf{H}^{#2}_{#1}}}
}
\newcommand*{\DDFidAI}[2]{
\ifthenelse{\equal{#1}{}}{\ensuremath{\mathbf{I}^{#2}_{\vphantom{|}}}}{\ensuremath{\mathbf{I}^{#2}_{#1}}}
}
\newcommand*\dctrl[1]{\ensuremath{\scptspc\Delta\ctrl{}}}

\pgfplotsset{compat=1.12,
  colormap/inferno/.style={
    colormap={inferno}{
      rgb=(0.00146159096,  0.000466127766,  0.0138655200)
      rgb=(0.00226726368,  0.00126992553,  0.0185703520)
      rgb=(0.00329899092,  0.00224934863,  0.0242390508)
      rgb=(0.00454690615,  0.00339180156,  0.0309092475)
      rgb=(0.00600552565,  0.00469194561,  0.0385578980)
      rgb=(0.00767578856,  0.00613611626,  0.0468360336)
      rgb=(0.00956051094,  0.00771344131,  0.0551430756)
      rgb=(0.0116634769,  0.00941675403,  0.0634598080)
      rgb=(0.0139950388,  0.0112247138,  0.0718616890)
      rgb=(0.0165605595,  0.0131362262,  0.0802817951)
      rgb=(0.0193732295,  0.0151325789,  0.0887668094)
      rgb=(0.0224468865,  0.0171991484,  0.0973274383)
      rgb=(0.0257927373,  0.0193306298,  0.105929835)
      rgb=(0.0294324251,  0.0215030771,  0.114621328)
      rgb=(0.0333852235,  0.0237024271,  0.123397286)
      rgb=(0.0376684211,  0.0259207864,  0.132232108)
      rgb=(0.0422525554,  0.0281385015,  0.141140519)
      rgb=(0.0469146287,  0.0303236129,  0.150163867)
      rgb=(0.0516437624,  0.0324736172,  0.159254277)
      rgb=(0.0564491009,  0.0345691867,  0.168413539)
      rgb=(0.0613397200,  0.0365900213,  0.177642172)
      rgb=(0.0663312620,  0.0385036268,  0.186961588)
      rgb=(0.0714289181,  0.0402939095,  0.196353558)
      rgb=(0.0766367560,  0.0419053329,  0.205798788)
      rgb=(0.0819620773,  0.0433278666,  0.215289113)
      rgb=(0.0874113897,  0.0445561662,  0.224813479)
      rgb=(0.0929901526,  0.0455829503,  0.234357604)
      rgb=(0.0987024972,  0.0464018731,  0.243903700)
      rgb=(0.104550936,  0.0470080541,  0.253430300)
      rgb=(0.110536084,  0.0473986708,  0.262912235)
      rgb=(0.116656423,  0.0475735920,  0.272320803)
      rgb=(0.122908126,  0.0475360183,  0.281624170)
      rgb=(0.129284984,  0.0472930838,  0.290788012)
      rgb=(0.135778450,  0.0468563678,  0.299776404)
      rgb=(0.142377819,  0.0462422566,  0.308552910)
      rgb=(0.149072957,  0.0454676444,  0.317085139)
      rgb=(0.155849711,  0.0445588056,  0.325338414)
      rgb=(0.162688939,  0.0435542881,  0.333276678)
      rgb=(0.169575148,  0.0424893149,  0.340874188)
      rgb=(0.176493202,  0.0414017089,  0.348110606)
      rgb=(0.183428775,  0.0403288858,  0.354971391)
      rgb=(0.190367453,  0.0393088888,  0.361446945)
      rgb=(0.197297425,  0.0384001825,  0.367534629)
      rgb=(0.204209298,  0.0376322609,  0.373237557)
      rgb=(0.211095463,  0.0370296488,  0.378563264)
      rgb=(0.217948648,  0.0366146049,  0.383522415)
      rgb=(0.224762908,  0.0364049901,  0.388128944)
      rgb=(0.231538148,  0.0364052511,  0.392400150)
      rgb=(0.238272961,  0.0366209949,  0.396353388)
      rgb=(0.244966911,  0.0370545017,  0.400006615)
      rgb=(0.251620354,  0.0377052832,  0.403377897)
      rgb=(0.258234265,  0.0385706153,  0.406485031)
      rgb=(0.264809649,  0.0396468666,  0.409345373)
      rgb=(0.271346664,  0.0409215821,  0.411976086)
      rgb=(0.277849829,  0.0423528741,  0.414392106)
      rgb=(0.284321318,  0.0439325787,  0.416607861)
      rgb=(0.290763373,  0.0456437598,  0.418636756)
      rgb=(0.297178251,  0.0474700293,  0.420491164)
      rgb=(0.303568182,  0.0493958927,  0.422182449)
      rgb=(0.309935342,  0.0514069729,  0.423720999)
      rgb=(0.316281835,  0.0534901321,  0.425116277)
      rgb=(0.322609671,  0.0556335178,  0.426376869)
      rgb=(0.328920763,  0.0578265505,  0.427510546)
      rgb=(0.335216916,  0.0600598734,  0.428524320)
      rgb=(0.341499828,  0.0623252772,  0.429424503)
      rgb=(0.347771086,  0.0646156100,  0.430216765)
      rgb=(0.354032169,  0.0669246832,  0.430906186)
      rgb=(0.360284449,  0.0692471753,  0.431497309)
      rgb=(0.366529195,  0.0715785403,  0.431994185)
      rgb=(0.372767575,  0.0739149211,  0.432400419)
      rgb=(0.379000659,  0.0762530701,  0.432719214)
      rgb=(0.385228383,  0.0785914864,  0.432954973)
      rgb=(0.391452659,  0.0809267058,  0.433108763)
      rgb=(0.397674379,  0.0832568129,  0.433182647)
      rgb=(0.403894278,  0.0855803445,  0.433178526)
      rgb=(0.410113015,  0.0878961593,  0.433098056)
      rgb=(0.416331169,  0.0902033992,  0.432942678)
      rgb=(0.422549249,  0.0925014543,  0.432713635)
      rgb=(0.428767696,  0.0947899342,  0.432411996)
      rgb=(0.434986885,  0.0970686417,  0.432038673)
      rgb=(0.441207124,  0.0993375510,  0.431594438)
      rgb=(0.447428382,  0.101597079,  0.431080497)
      rgb=(0.453650614,  0.103847716,  0.430497898)
      rgb=(0.459874623,  0.106089165,  0.429845789)
      rgb=(0.466100494,  0.108321923,  0.429124507)
      rgb=(0.472328255,  0.110546584,  0.428334320)
      rgb=(0.478557889,  0.112763831,  0.427475431)
      rgb=(0.484789325,  0.114974430,  0.426547991)
      rgb=(0.491022448,  0.117179219,  0.425552106)
      rgb=(0.497257069,  0.119379132,  0.424487908)
      rgb=(0.503492698,  0.121575414,  0.423356110)
      rgb=(0.509729541,  0.123768654,  0.422155676)
      rgb=(0.515967304,  0.125959947,  0.420886594)
      rgb=(0.522205646,  0.128150439,  0.419548848)
      rgb=(0.528444192,  0.130341324,  0.418142411)
      rgb=(0.534682523,  0.132533845,  0.416667258)
      rgb=(0.540920186,  0.134729286,  0.415123366)
      rgb=(0.547156706,  0.136928959,  0.413510662)
      rgb=(0.553391649,  0.139134147,  0.411828882)
      rgb=(0.559624442,  0.141346265,  0.410078028)
      rgb=(0.565854477,  0.143566769,  0.408258132)
      rgb=(0.572081108,  0.145797150,  0.406369246)
      rgb=(0.578303656,  0.148038934,  0.404411444)
      rgb=(0.584521407,  0.150293679,  0.402384829)
      rgb=(0.590733615,  0.152562977,  0.400289528)
      rgb=(0.596939751,  0.154848232,  0.398124897)
      rgb=(0.603138930,  0.157151161,  0.395891308)
      rgb=(0.609330184,  0.159473549,  0.393589349)
      rgb=(0.615512627,  0.161817111,  0.391219295)
      rgb=(0.621685340,  0.164183582,  0.388781456)
      rgb=(0.627847374,  0.166574724,  0.386276180)
      rgb=(0.633997746,  0.168992314,  0.383703854)
      rgb=(0.640135447,  0.171438150,  0.381064906)
      rgb=(0.646259648,  0.173913876,  0.378358969)
      rgb=(0.652369348,  0.176421271,  0.375586209)
      rgb=(0.658463166,  0.178962399,  0.372748214)
      rgb=(0.664539964,  0.181539111,  0.369845599)
      rgb=(0.670598572,  0.184153268,  0.366879025)
      rgb=(0.676637795,  0.186806728,  0.363849195)
      rgb=(0.682656407,  0.189501352,  0.360756856)
      rgb=(0.688653158,  0.192238994,  0.357602797)
      rgb=(0.694626769,  0.195021500,  0.354387853)
      rgb=(0.700575937,  0.197850703,  0.351112900)
      rgb=(0.706499709,  0.200728196,  0.347776863)
      rgb=(0.712396345,  0.203656029,  0.344382594)
      rgb=(0.718264447,  0.206635993,  0.340931208)
      rgb=(0.724102613,  0.209669834,  0.337423766)
      rgb=(0.729909422,  0.212759270,  0.333861367)
      rgb=(0.735683432,  0.215905976,  0.330245147)
      rgb=(0.741423185,  0.219111589,  0.326576275)
      rgb=(0.747127207,  0.222377697,  0.322855952)
      rgb=(0.752794009,  0.225705837,  0.319085410)
      rgb=(0.758422090,  0.229097492,  0.315265910)
      rgb=(0.764009940,  0.232554083,  0.311398734)
      rgb=(0.769556038,  0.236076967,  0.307485188)
      rgb=(0.775058888,  0.239667435,  0.303526312)
      rgb=(0.780517023,  0.243326720,  0.299522665)
      rgb=(0.785928794,  0.247055968,  0.295476756)
      rgb=(0.791292674,  0.250856232,  0.291389943)
      rgb=(0.796607144,  0.254728485,  0.287263585)
      rgb=(0.801870689,  0.258673610,  0.283099033)
      rgb=(0.807081807,  0.262692401,  0.278897629)
      rgb=(0.812239008,  0.266785558,  0.274660698)
      rgb=(0.817340818,  0.270953688,  0.270389545)
      rgb=(0.822385784,  0.275197300,  0.266085445)
      rgb=(0.827372474,  0.279516805,  0.261749643)
      rgb=(0.832299481,  0.283912516,  0.257383341)
      rgb=(0.837165425,  0.288384647,  0.252987700)
      rgb=(0.841968959,  0.292933312,  0.248563825)
      rgb=(0.846708768,  0.297558528,  0.244112767)
      rgb=(0.851383572,  0.302260213,  0.239635512)
      rgb=(0.855992130,  0.307038188,  0.235132978)
      rgb=(0.860533241,  0.311892183,  0.230606009)
      rgb=(0.865005747,  0.316821833,  0.226055368)
      rgb=(0.869408534,  0.321826685,  0.221481734)
      rgb=(0.873740530,  0.326906201,  0.216885699)
      rgb=(0.878000715,  0.332059760,  0.212267762)
      rgb=(0.882188112,  0.337286663,  0.207628326)
      rgb=(0.886301795,  0.342586137,  0.202967696)
      rgb=(0.890340885,  0.347957340,  0.198286080)
      rgb=(0.894304553,  0.353399363,  0.193583583)
      rgb=(0.898192017,  0.358911240,  0.188860212)
      rgb=(0.902002544,  0.364491949,  0.184115876)
      rgb=(0.905735448,  0.370140419,  0.179350388)
      rgb=(0.909390090,  0.375855533,  0.174563472)
      rgb=(0.912965874,  0.381636138,  0.169754764)
      rgb=(0.916462251,  0.387481044,  0.164923826)
      rgb=(0.919878710,  0.393389034,  0.160070152)
      rgb=(0.923214783,  0.399358867,  0.155193185)
      rgb=(0.926470039,  0.405389282,  0.150292329)
      rgb=(0.929644083,  0.411479007,  0.145366973)
      rgb=(0.932736555,  0.417626756,  0.140416519)
      rgb=(0.935747126,  0.423831237,  0.135440416)
      rgb=(0.938675494,  0.430091162,  0.130438175)
      rgb=(0.941521384,  0.436405243,  0.125409440)
      rgb=(0.944284543,  0.442772199,  0.120354038)
      rgb=(0.946964741,  0.449190757,  0.115272059)
      rgb=(0.949561766,  0.455659658,  0.110163947)
      rgb=(0.952075421,  0.462177656,  0.105030614)
      rgb=(0.954505523,  0.468743522,  0.0998735931)
      rgb=(0.956851903,  0.475356048,  0.0946952268)
      rgb=(0.959114397,  0.482014044,  0.0894989073)
      rgb=(0.961292850,  0.488716345,  0.0842893891)
      rgb=(0.963387110,  0.495461806,  0.0790731907)
      rgb=(0.965397031,  0.502249309,  0.0738591143)
      rgb=(0.967322465,  0.509077761,  0.0686589199)
      rgb=(0.969163264,  0.515946092,  0.0634881971)
      rgb=(0.970919277,  0.522853259,  0.0583674890)
      rgb=(0.972590351,  0.529798246,  0.0533237243)
      rgb=(0.974176327,  0.536780059,  0.0483920090)
      rgb=(0.975677038,  0.543797733,  0.0436177922)
      rgb=(0.977092313,  0.550850323,  0.0390500131)
      rgb=(0.978421971,  0.557936911,  0.0349306227)
      rgb=(0.979665824,  0.565056600,  0.0314091591)
      rgb=(0.980823673,  0.572208516,  0.0285075931)
      rgb=(0.981895311,  0.579391803,  0.0262497353)
      rgb=(0.982880522,  0.586605627,  0.0246613416)
      rgb=(0.983779081,  0.593849168,  0.0237702263)
      rgb=(0.984590755,  0.601121626,  0.0236063833)
      rgb=(0.985315301,  0.608422211,  0.0242021174)
      rgb=(0.985952471,  0.615750147,  0.0255921853)
      rgb=(0.986502013,  0.623104667,  0.0278139496)
      rgb=(0.986963670,  0.630485011,  0.0309075459)
      rgb=(0.987337182,  0.637890424,  0.0349160639)
      rgb=(0.987622296,  0.645320152,  0.0398857472)
      rgb=(0.987818759,  0.652773439,  0.0455808037)
      rgb=(0.987926330,  0.660249526,  0.0517503867)
      rgb=(0.987944783,  0.667747641,  0.0583286889)
      rgb=(0.987873910,  0.675267000,  0.0652570167)
      rgb=(0.987713535,  0.682806802,  0.0724892330)
      rgb=(0.987463516,  0.690366218,  0.0799897176)
      rgb=(0.987123759,  0.697944391,  0.0877314215)
      rgb=(0.986694229,  0.705540424,  0.0956941797)
      rgb=(0.986174970,  0.713153375,  0.103863324)
      rgb=(0.985565739,  0.720782460,  0.112228756)
      rgb=(0.984865203,  0.728427497,  0.120784651)
      rgb=(0.984075129,  0.736086521,  0.129526579)
      rgb=(0.983195992,  0.743758326,  0.138453063)
      rgb=(0.982228463,  0.751441596,  0.147564573)
      rgb=(0.981173457,  0.759134892,  0.156863224)
      rgb=(0.980032178,  0.766836624,  0.166352544)
      rgb=(0.978806183,  0.774545028,  0.176037298)
      rgb=(0.977497453,  0.782258138,  0.185923357)
      rgb=(0.976108474,  0.789973753,  0.196017589)
      rgb=(0.974637842,  0.797691563,  0.206331925)
      rgb=(0.973087939,  0.805409333,  0.216876839)
      rgb=(0.971467822,  0.813121725,  0.227658046)
      rgb=(0.969783146,  0.820825143,  0.238685942)
      rgb=(0.968040817,  0.828515491,  0.249971582)
      rgb=(0.966242589,  0.836190976,  0.261533898)
      rgb=(0.964393924,  0.843848069,  0.273391112)
      rgb=(0.962516656,  0.851476340,  0.285545675)
      rgb=(0.960625545,  0.859068716,  0.298010219)
      rgb=(0.958720088,  0.866624355,  0.310820466)
      rgb=(0.956834075,  0.874128569,  0.323973947)
      rgb=(0.954997177,  0.881568926,  0.337475479)
      rgb=(0.953215092,  0.888942277,  0.351368713)
      rgb=(0.951546225,  0.896225909,  0.365627005)
      rgb=(0.950018481,  0.903409063,  0.380271225)
      rgb=(0.948683391,  0.910472964,  0.395289169)
      rgb=(0.947594362,  0.917399053,  0.410665194)
      rgb=(0.946809163,  0.924168246,  0.426373236)
      rgb=(0.946391536,  0.930760752,  0.442367495)
      rgb=(0.946402951,  0.937158971,  0.458591507)
      rgb=(0.946902568,  0.943347775,  0.474969778)
      rgb=(0.947936825,  0.949317522,  0.491426053)
      rgb=(0.949544830,  0.955062900,  0.507859649)
      rgb=(0.951740304,  0.960586693,  0.524203026)
      rgb=(0.954529281,  0.965895868,  0.540360752)
      rgb=(0.957896053,  0.971003330,  0.556275090)
      rgb=(0.961812020,  0.975924241,  0.571925382)
      rgb=(0.966248822,  0.980678193,  0.587205773)
      rgb=(0.971161622,  0.985282161,  0.602154330)
      rgb=(0.976510983,  0.989753437,  0.616760413)
      rgb=(0.982257307,  0.994108844,  0.631017009)
      rgb=(0.988362068,  0.998364143,  0.644924005)
    }
  }
}
\pgfplotsset{compat=1.12,
  colormap/magma/.style={
    colormap={magma}{
      rgb=(0.00146159096,  0.000466127766,  0.0138655200)
      rgb=(0.00225764007,  0.00129495431,  0.0183311461)
      rgb=(0.00327943222,  0.00230452991,  0.0237083291)
      rgb=(0.00451230222,  0.00349037666,  0.0299647059)
      rgb=(0.00594976987,  0.00484285000,  0.0371296695)
      rgb=(0.00758798550,  0.00635613622,  0.0449730774)
      rgb=(0.00942604390,  0.00802185006,  0.0528443561)
      rgb=(0.0114654337,  0.00982831486,  0.0607496380)
      rgb=(0.0137075706,  0.0117705913,  0.0686665843)
      rgb=(0.0161557566,  0.0138404966,  0.0766026660)
      rgb=(0.0188153670,  0.0160262753,  0.0845844897)
      rgb=(0.0216919340,  0.0183201254,  0.0926101050)
      rgb=(0.0247917814,  0.0207147875,  0.100675555)
      rgb=(0.0281228154,  0.0232009284,  0.108786954)
      rgb=(0.0316955304,  0.0257651161,  0.116964722)
      rgb=(0.0355204468,  0.0283974570,  0.125209396)
      rgb=(0.0396084872,  0.0310895652,  0.133515085)
      rgb=(0.0438295350,  0.0338299885,  0.141886249)
      rgb=(0.0480616391,  0.0366066101,  0.150326989)
      rgb=(0.0523204388,  0.0394066020,  0.158841025)
      rgb=(0.0566148978,  0.0421598925,  0.167445592)
      rgb=(0.0609493930,  0.0447944924,  0.176128834)
      rgb=(0.0653301801,  0.0473177796,  0.184891506)
      rgb=(0.0697637296,  0.0497264666,  0.193735088)
      rgb=(0.0742565152,  0.0520167766,  0.202660374)
      rgb=(0.0788150034,  0.0541844801,  0.211667355)
      rgb=(0.0834456313,  0.0562249365,  0.220755099)
      rgb=(0.0881547730,  0.0581331465,  0.229921611)
      rgb=(0.0929486914,  0.0599038167,  0.239163669)
      rgb=(0.0978334770,  0.0615314414,  0.248476662)
      rgb=(0.102814972,  0.0630104053,  0.257854400)
      rgb=(0.107898679,  0.0643351102,  0.267288933)
      rgb=(0.113094451,  0.0654920358,  0.276783978)
      rgb=(0.118405035,  0.0664791593,  0.286320656)
      rgb=(0.123832651,  0.0672946449,  0.295879431)
      rgb=(0.129380192,  0.0679349264,  0.305442931)
      rgb=(0.135053322,  0.0683912798,  0.314999890)
      rgb=(0.140857952,  0.0686540710,  0.324537640)
      rgb=(0.146785234,  0.0687382323,  0.334011109)
      rgb=(0.152839217,  0.0686368599,  0.343404450)
      rgb=(0.159017511,  0.0683540225,  0.352688028)
      rgb=(0.165308131,  0.0679108689,  0.361816426)
      rgb=(0.171713033,  0.0673053260,  0.370770827)
      rgb=(0.178211730,  0.0665758073,  0.379497161)
      rgb=(0.184800877,  0.0657324381,  0.387972507)
      rgb=(0.191459745,  0.0648183312,  0.396151969)
      rgb=(0.198176877,  0.0638624166,  0.404008953)
      rgb=(0.204934882,  0.0629066192,  0.411514273)
      rgb=(0.211718061,  0.0619917876,  0.418646741)
      rgb=(0.218511590,  0.0611584918,  0.425391816)
      rgb=(0.225302032,  0.0604451843,  0.431741767)
      rgb=(0.232076515,  0.0598886855,  0.437694665)
      rgb=(0.238825991,  0.0595170384,  0.443255999)
      rgb=(0.245543175,  0.0593524384,  0.448435938)
      rgb=(0.252220252,  0.0594147119,  0.453247729)
      rgb=(0.258857304,  0.0597055998,  0.457709924)
      rgb=(0.265446744,  0.0602368754,  0.461840297)
      rgb=(0.271994089,  0.0609935552,  0.465660375)
      rgb=(0.278493300,  0.0619778136,  0.469190328)
      rgb=(0.284951097,  0.0631676261,  0.472450879)
      rgb=(0.291365817,  0.0645534486,  0.475462193)
      rgb=(0.297740413,  0.0661170432,  0.478243482)
      rgb=(0.304080941,  0.0678353452,  0.480811572)
      rgb=(0.310382027,  0.0697024767,  0.483186340)
      rgb=(0.316654235,  0.0716895272,  0.485380429)
      rgb=(0.322899126,  0.0737819504,  0.487408399)
      rgb=(0.329114038,  0.0759715081,  0.489286796)
      rgb=(0.335307503,  0.0782361045,  0.491024144)
      rgb=(0.341481725,  0.0805635079,  0.492631321)
      rgb=(0.347635742,  0.0829463512,  0.494120923)
      rgb=(0.353773161,  0.0853726329,  0.495501096)
      rgb=(0.359897941,  0.0878311772,  0.496778331)
      rgb=(0.366011928,  0.0903143031,  0.497959963)
      rgb=(0.372116205,  0.0928159917,  0.499053326)
      rgb=(0.378210547,  0.0953322947,  0.500066568)
      rgb=(0.384299445,  0.0978549106,  0.501001964)
      rgb=(0.390384361,  0.100379466,  0.501864236)
      rgb=(0.396466670,  0.102902194,  0.502657590)
      rgb=(0.402547663,  0.105419865,  0.503385761)
      rgb=(0.408628505,  0.107929771,  0.504052118)
      rgb=(0.414708664,  0.110431177,  0.504661843)
      rgb=(0.420791157,  0.112920210,  0.505214935)
      rgb=(0.426876965,  0.115395258,  0.505713602)
      rgb=(0.432967001,  0.117854987,  0.506159754)
      rgb=(0.439062114,  0.120298314,  0.506555026)
      rgb=(0.445163096,  0.122724371,  0.506900806)
      rgb=(0.451270678,  0.125132484,  0.507198258)
      rgb=(0.457385535,  0.127522145,  0.507448336)
      rgb=(0.463508291,  0.129892998,  0.507651812)
      rgb=(0.469639514,  0.132244819,  0.507809282)
      rgb=(0.475779723,  0.134577500,  0.507921193)
      rgb=(0.481928997,  0.136891390,  0.507988509)
      rgb=(0.488088169,  0.139186217,  0.508010737)
      rgb=(0.494257673,  0.141462106,  0.507987836)
      rgb=(0.500437834,  0.143719323,  0.507919772)
      rgb=(0.506628929,  0.145958202,  0.507806420)
      rgb=(0.512831195,  0.148179144,  0.507647570)
      rgb=(0.519044825,  0.150382611,  0.507442938)
      rgb=(0.525269968,  0.152569121,  0.507192172)
      rgb=(0.531506735,  0.154739247,  0.506894860)
      rgb=(0.537755194,  0.156893613,  0.506550538)
      rgb=(0.544015371,  0.159032895,  0.506158696)
      rgb=(0.550287252,  0.161157816,  0.505718782)
      rgb=(0.556570783,  0.163269149,  0.505230210)
      rgb=(0.562865867,  0.165367714,  0.504692365)
      rgb=(0.569172368,  0.167454379,  0.504104606)
      rgb=(0.575490107,  0.169530062,  0.503466273)
      rgb=(0.581818864,  0.171595728,  0.502776690)
      rgb=(0.588158375,  0.173652392,  0.502035167)
      rgb=(0.594508337,  0.175701122,  0.501241011)
      rgb=(0.600868399,  0.177743036,  0.500393522)
      rgb=(0.607238169,  0.179779309,  0.499491999)
      rgb=(0.613617209,  0.181811170,  0.498535746)
      rgb=(0.620005032,  0.183839907,  0.497524075)
      rgb=(0.626401108,  0.185866869,  0.496456304)
      rgb=(0.632804854,  0.187893468,  0.495331769)
      rgb=(0.639215638,  0.189921182,  0.494149821)
      rgb=(0.645632778,  0.191951556,  0.492909832)
      rgb=(0.652055535,  0.193986210,  0.491611196)
      rgb=(0.658483116,  0.196026835,  0.490253338)
      rgb=(0.664914668,  0.198075202,  0.488835712)
      rgb=(0.671349279,  0.200133166,  0.487357807)
      rgb=(0.677785975,  0.202202663,  0.485819154)
      rgb=(0.684223712,  0.204285721,  0.484219325)
      rgb=(0.690661380,  0.206384461,  0.482557941)
      rgb=(0.697097796,  0.208501100,  0.480834678)
      rgb=(0.703531700,  0.210637956,  0.479049270)
      rgb=(0.709961888,  0.212797337,  0.477201121)
      rgb=(0.716387038,  0.214981693,  0.475289780)
      rgb=(0.722805451,  0.217193831,  0.473315708)
      rgb=(0.729215521,  0.219436516,  0.471278924)
      rgb=(0.735615545,  0.221712634,  0.469179541)
      rgb=(0.742003713,  0.224025196,  0.467017774)
      rgb=(0.748378107,  0.226377345,  0.464793954)
      rgb=(0.754736692,  0.228772352,  0.462508534)
      rgb=(0.761077312,  0.231213625,  0.460162106)
      rgb=(0.767397681,  0.233704708,  0.457755411)
      rgb=(0.773695380,  0.236249283,  0.455289354)
      rgb=(0.779967847,  0.238851170,  0.452765022)
      rgb=(0.786212372,  0.241514325,  0.450183695)
      rgb=(0.792426972,  0.244242250,  0.447543155)
      rgb=(0.798607760,  0.247039798,  0.444848441)
      rgb=(0.804751511,  0.249911350,  0.442101615)
      rgb=(0.810854841,  0.252861399,  0.439304963)
      rgb=(0.816914186,  0.255894550,  0.436461074)
      rgb=(0.822925797,  0.259015505,  0.433572874)
      rgb=(0.828885740,  0.262229049,  0.430643647)
      rgb=(0.834790818,  0.265539703,  0.427671352)
      rgb=(0.840635680,  0.268952874,  0.424665620)
      rgb=(0.846415804,  0.272473491,  0.421631064)
      rgb=(0.852126490,  0.276106469,  0.418572767)
      rgb=(0.857762870,  0.279856666,  0.415496319)
      rgb=(0.863320397,  0.283729003,  0.412402889)
      rgb=(0.868793368,  0.287728205,  0.409303002)
      rgb=(0.874176342,  0.291858679,  0.406205397)
      rgb=(0.879463944,  0.296124596,  0.403118034)
      rgb=(0.884650824,  0.300530090,  0.400047060)
      rgb=(0.889731418,  0.305078817,  0.397001559)
      rgb=(0.894700194,  0.309773445,  0.393994634)
      rgb=(0.899551884,  0.314616425,  0.391036674)
      rgb=(0.904281297,  0.319609981,  0.388136889)
      rgb=(0.908883524,  0.324755126,  0.385308008)
      rgb=(0.913354091,  0.330051947,  0.382563414)
      rgb=(0.917688852,  0.335500068,  0.379915138)
      rgb=(0.921884187,  0.341098112,  0.377375977)
      rgb=(0.925937102,  0.346843685,  0.374959077)
      rgb=(0.929845090,  0.352733817,  0.372676513)
      rgb=(0.933606454,  0.358764377,  0.370540883)
      rgb=(0.937220874,  0.364929312,  0.368566525)
      rgb=(0.940687443,  0.371224168,  0.366761699)
      rgb=(0.944006448,  0.377642889,  0.365136328)
      rgb=(0.947179528,  0.384177874,  0.363701130)
      rgb=(0.950210150,  0.390819546,  0.362467694)
      rgb=(0.953099077,  0.397562894,  0.361438431)
      rgb=(0.955849237,  0.404400213,  0.360619076)
      rgb=(0.958464079,  0.411323666,  0.360014232)
      rgb=(0.960949221,  0.418323245,  0.359629789)
      rgb=(0.963310281,  0.425389724,  0.359469020)
      rgb=(0.965549351,  0.432518707,  0.359529151)
      rgb=(0.967671128,  0.439702976,  0.359810172)
      rgb=(0.969680441,  0.446935635,  0.360311120)
      rgb=(0.971582181,  0.454210170,  0.361030156)
      rgb=(0.973381238,  0.461520484,  0.361964652)
      rgb=(0.975082439,  0.468860936,  0.363111292)
      rgb=(0.976690494,  0.476226350,  0.364466162)
      rgb=(0.978209957,  0.483612031,  0.366024854)
      rgb=(0.979645181,  0.491013764,  0.367782559)
      rgb=(0.981000291,  0.498427800,  0.369734157)
      rgb=(0.982279159,  0.505850848,  0.371874301)
      rgb=(0.983485387,  0.513280054,  0.374197501)
      rgb=(0.984622298,  0.520712972,  0.376698186)
      rgb=(0.985692925,  0.528147545,  0.379370774)
      rgb=(0.986700017,  0.535582070,  0.382209724)
      rgb=(0.987646038,  0.543015173,  0.385209578)
      rgb=(0.988533173,  0.550445778,  0.388365009)
      rgb=(0.989363341,  0.557873075,  0.391670846)
      rgb=(0.990138201,  0.565296495,  0.395122099)
      rgb=(0.990871208,  0.572706259,  0.398713971)
      rgb=(0.991558165,  0.580106828,  0.402441058)
      rgb=(0.992195728,  0.587501706,  0.406298792)
      rgb=(0.992784669,  0.594891088,  0.410282976)
      rgb=(0.993325561,  0.602275297,  0.414389658)
      rgb=(0.993834412,  0.609643540,  0.418613221)
      rgb=(0.994308514,  0.616998953,  0.422949672)
      rgb=(0.994737698,  0.624349657,  0.427396771)
      rgb=(0.995121854,  0.631696376,  0.431951492)
      rgb=(0.995480469,  0.639026596,  0.436607159)
      rgb=(0.995809924,  0.646343897,  0.441360951)
      rgb=(0.996095703,  0.653658756,  0.446213021)
      rgb=(0.996341406,  0.660969379,  0.451160201)
      rgb=(0.996579803,  0.668255621,  0.456191814)
      rgb=(0.996774784,  0.675541484,  0.461314158)
      rgb=(0.996925427,  0.682827953,  0.466525689)
      rgb=(0.997077185,  0.690087897,  0.471811461)
      rgb=(0.997186253,  0.697348991,  0.477181727)
      rgb=(0.997253982,  0.704610791,  0.482634651)
      rgb=(0.997325180,  0.711847714,  0.488154375)
      rgb=(0.997350983,  0.719089119,  0.493754665)
      rgb=(0.997350583,  0.726324415,  0.499427972)
      rgb=(0.997341259,  0.733544671,  0.505166839)
      rgb=(0.997284689,  0.740771893,  0.510983331)
      rgb=(0.997228367,  0.747980563,  0.516859378)
      rgb=(0.997138480,  0.755189852,  0.522805996)
      rgb=(0.997019342,  0.762397883,  0.528820775)
      rgb=(0.996898254,  0.769590975,  0.534892341)
      rgb=(0.996726862,  0.776794860,  0.541038571)
      rgb=(0.996570645,  0.783976508,  0.547232992)
      rgb=(0.996369065,  0.791167346,  0.553498939)
      rgb=(0.996162309,  0.798347709,  0.559819643)
      rgb=(0.995932448,  0.805527126,  0.566201824)
      rgb=(0.995680107,  0.812705773,  0.572644795)
      rgb=(0.995423973,  0.819875302,  0.579140130)
      rgb=(0.995131288,  0.827051773,  0.585701463)
      rgb=(0.994851089,  0.834212826,  0.592307093)
      rgb=(0.994523666,  0.841386618,  0.598982818)
      rgb=(0.994221900,  0.848540474,  0.605695903)
      rgb=(0.993865767,  0.855711038,  0.612481798)
      rgb=(0.993545285,  0.862858846,  0.619299300)
      rgb=(0.993169558,  0.870024467,  0.626189463)
      rgb=(0.992830963,  0.877168404,  0.633109148)
      rgb=(0.992439881,  0.884329694,  0.640099465)
      rgb=(0.992089454,  0.891469549,  0.647116021)
      rgb=(0.991687744,  0.898627050,  0.654201544)
      rgb=(0.991331929,  0.905762748,  0.661308839)
      rgb=(0.990929685,  0.912915010,  0.668481201)
      rgb=(0.990569914,  0.920048699,  0.675674592)
      rgb=(0.990174637,  0.927195612,  0.682925602)
      rgb=(0.989814839,  0.934328540,  0.690198194)
      rgb=(0.989433736,  0.941470354,  0.697518628)
      rgb=(0.989077438,  0.948604077,  0.704862519)
      rgb=(0.988717064,  0.955741520,  0.712242232)
      rgb=(0.988367028,  0.962878026,  0.719648627)
      rgb=(0.988032885,  0.970012413,  0.727076773)
      rgb=(0.987690702,  0.977154231,  0.734536205)
      rgb=(0.987386827,  0.984287561,  0.742001547)
      rgb=(0.987052509,  0.991437853,  0.749504188)
    }
  }
}
\pgfplotsset{compat=1.12,
  colormap/plasma/.style={
    colormap={plasma}{
      rgb=(0.0503832136,  0.0298028976,  0.527974883)
      rgb=(0.0635363639,  0.0284259729,  0.533123681)
      rgb=(0.0753531234,  0.0272063728,  0.538007001)
      rgb=(0.0862217979,  0.0261253206,  0.542657691)
      rgb=(0.0963786097,  0.0251650976,  0.547103487)
      rgb=(0.105979704,  0.0243092436,  0.551367851)
      rgb=(0.115123641,  0.0235562500,  0.555467728)
      rgb=(0.123902903,  0.0228781011,  0.559423480)
      rgb=(0.132380720,  0.0222583774,  0.563250116)
      rgb=(0.140603076,  0.0216866674,  0.566959485)
      rgb=(0.148606527,  0.0211535876,  0.570561711)
      rgb=(0.156420649,  0.0206507174,  0.574065446)
      rgb=(0.164069722,  0.0201705326,  0.577478074)
      rgb=(0.171573925,  0.0197063415,  0.580805890)
      rgb=(0.178950212,  0.0192522243,  0.584054243)
      rgb=(0.186212958,  0.0188029767,  0.587227661)
      rgb=(0.193374449,  0.0183540593,  0.590329954)
      rgb=(0.200445260,  0.0179015512,  0.593364304)
      rgb=(0.207434551,  0.0174421086,  0.596333341)
      rgb=(0.214350298,  0.0169729276,  0.599239207)
      rgb=(0.221196750,  0.0164970484,  0.602083323)
      rgb=(0.227982971,  0.0160071509,  0.604867403)
      rgb=(0.234714537,  0.0155015065,  0.607592438)
      rgb=(0.241396253,  0.0149791041,  0.610259089)
      rgb=(0.248032377,  0.0144393586,  0.612867743)
      rgb=(0.254626690,  0.0138820918,  0.615418537)
      rgb=(0.261182562,  0.0133075156,  0.617911385)
      rgb=(0.267702993,  0.0127162163,  0.620345997)
      rgb=(0.274190665,  0.0121091423,  0.622721903)
      rgb=(0.280647969,  0.0114875915,  0.625038468)
      rgb=(0.287076059,  0.0108554862,  0.627294975)
      rgb=(0.293477695,  0.0102128849,  0.629490490)
      rgb=(0.299855122,  0.00956079551,  0.631623923)
      rgb=(0.306209825,  0.00890185346,  0.633694102)
      rgb=(0.312543124,  0.00823900704,  0.635699759)
      rgb=(0.318856183,  0.00757551051,  0.637639537)
      rgb=(0.325150025,  0.00691491734,  0.639512001)
      rgb=(0.331425547,  0.00626107379,  0.641315649)
      rgb=(0.337683446,  0.00561830889,  0.643048936)
      rgb=(0.343924591,  0.00499053080,  0.644710195)
      rgb=(0.350149699,  0.00438202557,  0.646297711)
      rgb=(0.356359209,  0.00379781761,  0.647809772)
      rgb=(0.362553473,  0.00324319591,  0.649244641)
      rgb=(0.368732762,  0.00272370721,  0.650600561)
      rgb=(0.374897270,  0.00224514897,  0.651875762)
      rgb=(0.381047116,  0.00181356205,  0.653068467)
      rgb=(0.387182639,  0.00143446923,  0.654176761)
      rgb=(0.393304010,  0.00111388259,  0.655198755)
      rgb=(0.399410821,  0.000859420809,  0.656132835)
      rgb=(0.405502914,  0.000678091517,  0.656977276)
      rgb=(0.411580082,  0.000577101735,  0.657730380)
      rgb=(0.417642063,  0.000563847476,  0.658390492)
      rgb=(0.423688549,  0.000645902780,  0.658956004)
      rgb=(0.429719186,  0.000831008207,  0.659425363)
      rgb=(0.435733575,  0.00112705875,  0.659797077)
      rgb=(0.441732123,  0.00153984779,  0.660069009)
      rgb=(0.447713600,  0.00207954744,  0.660240367)
      rgb=(0.453677394,  0.00275470302,  0.660309966)
      rgb=(0.459622938,  0.00357374415,  0.660276655)
      rgb=(0.465549631,  0.00454518084,  0.660139383)
      rgb=(0.471456847,  0.00567758762,  0.659897210)
      rgb=(0.477343929,  0.00697958743,  0.659549311)
      rgb=(0.483210198,  0.00845983494,  0.659094989)
      rgb=(0.489054951,  0.0101269996,  0.658533677)
      rgb=(0.494877466,  0.0119897486,  0.657864946)
      rgb=(0.500677687,  0.0140550640,  0.657087561)
      rgb=(0.506454143,  0.0163333443,  0.656202294)
      rgb=(0.512206035,  0.0188332232,  0.655209222)
      rgb=(0.517932580,  0.0215631918,  0.654108545)
      rgb=(0.523632990,  0.0245316468,  0.652900629)
      rgb=(0.529306474,  0.0277468735,  0.651586010)
      rgb=(0.534952244,  0.0312170300,  0.650165396)
      rgb=(0.540569510,  0.0349501310,  0.648639668)
      rgb=(0.546157494,  0.0389540334,  0.647009884)
      rgb=(0.551715423,  0.0431364795,  0.645277275)
      rgb=(0.557242538,  0.0473307585,  0.643443250)
      rgb=(0.562738096,  0.0515448092,  0.641509389)
      rgb=(0.568201372,  0.0557776706,  0.639477440)
      rgb=(0.573631859,  0.0600281369,  0.637348841)
      rgb=(0.579028682,  0.0642955547,  0.635126108)
      rgb=(0.584391137,  0.0685790261,  0.632811608)
      rgb=(0.589718606,  0.0728775875,  0.630407727)
      rgb=(0.595010505,  0.0771902878,  0.627916992)
      rgb=(0.600266283,  0.0815161895,  0.625342058)
      rgb=(0.605485428,  0.0858543713,  0.622685703)
      rgb=(0.610667469,  0.0902039303,  0.619950811)
      rgb=(0.615811974,  0.0945639838,  0.617140367)
      rgb=(0.620918555,  0.0989336721,  0.614257440)
      rgb=(0.625986869,  0.103312160,  0.611305174)
      rgb=(0.631016615,  0.107698641,  0.608286774)
      rgb=(0.636007543,  0.112092335,  0.605205491)
      rgb=(0.640959444,  0.116492495,  0.602064611)
      rgb=(0.645872158,  0.120898405,  0.598867442)
      rgb=(0.650745571,  0.125309384,  0.595617300)
      rgb=(0.655579615,  0.129724785,  0.592317494)
      rgb=(0.660374266,  0.134143997,  0.588971318)
      rgb=(0.665129493,  0.138566428,  0.585582301)
      rgb=(0.669845385,  0.142991540,  0.582153572)
      rgb=(0.674522060,  0.147418835,  0.578688247)
      rgb=(0.679159664,  0.151847851,  0.575189431)
      rgb=(0.683758384,  0.156278163,  0.571660158)
      rgb=(0.688318440,  0.160709387,  0.568103380)
      rgb=(0.692840088,  0.165141174,  0.564521958)
      rgb=(0.697323615,  0.169573215,  0.560918659)
      rgb=(0.701769334,  0.174005236,  0.557296144)
      rgb=(0.706177590,  0.178437000,  0.553656970)
      rgb=(0.710548747,  0.182868306,  0.550003579)
      rgb=(0.714883195,  0.187298986,  0.546338299)
      rgb=(0.719181339,  0.191728906,  0.542663338)
      rgb=(0.723443604,  0.196157962,  0.538980786)
      rgb=(0.727670428,  0.200586086,  0.535292612)
      rgb=(0.731862231,  0.205013174,  0.531600995)
      rgb=(0.736019424,  0.209439071,  0.527908434)
      rgb=(0.740142557,  0.213863965,  0.524215533)
      rgb=(0.744232102,  0.218287899,  0.520523766)
      rgb=(0.748288533,  0.222710942,  0.516834495)
      rgb=(0.752312321,  0.227133187,  0.513148963)
      rgb=(0.756303937,  0.231554749,  0.509468305)
      rgb=(0.760263849,  0.235975765,  0.505793543)
      rgb=(0.764192516,  0.240396394,  0.502125599)
      rgb=(0.768090391,  0.244816813,  0.498465290)
      rgb=(0.771957916,  0.249237220,  0.494813338)
      rgb=(0.775795522,  0.253657797,  0.491170517)
      rgb=(0.779603614,  0.258078397,  0.487539124)
      rgb=(0.783382636,  0.262499662,  0.483917732)
      rgb=(0.787132978,  0.266921859,  0.480306702)
      rgb=(0.790855015,  0.271345267,  0.476706319)
      rgb=(0.794549101,  0.275770179,  0.473116798)
      rgb=(0.798215577,  0.280196901,  0.469538286)
      rgb=(0.801854758,  0.284625750,  0.465970871)
      rgb=(0.805466945,  0.289057057,  0.462414580)
      rgb=(0.809052419,  0.293491117,  0.458869577)
      rgb=(0.812611506,  0.297927865,  0.455337565)
      rgb=(0.816144382,  0.302368130,  0.451816385)
      rgb=(0.819651255,  0.306812282,  0.448305861)
      rgb=(0.823132309,  0.311260703,  0.444805781)
      rgb=(0.826587706,  0.315713782,  0.441315901)
      rgb=(0.830017584,  0.320171913,  0.437835947)
      rgb=(0.833422053,  0.324635499,  0.434365616)
      rgb=(0.836801237,  0.329104836,  0.430905052)
      rgb=(0.840155276,  0.333580106,  0.427454836)
      rgb=(0.843484103,  0.338062109,  0.424013059)
      rgb=(0.846787726,  0.342551272,  0.420579333)
      rgb=(0.850066132,  0.347048028,  0.417153264)
      rgb=(0.853319279,  0.351552815,  0.413734445)
      rgb=(0.856547103,  0.356066072,  0.410322469)
      rgb=(0.859749520,  0.360588229,  0.406916975)
      rgb=(0.862926559,  0.365119408,  0.403518809)
      rgb=(0.866077920,  0.369660446,  0.400126027)
      rgb=(0.869203436,  0.374211795,  0.396738211)
      rgb=(0.872302917,  0.378773910,  0.393354947)
      rgb=(0.875376149,  0.383347243,  0.389975832)
      rgb=(0.878422895,  0.387932249,  0.386600468)
      rgb=(0.881442916,  0.392529339,  0.383228622)
      rgb=(0.884435982,  0.397138877,  0.379860246)
      rgb=(0.887401682,  0.401761511,  0.376494232)
      rgb=(0.890339687,  0.406397694,  0.373130228)
      rgb=(0.893249647,  0.411047871,  0.369767893)
      rgb=(0.896131191,  0.415712489,  0.366406907)
      rgb=(0.898983931,  0.420391986,  0.363046965)
      rgb=(0.901807455,  0.425086807,  0.359687758)
      rgb=(0.904601295,  0.429797442,  0.356328796)
      rgb=(0.907364995,  0.434524335,  0.352969777)
      rgb=(0.910098088,  0.439267908,  0.349610469)
      rgb=(0.912800095,  0.444028574,  0.346250656)
      rgb=(0.915470518,  0.448806744,  0.342890148)
      rgb=(0.918108848,  0.453602818,  0.339528771)
      rgb=(0.920714383,  0.458417420,  0.336165582)
      rgb=(0.923286660,  0.463250828,  0.332800827)
      rgb=(0.925825146,  0.468103387,  0.329434512)
      rgb=(0.928329275,  0.472975465,  0.326066550)
      rgb=(0.930798469,  0.477867420,  0.322696876)
      rgb=(0.933232140,  0.482779603,  0.319325444)
      rgb=(0.935629684,  0.487712357,  0.315952211)
      rgb=(0.937990034,  0.492666544,  0.312575440)
      rgb=(0.940312939,  0.497642038,  0.309196628)
      rgb=(0.942597771,  0.502639147,  0.305815824)
      rgb=(0.944843893,  0.507658169,  0.302433101)
      rgb=(0.947050662,  0.512699390,  0.299048555)
      rgb=(0.949217427,  0.517763087,  0.295662308)
      rgb=(0.951343530,  0.522849522,  0.292274506)
      rgb=(0.953427725,  0.527959550,  0.288883445)
      rgb=(0.955469640,  0.533093083,  0.285490391)
      rgb=(0.957468770,  0.538250172,  0.282096149)
      rgb=(0.959424430,  0.543431038,  0.278700990)
      rgb=(0.961335930,  0.548635890,  0.275305214)
      rgb=(0.963202573,  0.553864931,  0.271909159)
      rgb=(0.965023656,  0.559118349,  0.268513200)
      rgb=(0.966798470,  0.564396327,  0.265117752)
      rgb=(0.968525639,  0.569699633,  0.261721488)
      rgb=(0.970204593,  0.575028270,  0.258325424)
      rgb=(0.971835007,  0.580382015,  0.254931256)
      rgb=(0.973416145,  0.585761012,  0.251539615)
      rgb=(0.974947262,  0.591165394,  0.248151200)
      rgb=(0.976427606,  0.596595287,  0.244766775)
      rgb=(0.977856416,  0.602050811,  0.241387186)
      rgb=(0.979232922,  0.607532077,  0.238013359)
      rgb=(0.980556344,  0.613039190,  0.234646316)
      rgb=(0.981825890,  0.618572250,  0.231287178)
      rgb=(0.983040742,  0.624131362,  0.227937141)
      rgb=(0.984198924,  0.629717516,  0.224595006)
      rgb=(0.985300760,  0.635329876,  0.221264889)
      rgb=(0.986345421,  0.640968508,  0.217948456)
      rgb=(0.987332067,  0.646633475,  0.214647532)
      rgb=(0.988259846,  0.652324832,  0.211364122)
      rgb=(0.989127893,  0.658042630,  0.208100426)
      rgb=(0.989935328,  0.663786914,  0.204858855)
      rgb=(0.990681261,  0.669557720,  0.201642049)
      rgb=(0.991364787,  0.675355082,  0.198452900)
      rgb=(0.991984990,  0.681179025,  0.195294567)
      rgb=(0.992540939,  0.687029567,  0.192170500)
      rgb=(0.993031693,  0.692906719,  0.189084459)
      rgb=(0.993456302,  0.698810484,  0.186040537)
      rgb=(0.993813802,  0.704740854,  0.183043180)
      rgb=(0.994103226,  0.710697814,  0.180097207)
      rgb=(0.994323596,  0.716681336,  0.177207826)
      rgb=(0.994473934,  0.722691379,  0.174380656)
      rgb=(0.994553260,  0.728727890,  0.171621733)
      rgb=(0.994560594,  0.734790799,  0.168937522)
      rgb=(0.994494964,  0.740880020,  0.166334918)
      rgb=(0.994355411,  0.746995448,  0.163821243)
      rgb=(0.994140989,  0.753136955,  0.161404226)
      rgb=(0.993850778,  0.759304390,  0.159091984)
      rgb=(0.993482190,  0.765498551,  0.156890625)
      rgb=(0.993033251,  0.771719833,  0.154807583)
      rgb=(0.992505214,  0.777966775,  0.152854862)
      rgb=(0.991897270,  0.784239120,  0.151041581)
      rgb=(0.991208680,  0.790536569,  0.149376885)
      rgb=(0.990438793,  0.796858775,  0.147869810)
      rgb=(0.989587065,  0.803205337,  0.146529128)
      rgb=(0.988647741,  0.809578605,  0.145357284)
      rgb=(0.987620557,  0.815977942,  0.144362644)
      rgb=(0.986509366,  0.822400620,  0.143556679)
      rgb=(0.985314198,  0.828845980,  0.142945116)
      rgb=(0.984031139,  0.835315360,  0.142528388)
      rgb=(0.982652820,  0.841811730,  0.142302653)
      rgb=(0.981190389,  0.848328902,  0.142278607)
      rgb=(0.979643637,  0.854866468,  0.142453425)
      rgb=(0.977994918,  0.861432314,  0.142808191)
      rgb=(0.976264977,  0.868015998,  0.143350944)
      rgb=(0.974443038,  0.874622194,  0.144061156)
      rgb=(0.972530009,  0.881250063,  0.144922913)
      rgb=(0.970532932,  0.887896125,  0.145918663)
      rgb=(0.968443477,  0.894563989,  0.147014438)
      rgb=(0.966271225,  0.901249365,  0.148179639)
      rgb=(0.964021057,  0.907950379,  0.149370428)
      rgb=(0.961681481,  0.914672479,  0.150520343)
      rgb=(0.959275646,  0.921406537,  0.151566019)
      rgb=(0.956808068,  0.928152065,  0.152409489)
      rgb=(0.954286813,  0.934907730,  0.152921158)
      rgb=(0.951726083,  0.941670605,  0.152925363)
      rgb=(0.949150533,  0.948434900,  0.152177604)
      rgb=(0.946602270,  0.955189860,  0.150327944)
      rgb=(0.944151742,  0.961916487,  0.146860789)
      rgb=(0.941896120,  0.968589814,  0.140955606)
      rgb=(0.940015097,  0.975158357,  0.131325517)
    }
  }
}
\pgfplotsset{compat=1.12,
  colormap/viridis/.style={
    colormap={viridis}{
      rgb=(0.26700401,  0.00487433,  0.32941519)
      rgb=(0.26851048,  0.00960483,  0.33542652)
      rgb=(0.26994384,  0.01462494,  0.34137895)
      rgb=(0.27130489,  0.01994186,  0.34726862)
      rgb=(0.27259384,  0.02556309,  0.35309303)
      rgb=(0.27380934,  0.03149748,  0.35885256)
      rgb=(0.27495242,  0.03775181,  0.36454323)
      rgb=(0.27602238,  0.04416723,  0.37016418)
      rgb=(0.2770184 ,  0.05034437,  0.37571452)
      rgb=(0.27794143,  0.05632444,  0.38119074)
      rgb=(0.27879067,  0.06214536,  0.38659204)
      rgb=(0.2795655 ,  0.06783587,  0.39191723)
      rgb=(0.28026658,  0.07341724,  0.39716349)
      rgb=(0.28089358,  0.07890703,  0.40232944)
      rgb=(0.28144581,  0.0843197 ,  0.40741404)
      rgb=(0.28192358,  0.08966622,  0.41241521)
      rgb=(0.28232739,  0.09495545,  0.41733086)
      rgb=(0.28265633,  0.10019576,  0.42216032)
      rgb=(0.28291049,  0.10539345,  0.42690202)
      rgb=(0.28309095,  0.11055307,  0.43155375)
      rgb=(0.28319704,  0.11567966,  0.43611482)
      rgb=(0.28322882,  0.12077701,  0.44058404)
      rgb=(0.28318684,  0.12584799,  0.44496   )
      rgb=(0.283072  ,  0.13089477,  0.44924127)
      rgb=(0.28288389,  0.13592005,  0.45342734)
      rgb=(0.28262297,  0.14092556,  0.45751726)
      rgb=(0.28229037,  0.14591233,  0.46150995)
      rgb=(0.28188676,  0.15088147,  0.46540474)
      rgb=(0.28141228,  0.15583425,  0.46920128)
      rgb=(0.28086773,  0.16077132,  0.47289909)
      rgb=(0.28025468,  0.16569272,  0.47649762)
      rgb=(0.27957399,  0.17059884,  0.47999675)
      rgb=(0.27882618,  0.1754902 ,  0.48339654)
      rgb=(0.27801236,  0.18036684,  0.48669702)
      rgb=(0.27713437,  0.18522836,  0.48989831)
      rgb=(0.27619376,  0.19007447,  0.49300074)
      rgb=(0.27519116,  0.1949054 ,  0.49600488)
      rgb=(0.27412802,  0.19972086,  0.49891131)
      rgb=(0.27300596,  0.20452049,  0.50172076)
      rgb=(0.27182812,  0.20930306,  0.50443413)
      rgb=(0.27059473,  0.21406899,  0.50705243)
      rgb=(0.26930756,  0.21881782,  0.50957678)
      rgb=(0.26796846,  0.22354911,  0.5120084 )
      rgb=(0.26657984,  0.2282621 ,  0.5143487 )
      rgb=(0.2651445 ,  0.23295593,  0.5165993 )
      rgb=(0.2636632 ,  0.23763078,  0.51876163)
      rgb=(0.26213801,  0.24228619,  0.52083736)
      rgb=(0.26057103,  0.2469217 ,  0.52282822)
      rgb=(0.25896451,  0.25153685,  0.52473609)
      rgb=(0.25732244,  0.2561304 ,  0.52656332)
      rgb=(0.25564519,  0.26070284,  0.52831152)
      rgb=(0.25393498,  0.26525384,  0.52998273)
      rgb=(0.25219404,  0.26978306,  0.53157905)
      rgb=(0.25042462,  0.27429024,  0.53310261)
      rgb=(0.24862899,  0.27877509,  0.53455561)
      rgb=(0.2468114 ,  0.28323662,  0.53594093)
      rgb=(0.24497208,  0.28767547,  0.53726018)
      rgb=(0.24311324,  0.29209154,  0.53851561)
      rgb=(0.24123708,  0.29648471,  0.53970946)
      rgb=(0.23934575,  0.30085494,  0.54084398)
      rgb=(0.23744138,  0.30520222,  0.5419214 )
      rgb=(0.23552606,  0.30952657,  0.54294396)
      rgb=(0.23360277,  0.31382773,  0.54391424)
      rgb=(0.2316735 ,  0.3181058 ,  0.54483444)
      rgb=(0.22973926,  0.32236127,  0.54570633)
      rgb=(0.22780192,  0.32659432,  0.546532  )
      rgb=(0.2258633 ,  0.33080515,  0.54731353)
      rgb=(0.22392515,  0.334994  ,  0.54805291)
      rgb=(0.22198915,  0.33916114,  0.54875211)
      rgb=(0.22005691,  0.34330688,  0.54941304)
      rgb=(0.21812995,  0.34743154,  0.55003755)
      rgb=(0.21620971,  0.35153548,  0.55062743)
      rgb=(0.21429757,  0.35561907,  0.5511844 )
      rgb=(0.21239477,  0.35968273,  0.55171011)
      rgb=(0.2105031 ,  0.36372671,  0.55220646)
      rgb=(0.20862342,  0.36775151,  0.55267486)
      rgb=(0.20675628,  0.37175775,  0.55311653)
      rgb=(0.20490257,  0.37574589,  0.55353282)
      rgb=(0.20306309,  0.37971644,  0.55392505)
      rgb=(0.20123854,  0.38366989,  0.55429441)
      rgb=(0.1994295 ,  0.38760678,  0.55464205)
      rgb=(0.1976365 ,  0.39152762,  0.55496905)
      rgb=(0.19585993,  0.39543297,  0.55527637)
      rgb=(0.19410009,  0.39932336,  0.55556494)
      rgb=(0.19235719,  0.40319934,  0.55583559)
      rgb=(0.19063135,  0.40706148,  0.55608907)
      rgb=(0.18892259,  0.41091033,  0.55632606)
      rgb=(0.18723083,  0.41474645,  0.55654717)
      rgb=(0.18555593,  0.4185704 ,  0.55675292)
      rgb=(0.18389763,  0.42238275,  0.55694377)
      rgb=(0.18225561,  0.42618405,  0.5571201 )
      rgb=(0.18062949,  0.42997486,  0.55728221)
      rgb=(0.17901879,  0.43375572,  0.55743035)
      rgb=(0.17742298,  0.4375272 ,  0.55756466)
      rgb=(0.17584148,  0.44128981,  0.55768526)
      rgb=(0.17427363,  0.4450441 ,  0.55779216)
      rgb=(0.17271876,  0.4487906 ,  0.55788532)
      rgb=(0.17117615,  0.4525298 ,  0.55796464)
      rgb=(0.16964573,  0.45626209,  0.55803034)
      rgb=(0.16812641,  0.45998802,  0.55808199)
      rgb=(0.1666171 ,  0.46370813,  0.55811913)
      rgb=(0.16511703,  0.4674229 ,  0.55814141)
      rgb=(0.16362543,  0.47113278,  0.55814842)
      rgb=(0.16214155,  0.47483821,  0.55813967)
      rgb=(0.16066467,  0.47853961,  0.55811466)
      rgb=(0.15919413,  0.4822374 ,  0.5580728 )
      rgb=(0.15772933,  0.48593197,  0.55801347)
      rgb=(0.15626973,  0.4896237 ,  0.557936  )
      rgb=(0.15481488,  0.49331293,  0.55783967)
      rgb=(0.15336445,  0.49700003,  0.55772371)
      rgb=(0.1519182 ,  0.50068529,  0.55758733)
      rgb=(0.15047605,  0.50436904,  0.55742968)
      rgb=(0.14903918,  0.50805136,  0.5572505 )
      rgb=(0.14760731,  0.51173263,  0.55704861)
      rgb=(0.14618026,  0.51541316,  0.55682271)
      rgb=(0.14475863,  0.51909319,  0.55657181)
      rgb=(0.14334327,  0.52277292,  0.55629491)
      rgb=(0.14193527,  0.52645254,  0.55599097)
      rgb=(0.14053599,  0.53013219,  0.55565893)
      rgb=(0.13914708,  0.53381201,  0.55529773)
      rgb=(0.13777048,  0.53749213,  0.55490625)
      rgb=(0.1364085 ,  0.54117264,  0.55448339)
      rgb=(0.13506561,  0.54485335,  0.55402906)
      rgb=(0.13374299,  0.54853458,  0.55354108)
      rgb=(0.13244401,  0.55221637,  0.55301828)
      rgb=(0.13117249,  0.55589872,  0.55245948)
      rgb=(0.1299327 ,  0.55958162,  0.55186354)
      rgb=(0.12872938,  0.56326503,  0.55122927)
      rgb=(0.12756771,  0.56694891,  0.55055551)
      rgb=(0.12645338,  0.57063316,  0.5498411 )
      rgb=(0.12539383,  0.57431754,  0.54908564)
      rgb=(0.12439474,  0.57800205,  0.5482874 )
      rgb=(0.12346281,  0.58168661,  0.54744498)
      rgb=(0.12260562,  0.58537105,  0.54655722)
      rgb=(0.12183122,  0.58905521,  0.54562298)
      rgb=(0.12114807,  0.59273889,  0.54464114)
      rgb=(0.12056501,  0.59642187,  0.54361058)
      rgb=(0.12009154,  0.60010387,  0.54253043)
      rgb=(0.11973756,  0.60378459,  0.54139999)
      rgb=(0.11951163,  0.60746388,  0.54021751)
      rgb=(0.11942341,  0.61114146,  0.53898192)
      rgb=(0.11948255,  0.61481702,  0.53769219)
      rgb=(0.11969858,  0.61849025,  0.53634733)
      rgb=(0.12008079,  0.62216081,  0.53494633)
      rgb=(0.12063824,  0.62582833,  0.53348834)
      rgb=(0.12137972,  0.62949242,  0.53197275)
      rgb=(0.12231244,  0.63315277,  0.53039808)
      rgb=(0.12344358,  0.63680899,  0.52876343)
      rgb=(0.12477953,  0.64046069,  0.52706792)
      rgb=(0.12632581,  0.64410744,  0.52531069)
      rgb=(0.12808703,  0.64774881,  0.52349092)
      rgb=(0.13006688,  0.65138436,  0.52160791)
      rgb=(0.13226797,  0.65501363,  0.51966086)
      rgb=(0.13469183,  0.65863619,  0.5176488 )
      rgb=(0.13733921,  0.66225157,  0.51557101)
      rgb=(0.14020991,  0.66585927,  0.5134268 )
      rgb=(0.14330291,  0.66945881,  0.51121549)
      rgb=(0.1466164 ,  0.67304968,  0.50893644)
      rgb=(0.15014782,  0.67663139,  0.5065889 )
      rgb=(0.15389405,  0.68020343,  0.50417217)
      rgb=(0.15785146,  0.68376525,  0.50168574)
      rgb=(0.16201598,  0.68731632,  0.49912906)
      rgb=(0.1663832 ,  0.69085611,  0.49650163)
      rgb=(0.1709484 ,  0.69438405,  0.49380294)
      rgb=(0.17570671,  0.6978996 ,  0.49103252)
      rgb=(0.18065314,  0.70140222,  0.48818938)
      rgb=(0.18578266,  0.70489133,  0.48527326)
      rgb=(0.19109018,  0.70836635,  0.48228395)
      rgb=(0.19657063,  0.71182668,  0.47922108)
      rgb=(0.20221902,  0.71527175,  0.47608431)
      rgb=(0.20803045,  0.71870095,  0.4728733 )
      rgb=(0.21400015,  0.72211371,  0.46958774)
      rgb=(0.22012381,  0.72550945,  0.46622638)
      rgb=(0.2263969 ,  0.72888753,  0.46278934)
      rgb=(0.23281498,  0.73224735,  0.45927675)
      rgb=(0.2393739 ,  0.73558828,  0.45568838)
      rgb=(0.24606968,  0.73890972,  0.45202405)
      rgb=(0.25289851,  0.74221104,  0.44828355)
      rgb=(0.25985676,  0.74549162,  0.44446673)
      rgb=(0.26694127,  0.74875084,  0.44057284)
      rgb=(0.27414922,  0.75198807,  0.4366009 )
      rgb=(0.28147681,  0.75520266,  0.43255207)
      rgb=(0.28892102,  0.75839399,  0.42842626)
      rgb=(0.29647899,  0.76156142,  0.42422341)
      rgb=(0.30414796,  0.76470433,  0.41994346)
      rgb=(0.31192534,  0.76782207,  0.41558638)
      rgb=(0.3198086 ,  0.77091403,  0.41115215)
      rgb=(0.3277958 ,  0.77397953,  0.40664011)
      rgb=(0.33588539,  0.7770179 ,  0.40204917)
      rgb=(0.34407411,  0.78002855,  0.39738103)
      rgb=(0.35235985,  0.78301086,  0.39263579)
      rgb=(0.36074053,  0.78596419,  0.38781353)
      rgb=(0.3692142 ,  0.78888793,  0.38291438)
      rgb=(0.37777892,  0.79178146,  0.3779385 )
      rgb=(0.38643282,  0.79464415,  0.37288606)
      rgb=(0.39517408,  0.79747541,  0.36775726)
      rgb=(0.40400101,  0.80027461,  0.36255223)
      rgb=(0.4129135 ,  0.80304099,  0.35726893)
      rgb=(0.42190813,  0.80577412,  0.35191009)
      rgb=(0.43098317,  0.80847343,  0.34647607)
      rgb=(0.44013691,  0.81113836,  0.3409673 )
      rgb=(0.44936763,  0.81376835,  0.33538426)
      rgb=(0.45867362,  0.81636288,  0.32972749)
      rgb=(0.46805314,  0.81892143,  0.32399761)
      rgb=(0.47750446,  0.82144351,  0.31819529)
      rgb=(0.4870258 ,  0.82392862,  0.31232133)
      rgb=(0.49661536,  0.82637633,  0.30637661)
      rgb=(0.5062713 ,  0.82878621,  0.30036211)
      rgb=(0.51599182,  0.83115784,  0.29427888)
      rgb=(0.52577622,  0.83349064,  0.2881265 )
      rgb=(0.5356211 ,  0.83578452,  0.28190832)
      rgb=(0.5455244 ,  0.83803918,  0.27562602)
      rgb=(0.55548397,  0.84025437,  0.26928147)
      rgb=(0.5654976 ,  0.8424299 ,  0.26287683)
      rgb=(0.57556297,  0.84456561,  0.25641457)
      rgb=(0.58567772,  0.84666139,  0.24989748)
      rgb=(0.59583934,  0.84871722,  0.24332878)
      rgb=(0.60604528,  0.8507331 ,  0.23671214)
      rgb=(0.61629283,  0.85270912,  0.23005179)
      rgb=(0.62657923,  0.85464543,  0.22335258)
      rgb=(0.63690157,  0.85654226,  0.21662012)
      rgb=(0.64725685,  0.85839991,  0.20986086)
      rgb=(0.65764197,  0.86021878,  0.20308229)
      rgb=(0.66805369,  0.86199932,  0.19629307)
      rgb=(0.67848868,  0.86374211,  0.18950326)
      rgb=(0.68894351,  0.86544779,  0.18272455)
      rgb=(0.69941463,  0.86711711,  0.17597055)
      rgb=(0.70989842,  0.86875092,  0.16925712)
      rgb=(0.72039115,  0.87035015,  0.16260273)
      rgb=(0.73088902,  0.87191584,  0.15602894)
      rgb=(0.74138803,  0.87344918,  0.14956101)
      rgb=(0.75188414,  0.87495143,  0.14322828)
      rgb=(0.76237342,  0.87642392,  0.13706449)
      rgb=(0.77285183,  0.87786808,  0.13110864)
      rgb=(0.78331535,  0.87928545,  0.12540538)
      rgb=(0.79375994,  0.88067763,  0.12000532)
      rgb=(0.80418159,  0.88204632,  0.11496505)
      rgb=(0.81457634,  0.88339329,  0.11034678)
      rgb=(0.82494028,  0.88472036,  0.10621724)
      rgb=(0.83526959,  0.88602943,  0.1026459 )
      rgb=(0.84556056,  0.88732243,  0.09970219)
      rgb=(0.8558096 ,  0.88860134,  0.09745186)
      rgb=(0.86601325,  0.88986815,  0.09595277)
      rgb=(0.87616824,  0.89112487,  0.09525046)
      rgb=(0.88627146,  0.89237353,  0.09537439)
      rgb=(0.89632002,  0.89361614,  0.09633538)
      rgb=(0.90631121,  0.89485467,  0.09812496)
      rgb=(0.91624212,  0.89609127,  0.1007168 )
      rgb=(0.92610579,  0.89732977,  0.10407067)
      rgb=(0.93590444,  0.8985704 ,  0.10813094)
      rgb=(0.94563626,  0.899815  ,  0.11283773)
      rgb=(0.95529972,  0.90106534,  0.11812832)
      rgb=(0.96489353,  0.90232311,  0.12394051)
      rgb=(0.97441665,  0.90358991,  0.13021494)
      rgb=(0.98386829,  0.90486726,  0.13689671)
      rgb=(0.99324789,  0.90615657,  0.1439362 )
    }
  }
}

\newcommand*\circled[1]{\tikz[baseline=(char.base)]{\node[shape=circle,thick,draw,inner sep=1.5pt,black!75] (char) {#1};}}
\newcommand*\plotlabel[1]{{\scshape\circled{{#1}}}}
\setlist[enumerate]{itemsep=1.0em,topsep=1em}
\setlist[enumerate,1]{label=\protect\circled{\arabic*}}
\setlist[enumerate,2]{topsep=0.5em,label=\protect{\small\Roman*.},ref=\theenumi.\small\Roman*}

\makeatletter
\renewcommand{\@chapapp}{}
\newenvironment{chapquote}[2][2em]
  {\setlength{\@tempdima}{#1}%
   \def\chapquote@author{#2}%
   \parshape 1 \@tempdima \dimexpr\textwidth-2\@tempdima\relax%
   \itshape}
  {\par\normalfont\hfill--\ \chapquote@author\hspace*{\@tempdima}\par\bigskip}
\makeatother

\newcommand{\pgfplotsdir}{./figures/}

\newcommand{\datadir}{./data/}
\removecolourlinks

\graphicspath{{\pgfplotsdir}}   
\tikzsetnextfilename{\tmp@a}

\definecolor{MatlabBlue}{rgb}{0, 0.4470, 0.7410}
\definecolor{MatlabGreen}{rgb}{0.4660, 0.6740, 0.1880}
\definecolor{MatlabRed}{rgb}{0.6350, 0.0780, 0.1840}
\definecolor{MatlabCyan}{rgb}{0.3010, 0.7450, 0.9330}
\definecolor{MatlabMagenta}{rgb}{0.4940, 0.1840, 0.5560}
\definecolor{MatlabYellow}{rgb}{0.9290, 0.6940, 0.1250}
\definecolor{MatlabOrange}{rgb}{0.8500, 0.3250, 0.0980}

\pgfplotsset{compat=newest,
colormap default colorspace=rgb,
every axis/.append style={
  width=0.45\columnwidth,
  font=\scshape\footnotesize,
  axis x line=middle,
  axis y line=middle,
  axis z line=middle,
  },
every axis label/.append style={
  font=\scshape\footnotesize,
  },
every tick label/.append style={font=\scriptsize,},
minor grid style={very thin,black!5},
major grid style={thin,black!15},
every non boxed x axis/.style={},
    matrix plot/.style={
        axis on top,
        clip marker paths=true,
        scale only axis,
        height=\matrixrows/\matrixcols*\pgfkeysvalueof{/pgfplots/width},
        enlarge x limits={rel=0.5/\matrixcols},
        enlarge y limits={rel=0.5/\matrixrows},
        scatter/use mapped color={draw=mapped color, fill=mapped color},
        scatter,
        point meta=explicit,
        mark=square*,
        cycle list={
            mark size=0.5*\pgfkeysvalueof{/pgfplots/width}/\matrixcols
        }
    },
matrix rows/.store in=\matrixrows,
matrix cols/.store in=\matrixcols,
cycle list set/.initial=,
}

\tikzset{%
    legendline/.style n args={2}{%
        draw,ultra thick,line cap=round,color=#1,fill=#1,draw opacity=#2
    },
mixer/.style n args={4}{regular polygon, regular polygon sides=360,
    minimum width=3em,
    path picture={
        \draw[black] 
            (path picture bounding box.south east) -- (path picture bounding box.north west)
            (path picture bounding box.south west) -- (path picture bounding box.north east);
        \node at ($(path picture bounding box.south)+(0,0.25)$)     {#1};
        \node at ($(path picture bounding box.west)+(0.25,0)$)      {#2};
        \node at ($(path picture bounding box.north)+(0,-0.25)$)    {#3};
        \node at ($(path picture bounding box.east)+(-0.25,0)$)     {#4};
        }
    },
adder/.style n args={6}{regular polygon, regular polygon sides=4,
    minimum width=5em,
    minimum height=5em,
    path picture={
        \node at ($(path picture bounding box.west)+(0.25,+0.4)$)     {#1};
        \node at ($(path picture bounding box.west)+(0.25,0.00)$)      {#2};
        \node at ($(path picture bounding box.west)+(0.25,-0.4)$)    {#3};
        \node at ($(path picture bounding box.east)+(-0.25,+0.4)$)     {#4};
        \node at ($(path picture bounding box.east)+(-0.25,0.00)$)     {#5};
        \node at ($(path picture bounding box.east)+(-0.25,-0.4)$)     {#6};
        }
    },
tri-buffer/.style={regular polygon, regular polygon sides=3,
    minimum width=4em,
    minimum height=4em,
    rotate=-90
    },
awg/.style={regular polygon, regular polygon sides=4,
    minimum size=6em,
    yscale=2,
    path picture={
        \node at ($(path picture bounding box.south)+(0,0.25)$)     {AWG};
        \node at ($(path picture bounding box.north)+(-0.33,-0.15)$)      {\scriptsize{I}};
        \node at ($(path picture bounding box.north)+(0.33,-0.15)$)      {\scriptsize{Q}};
        }
    },
dir-couple/.style args={#1}{regular polygon, regular polygon sides=4,
    minimum size=4em,
    xscale=1.5,
    path picture={
        \draw[black,-latex] 
            ($(path picture bounding box.north west)!0.25!(path picture bounding box.south west)+(0.05,0)$) -- ($(path picture bounding box.north east)!0.25!(path picture bounding box.south east)+(-0.05,0)$);
        \draw ($(path picture bounding box.north)!0.25!(path picture bounding box.south)-(0.1,0)$) edge[out=0,in=90,-latex] ($(path picture bounding box.south east)!0.25!(path picture bounding box.south west)+(0,0.05)$);
        \node at ($(path picture bounding box.center)-(0.1,0.1)$)     {#1};
        }
    },
legendline/.style n args={2}{%
        draw,ultra thick,line cap=round,color=#1,fill=#1,draw opacity=#2
    },
legenddash/.style n args={2}{%
        draw,ultra thick, dashed,line cap=round,color=#1,fill=#1,draw opacity=#2
    },
toprule/.style={
	  execute at end cell={
	      \draw[line cap=rect,#1] (\tikzmatrixname-\the\pgfmatrixcurrentrow-\the\pgfmatrixcurrentcolumn.north west) -- (\tikzmatrixname-\the\pgfmatrixcurrentrow-\the\pgfmatrixcurrentcolumn.north east);
	      }
	  },
toprule/.style={
	  execute at end cell={
	      \draw[line cap=rect,#1] (\tikzmatrixname-\the\pgfmatrixcurrentrow-\the\pgfmatrixcurrentcolumn.north west) -- (\tikzmatrixname-\the\pgfmatrixcurrentrow-\the\pgfmatrixcurrentcolumn.north east);
	      }
	  },
}

\makeatletter
\pgfplotsset{colormap={reverseviridis}{
        indices of colormap={
            \pgfplotscolormaplastindexof{viridis},...,0 of viridis}
            },
    colormap/viridisfive/.style={
    colormap={viridisfive}{samples of colormap=(5 of reverseviridis)}},
    colormap/viridisfourteen/.style={
    colormap={viridisfourteen}{samples of colormap=(14 of reverseviridis)}},
    groupplot xlabel/.initial={},
    every groupplot x label/.style={
        at={($({\pgfplots@group@name\space c1r\pgfplots@group@rows.west}|-{\pgfplots@group@name\space c1r\pgfplots@group@rows.outer south})!0.5!({\pgfplots@group@name\space c\pgfplots@group@columns r\pgfplots@group@rows.east}|-{\pgfplots@group@name\space c\pgfplots@group@columns r\pgfplots@group@rows.outer south})$)},
        anchor=north,
    },
    groupplot xlabel2/.initial={},
    every groupplot x label2/.style={
        at={($({\pgfplots@group@name\space c1r\pgfplots@group@rows.west}|-{\pgfplots@group@name\space c1r\pgfplots@group@rows.outer north})!0.5!({\pgfplots@group@name\space c\pgfplots@group@columns r\pgfplots@group@rows.east}|-{\pgfplots@group@name\space c\pgfplots@group@columns r\pgfplots@group@rows.outer north})$)},
        anchor=south,
    },
    groupplot ylabel/.initial={},
    every groupplot y label/.style={
            rotate=90,
        at={($({\pgfplots@group@name\space c1r1.north}-|{\pgfplots@group@name\space c1r1.outer
west})!0.5!({\pgfplots@group@name\space c1r\pgfplots@group@rows.south}-|{\pgfplots@group@name\space c1r\pgfplots@group@rows.outer west})$)},
        anchor=center,
    },
    execute at end groupplot/.code={%
      \node [/pgfplots/every groupplot x label]
{\pgfkeysvalueof{/pgfplots/groupplot xlabel}};  
      \node [/pgfplots/every groupplot x label2]
{\pgfkeysvalueof{/pgfplots/groupplot xlabel2}};  
      \node [/pgfplots/every groupplot y label] 
{\pgfkeysvalueof{/pgfplots/groupplot ylabel}};  
    }
}
\def\endpgfplots@environment@groupplot{%
    \endpgfplots@environment@opt%
    \pgfkeys{/pgfplots/execute at end groupplot}%
    \endgroup%
}
\makeatother

\pgfplotsset{conv style/.style=
  { compat=newest,
    font=\small,
    axis x line=bottom,
    axis y line=left,
    ylabel style={align=center,anchor=center},
    axis line style=-latex,
    grid=both,
    tick align=outside,
    scaled ticks=false,
    mark=none,
    thick,
    solid,
  }
}

\pgfplotsset{echo style/.style=
  { compat=1.13,
    font=\small,
    axis x line=bottom,
    axis y line=left,
    axis line style=-latex,
    grid=both,
    tick align=outside,
    scaled ticks=false,
    mark=none,
    thick,
    solid,
    colormap/viridisfourteen,
    colormap access=piecewise constant,
    colorbar sampled,
    colorbar right,
    point meta min=0,
    point meta max=14,
    y tick label style={
        /pgf/number format/.cd,
            fixed,
            fixed zerofill,
            precision=1,
        /tikz/.cd
    },
    x tick label style={
        /pgf/number format/.cd,
            fixed,
            fixed zerofill,
            precision=0,
        /tikz/.cd
    },
    yticklabel={\ifdim\tick pt=0pt$0$\else\axisdefaultticklabel\fi},
    xticklabel={\ifdim\tick pt=0pt$0$\else\axisdefaultticklabel\fi},
    colorbar style={
      at={(1.2,0)},anchor=south west,
      ytick align=center,
      yticklabel style={text=white,anchor=center},
      yticklabel shift=-5.5pt,
      samples=15,
      ytick={0.5,1.5,...,13.5},
      ytick style={draw=none},
      ylabel={Iteration},
      ylabel style={align=center,anchor=south,yshift=-0.5ex},
      minor y tick num = 0, 
      axis y line=left,
      axis line style={ -,draw opacity=0 },
      y unit={},
      x unit={},
      yticklabels={1,2,...,14},
      tick align=inside,
      x post scale=0.775,
      y post scale=1,
    },
  }
}

\pgfplotsset{
  eseem style/.style=
  { compat=1.13,
    font=\small,
    axis x line=bottom,
    axis y line=left,
    axis line style=-latex,
    grid=both,
    tick align=outside,
    scaled ticks=false,
    mark=none,
    thick,
    solid,
    colormap/viridisfive,
    colormap access=piecewise constant,
    colorbar sampled,
    colorbar right,
    point meta min=0,
    point meta max=5,
    y tick label style={
        /pgf/number format/.cd,
            fixed,
            fixed zerofill,
            precision=1,
        /tikz/.cd
    },
    x tick label style={
        /pgf/number format/.cd,
            fixed,
            fixed zerofill,
            precision=2,
        /tikz/.cd
    },
    yticklabel={\ifdim\tick pt=0pt$0$\else\axisdefaultticklabel\fi},
    xticklabel={\ifdim\tick pt=0pt$0$\else\axisdefaultticklabel\fi},
    colorbar style={
      at={(1.1,0)},anchor=south west,
      ytick align=center,
      yticklabel style={text=white,anchor=center},
      yticklabel shift=-5.5pt,
      samples=6,
      ytick={0.5,1.5,...,4.5},
      ytick style={draw=none},
      ylabel={Iteration},
      ylabel style={align=center,anchor=south,yshift=-0.5ex},
      minor y tick num = 0, 
      axis y line=left,
      axis line style={ -,draw opacity=0 },
      y unit={},
      x unit={},
      yticklabels={1,2,...,5},
      tick align=inside,
      x post scale=0.775,
      y post scale=1,
    },
  }
}

\pgfplotsset{
  amplitude plot/.style=
  { title style={at={(0.5,1)},anchor=center,align=center},
    grid=none,
    axis x line=middle,
    xtick align=center,
    ytick distance=1,
    minor y tick num = 0,
    ymin=0, ymax=1,
    y axis line style=|-|,
    x axis line style=-latex,
    yticklabels={,$0$,$+1$},
    ylabel style={align=center,anchor=south},
    ylabel shift=-2ex,
  }
}

\pgfplotsset{
  amplitude plot r/.style=
  { title style={at={(0.5,1)},anchor=center,align=center},
    grid=none,
    axis x line=middle,
    xtick align=center,
    ytick distance=1,
    minor y tick num = 0,
    ymin=0, ymax=1,
    y axis line style=|-|,
    x axis line style=|-latex,
    yticklabels={,$0$,$+1$},
    ylabel style={align=center,anchor=south},
    ylabel shift=-2ex,
  }
}

\pgfplotsset{
  phase plot/.style=
  { title style={at={(0.5,1)},anchor=center,align=center},
    grid=none,
    axis x line=middle,
    xtick align=center,
    ytick distance={pi},
    minor y tick num = 1,
    y axis line style=latex-latex,
    x axis line style=|-latex,
  }
}

\pgfplotsset{
  phase plot r/.style=
  { title style={at={(0.5,1)},anchor=center,align=center},
    grid=none,
    axis x line=middle,
    xtick align=center,
    axis y line=right,
    ytick distance={pi},
    minor y tick num = 1,
    y axis line style=latex-latex,
    x axis line style=|-latex,
    ylabel style={align=center,anchor=center},
  }
}

\def\initmd(#1,#2,#3){
    \coordinate (A) at (#1-#3,#2-#3);
    \coordinate (B) at (#1+#3,#2-#3);
    \coordinate (W) at (#1,#2+#3);

    \draw [black, ultra thin, fill=gray!10,fill opacity=0.5] (A) -- (B) -- (W) -- cycle;
    \filldraw[black] (W) circle (2*#3pt);
    \filldraw[black] (A) circle (2*#3pt);
    \filldraw[MatlabYellow] (B) circle (2.5*#3pt);
    }
    
\def\reflmd(#1,#2,#3){
    \coordinate (A) at (#1-#3,#2-#3);
    \coordinate (B) at (#1+#3,#2-#3);
    \coordinate (W) at (#1,#2+#3);
    
    \coordinate (R1) at (#1+3*#3,#2-#3);
    \coordinate (R2) at (#1+2*#3,#2-3*#3);

    \draw [black, ultra thin, fill=gray!10,fill opacity=0.5] (A) -- (B) -- (W) -- cycle;
    \filldraw[MatlabBlue!50,opacity=0.5] (B) -- (R1) -- (R2) -- cycle;
    \filldraw[black] (W) circle (2*#3pt);
    \filldraw[black] (A) circle (2*#3pt);
    \filldraw[MatlabYellow] (B) circle (2.5*#3pt);
    \filldraw[MatlabCyan] (R1) circle (2.5*#3pt);
    \filldraw[MatlabCyan] (R2) circle (2.5*#3pt);
    }

\def\expdmd(#1,#2,#3){
    \coordinate (A) at (#1-#3,#2-#3);
    \coordinate (B) at (#1+#3,#2-#3);
    \coordinate (W) at (#1,#2+#3);
    
    \coordinate (R1) at (#1+3*#3,#2-#3);
    \coordinate (R2) at (#1+2*#3,#2-3*#3);
    \coordinate (E1) at (#1+5*#3,#2-#3);
    \coordinate (E2) at (#1+3*#3,#2-5*#3);

    \draw [black, ultra thin, fill=gray!10,fill opacity=0.5] (A) -- (B) -- (W) -- cycle;
    \filldraw[MatlabBlue!50,opacity=0.5] (B) -- (R1) -- (R2) -- cycle;
    \filldraw[MatlabBlue!50,opacity=0.5] (B) -- (E1) -- (E2) -- cycle;
    \filldraw[black] (W) circle (2*#3pt);
    \filldraw[black] (A) circle (2*#3pt);
    \filldraw[MatlabYellow] (B) circle (2.5*#3pt);
    \filldraw[MatlabCyan] (R1) circle (2.5*#3pt);
    \filldraw[MatlabCyan] (R2) circle (2.5*#3pt);
    \filldraw[MatlabCyan] (E1) circle (2.5*#3pt);
    \filldraw[MatlabCyan] (E2) circle (2.5*#3pt);
    }
    
\def\contmd(#1,#2,#3){
    \coordinate (A) at (#1-#3,#2-#3);
    \coordinate (B) at (#1+#3,#2-#3);
    \coordinate (W) at (#1,#2+#3);

    \draw [black, ultra thin, fill=gray!10,fill opacity=0.5] (A) -- (B) -- (W) -- cycle;
    \filldraw[MatlabBlue!50,opacity=0.5] (#1,#2-#3) -- (B) -- (#1+#3/2,#2) -- cycle;
    \filldraw[black] (W) circle (2*#3pt);
    \filldraw[black] (A) circle (2*#3pt);
    \filldraw[MatlabYellow] (B) circle (2.5*#3pt);
    \filldraw[MatlabCyan] ($(A)!0.5!(B)$) circle (2.5*#3pt);
    \filldraw[MatlabCyan] ($(W)!0.5!(B)$) circle (2.5*#3pt);
    }
    
\def\initnm(#1,#2,#3){
\coordinate (A) at (#1-#3,#2-#3);
    \coordinate (B) at (#1+#3,#2-#3);
    \coordinate (W) at (#1,#2+#3);

    \draw [black, ultra thin, fill=gray!10,fill opacity=0.5] (A) -- (B) -- (W) -- cycle;
    \filldraw[MatlabOrange] (W) circle (2.5*#3pt);
    \filldraw[black] (A) circle (2*#3pt);
    \filldraw[black] (B) circle (2*#3pt);
    }
    
\def\reflnm(#1,#2,#3){
     \coordinate  (A) at (#1-#3,#2-#3);
    \coordinate (B) at (#1+#3,#2-#3);
    \coordinate (W) at (#1,#2+#3);
    
    \coordinate (R) at (#1,#2-3*#3);

    \draw [black, ultra thin, fill=gray!10,fill opacity=0.5] (A) -- (B) -- (W) -- cycle;
    \filldraw[MatlabBlue!50,opacity=0.5] (A) -- (B) -- (R) -- cycle;
    \filldraw[MatlabOrange] (W) circle (2.5*#3pt);
    \filldraw[MatlabCyan] (R) circle (2.5*#3pt);
    \filldraw[black] (A) circle (2*#3pt);
    \filldraw[black] (B) circle (2*#3pt);
    }

\def\expdnm(#1,#2,#3){
     \coordinate  (A) at (#1-#3,#2-#3);
    \coordinate (B) at (#1+#3,#2-#3);
    \coordinate (W) at (#1,#2+#3);
    
    \coordinate (R) at (#1,#2-3*#3);
    \coordinate (E) at (#1,#2-5*#3);

    \draw [black, ultra thin, fill=gray!10,fill opacity=0.5] (A) -- (B) -- (W) -- cycle;
    \filldraw[MatlabBlue!50,opacity=0.5] (A) -- (B) -- (R) -- cycle;
    \filldraw[MatlabBlue!50,opacity=0.5] (A) -- (B) -- (E) -- cycle;
    \filldraw[MatlabOrange] (W) circle (2.5*#3pt);
    \filldraw[MatlabCyan] (R) circle (2.5*#3pt);
    \filldraw[MatlabCyan] (E) circle (2.5*#3pt);
    \filldraw[black] (A) circle (2*#3pt);
    \filldraw[black] (B) circle (2*#3pt);
    }

\def\contnm(#1,#2,#3){
     \coordinate  (A) at (#1-#3,#2-#3);
    \coordinate (B) at (#1+#3,#2-#3);
    \coordinate (W) at (#1,#2+#3);

    \coordinate (C) at (#1,#2);

    \draw [black, ultra thin, fill=gray!10,fill opacity=0.5] (A) -- (B) -- (W) -- cycle;
    \filldraw[MatlabBlue!50,opacity=0.5] (A) -- (B) -- (C) -- cycle;
    \filldraw[MatlabOrange] (W) circle (2.5*#3pt);
    \filldraw[MatlabCyan] ($(A)!0.5!(B)!0.5!(W)$) circle (2*#3pt);
    \filldraw[black] (A) circle (2*#3pt);
    \filldraw[black] (B) circle (2*#3pt);
    }

    \makeatletter
\def\pgfplots@getautoplotspec into#1{%
    \begingroup
    \let#1=\pgfutil@empty
    \pgfkeysgetvalue{/pgfplots/cycle multi list/@dim}\pgfplots@cycle@dim
    \let\pgfplots@listindex=\pgfplots@numplots
    \pgfkeysgetvalue{/pgfplots/cycle list set}\pgfplots@listindex@set
    \ifx\pgfplots@listindex@set\pgfutil@empty
    \else 
        \c@pgf@counta=\pgfplots@listindex
        \c@pgf@countb=\pgfplots@listindex@set
        \advance\c@pgf@countb by -\c@pgf@counta
        \globaldefs=1\relax
        \edef\setshift{%
            \noexpand\pgfkeys{
                /pgfplots/cycle list shift=\the\c@pgf@countb,
                /pgfplots/cycle list set=
            }
        }%
        \setshift%
    \fi
    \pgfkeysgetvalue{/pgfplots/cycle list shift}\pgfplots@listindex@shift
    \ifx\pgfplots@listindex@shift\pgfutil@empty
    \else
        \c@pgf@counta=\pgfplots@listindex\relax
        \advance\c@pgf@counta by\pgfplots@listindex@shift\relax
        \ifnum\c@pgf@counta<0
            \c@pgf@counta=-\c@pgf@counta
        \fi
        \edef\pgfplots@listindex{\the\c@pgf@counta}%
    \fi
    \ifnum\pgfplots@cycle@dim>0
        \c@pgf@counta=\pgfplots@cycle@dim\relax
        \c@pgf@countb=\pgfplots@listindex\relax
        \advance\c@pgf@counta by-1
        \pgfplotsloop{%
            \ifnum\c@pgf@counta<0
                \pgfplotsloopcontinuefalse
            \else
                \pgfplotsloopcontinuetrue
            \fi
        }{%
            \pgfkeysgetvalue{/pgfplots/cycle multi list/@N\the\c@pgf@counta}\pgfplots@cycle@N
            \pgfplotsmathmodint{\c@pgf@countb}{\pgfplots@cycle@N}%
            \divide\c@pgf@countb by \pgfplots@cycle@N\relax
            \expandafter\pgfplots@getautoplotspec@
                \csname pgfp@cyclist@/pgfplots/cycle multi list/@list\the\c@pgf@counta @\endcsname
                {\pgfplots@cycle@N}%
                {\pgfmathresult}%
            \t@pgfplots@toka=\expandafter{#1,}%
            \t@pgfplots@tokb=\expandafter{\pgfplotsretval}%
            \edef#1{\the\t@pgfplots@toka\the\t@pgfplots@tokb}%
            \advance\c@pgf@counta by-1
        }%
    \else
        \pgfplotslistsize\autoplotspeclist\to\c@pgf@countd
        \pgfplots@getautoplotspec@{\autoplotspeclist}{\c@pgf@countd}{\pgfplots@listindex}%
        \let#1=\pgfplotsretval
    \fi
    \pgfmath@smuggleone#1%
    \endgroup
}

\pgfplotsset{
    cycle list set/.initial=
}
\makeatother
\newcommand\scalemath[2]{\scalebox{#1}{\mbox{\ensuremath{\displaystyle #2}}}}
\newcommand*{\CurrentPath}{.}

\usepackage[T1]{fontenc}
\usepackage{lmodern}
\usepackage{textcomp}

\let\cftchapfontorig\cftchapfont
\let\cftchappagefontorig\cftchappagefont
\renewcommand{\cftchapfont}{\itshape}
\renewcommand{\cftchappagefont}{\itshape}%

\tikzexternaldisable

\title        {Advanced Optimal Control Methods for Spin Systems}
\authors      {David L. Goodwin}
\supervisor   {Dr. Ilya Kuprov}
\examiner     {Prof. Stefan Glaser \& Dr. Tim Freegarde}
\degree       {Philosophy Doctorate}
\authors      {David L. Goodwin}
\university   {University of Southampton}
\department   {School of Chemistry}
\group        {Spin Dynamics Group}
\faculty      {Faculty of Natural and Environmental Sciences}
\addresses    {}
\subject      {Magnetic Resonance}
\documenttype {\textit{A thesis submitted for the degree of\\ Doctor of Philosophy}}
\date         {October 2017}

\begin{document}
\frontmatter
\maketitle
\begin{abstract}
Work within this thesis advances optimal control algorithms for application to magnetic resonance systems. Specifically, presenting a quadratically convergent version of the \textit{gradient ascent pulse engineering} method. The work is formulated in a superoperator representation of the Liouville-von Neumann equation.

A Newton-\textsc{grape} method is developed using efficient calculation of analytical second directional derivatives. The method is developed to scale with the same complexity as methods that use only first directional derivatives. Algorithms to ensure a well-conditioned and positive definite matrix of second directional derivatives are used so the sufficient conditions of gradient-based numerical optimisation are met.

A number of applications of optimal control in magnetic resonance are investigated: solid-state nuclear magnetic resonance, magnetisation-to-singlet pulses, and electron spin resonance experiments.
\end{abstract}

\subsection*{\hfill Declaration of Authorship \hfill}

{\normalsize I, David L.~Goodwin, declare that the thesis entitled \emph{Advanced Optimal Control Methods for Spin Systems} and the work presented in the thesis are both my own, and have been generated by me as the result of my own original research. I confirm that:

\begin{itemize}\setlength\itemsep{0.15em}
\item this work was done wholly or mainly while in candidature for a research degree
at this University;
\item where any part of this thesis has previously been submitted for a degree or any
other qualification at this University or any other institution, this has been clearly
stated;
\item where I have consulted the published work of others, this is always clearly
attributed;
\item where I have quoted from the work of others, the source is always given. With
the exception of such quotations, this thesis is entirely my own work;
\item with the oversight of my main supervisor, editorial advice has been sought. 
No changes of intellectual content were made as a result of this advice.
\item I have acknowledged all main sources of help;
\item where the thesis is based on work done by myself jointly with others, I have
made clear exactly what was done by others and what I have contributed myself;
\item parts of this work have been published as \cite{GOODWIN15,GOODWIN16,GOODWIN17,GOODWIN18}
\end{itemize}

\null\vfill\null

\begin{tabular}{l c l l c}
Signed: & & \begin{minipage}{2.5cm}\includegraphics[width=\linewidth]{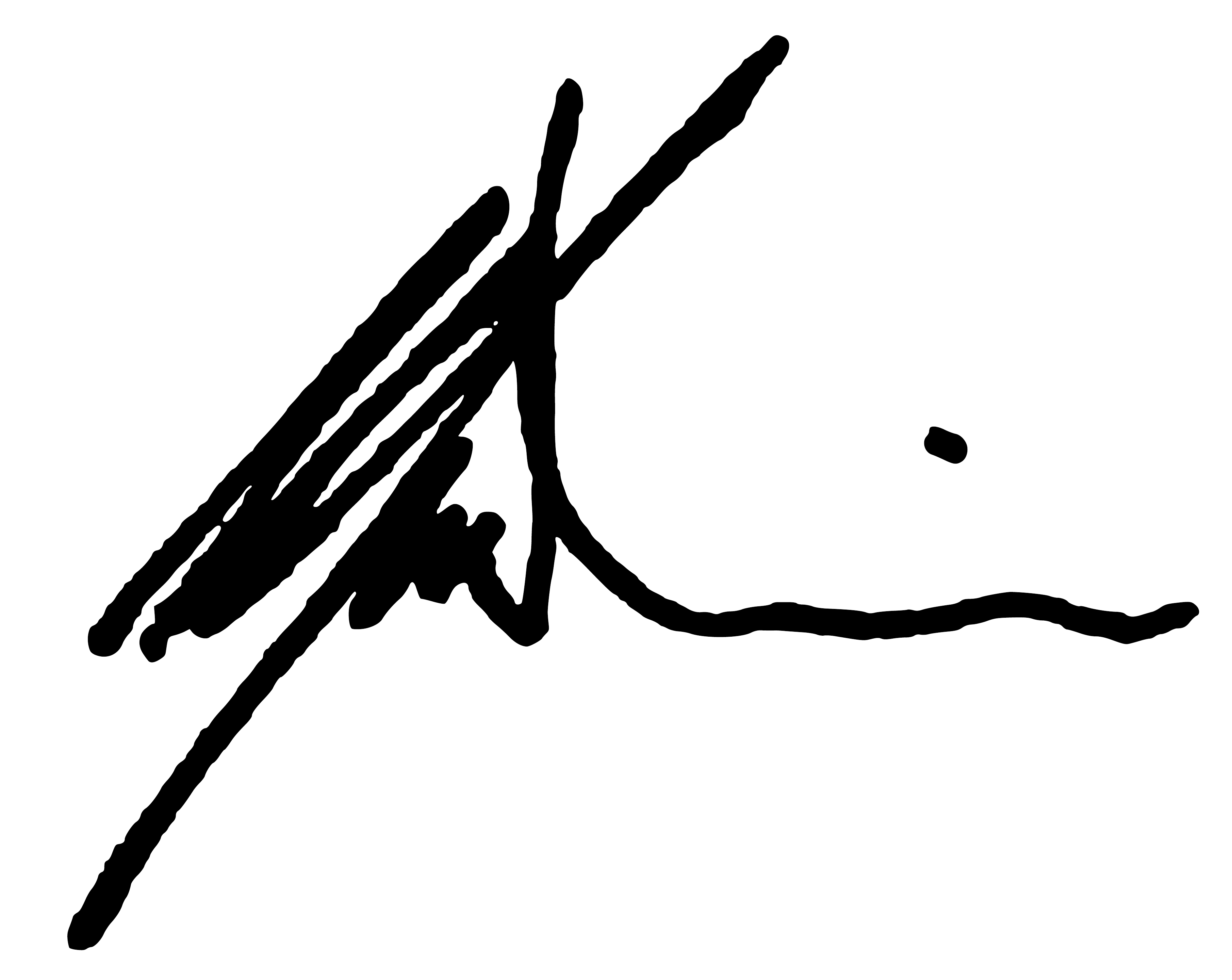}\end{minipage} & & \\
Date: & &\textit{1\(^\text{st}\) March 2018} & & \\
 & & & & \\
Supervisor: & & \textit{Dr. Ilya Kuprov} & & \\
Co-supervisor: & & \textit{Dr. Marina Carravetta} & & \\
 & & & & \\
External examiner: & & \textit{Prof. Steffen Glaser} & & \\
Internal examiner: & & \textit{Dr. Tim Freegarde} & & \\
\end{tabular}

\cleardoublepage
\label{Table of Contents}
\pdfbookmark{\contentsname}{Contents}
\tableofcontents
\cleardoublepage

\phantomsection
\label{List of Figures}
\addcontentsline{toc}{chapter}{\listfigurename}
\listoffigures
\cleardoublepage

\begin{doublespacing}
\begin{small}
\listofsymbols{@{}l@{}>{\raggedleft}p{0.45\linewidth}l@{}r@{}}{
\hline\hline
\textbf{Symbol} & \textbf{Description} & \textbf{Notes} & \textbf{Chapter}\\
\hline
\endfirsthead 
\textbf{Symbol} & \textbf{Description} & \textbf{Notes} & \textbf{Chapter}\\
\hline
\endhead 
\endfoot
\hline\hline
\endlastfoot
\multicolumn{4}{@{}l}{} \vspace{-1em}\\
\multicolumn{4}{@{}l}{\textsc{Magnetic Resonance:}} \\
\(\langle\,\cdot\,|\,\cdot\,\rangle\) & inner product & \hfill & \cref{Chapter:Introduction,Chapter:MagResTheory,Chapter:GRAPE,Chapter:Hessian,Chapter:Penalty,Chapter:ssNMR} \\
\(t\) & time interval & \hfill & \cref{Chapter:MagResTheory,Chapter:GRAPE,Chapter:AuxMat,Chapter:ssNMR}\\
\(i\) & imaginary number & \(=\sqrt{-1}=\frac{-1}{\sqrt{-1}}\) \hfill & \cref{Chapter:MagResTheory,Chapter:GRAPE,Chapter:Hessian,Chapter:AuxMat,Chapter:ssNMR}\\
\(\big|\Psi(t)\big\rangle\) & state of a system at a time \(t\) & \hfill & \cref{Chapter:MagResTheory}\\
\(\big|\Psi_i\big\rangle\)  & complete orthonormal basis states & (Hilbert space) \hfill & \cref{Chapter:MagResTheory}\\
\(c_i^{}(t)\)  & time dependent coefficient of state & \hfill & \cref{Chapter:MagResTheory}\\
\(\hat{\Ham}\)  & Hamiltonian operator & \hfill & \cref{Chapter:MagResTheory}\\
\(\hat{R}\) & rotation operation & \hfill & \cref{Chapter:MagResTheory}\\
\(\mA\) & (bold typeface) a matrix & \hfill & \cref{Chapter:MagResTheory,Chapter:GRAPE,Chapter:Hessian,Chapter:AuxMat,Chapter:Penalty}\\
\(\mA_{}^{T}\) & transpose of a matrix \(\mA\) & \(A_{ij}^{T}=A_{ji}^{}\) \hfill & \cref{Chapter:MagResTheory,Chapter:Hessian}\\
\(\mathbf{A}_{}^{\ast}\) & complex conjugate of \(\mA\) & \(=\Re{(\mA)}-\Im{(\mA)}\) \hfill & \cref{Chapter:MagResTheory,Chapter:GRAPE,Chapter:Penalty}\\
\(\mA_{}^{\dagger}\) & Hermitian conjugate of \(\mA\) & \(=\big(\mA_{}^{\ast}\big)_{}^{\!T}\) \hfill & \cref{Chapter:MagResTheory,Chapter:GRAPE}\\
\(\mA_{}^{-1}\) & matrix inverse of \(\mA\) & \(\mA=\big(\mA_{}^{-1}\big)_{}^{\!-1}\) \hfill & \cref{Chapter:MagResTheory,Chapter:GRAPE,Chapter:Hessian,Chapter:AuxMat,Chapter:ssNMR}\\
\(\big[\mA,\mB\big]\) & commutator of matrices \(\mA\) and \(\mB\) & \(=\mA\mB-\mB\mA\) & \cref{Chapter:MagResTheory,Chapter:Hessian,Chapter:AuxMat}\\
\(\big|\Psi\big\rangle\) & arbitrary state vector & \hfill & \cref{Chapter:MagResTheory,Chapter:GRAPE,Chapter:Hessian}\\
\(\big|\Psi_{}^{\prime}\big\rangle\) & transformed state vector \(\big|\Psi\big\rangle\) & \hfill & \cref{Chapter:MagResTheory}\\
\(\big\langle\Psi\big|\) & Hermitian conjugate of \(\big|\Psi\big\rangle\) & \(\big\langle\Psi\big|=\big|\Psi\big\rangle_{}^{\!\dagger}\) \hfill & \cref{Chapter:MagResTheory,Chapter:GRAPE,Chapter:Hessian}\\
\(\hat{r}\) & displacement operator & \hfill & \cref{Chapter:MagResTheory}\\
\(\hat{p}\) & momentum operator & \(=-i\hbar\nabla\) \hfill & \cref{Chapter:MagResTheory}\\
\(\nabla\) & gradient operator & \(=\Big(\frac{\partial}{\partial x},\frac{\partial}{\partial y},\frac{\partial}{\partial z}\Big)_{}^{\!\!T}\) \hfill & \cref{Chapter:MagResTheory,Chapter:GRAPE,Chapter:Hessian,Chapter:AuxMat,Chapter:Penalty,Chapter:ssNMR}\\
\(\hbar\) & reduced Planck constant (Js(rad)\(^{-1}\)) & \(\approx1.05\times 10^{-34}\) \hfill & \cref{Chapter:MagResTheory}\\
\(\Big(x,y,z\Big)\) & Cartesian coordinates & \hfill & \cref{Chapter:MagResTheory,Chapter:Penalty}\\
\(\Big(r,\theta,\varphi\Big)\) & spherical coordinates & \hfill & \cref{Chapter:MagResTheory,Chapter:Penalty}\\
\(r\) & radial distance & \(=\sqrt{x^2+y^2+z^2}\) \hfill & \cref{Chapter:MagResTheory,Chapter:Penalty}\\
\(\theta\) & inclination/polar angle & \(=\arccos{\frac{z}{r}}\) \hfill & \cref{Chapter:MagResTheory}\\
\(\varphi\) & azimuthal angle & \(=\atan2{\frac{y}{x}}\) \hfill & \cref{Chapter:MagResTheory,Chapter:Hessian,Chapter:Penalty}\\
\(\hat{L}\) & orbital angular momentum operator & \(=\hat{r}\times\hat{p}\) \hfill & \cref{Chapter:MagResTheory}\\
\(\hat{L}_x^{}\) & \(x\)-component of angular momentum operator & \(=-i\big[\hat{L}_y^{},\hat{L}_z^{}\big]\) \hfill & \cref{Chapter:MagResTheory}\\
\(\hat{L}_y^{}\) & \(y\)-component of angular momentum operator & \(=-i\big[\hat{L}_z^{},\hat{L}_x^{}\big]\) \hfill & \cref{Chapter:MagResTheory}\\
\(\hat{L}_z^{}\) & \(z\)-component of angular momentum operator & \(=-i\big[\hat{L}_x^{},\hat{L}_y^{}\big]\) \hfill & \cref{Chapter:MagResTheory}\\
\(\hat{L}_{}^{2}\) & Casimir operator & \(=\hat{L}_x^{2}+\hat{L}_y^{2}+\hat{L}_z^{2}\) \hfill & \cref{Chapter:MagResTheory}\\
\(\hat{L}_+^{}\) & raising operator & \(=\hat{L}_x^{}+i\hat{L}_y^{}\) \hfill & \cref{Chapter:MagResTheory}\\
\(\hat{L}_-^{}\) & lowering operator & \(=\hat{L}_x^{}-i\hat{L}_y^{}\) \hfill & \cref{Chapter:MagResTheory}\\
\(\Delta x\) & change in the variable \(x\) & \hfill & \cref{Chapter:MagResTheory}\\
\(\big\langle A\big\rangle\) & expected value of \(A\) & \hfill & \cref{Chapter:MagResTheory}\\
\(Y_{\!\ell}^{m_{}^{\vphantom{\prime}}}\!(\theta,\varphi)\) & spherical harmonics & \hfill & \cref{Chapter:MagResTheory}\\
\(\mathbb{N}\) & natural numbers & \(\dots,-2,-1,0,1,2,\dots\) \hfill & \cref{Chapter:MagResTheory}\\
\(\ell\) & angular momentum quantum number & \(\ell\in\mathbb{N}\) \hfill & \cref{Chapter:MagResTheory}\\
\(m\) & magnetic quantum number & \(m=-\ell,\dots,\ell\) \hfill & \cref{Chapter:MagResTheory}\\
\(P_{\!\ell}^{}(\xi)\) & Legendre polynomials & \(=\frac{1}{2^\ell \ell!}\frac{{\dd}^{\:\ell}}{\dd \xi^\ell}\Big[\big( \xi^2-1 \big)^{\!\ell}\Big]\) \hfill & \cref{Chapter:MagResTheory}\\
\(\big|\Psi(\vec{r},s)\big\rangle\) & spinor & \(=\!\sum\limits_{n,k}\!\big|\Psi_{\!n}^{}(\vec{r}\,)\big\rangle\!\otimes\!\big|\Psi_{\!k}^{}(s)\big\rangle\)\hfill & \cref{Chapter:MagResTheory}\\
\(s\) & spin quantum number & \(s=0,\frac{1}{2},1,\frac{3}{2},2,\dots\) \hfill & \cref{Chapter:MagResTheory}\\
\(\Unit\) & identity matrix & \hfill & \cref{Chapter:MagResTheory,Chapter:GRAPE,Chapter:AuxMat,Chapter:Penalty}\\
\(\Zero\) & vector or matrix of zeros & \hfill & \cref{Chapter:MagResTheory,Chapter:GRAPE}\\
\(\hat{S}\) & intrinsic angular momentum (spin) & \hfill & \cref{Chapter:MagResTheory}\\
\(\hat{S}_x^{}\) & \(x\)-component of spin operator & \(=-i\big[\hat{S}_y^{},\hat{S}_z^{}\big]\) \hfill & \cref{Chapter:MagResTheory}\\
\(\hat{S}_y^{}\) & \(y\)-component of spin operator & \(=-i\big[\hat{S}_z^{},\hat{S}_x^{}\big]\) \hfill & \cref{Chapter:MagResTheory}\\
\(\hat{S}_z^{}\) & \(z\)-component of spin operator & \(=-i\big[\hat{S}_x^{},\hat{S}_y^{}\big]\) \hfill & \cref{Chapter:MagResTheory}\\
\(\hat{S}_+^{}\) & spin raising operator & \(=\hat{S}_x^{}+i\hat{S}_y^{}\) \hfill & \cref{Chapter:MagResTheory}\\
\(\hat{S}_-^{}\) & spin lowering operator & \(=\hat{S}_x^{}-i\hat{S}_y^{}\) \hfill & \cref{Chapter:MagResTheory}\\
\(\big|\alpha\big\rangle\) & \(\alpha\) spin state & \(=\Big(1,0\Big)_{}^{\!\!T}\) \hfill & \cref{Chapter:MagResTheory}\\
\(\big|\beta\big\rangle\) & \(\beta\) spin state & \(=\Big(0,1\Big)_{}^{\!\!T}\) \hfill & \cref{Chapter:MagResTheory}\\
\(T\) & total time & \hfill & \cref{Chapter:MagResTheory,Chapter:GRAPE}\\
\(\HamH(t)\) & time-dependent Hamiltonian & \hfill & \cref{Chapter:MagResTheory,Chapter:GRAPE}\\
\(\big|\psi(t)\big\rangle\) & time dependent state & \hfill & \cref{Chapter:MagResTheory}\\
\(\exp_{(\text{o})}\) & time-ordered exponential & \hfill & \cref{Chapter:MagResTheory,Chapter:GRAPE}\\
\(\hat{\rho}(t)\) & time-dependent density operator & \(=\big|\psi(t)\big\rangle\!\big\langle\psi(t)\big| \)  \hfill & \cref{Chapter:MagResTheory,Chapter:GRAPE,Chapter:Penalty,Chapter:Penalty}\\
\(p_{ij}^{}\) & probability associated with states \(\psi_i\) and \(\psi_j\) & \hfill & \cref{Chapter:MagResTheory}\\
\(\hat{\mathcal{U}}(T)\) & time propagator over the interval \([0,T]\) & \hfill & \cref{Chapter:MagResTheory,Chapter:ssNMR}\\
\(\hat{\rho}_\textsc{1,2}^{}\) & \(2\) non-interacting, uncorrelated subsystems & \(=\hat{\rho}_\textsc{2}^{}\!\otimes\!\hat{\rho}_\textsc{2}^{}\) \hfill & \cref{Chapter:MagResTheory}\\
\(\hat{\rho}_{1,2,\dots k}^{}\) & \(k\) non-interacting, uncorrelated subsystems & \(=\hat{\rho}_\textsc{2}^{}\!\otimes\!\hat{\rho}_\textsc{2}^{}\!\otimes\!\cdots\!\otimes\!\hat{\rho}_\textsc{k}^{}\) \hfill & \cref{Chapter:MagResTheory}\\
\(\hat{S}_{}^{(k)}\) & spin operator for the \(k^\text{th}\) spin & \hfill & \cref{Chapter:MagResTheory}\\
\(\HamH_{1,2}^{}\) & Hamiltonian of \(2\) non-interacting subsystems & \(=\HamH_1^{}\!\otimes\!\Unit_2^{}\!+\!\Unit_1^{}\!\otimes\!\HamH_2^{}\) \hfill & \cref{Chapter:MagResTheory}\\
\(\hat{T}_{\ell m}^{(12)}\) & \(2\)--particle tensor operators & \hfill & \cref{Chapter:MagResTheory,Chapter:Hessian}\\
\(\mathcal{C}_{m_1^{}m_2^{}m}^{\ell_1^{}\ell_2^{}\ell}\) & Clebsch-Gordan coefficients & \hfill & \cref{Chapter:MagResTheory}\\
\(\HamHH\) & Hamiltonian superoperator & \(=\Unit\!\otimes\!\HamH\! - \!\HamH_{}^{\dagger}\!\otimes\!\Unit\) \hfill & \cref{Chapter:MagResTheory,Chapter:GRAPE,Chapter:ssNMR}\\
\(\hhat{\mathcal{R}}\) & relaxation superoperator & \hfill & \cref{Chapter:MagResTheory,Chapter:GRAPE}\\
\(\big|\state(t)\big\rangle\) & time-dependent, vectorised density matrix & \hfill & \cref{Chapter:MagResTheory,Chapter:GRAPE,Chapter:ssNMR}\\
\(\B{0}\) & static magnetic field in the \(z\)-direction & \hfill & \cref{Chapter:MagResTheory,Chapter:Penalty}\\
\(\B{1}\) & pulsed magnetic field & \hfill & \cref{Chapter:MagResTheory,Chapter:Penalty}\\
\(\vec{B}\) & external magnetic field vector & \hfill & \cref{Chapter:MagResTheory}\\
\(\vhat{S}\) & vector of Cartesian spin operators & \(=\Big(\hat{S}_x^{}, \hat{S}_y^{}, \hat{S}_z^{}\Big)\) \hfill & \cref{Chapter:MagResTheory,Chapter:Hessian}\\
\(\HamZ\) & Zeeman interaction Hamiltonian & \(=\vhat{S}\ccdot\bm{A}\ccdot\vec{B}\) \hfill & \cref{Chapter:MagResTheory}\\
\(\bm{A}\) & Zeeman interaction tensor & \hfill & \cref{Chapter:MagResTheory}\\
\(\HamH_{\mathcal{CS}}\) & chemical shielding Hamiltonian & \(=\vhat{S}\ccdot\bm{\delta}\ccdot\vec{B}\) \hfill & \cref{Chapter:MagResTheory}\\
\(\bm{\delta}\) & chemical shielding tensor & \hfill & \cref{Chapter:MagResTheory}\\
\(\HamH_{\mathcal{G}}\) & electron Zeeman Hamiltonian & \(=\mu_B\vhat{L}\ccdot\bm{g}\ccdot\vec{B}\) \hfill & \cref{Chapter:MagResTheory}\\
\(\mu_B\) & Bohr magneton & \(=\frac{e\hbar}{2m_e}\) \hfill & \cref{Chapter:MagResTheory}\\
\(m_e\) & electron mass (kg) & \(\approx 9.11\times10^{−31}\) \hfill & \cref{Chapter:MagResTheory}\\
\(\bm{g}\) & electron g-tensor & \hfill & \cref{Chapter:MagResTheory}\\
\(\HamNN\) & inter-nuclear interaction Hamiltonian & \hfill & \cref{Chapter:MagResTheory}\\
\(\HamH_{\mathcal{DD}}\) & dipole-dipole interaction Hamiltonian & \hfill & \cref{Chapter:MagResTheory}\\
\(\gamma\) & magnetogyric ratio & \hfill & \cref{Chapter:MagResTheory}\\
\(\HamH_{\mathcal{SR}}\) & spin-rotation coupling Hamiltonian & \(=-\sum\limits_k\vhat{L}\ccdot\bm{A}_k\ccdot\vhat{S}_k\) \hfill & \cref{Chapter:MagResTheory}\\
\(\bm{A}_k\) & spin-rotation coupling tensors & \hfill & \cref{Chapter:MagResTheory}\\
\(\HamH_{\mathcal{Q}}\) & quadrupolar coupling Hamiltonian & \hfill & \cref{Chapter:MagResTheory,Chapter:Hessian}\\
\(\HamH_{\mathcal{J}}\) & J-coupling Hamiltonian & \(a\big(\vhat{L}\ccdot\vhat{S}\big)\) \hfill & \cref{Chapter:MagResTheory,Chapter:Hessian}\\
\(\HamNE\) & electron-nuclear interaction Hamiltonian & \hfill & \cref{Chapter:MagResTheory}\\
\(\HamEE\) & inter-electron interaction Hamiltonian & \hfill & \cref{Chapter:MagResTheory}\\
\(H_n^{}\) & Hermitian matrices of periodic time-dependent functions & \hfill & \cref{Chapter:ssNMR}\\
\(s(t)\) & measured signal as a function of time & \hfill & \cref{Chapter:Penalty}\\
\(\vec{s}\) & measured signal over time & \hfill & \cref{Chapter:Penalty}\\
\(\vec{r}\) & reference signal measured from an experiment with hard pulses & \hfill & \cref{Chapter:Penalty}\\
\(Q\) & characteristic exponents & \hfill & \cref{Chapter:ssNMR}\\
\(\hat{F}_n\) & ladder operators in Fourier space & \hfill & \cref{Chapter:ssNMR}\\
\(\hat{N}\) & number operator & \hfill & \cref{Chapter:ssNMR}\\
\multicolumn{4}{@{}l}{} \vspace{-1em}\\
\multicolumn{4}{@{}l}{\textsc{Numerical Optimisation:}} \\
\(\Obj{\objv{}}\) & scalar objective function & \hfill & \cref{Chapter:GRAPE,Chapter:AWG}\\
\(\objv{}\) & objective variable & \(=\Big(c_1, c_2,\dots,  c_n\Big)_{}^{\!\!T}\) \hfill & \cref{Chapter:GRAPE,Chapter:AWG}\\
\(\Real^n\) & \(n\)D space of real numbers & \hfill & \cref{Chapter:GRAPE,Chapter:Hessian,Chapter:AuxMat}\\
\(\Complex^n\) & \(n\)D space of complex numbers & \hfill & \cref{Chapter:AuxMat}\\
\(\min\limits_{\objv{}}\Big\{\!\Obj{\objv{}}\Big\}\) & minimum of a function \(\Obj{\objv{}}\) & \hfill & \cref{Chapter:GRAPE,Chapter:Hessian,Chapter:AuxMat}\\
\(\max\limits_{\objv{}}\Big\{\!\Obj{\objv{}}\Big\}\) & maximum of a function \(\Obj{\objv{}}\) & \hfill & \cref{Chapter:GRAPE,Chapter:AuxMat,Chapter:Penalty}\\
\(\bm{h}\) & small displacement vector & \hfill & \cref{Chapter:GRAPE}\\
\(\Grad{}\) & gradient vector of objective & \hfill & \cref{Chapter:GRAPE,Chapter:Hessian}\\
\(\Hess{}\) & Hessian matrix of objective & \hfill & \cref{Chapter:GRAPE,Chapter:Hessian}\\
\(\Gradop\) & del operator & \(=\Big(\frac{\partial}{\partial c_1},\frac{\partial}{\partial c_2},\dots,\frac{\partial}{\partial c_n}\Big)_{}^{\!\!T}\)\hfill & \cref{Chapter:GRAPE,Chapter:Hessian,Chapter:AuxMat,Chapter:Penalty}\\
\(\Hessop\) & Hessian operator & \(=\big|\Gradop\big\rangle\!\big\langle\Gradop\big|\)\hfill & \cref{Chapter:GRAPE,Chapter:Hessian,Chapter:AuxMat,Chapter:Penalty}\\
\(\nabla_{}^r\) & \(r^\text{th}\) order derivative & \hfill & \cref{Chapter:GRAPE,Chapter:AuxMat}\\
\(\itr\) & optimisation iteration & \hfill & \cref{Chapter:GRAPE,Chapter:Hessian}\\
\(\objv{\itr}\) & objective function at an iteration \(\itr\) & \hfill & \cref{Chapter:GRAPE}\\
\(\objv{\ast}\) & minimiser/maximiser of objective function & \hfill & \cref{Chapter:GRAPE}\\
\(\big|\objv{\itr}\big\rangle\) & direction at a point of surface, \(\objv{\itr}\) & \hfill & \cref{Chapter:GRAPE}\\
\(\big\|\cdot\big\|_{n}^{}\) & \(n\)-norm & \hfill & \cref{Chapter:Hessian,Chapter:AuxMat,Chapter:Penalty}\\
\(\big\|\cdot\big\|_{\infty}^{}\) & infinity norm & \hfill & \cref{Chapter:GRAPE,Chapter:Hessian}\\
\(\big\|\cdot\big\|_{\text{F}}^{}\) & Frobenius norm & \hfill & \cref{Chapter:Hessian}\\
\(\lambda\) & eigenvalues & \hfill & \cref{Chapter:GRAPE,Chapter:Hessian}\\
\(\sdir{}\) & line search direction & \hfill & \cref{Chapter:GRAPE}\\
\(\sdir{\itr}\) & line search direction at iteration \(\itr\) & \hfill & \cref{Chapter:GRAPE}\\
\(\slen{}\) & line search step length & \hfill & \cref{Chapter:GRAPE,Chapter:Hessian}\\
\(\snewt\) & Newton step & \hfill & \cref{Chapter:GRAPE}\\
\(\BigO(\cdot)\) & error term & \hfill & \cref{Chapter:GRAPE}\\
\(\eps\) & double-precision floating point accuracy & \(\approx 2.22\times 10^{-16}\) \hfill & \cref{Chapter:GRAPE,Chapter:Hessian}\\
\(\dobjvar{\itr}\) & change in objective variable & \(=\objv{\itr+1}-\objv{\itr}\) \hfill & \cref{Chapter:GRAPE}\\
\(\dGrad{\itr}\) & change in gradient & \(=\Grad{\itr+1}\!-\!\Grad{\itr}\) \hfill & \cref{Chapter:GRAPE}\\
\(\HessA{}{}\) & approximation to the Hessian matrix & \hfill & \cref{Chapter:GRAPE}\\
\(\HessAI{}{}\) & approximation to the inverse of the Hessian matrix & \hfill & \cref{Chapter:GRAPE}\\
\(\dOBJVAR{}_{\!\itr}^{}\) & store of most recent \(\dobjvar{\itr}\) & \hfill & \cref{Chapter:GRAPE}\\
\(\dGRAD{}_{\!\itr}^{}\) & store of most recent \(\dGrad{\itr}\) & \hfill & \cref{Chapter:GRAPE}\\
\(\mL\) & lower triangular matrix & \hfill & \cref{Chapter:Hessian}\\
\(\mLbd\) & diagonal matrix of eigenvalues & \hfill & \cref{Chapter:Hessian}\\
\(\mQ\) & matrix with columns of eigenvectors & \hfill & \cref{Chapter:Hessian}\\
\(\mS\) & scaling matrix & \hfill & \cref{Chapter:Hessian}\\
\(D_{\mB}^{}(f(\mA))\) & derivative of the function \(f\) at \(\mA\) in the direction \(\mB\) & \hfill & \cref{Chapter:AuxMat}\\
\(D_{\mB}^{r}(f(\mA))\) & \(r^\text{th}\)-derivative of the function \(f\) at \(\mA\) in the direction \(\mB\) & \hfill & \cref{Chapter:AuxMat}\\
\(\beta\) & simplex expansion factor & \hfill & \cref{Chapter:AWG}\\
\(\gamma\) & simplex contraction factor & \hfill & \cref{Chapter:AWG}\\
\(\objv{j}\) & simplex vertex & \hfill & \cref{Chapter:AWG}\\
\(\objv{0}\) & initial simplex vertex & \hfill & \cref{Chapter:AWG}\\
\(\objv{c}\) & simplex centroid & \hfill & \cref{Chapter:AWG}\\
\(\mX\) & simplex & \hfill & \cref{Chapter:AWG}\\
\(\mR\) & reflected simplex & \hfill & \cref{Chapter:AWG}\\
\(\mE\) & expanded simplex & \hfill & \cref{Chapter:AWG}\\
\(\mC\) & contracted simplex & \hfill & \cref{Chapter:AWG}\\
\(\hat{\mathbf{\Delta}}\) & Finite difference matrix/differentiation approximation & \hfill & \cref{Chapter:Penalty}\\
\multicolumn{4}{@{}l}{} \vspace{-1em}\\
\multicolumn{4}{@{}l}{\textsc{Optimal Control:}} \\
\(\HamC\) & Control operator & \hfill & \cref{Chapter:MagResTheory,Chapter:GRAPE}\\
\(\HamH_x^{(k)},\HamHH_x^{(k)}\) & \(k^\text{th}\) control operator in the \(x\)-direction & \hfill & \cref{Chapter:MagResTheory,Chapter:GRAPE,Chapter:Hessian,Chapter:Penalty}\\
\(\HamH_y^{(k)},\HamHH_y^{(k)}\) & \(k^\text{th}\) control operator in the \(y\)-direction & \hfill & \cref{Chapter:MagResTheory,Chapter:GRAPE,Chapter:Hessian,Chapter:Penalty}\\
\(\ctrl{k}(t)\) & time-dependent control amplitudes & \hfill & \cref{Chapter:GRAPE,Chapter:Penalty}\\
\(K\) & number of control operators & \hfill & \cref{Chapter:GRAPE,Chapter:Hessian,Chapter:AuxMat}\\
\(\big|\init\big\rangle\) & initial state of system & \hfill & \cref{Chapter:GRAPE}\\
\(\big|\targ\big\rangle\) & desired state of system & \hfill & \cref{Chapter:GRAPE,Chapter:Penalty}\\
\(\Delta t\) & small time slice & \hfill & \cref{Chapter:GRAPE,Chapter:AuxMat}\\
\(N\) & number of time slices & \hfill & \cref{Chapter:GRAPE,Chapter:Hessian,Chapter:AuxMat,Chapter:Penalty}\\
\(t_n^{}\) & \(n^\text{th}\) time interval & \hfill & \cref{Chapter:GRAPE}\\
\(\ctrlv{}\) & 2D vector of \(K\times N\) control amplitudes & \hfill & \cref{Chapter:GRAPE,Chapter:Hessian,Chapter:Penalty}\\
\(\ctrlv{k}\) & row vector of \(N\) control amplitudes & \hfill & \cref{Chapter:GRAPE,Chapter:Penalty}\\
\(\ctrlv{n}\) & column vector of \(K\) control amplitudes & \hfill & \cref{Chapter:GRAPE}\\
\(\ctrl{k,n}\) & individual element of \(\ctrlv{}\) & \hfill & \cref{Chapter:GRAPE,Chapter:Hessian,Chapter:AuxMat}\\
\(\dot{\state}(t)\) & time derivative of \(\state\) & \hfill & \cref{Chapter:GRAPE}\\
\(\mathcal{L}\big(\ctrl{}(t)\big)\) & Lagrangian as a function of control \(\ctrl{}(t)\) & \hfill & \cref{Chapter:GRAPE}\\
\(\Fid{}\big(\state(T)\big)\) & fidelity: terminal cost of Lagrangian & \hfill & \cref{Chapter:GRAPE,Chapter:Hessian,Chapter:AuxMat,Chapter:Penalty}\\
\(\Pen\) & penalty: running cost of Lagrangian & \hfill & \cref{Chapter:GRAPE,Chapter:Penalty}\\
\(\big|\Astate(t)\big\rangle\) & costate/adjoint state of system & \(\big|\Astate(T)\big\rangle=\big|\targ\big\rangle\) \hfill & \cref{Chapter:GRAPE}\\
\(\big|\state_n\big\rangle\) & state at a time interval \(n\) & \hfill & \cref{Chapter:GRAPE,Chapter:Hessian,Chapter:AuxMat}\\
\(\big|\Astate_n\big\rangle\) & adjoint state at a time interval \(n\) & \hfill & \cref{Chapter:GRAPE,Chapter:Hessian,Chapter:AuxMat}\\
\(\mathcal{H}_\textsc{p}^{}\) & pseudo-Hamiltonian & \hfill & \cref{Chapter:GRAPE}\\
\(\Prop_n^{}(\ctrlv{n})\) & time propagator of controls & \hfill & \cref{Chapter:GRAPE,Chapter:Hessian}\\
\(\ctrlvn{}\) & normalised control amplitude vector & \hfill & \cref{Chapter:GRAPE,Chapter:Penalty}\\
\(\pwr\) & nominal power level of control pulses & \hfill & \cref{Chapter:GRAPE}\\
\(\ctrlv{}^{(\itr)}\) & control vector at an optimisation iteration \(\itr\) & \hfill & \cref{Chapter:Hessian}\\
\(\mD_{m,n}\) & Hessian matrix of controls at a single time point, \(K\times K\) & \hfill & \cref{Chapter:AuxMat}\\
\(\PropB_{n}^{}\) & time propagator matrix& \hfill & \cref{Chapter:AuxMat}\\
\(\HamB\) & Hamiltonian matrix & \hfill & \cref{Chapter:AuxMat}\\
}\end{small}\end{doublespacing}
\cleardoublepage

\chapter{Preface}
\begin{chapquote}{Marcus Aurelius, \textsc{Meditations IV}}
Do not think that what is hard for you to master is humanly impossible; but if a thing is humanly possible, consider it to be within your reach.
\end{chapquote}

\section*{Resources}

The main resource used within this thesis is the \textit{Spinach} software library\footnote{Available at \url{www.spindynamics.org}} \cite{HOGBEN11}, developed by Ilya Kuprov, Hannah Hogben, Luke Edwards, Matthew Krzystyniak, Peter J. Hore, Gareth T.P. Charnock, Dmitry Savostyanov, Sergey Dolgov, Liza Sutarina, and Zenawi Welderufael.

The majority of numerical simulation in this work was completed with Matlab 2016-2017a/b. All random numbers were seeded with \texttt{rng(1)} from within a Matlab session, unless otherwise stated.

\textsc{Esr} spectroscopy measurements were performed in the Centre for Advanced \textsc{esr} (CAESR) in the Department of Chemistry, University of Oxford, using a Bruker BioSpin EleXSys II E580 spectrometer operating with a SpinJet arbitrary waveform generator (\textsc{awg}) that is based on a SPDevices SDR14 PCI board, with a 0.625~ns time base. Samples were held at 85~K in a Oxford Instruments CF935O cryostat under a flow of cold N2 gas, controlled by an Oxford Instruments Mercury instrument temperature controller. At X-band, the Bruker Biospin ER4118-MD5-W1 resonator was used, which is a sapphire dielectric resonator of dimensions 5~mm ID, 10~mm OD and 13~mm height, that was overcoupled for pulsed measurements to a Q-value of about 200. At Q-band the resonator was a ER5107D2 of typically 80-100~MHz bandwidths.

Photoexcitation of the \textsc{oop-eseem} sample was accomplished with a Continuum Shurlite Nd:YAG emitting a 7~ns \textsc{laser} pulse at the 3rd harmonic of 1064~nm, 355~nm, at a rate of 10~Hz, attenuated with a \(\lambda\)/2 plate and finally depolarized to a 1~mJ pulse energy. The beam was found to match the 5~mm cryostat window and was not further focused on the sample. Synchronisation of the \textsc{laser} and \textsc{oop-eseem} measurement involved using the E580 console PatternJet board user-defined channel as an external trigger to the Stanford Research DG645 delay generator.

The author acknowledges the use of the \textit{iridis} High Performance Computing Facility, and associated support services at the University of Southampton, in the completion of some parts of this work.

This report was typeset by the author using \LaTeX{} and the \TeX{} mathematical typesetting system of Donald Knuth\footnote{D.E.Knuth, \textsc{tex} and \textsc{metafont}, New Directions in Typesetting, American Mathematical Society and Digital Press, Bedford, Massachusetts (1979)}, and vector graphics were generated with 

\section*{Acknowledgements}
A number of people have influenced this work, particularly my supervisor Prof. Ilya Kuprov; without his advice, I would not have had confidence to undertake this project. 

Academics working in the area of optimal control that have had a direct involvement in this work through collaboration and conversation are Mads Sloth Vinding, Thomas Schulte-Herbr\"{u}ggen, Tony Reinsperger, Robert Zeier, David Tannor, Reuven Eitan, Sean Wright, Jack Saywell. Less direct involvement through discussion, although sometimes more substantial; Thomas Heydenreich, Wolfgang Kallies, Sophie Schirmer, Frank Langbein, Martin Koos, Burkhard Luy, Daniel Wisniewski, Shai Machines, and Tim Freegarde.

Magnetisation-to-singlet state work was an outcome of interesting conversation with Stuart Elliott, Chiristian Bengs, Gabriele Stevanato, Giuseppe Pileio, and Malcolm Levitt.

Work on solid state \textsc{nmr} was a collaboration with James Jarvis, Phil Williamson, Marina Carravetta, and Maria Concistre.

In developing the work using the \textsc{esr} \textsc{awg}, I am most grateful to Will Myers for his extensive knowledge of all hardware under his command, but mostly for his patience -- I'm sure he saw emptiness behind my eyes while he tried to explain how to configure the spectrometer to perform basic Hahn echo to me. I don't know if he knows that this was the only physical experiment I have done since my undergraduate days with physics experiments in 2001--2004. In enabling me to do this work I am also grateful to Christiane Timmel, Arzhang Ardavan, Alice Bowen, Thomas Prisner, Anton Tcholakov, Gavin Morley, and Gunnar Jeschke for insightful discussions, and technical support of Bruker through Peter H\"{o}fer and Patrick Carl.

Academics to influence me and allow me to believe in myself are Phil Dawson, Tobias Brandes, Mike Godfrey, Clive Saunders, Klaus Gernoth (whose undergraduate course in advanced quantum mechanics and path integrals seduced me into enjoying difficult, and possibly insoluble, problems). Also, Jason Ralph, Neil Oxtoby, and Charles Hill for allowing me to work in self imposed solitude; a mistake I needed to make at that time.

Finally, personal acknowledgement is given to my wife Katarzyna -- who helped me in an uncountable number of ways -- my father Martin, and my mother Jane. These people have become a bedrock for supporting me and accepting my eccentricities without question. Further acknowledgement is also given to Martin Goodwin and Emmaleen Goodwin, who helped in proofreading this document.

This thesis is dedicated to my son Ruadh\'{a}n, born three months before submission of this Ph.D. thesis, and who I hope will surpass all of my achievements.
\vspace{2em}\\
Southampton\hfill David L. Goodwin\linebreak
October 2017

\cleardoublepage
\addtocontents{toc}{\protect\renewcommand{\cftchapfont}{\cftchapfontorig}}
\addtocontents{toc}{\protect\renewcommand{\cftchappagefont}{\cftchappagefontorig}}
\mainmatter

\chapter{Introduction} \label{Chapter:Introduction} 

\begin{chapquote}{Terry Pratchett, \textsc{Lords and Ladies}}
In fact, the mere act of opening the box will determine the state of the cat, although in this case there were three determinate states the cat could be in: these being Alive, Dead, and Bloody Furious.
\end{chapquote}
\renewcommand*{\CurrentPath}{./Chapter_1}


The subject of this thesis is \textit{optimal control} and its aim is advancing existing methods of optimal control for spin systems of magnetic resonance. A novel method with improved convergence properties will be presented with a focus on computational efficiency and avoiding numerical problems. The computationally expensive task of calculating the exponential of a matrix will be reviewed and concluded with particularly useful numerical efficiencies which decrease simulation time. The thesis will be completed by investigating realistic applications of optimal control solutions, including methods to overcome experimental limitations.

Many descriptions of optimisation use the analogy of hiking in a mountainous terrain to highlight the character of numerical optimisation. The goal of the hiking experience could be to get to the highest peak within a range of mountains, or descend to the lowest altitude. The first of these goals is finding the maximum altitude within the landscape, and the second is finding the minimum; here we optimise the altitude as a function of the coordinates. Although this analogy is easy to relate to, it is not a good example of optimisation: In navigating the mountain we have a map and possibly a compass, more importantly we have a clear view of the mountain and can pick our path to a visible peak; we know the easiest path to the maximum on the outset. Furthermore, the mountain range has many peaks -- an initial path may only find a low-lying peak which becomes apparent from the summit.

\subsubsection{Example of numerical optimisation}

A better example is to follow that of Isaac Newton, finding the roots to a simple polynomial function of a single variable, \(f(x)\). Assuming the function is well behaved, smooth with no discontinuities, the maximum and minimum should reside in places where the gradient of the function is equal to zero. As a first step in finding these solutions is to use an algorithm to find iteratively better estimates to the roots of a function, the places where \(f(x)=0\): This algorithm is known as the Newton-Raphson method (also known as the Newton method) and is attributed to Isaac Newton \cite{NEWTON1671} and Joseph Raphson \cite{RAPHSON1690}:
\begin{equation*}
 x_{n+1}=x_n-\frac{f(x_n)}{f^{\prime}(x_n)}
\end{equation*}
where \(f(x_n)\) if the function of a real variable \(x_n\) at the iteration \(n\). \(f^{\prime}(x_n)=\tfrac{\dd f}{\dd x}\big|_{x_n}\) is the tangent to the function \(f(x)\) at \(x_n\). It is instructive to show the algorithm with an example\footnote{The function used by Newton in his original example \cite{NEWTON1671} was \(f(x)=x^3-2x+5\).}. For the function \(f(x)=\tfrac{1}{3}x^3-x+1\), the tangent to the function can be found analytically by differentiation.

\begin{figure}
\centering{\includegraphics{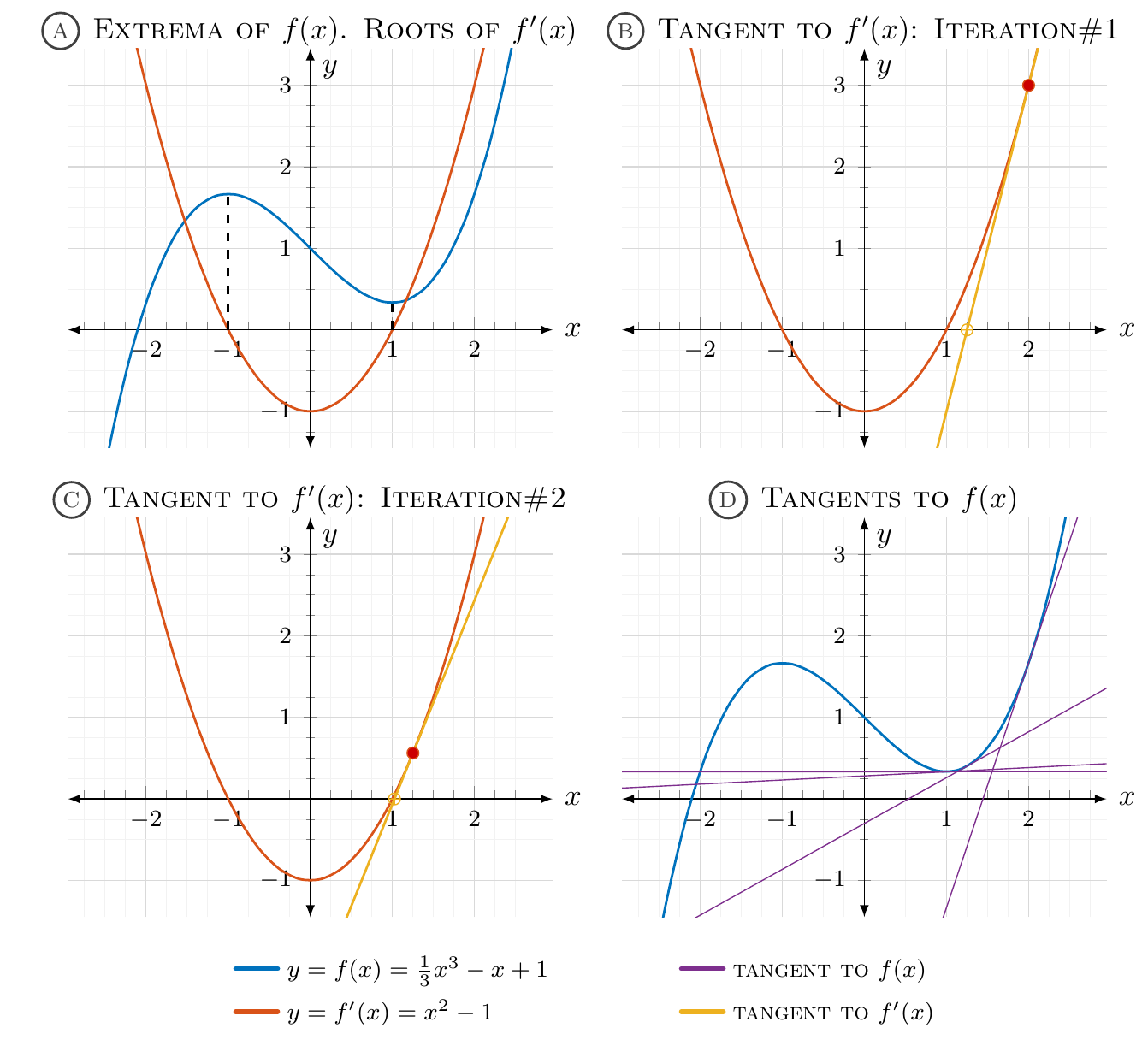}}
\caption[Finding a minimum of a function using tangents of its gradient plot]{Finding a minimum of a function using tangents of the gradient from an initial trial point. \plotlabel{a} the function \(f(x)=\tfrac{1}{3}x^3-x+1\) and its gradient \(x^2-1\) with two extremal solutions at \(x=\pm1\). \plotlabel{b} First iteration: An initial guess to the extrema at \(x=2\). The tangent to the gradient shows where to look for a better point, at \(x=1.25\). \plotlabel{c} Second iteration: tangent of the gradient at \(x=1.25\), finding a better point at \(x=1.025\). \plotlabel{d} Iterations shown as the tangents on the functional, converging closer to a tangential solution with zero gradient.}\label{Fig:NewtonIterates}
\end{figure}

However, the aim is not to find the roots of the function, the aim is to find the roots of the derivative of the function: the place where the gradient is \(=0\)\footnote{Pre-dating Isaac Newton's and Gottfried Leibniz' work on calculus, in the middle of the seventeenth century, de Fermat's work in analytical geometry used tangents to show that at the extreme points of various curves, the gradient becomes zero.}. Substituting this into the previous equation we find an iterative algorithm to find the extrema of a function \(f(x)\):
\begin{equation*}
 x_{n+1}=x_n-\frac{f^{\prime}(x_n)}{f^{\prime\prime}(x_n)}
\end{equation*}
where the second derivative, the derivative of the gradient is \(f^{\prime\prime}(x_n)=\tfrac{\dd }{\dd x}\tfrac{\dd f}{\dd x}\big|_{x_n}=\tfrac{\dd^2f}{\dd x^2}\big|_{x_n}\) evaluated at \(x_n\). The Newton-Raphson method can be used to find these roots to the gradient of the function, and the method is shown for the function \(f(x)=\tfrac{1}{3}x^3-x+1\) in \cref{Fig:NewtonIterates}.

The algorithm converges to an extremum at \(x=+1\); within an acceptable level of accuracy at \(4\) iterations and further than the machine precision\footnote{this is floating-point arithmetic, also referred to as \textit{eps}, is about 16 decimal places on a 64-bit computer} after \(5\) iterations. Notice that starting from a guess \(x_0<0\) converges to a minimum at \(-1\), a starting guess of \(x_0=0\) will be undefined at the first iteration, and a starting guess \(x_0>0\) converges to a maximum at \(+1\). The reason is clearly seen with a graphical representation of the algorithm in \cref{Fig:NewtonIterates}.

\subsubsection{Example of control theory}

Optimal control theory is more subtle. Although optimisation is a part of optimal control theory, the functional of the optimisation is more than navigating an \(n\)-dimensional space. Control theory separates the controllable and uncontrollable physics of the system, allowing the controllable physics to become part of the optimisation functional, and the uncontrollable physics of the system become inherent ``drift''.

\begin{figure}
\centering{\includegraphics{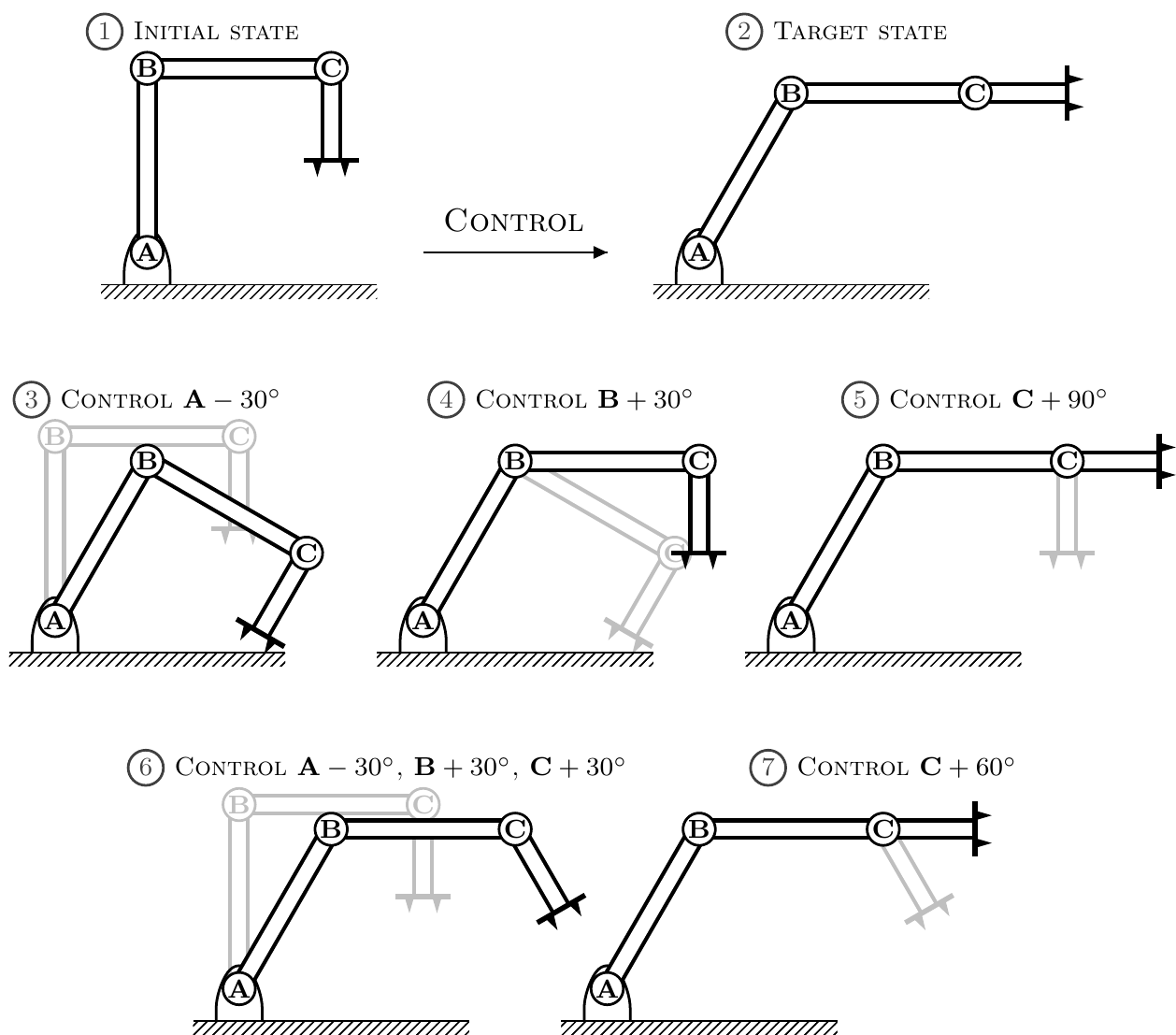}}
\caption[Robot manipulator]{Diagram of a robot manipulator control task. Upper row is the control task of taking the manipulator from \circled{1} an initial state to  \circled{2} a defined target state. Central row shows a control strategy, moving each control motor in turn, \circled{3}, \circled{4}, then \circled{5}. Lower row shows an efficient control strategy, allowing all control motors to be moved simultaneously where possible, \circled{6}, then a final single control \circled{7}. Assuming all motors move at the same speed, the control strategy of the lower low takes 60\% of the time it takes the central row control task.}\label{fig:robot_arm}
\end{figure}

A classical example of a control theory is shown in \cref{fig:robot_arm}, where the task of taking a robot manipulator from an initial position to a target position can be more efficient if control theory is used. The robot manipulator in \cref{fig:robot_arm} has three degrees of freedom, three controllable motors to position the grippers. The easiest way to control the robot manipulator is to move each pivot in turn, controlling each motor to the desired final position in turn. However, if motors are allowed to be controlled simultaneously, the task is performed quicker. This can be viewed as an application of control theory, although simplistic, reducing the amount of time the control task takes.

A proper use of control theory would take into account the weight of each section of the manipulator, the weight of any potential load at the end of the manipulator, the surrounding environment, and ensuring movement of the manipulator is smooth with motors started slowly and finishing slowly. This information, the physics, is easily programmed into a robot manipulator which performs repetitive tasks.

The task of finding the most efficient, the most optimal, controls to move from the initial state to the target state is the task of optimal control.

\section{Thesis Outline}\label{Section:IntroOutline}

\begin{description}
\item[Chapter \ref{Chapter:Introduction}] \hfill\\
An introduction to optimal control, broadly reviewing the literature of the last fifteen years, where optimal control methods have been applied to the area of magnetic resonance. The review focuses on the optimal control method know as \textsc{grape}, but includes a number of other methods also used for optimal control, some well established and some newer.

\item[Chapter \ref{Chapter:MagResTheory}] \hfill\\
This chapter will give an overview of magnetic resonance theory. It will form a particularly theoretical chapter that will be referred to in later chapters, and also serve the purpose of setting out mathematical formalisms that are used in the software toolbox \textit{Spinach} -- the software used in all numerical simulation within this thesis.

\item[Chapter \ref{Chapter:GRAPE}] \hfill\\
A presentation of the \textsc{grape} method in the context of magnetic resonance. The chapter starts from the mathematical area of numerical optimisation, progressing through control theory to optimal control theory. The \textsc{grape} method in this thesis uses a superoperator algebra representation of the Liouville-von Neumann equation and the mathematical derivations of the chapter are all in the context of this formalism.

\item[Chapter \ref{Chapter:Hessian}] \hfill\\
Work developed by the author for this thesis and published in \cite{GOODWIN16}: advancing the \textsc{grape} method from the previous best superlinearly convergent method \cite{FOUQUIERES11} to quadratic convergence. The method presented is shown to scale linearly with a similar computational complexity to previous methods. The Newton-\textsc{grape} method, unique work for this thesis, will be presented along with well established numerical procedures to avoid predicted singularities in the second-order optimal control method. Convergence analysis will be shown for simulations of magnetic resonance pulse experiments using this optimal control method.

\item[Chapter \ref{Chapter:AuxMat}] \hfill\\
The matrix exponential is far from a trivial mathematical operator. This chapter reviews a number series methods from the ``\textit{nineteen dubious ways to compute the exponential of a matrix}'' \cite{MOLER78,MOLER03}, giving recommendation on which method to use for simulation in magnetic resonance systems -- the method used in the software toolbox \textit{Spinach}. The chapter then goes on to present work unique to this thesis, published in \cite{GOODWIN15}, of analytical second directional derivative for the \textsc{grape} method. \cref{Chapter:Hessian,Chapter:AuxMat} set out the focus of this thesis -- the Newton-\textsc{grape} method of optimal control.

\item[Chapter \ref{Chapter:Penalty}] \hfill\\
An overview of popular quadratic penalty methods is presented, optimal control of phase in an amplitude-phase representation of control pulses, and smoothed and shaped pulse solutions with application to finding smooth and robust pulses to generate the singlet state.

\item[Chapter \ref{Chapter:ssNMR}] \hfill\\
This chapter sets out the application of optimal control to solid-state \textsc{nmr}, simulating the optimisation algorithm over the crystalline orientations in a powder sample, and also for magic angle spinning experiments. The chapter builds to finding optimal control pulses to excite the overtone transition in an \textsc{hmqc}-type experiment.

\item[Chapter \ref{Chapter:AWG}] \hfill\\
Work on optimal control in the area of electron spin resonance is far from trivial -- the pulses sent through the spectrometer are not the same shape as those seen by the sample. This is not a problem for shapes that do not vary rapidly -- but optimal control pulses tend to be very specific, with high frequency components that are very specific to the outcome of the experiment. To investigate this, a real-time feedback control loop is set up and experiments are optimised in a brute-force way. Experimental results are shown for Hahn echo experiments with shaped pulses produced from this type of real-time feedback loop.

\item[Chapter \ref{Chapter:Conclusions}] \hfill\\
Summary of main findings of the thesis.
\end{description}

The remainder of this introductory chapter will outline a survey of published literature in the area of optimal control. Although not extensive in its applications, the story of the development of optimal control theory in the context of finding optimal pulses for magnetic resonance systems is fully considered. The reader is directed to a more detailed perspective of modern optimal control theory in the report \textit{Training Schr\"{o}dinger's cat: Quantum optimal control: Strategic report on current status, visions and goals for research in Europe} \cite{GLASER15}.

\section{Early optimal control}\label{Section:IntroEarlyOC}

Optimal control theory in magnetic resonance started at a similar time to hardware implementation of arbitrarily shaped pulses \cite{SILVER84,TEMPS84}. Although not optimal control theory in itself, numerical optimisation was used to design highly selective excitation pulses \cite{LURIE85} and population inversion pulses \cite{WARREN84}. A simple approach was to minimise the square of the difference between an ideal signal and a measured signal,
\begin{equation*}
 \min_{\text{pulse shape}}\Big(\textsc{Ideal Signal} - \textsc{Measured Signal}\Big)^2
\end{equation*}
using a Nelder-Mead simplex method \cite{NELDER65} varying parameters describing a shaped pulse \cite{LURIE85}. This approach to optimisation will be revisited in \cref{Chapter:AWG}.

The development of appropriate hardware, and the early successes of experiments using non-rectangular pulses, led researchers to explore the use of optimal control theory to design ``better'' pulses for magnetic resonance experiments. Selective excitation problems using numerically designed \(\sfrac{\pi}{2}\) and \(\pi\) pulses used optimal control theory to solve Bloch equations\footnote{Bloch equations describe magnetisation of isolated spins.} with a piecewise constant approximation of an arbitrary pulse shape \cite{CONOLLY86}. The optimisation metric used was defined as a \textit{minimum-distance} measure between the current magnetisation vector and a desired magnetisation vector, then solving a Lagrange equation to find an optimum. Later, a modification to the \textit{minimum-distance} measure to include a measure of adiabaticity gave adiabatic and band selective pulse shape solutions \cite{ROSENFELD96}.

In magnetic resonance systems an ensemble of \textit{spins} is manipulated to achieve a desired ensemble state by the application of \textit{unitary operations} \cite{ERNST87}. Investigation into the \textit{controllability} of quantum systems showed that, given a defined initial state and enough pulsing time, any eigenstate of the system can be populated when \(\HamHH_x^{}\) and \(\HamHH_y^{}\) controls are available, and their commutators with the drift span the active space \cite{JUDSON90}\footnote{Unitary operations, \(\HamHH_x^{}\) controls, and \(\HamHH_y^{}\) controls will be introduced in \cref{Chapter:MagResTheory}}.

\section{Analytical optimal control}\label{Section:IntroAnalyricalOC}

One of the first successful applications of optimal control theory in \textsc{nmr} used an existing deuterium decoupling pulse sequence \cite{SCHENKER87} as an initial guess to a gradient-based numerical optimisation algorithm \cite{LEVANTE96}. The optimisation metric was defined by the eigenvalues and eigenvectors of the propagator from an average Hamiltonian. The analytical derivatives of the linear operators \cite{AIZU63} used for the gradient-based optimisation were derived for a piecewise constant approximation of an arbitrary pulse sequence.

Using a method developed for laser spectroscopy \cite{JUDSON92,WARREN93}, analytical design of optimal unitary operations for use in \textsc{nmr} proved to be a difficult task for complicated, coherence-order selective experiments \cite{UNTIDT98,UNTIDT99}. However, the maximal unitary bounds on general \textsc{nmr} polarisation transfer experiments were shown to to be much larger than the apparent limits of \textsc{nmr} polarisation transfer using state-of-the-art experiments of that time \cite{GLASER98,UNTIDT99}. The search for solutions nearer to these bounds, finding pulsed solutions to give better experimental results, and the search for solutions at these bounds, would become the work of the following decade.

Given that a system is controllable and can be steered from one defined state to another \cite{JUDSON90}, optimal solutions must exist. With the aim of designing optimally short pulses, to minimise the effects of relaxation, unitary transforms are found analytically by separating out a Lie subgroup corresponding to controllable operations, then deducing pulses from the unitary operator \cite{KHANEJA01}. Pulses with the shortest possible time were found for coherence transfer experiments in heteronuclear two-spin systems \cite{KHANEJA01} and linear three-spin systems \cite{KHANEJA02}.

\section{Broadband optimal pulses}\label{Section:BroadbandPulses}

Optimal control pulses can be designed to be broadband, robust to resonance offsets, by simulating the optimal control algorithm over a range of offsets \cite{SKINNER03}. These pulses are called Broadband Excitation By Optimised Pulses (\textsc{bebop}). Further to this, pulses designed by \textsc{bebop} were designed to have a shorted duration and less power \cite{SKINNER04}. 

A similar method was used to create Broadband Inversion By Optimised Pulse (\textsc{bibop}) \cite{KOBZAR04}. This optimal control problem is simulated over a range of control pulse power levels -- creating pulses robust to miscalibration. Allowing only the phase of the pulses to vary, keeping a constant amplitude, allows pulses designed with \textsc{bebop} to be calibration-free \cite{SKINNER05,SKINNER06}.

\section{Gradient based optimal control}\label{Section:IntroGradOC}

The original paper setting out the theory for \textit{Gradient Ascent Pulse Engineering}\footnote{also termed \textit{Gradient Assisted Pulse Engineering}} ({\sc grape}) followed the work on analytical optimisation \cite{KHANEJA01} - setting out an elegant base for Newton-type optimisation methods applied to quantum systems \cite{KHANEJA05}. The essence of the method involved discretising the system into slices of time, each with a constant Hamiltonian \cite{CONOLLY86}, and splitting the system into an uncontrollable drift Hamiltonian, and control operators with associated amplitude constants. The validity of the approximation defines the validity of {\sc grape}. Along with similar methods, the measure to maximise -- to make the highest possible value -- is the overlap between two states termed the fidelity of the two states \cite{GLASER98}:
\begin{equation*}
\max_{\text{pulse shape}}\Big\langle \textsc{Target State } \Big| \textsc{ Current State}\Big\rangle
\end{equation*}
The angled brackets, \(\langle\,\cdot\,|\,\cdot\,\rangle\), are a formal definition of the overlap between two vectors, denoting the \textit{inner product}\footnote{The inner product, \(\langle\,\cdot\,|\,\cdot\,\rangle\), is a generalisation of the scalar product between two vectors. In this context, \(|\,\cdot\,\rangle\) is a row vector and \(\langle\,\cdot\,|\) is a column vector, using the Dirac bra-ket notation \cite{DIRAC30}.} of the desired \textsc{Target state} vector and the simulated \textsc{Current state} of the system.

Following its publication in 2005, \textsc{grape} immediately found applications in creating broadband pulses designed with \textsc{bebop}, as any optimal control method should, to systems requiring robust excitation pulses -- pulses that are tolerant to miscalibration of the nominal pulse power. The \textsc{grape} method of optimal control can easily be modified to find broadband pulse excitation with tolerance to radio-frequency (\B{1}) miscalibration/inhomogeneity \cite{KOBZAR05}. Although the system and optimal control task are simple, transferring longitudinal magnetisation of an uncoupled spin to transverse magnetisation, the example shows the possible flexibility of \textsc{grape}, where rectangular pulses alone would fail to be adequately robust.

Transferring a single initial to final magnetisation component, point-to-point state transfer, may not entirely describe an adequate desired final state of the system \cite{KOBZAR04,KOBZAR08}. Simple \textit{point-to-point}\footnote{also termed \textit{state-to-state} problems; these define an initial state and a target state each as single, or linear combination of, members of the basis set} optimal control solutions can be used within the framework of finding \textit{universal rotation} pulses, required for mixing or refocusing pulses \cite{LUY05}. \textit{Universal rotation} pulses are solutions to problems where each of the three magnetisation vectors of the Bloch equation must have a defined rotation \cite{MACHNES11,SCHULTE_HERBRUEGGEN11}. When these universal rotations are the target state of a \textsc{grape} simulation, the resulting pulse shapes are found to exhibit symmetry around the centre of the pulse \cite{KOBZAR12}. Designing universal rotation pulses to be robust is called Broadband Universal Rotation By Optimised Pulses (\textsc{burbop}) \cite{JANICH11,JANICH12,SKINNER12a}.

The \textsc{cpmg} sequence (Carr-Purcell-Meiboom-Gill \cite{CARR54,MEIBOOM58}), useful in investigation of relaxation rates\footnote{Relaxation is called \textit{decoherence} in some literature}, was redesigned with \textsc{grape} resulting in universal rotations. This application extended the \textsc{cpmg} robustness to a larger range of static \B{0} resonance offsets and \B{1} pulse power miscalibration \cite{BORNEMAN10}. Robust shaped pulses, intended to be direct replacements for hard \(\sfrac{\pi}{2}\) and \(\pi\) pulses, were designed using the \textsc{grape} method \cite{NIMBALKAR13}. Furthermore, broadband optimal solutions for heteronuclear spin systems can be designed to compensate for coupling evolution over the duration of a pulse shape, with the strategy of using a point-to-point optimisation with its time reversal for finding universal rotation pulses \cite{EHNI13}. 

Spin systems simulated with \textsc{grape} speed up calculation of coherence transfer in larger spin and pseudospin chains \cite{SCHULTE_HERBRUEGGEN05}. Further to this, a method of using \textsc{grape} with partitioned subsystems used to describe a large spin system proves effective in finding optimal control solutions for very large spin systems \cite{RYAN08}. The long-range state transfer can be improved with the temporary occupation of higher order states for larger systems \cite{KHANI09}.

Pulses produced by \textsc{grape} may be difficult to implement on real hardware. Hardware constraints may interpolate a shape to a smooth realisation of pulses, rather than the square pulses produced by the \textsc{grape} method. Smooth, shaped pulses can be produced as analytic functions with a modification to the fidelity functional \cite{SKINNER10}. Alternatively, this hardware constraint can be accounted for in simulations by increasing the discretisation of time evolution above that of the discrete control pulses \cite{MOTZOI11}, using experimentally obtained transfer matrices within \textsc{grape} \cite{BORNEMAN12}, or by employing a feedback control loop after a designed pulse sequence is implemented \cite{EGGER14}.

\subsection{Solid-state nuclear magnetic resonance}\label{Section:IntroNMR}

When \textsc{nmr} is performed on a solid powder, the anisotropic spin interactions result in large line broadening of the spectrum. Spinning the sample at the magic angle\footnote{the magic angle is \(\theta_\text{magic}=\arccos{\tfrac{1}{\sqrt{3}}}=54.7^{\circ}\)} relative to the static \B{0} axis averages out many of these interactions, resulting in sharper spectral line shapes \cite{ERNST87}. This method is termed \textsc{mas} (Magic Angle Spinning), and can be useful for investigating dipolar and quadrupolar systems \cite{FRYDMAN95}.

In the same year as the publication of the \textsc{grape} method, it found application to solid-state \textsc{mas} \textsc{nmr},  accounting for anisotropic components of spin interactions, sample spinning and specifically addressing problems of \B{1} inhomogeneity \cite{VOSEGAARD05} within simulation. Design of pulses robust to hardware miscalibration of \B{1} averages the \textsc{grape} method over a range of \B{1} power levels, in effect, performing a \textsc{grape} simulation at each power level, then averaging the fidelities and the gradients \cite{VOSEGAARD05}. A range of chemical shift offsets can be treated in a similar way to \B{1} miscalibration \cite{SKINNER06}. Chemical shift offsets, \B{1} inhomogeneity, and dispersion from powder averaging, can also be simulated with the \textsc{grape} framework \cite{TOSNER06}, in this case using effective Hamiltonians. A combination of optimal control pulses and average Hamiltonian theory can be used to analyse and then remove unwanted second and third order coupling terms produced by optimal control solutions \cite{BJERRING13}.

A method of solid-state \textsc{nmr} can take advantage of the \textit{Overhauser effect} \cite{OVERHAUSER53} to transfer polarisation from unpaired electrons to nearby nuclei, in a method termed \textsc{dnp} (Dynamic Nuclear Polarisation). The framework to control \textsc{dnp} systems \cite{GRIESINGER12} is set out in \cite{KHANEJA07b,MAXIMOV08}, and simple \textsc{dnp} systems can be handled with \textsc{grape} to give optimal solutions \cite{HODGES08} in the presence of relaxation \cite{POMPLUN08}. Optimal control proves useful in combining incoherent and coherent transfer schemes, investigated in a system of one electron and two nuclear spins \cite{POMPLUN10}.

Again in the context of solid-state \textsc{nmr}, optimal pulses found use in experiments with labelled proteins \cite{KEHLET07,HANSEN07}. Biomolecular samples, where high power control pulses can overheat the sample, dictate the need for low power pulse solutions. These can be designed with \textsc{grape} by penalising high power pulse solutions \cite{KEHLET05,NIELSEN09}, and can be extended to \(\mathrm{{}_{}^{2}H - {}_{}^{13}C}\) cross-polarisation \textsc{mas} experiments \cite{WEI11}. The quadratic power penalty is also used in calculating optimal pulses for conversion of parahydrogen induced longitudinal two-spin order to single spin order \cite{BRETSCHNEIDER12}, and in designing robust pulses for transfer of magnetisation to singlet order \cite{LAUSTSEN14}. Further radio frequency power and amplitude restriction are investigated for broadband excitation pulses \cite{KOOS15}. 

Methods of simulating solid-state \textsc{mas} to find optimal control pulses will be given in \cref{Chapter:ssNMR} and an overview of useful penalty methods used in conjunction with \textsc{grape} will be given \cref{Chapter:Penalty}.

\subsection{Nuclear magnetic resonance imaging}\label{Section:IntroMRI}

A useful application of \textsc{nmr} in medicine is non-invasive, macroscopic imaging of the human body \cite{ERNST87}. Termed \textsc{mri} (Magnetic Resonance Imaging), the method uses magnetic field gradients to record a projection of proton density such that resonant frequencies are a linear functions of spatial coordinates \cite{LAUTERBUR73}. Extending \textsc{grape} into an average optimisation over an ensemble of Hamiltonians can compensate for the dispersion in the system dynamics in \textsc{mri} simulations \cite{LI06,KHANI12}.

When radio-frequency pulse durations are similar to relaxation rates of quadrupolar nuclei, as in the case of \textsc{mri} applications, previous work on optimal control with \textsc{mas} \cite{VOSEGAARD05} fails to be relevant. This was investigated, resulting in a novel method of using \textsc{grape} \cite{LEE08}. The idea is to find optimal solutions with \textsc{grape} for long total pulse duration, then decreasing the total pulse duration but using the same shape. When the total pulse duration was reduced to less than the inverse of the quadrupolar frequency, the resulting pulse shapes were simple, resembling two discrete pulses separated by a delay. Results from that work showed an increase in signal intensity of a \spinfrac{3}{2} central transition while suppressing its satellite transitions. In the context of \textsc{mri}, this is useful in separating the signal of \(\mathrm{{}_{}^{23}Na}\) in cartilage from that of free \(\mathrm{{}_{}^{23}Na}\) within the image. Although found by optimal control, the solutions to the problem indicated simple pulse shapes that do not need optimal control \cite{LEE09}. Further investigation proved optimal control theory found its use in designing robust pulses \cite{LEE10}, where simple pulse shapes could not. Robust pulses were designed with \textsc{grape} to excite the central transition of arbitrary \spinfrac{3}{2} systems with static powder patterns \cite{ODELL10}. \textsc{Dft} (Density Functional Theory)\footnote{a popular computational method to investigate electronic structure.} calculations can be used to predict the necessary interaction parameters for these \textsc{grape} simulations \cite{ODELL11}, giving a very general approach to numerically finding optimal control pulses.

A practicable realisation of an efficient optimal control algorithm for \textsc{mri} purposes should be fast, working in a small window of time, to be able to take patient specific distortions into account. The optimisation should be performed as the patient waits in the instrument, and timing is critical -- should the patient move during the optimal control simulation, the distortions also change and the optimal control solution becomes ineffective \cite{VINDING12}. Furthermore, advanced arrangements of control coils used in modern \textsc{mri}\footnote{the arrangement is called pTX (parallel transmission), and can have an arrangement of up to 6 control coils, rather the the conventional 2 in \textsc{mri}.} need to be taken into account when designing optimal pulses \cite{MASSIRE13}. In comparison to conventional adiabatic pulses, optimal control pulses can achieve similar results with reduced \textsc{sar}\footnote{Patient safety guidelines have strict limitations on \textsc{sar} (specific-absorption-rate) -- local hot-spots of temperature, due to the distribution of \textsc{sar} when using pTX, will be different for every patient and every scan.} \cite{LEE16} and within local and global \textsc{sar} limits \cite{VINDING17a,VINDING17b}.

\subsection{Quantum information}\label{Section:IntroQIP}

The research area of quantum information processing is based using two-level quantum systems, and can be thought of as extending the language of classical computation to these quantum bits \cite{NIELSEN00,WISEMAN10}. Just as the language of classical computation is based on circuitry and logic gates, so too is the area of quantum information processing. The difference between classical bits and quantum bits is that quantum bits can exist in a superposition of the classical binary states.

Although optimal control was applied to the area of quantum information before the advent of \textsc{grape}, and there have been many applications of optimal control to this area, this review is intended to show applications of \textsc{grape} to magnetic resonance systems -- specifically to ensembles of spins\footnote{Magnetic resonance on single spins is an emerging area, but beyond the scope of this work.}. As such, only a modest number of applications of \textsc{grape} to this area will be mentioned; those the author judges general enough to have use in optimal control of an ensemble system in magnetic resonance.

Compared with the previous best methods for control of superconducting two-level systems, \textsc{grape} can achieve lower information leakage, higher fidelity, and faster gates \cite{SPORL07}. A strategy for reducing information leakage entails using an extra \textit{control channel}, proportional to the derivative of the main control, within \textsc{grape} \cite{MOTZOI09}. A chain of 3 coupled superconducting two-level systems can use \textsc{grape} to reduce unwanted ``crosstalk'' between the two outer two-level systems, so mediating transfer in a serial manner \cite{GROSZKOWSKI11}. High fidelity spin entanglement can be achieved with universal rotations designed with \textsc{grape}, demonstrated experimentally for nitrogen-vacancy centres in diamond \cite{DOLDE14}.

The robust nature of \textsc{grape} has been shown to be tolerant to errors in pulse length and off-resonance pulses for nitrogen-vacancy centres in diamond \cite{SAID09,DOLDE14}, atomic vapour ensembles \cite{SCHONFELDT09}, and design of fault tolerant quantum gates using a modified \textsc{grape} optimisation update rule \cite{NIGMATULLIN09,SCHIRMER09a}. Time optimal solutions \cite{KHANEJA01} giving the minimal pulse length to achieve high fidelity transfer of Ising coupled cluster states are verified with \textsc{grape} simulated at different pulse lengths \cite{FISHER09}, with exponential falloff below a threshold pulse length \cite{KHANI09}. Realistic pulse implementation can be accounted for in the form of power and rise-time limits \cite{FISHER10}. Optimal \textsc{grape} solutions for these realistic pulses are realised by averaging optimisation over individual detunings of resonance frequencies (with longitudinal magnetisation control), with optimal pulses using only a single \(x\)-magnetisation control.

Although the \textsc{grape} method is designed for closed systems, in the case of coupling to an environment the method was advanced to open non-Markovian \cite{REBENTROST09}, and open Markovian systems \cite{SCHULTE_HERBRUEGGEN11}. Control pulses that maximise transfer starting with a high-temperature thermal ensemble are automatically optimal for lower-temperature initial ensembles \cite{WANG10}. Further, the control landscape of an open system is similar to a large closed system \cite{WU12} and there is evidence that noise can be suppressed with control \cite{ARENZ14}.

\subsection{Electron spin resonance}\label{Section:IntroESR}

The particularly challenging application of optimal control to \textsc{esr} (Electron Spin Resonance)\footnote{also termed \textsc{epr} (Electron Paramagnetic Resonance). \textsc{Esr} is used in this thesis to avoid conflict with the quantum mechanics term ``\textsc{epr}-pair'' derived from the Einstein-Podolsky-Rosen thought experiment \cite{EINSTEIN35}} includes the problem of the pulse response function of the microwave setup -- the pulses seen by the sample differ from those sent to the pulse shaping hardware. The transfer matrix (response function) may be measured either by adding an antenna to the resonator \cite{SPINDLER12}, or by using a sample with a narrow \textsc{esr} line to pick up the intensity of each spectral component \cite{KAUFMANN13}. Quasi-linear responses, such as the phase variation across the excitation bandwidth in nutation frequency experiments, can be described with additional transfer matrices \cite{DOLL13}. Once measured, the transfer matrix can be used within an optimal control algorithm to transform pulse shapes before trajectory time evolution is calculated, then be used to design broadband \textsc{esr} excitation pulses \cite{SPINDLER12}.

This relatively new area of applying optimal control pulses to \textsc{esr} systems will be investigated in \cref{Chapter:AWG}, finding a complimentary approach to those mentioned in the previous paragraph.

\section{Modified methods}\label{Section:IntroModMethods}

Effective Hamiltonian methods \cite{TOSNER06} and the \textsc{grape} algorithm \cite{KHANEJA05} have showed great success in finding optimal control solutions. However, more powerful algorithms have the potential to take full advantage of optimal control. Further modifications of the \textsc{grape} method spawned two useful optimal control methods, \textit{optimal tracking} and \textit{cooperative pulses}.

\subsection{Optimal tracking}\label{Section:IntroTracking}

\textit{Optimal tracking} is a modified \textsc{grape} method that takes an input control field that will approximate a desired output trajectory rather than a state \cite{NEVES09}. Heteronuclear decoupling is an example of a limitation of the original \textsc{grape} algorithm. The control problem involves taking a system to a desired intermediate state then back to the original state. In this context, defining the cost functional as the overlap between the initial and final states is not enough, since the initial and desired final states are the same -- the system may simply do nothing and produce a maximal overlap while not penetrating the intermediate state. As an extension of the \textsc{grape} algorithm, \textit{optimal tracking} \cite{NEVES09} propagates the system to an intermediate desired state while simultaneously propagating backward from the final state to this desired intermediate state -- so ensuring that the initial and final states are always the same. \textit{Optimal tracking} used for heteronuclear decoupling sequences, with linear offset-dependent scaling factors, can encode chemical shift correlations from an indirect dimension \cite{SCHILLING12,ZHANG13}.

\subsection{Cooperative pulses}\label{Section:IntroCoop}

\textit{Cooperative pulses} take into account multi-pulse experiments, generalising the experimental method of phase cycling and cooperatively compensating for each pulse imperfection \cite{BRAUN10}. By generalising the \textsc{grape} method, it can be applied to broadband and band-selective experiments \cite{BRAUN10}, quantum filter type experiments \cite{BRAUN14}, and pseudo-pure state preparation in the context of quantum information processing \cite{WEI14}.

\section{Other optimal control methods}\label{Section:IntroMiscOC}

Krotov type methods are algorithms of optimal control \cite{KROTOV96} that were developed in parallel to the \textsc{grape} method reviewed above. The methods were successfully developed for application to laser spectroscopy \cite{TANNOR92} and general two-level quantum systems \cite{ZHU98}. Krotov methods recursively find pulses necessary to perform a state-transfer problem by solving the Lagrange multiplier problem \cite{PEIRCE88}. The algorithm was recently formulated in the context of magnetic resonance \cite{MAXIMOV08} and is an alternative to \textsc{grape}. It was further developed to a hybrid type method, including a quasi-Newton character to improve convergence \cite{EITAN11}. The total pulse time of this method can be critical in finding high fidelity solutions to optimal control problems, and a characteristic threshold must be passed to find these solutions using Krotov type methods \cite{CANEVA09}. Krotov based methods can appear unattractive in that they need a seemingly arbitrary constant in the form of a Lagrange multiplier -- and although the constant is the same for similar systems, any new system needs investigation to find its own Lagrange multiplier.

The \textsc{crab} (Chopped Random Basis) algorithm shows promise in finding optimal control solutions to many-body problems \cite{DORIA11,CANEVA11}. Specifically, the method lends itself to \textit{tensor-network} descriptions of the system, also known as time-dependent \textit{density matrix renormalisation group} \cite{SCHOLLWOCK05} or the \textit{tensor train formalism} \cite{SAVOSTYANOV14} -- an almost approximation-free description.  An extension of the algorithm avoids local extrema by reformulating pulse bounds \cite{RACH15}. Pulse shapes are described by Fourier harmonics and these become the variables of optimisation. Calculating the gradient elements in this formalism was expected to be prohibitively expensive, therefore the method uses a gradient-free optimisation method such as the Nelder-Mead simplex method \cite{NELDER65}. A further modification includes gradient calculations in the \textsc{goat} method (Gradient Optimisation of Analytical controls) \cite{MACHNES15} giving improved convergence. 

A method for efficiently calculating gradients will be presented in \cref{Chapter:AuxMat}.
%
%
%

\section{Numerical computation}\label{Section:IntroNumComp}

A number of public software packages include functionality to perform optimal control calculations for magnetic resonance problems. A selection is listed below:
\begin{description}
\item[Dynamo] \cite{MACHNES11,SCHULTE_HERBRUEGGEN12} \hfill\\
Using Matlab, includes \textsc{grape}, Krotov, and hybrid methods for quantum optimal control simulations of open and closed systems.
\item[QuTiP] \cite{JOHANSSON12,JOHANSSON13} \hfill\\
Using Python, includes the \textsc{grape} and \textsc{crab} algorithms using \textsc{l-bfgs} and Nelder-Mead algorithms, respectively.
\item[Simpson] \cite{BAK00} \hfill\\
The release of the magnetic resonance simulation package \textsc{simpson} progressed to include optimal control in the form of the \textsc{grape} method \cite{TOSNER09,TOSNER14}. Includes the ability to simulate liquid- and solid-state \textsc{nmr}, \textsc{mri}, and quantum computation. Simpson is coded in C with the Tcl scripting interface.
\item[Spinach] \cite{HOGBEN11} \hfill\\
In the context of optimal control, \textit{Spinach} solves the Liouville-von Neumann equation using state-space restriction techniques in an irreducible spherical tensor basis using Matlab. \textsc{Grape}-\textsc{l-bfgs} \cite{FOUQUIERES11} and Krotov methods were included before the author became involved in the project\footnote{The \textit{Spinach} software package is used and developed by the author of this document.}.
\end{description}

The first systematic look at the algorithmic complexity of the \textsc{grape} algorithm \cite{KHANEJA05} was published in \cite{GRADL06}. Efficient implementation would take advantage of inherent parallelisation of the \textsc{grape} algorithm. The bottleneck in computation was identified as performing a matrix exponential -- a number of approaches to numerically compute this were compared: scaling and squaring based on the Pad\'{e} approximation and series expansion with either the Taylor expansion or Chebyshev polynomials \cite{SCHULTE_HERBRUEGGEN09,AUCKENTHALER10}.

The obvious way to create a gradient for a gradient ascent algorithm is to define it by a central finite-difference approximation -- however, for anything but small systems, this gives a computationally expensive calculation\footnote{A two point finite difference requires two functional evaluations, a four point requires four etc. whereas an analytic gradient would require one functional evaluation by definition.} -- making optimal control of larger systems impractical. The need for analytic gradients is paramount for any gradient ascent method to become useful in larger systems \cite{KUPROV09}. Based on an auxiliary matrix formalism, analytic gradient elements can be calculated by taking the exponential of a \(2\times 2\) triangular block matrix, with the drift Hamiltonian on the diagonal and the control operator on the off-diagonal \cite{FLOETHER12}.

A Quasi-Newton optimisation of \textsc{grape}, using the \(\ell\)\textsc{-bfgs} algorithm, was developed to make full use of the gradient history to give second-order derivative information and improve convergence to superlinear \cite{FOUQUIERES11}. A Newton-Raphson root finding approach, using the conditioned Jacobian matrix of first-order partial derivatives as a least-squares optimisation problem \cite{FOUQUIERES12}, improved convergence to quadratic for a chain of five coupled two-level systems using two controls. The Jacobian matrix is also used in Krylov-Newton methods for closed spin systems, showing impressive convergence without calculation of a full Hessian \cite{CIARAMELLA15}.

Many examples in the literature above, when simulating high dimensional spin systems, produce optimal control solutions which can be described to appear like ``noise''. Trajectory analysis of the ``noisy'' pulse sequence evolution shows an underlying order to the spin dynamics \cite{KUPROV13}, and a time-frequency representation of the pulse shapes reveals underlying patterns which are not apparent in a conventional Cartesian representation of pulses \cite{KOCHER14}.

\section[Summary]{Summary}\label{Section:IntroSummary}

Optimal control has become a useful tool in exploring the limits of, among other technologies, magnetic resonance spectroscopy and imaging. The design of pulsed magnetic resonance experiments with either composite pulses, shaped pulses, or adiabatic pulses has shown improvement when designed with optimal control methods.

The \textsc{grape} method of optimal control has become one of the leaders in its field for finding computationally efficient solutions, able to find state-of-the-art pulse sequences for large and complicated spin systems. Furthermore, pulse sequences produced can be made robust to experimental variability.

In addition to becoming an indispensable tool of chemical spectroscopy, the emerging area of optimal control of \textsc{mri} shows promise in reducing medical costs at the same time as increasing diagnostic abilities. And it is this last point that becomes the motivation of the work presented in this thesis -- finding more accurate optimal control solutions with decreased computational complexity -- with the vision of large, highly parallel computational infrastructure installed at medical facilities, real-time, bespoke optimal control solutions will become affordable. With reduced experiment time on the expensive apparatus, and increased throughput: the dream of scientific facilitators working in business and finance will become more real.

\chapter{Magnetic Resonance Theory} \label{Chapter:MagResTheory}

\begin{chapquote}{Joannes Stobaeus, \textsc{Physical and Moral Extracts}}
A youth who had begun to read geometry with Euclid, when he had learnt the first proposition, inquired, ``What do I get by learning these things?'' So Euclid called a slave and said ``Give him three pence, since he must make a gain out of what he learns.''
\end{chapquote}
\renewcommand*{\CurrentPath}{./Chapter_2}


Optimal control can be thought of as the study of controlling a system to find an optimally desired solution. A formal definition of optimal control will be given in \cref{Chapter:GRAPE}, culminating in the \textsc{grape} method. \cref{Chapter:Introduction} was a review of published literature on the subject of optimal control applied to magnetic resonance systems. Normally, when studying magnetic resonance systems, an ensemble of particle \textit{spins} are manipulated to give molecular information of how these \textit{spins} interact with one another.


Magnetic resonance spectroscopy classically uses rectangular pulses of electromagnetic radiation to manipulate spins located within a static magnetic field. Non-rectangular shaped pulses are also used in magnetic resonance with a review given in \cite{FREEMAN98}. The idea of applying optimal control to magnetic resonance is to optimise desired experimental outcomes with respect to the pulse shape.


This chapter will introduce features of magnetic resonance theory that will be used within this thesis. Much of this chapter is taken from standard textbooks on quantum mechanics \cite{LANDAU58,MERZBACHER61,ARFKEN99,TANNOR07}, nuclear magnetic resonance (\textsc{nmr}) \cite{ABRAGAM61,ERNST87,HORE95} and electron spin resonance (\textsc{esr}) \cite{SCHWEIGER01,BRUSTOLON09}.

\section{Angular momentum}\label{Section:AngularMomentum}

Angular momentum in quantum mechanics can be derived from many starting points. Here, the starting point is the Schr\"{o}dinger equation and Euclidean geometry.

\subsection{Schr\"{o}dinger's equation}\label{Section:Schrodinger}

In 1926 Erwin Schr\"{o}dinger formulated the famous Schr\"{o}dinger equation \cite{SCHRODINGER26} in an attempt to find solutions to a Hamiltonian equation using Louis de Broglie's wave-particle duality \cite{DEBROGLIE25}. The Hamiltonian, \(\HamH\), is an energy operator which dictates dynamics in the equation of motion of that system in the form of a Hamiltonian equation. If \(|\Psi(t)\rangle\) describes the state of a system at a particular time \(t\), then the time-dependent Schr\"{o}dinger equation is
\begin{equation}
 \frac{\partial}{\partial t}\big|\Psi(t)\big\rangle=-i\HamH\big|\Psi(t)\big\rangle\label{eqn:schrodinger}
\end{equation}
It is common to measure energy eigenvalues in angular frequency units within the area of magnetic resonance, and the usual factor of \(\hbar\) is therefore dropped from \cref{eqn:schrodinger}. The state \(|\Psi(t)\rangle\) can be expanded to a complete orthonormal basis
\begin{align}
 && \big|\Psi(t)\big\rangle=\sum\limits_{i=1}^n c_i^{}(t)\big|\Psi_i\big\rangle, && i=1,2,\dots,n
\end{align}
where the time dependence is now described with the coefficients \(c_i^{}(t)\) of the complete orthonormal basis states \(\Psi_i\). This basis spans a \textit{Hilbert space} of dimension \(n\).

\subsection{Euclidean principle of relativity}\label{Section:Relativity}

Quantum mechanics is built on the principle that space is subject to the laws of Euclidean geometry -- and that space itself is isotropic and homogeneous. This assumption is called the \textit{Euclidean principle of relativity}. A physical interpretation is -- in the absence of external fields and perturbations the effect of moving a physical system, such as experimental apparatus, should not result in measurements that depend on the location or orientation of the experiment. Or more precisely; a translation or rotation of the coordinate system should not change the Hamiltonian describing the energy of particles within the system. The Hamiltonian is invariant under rotation and translation. 

For a rotation operation denoted by \(\hat{R}\) the mapping of a state vector \(\Psi\) to \(\Psi^{\prime}\) is
\begin{align}
 && \big|\Psi\big\rangle \mapsto  \hat{R}\big|\Psi\big\rangle=\big|\Psi^{\prime}\big\rangle && \Rightarrow && \big\langle\Psi\big|\hat{R}^{\dagger}\HamH\hat{R}\big|\Psi\big\rangle=\big\langle\Psi\big|\HamH\big|\Psi\big\rangle
\end{align} 
This relation must hold for any state vectors, and therefore $\hat{R}$ must commute with the system Hamiltonian \(\HamH\)
\begin{align}
 && \hat{R}^{\dagger}\HamH\hat{R}=\HamH && \Rightarrow && \hat{R}^{-1}\!\HamH\hat{R}=\HamH && \Rightarrow && \big[\!\HamH,\hat{R}\big]=0
\end{align} 
Using the Schr\"{o}dinger \cref{eqn:schrodinger}, and the measurement of an observable\footnote{An Hermitian operator that possesses a complete set of eigenfunctions is an \textit{observable}.} property of the system being its expected value\footnote{The expected value of an hermitian operator is \(\big\langle\Psi\big|\hat{A}\big|\Psi\big\rangle\)} -- the following relation can be derived
\begin{equation}
 -i\Big\langle\Psi\Big|\big[\HamH,\hat{R}\big]\Big|\Psi\Big\rangle=0
\end{equation}
showing the conservation of the observable property associated with the rotation \(\hat{R}\).

\subsection{Orbital angular momentum}\label{Section:AngularMops}

The property conserved for the above rotation, from the assumption of isotropic space, is \textit{angular momentum} . The quantum mechanical angular momentum operator, \(\hat{L}\), is analogous to its classical vector counterpart \cite{LANDAU60}
\begin{align}
 \hat{L}=\hat{r}\times\hat{p}=\hat{r}\times\frac{\hbar}{i}\nabla, && \nabla\triangleq\begin{pmatrix}
  \frac{\partial}{\partial x^{}} \\
  \frac{\partial}{\partial y^{}} \\
  \frac{\partial}{\partial z^{}} \\
 \end{pmatrix}
\end{align}
The Cartesian components of the angular momentum operator can be derived by considering infinitesimal rotations giving
\begin{align}
 i\hat{L}_x=\bigg( y\frac{\partial}{\partial z} - z\frac{\partial}{\partial y} \bigg),&&
 i\hat{L}_y=\bigg( z\frac{\partial}{\partial x} - x\frac{\partial}{\partial z} \bigg),&&
 i\hat{L}_z=\bigg( x\frac{\partial}{\partial y} - y\frac{\partial}{\partial x} \bigg)\label{eqn:Lcommrel}
\end{align}
where the constant unit of \(\hbar\) is dropped, as is normal in the subject of magnetic resonance. The solution to a differential equation for the infinitesimal rotation \(\varphi\) is
\begin{align}
&& \frac{\dd}{\dd\varphi}\hat{R}(\varphi)=-i\hat{L}_z^{}\hat{R} && \Rightarrow && \hat{R}(\varphi)=\e^{-i\hat{L}_z^{}\varphi}
\end{align} 
This can be extended to show all rotation operators are exponentials of the angular momentum operators, and the angular momentum operators are the basis of the Lie algebra corresponding to the Lie group of rotations. By inspection of \cref{eqn:Lcommrel}, the following commutation relations can be formed:
\begin{align}
 i\hat{L}_x=\big[\hat{L}_y^{},\hat{L}_z^{}\big], &&
 i\hat{L}_y=\big[\hat{L}_z^{},\hat{L}_x^{}\big], &&
 i\hat{L}_z=\big[\hat{L}_x^{},\hat{L}_y^{}\big]
\end{align}
where the commutator of two elements \(a\) and \(b\) defines a Lie algebra, with \([a,b]=ab-ba\). The square of the angular momentum, the modulus, can be constructed from the angular momentum operator components:
\begin{equation}
 \hat{L}_{}^2=\hat{L}_{x}^2+\hat{L}_{y}^2+\hat{L}_{z}^2
\end{equation}
and this operator, called the Casimir operator, does commute with each of the angular momentum operators
\begin{align}
 \big[\hat{L}_{}^{2},\hat{L}_x^{}\big]=0, &&
 \big[\hat{L}_{}^{2},\hat{L}_y^{}\big]=0, &&
 \big[\hat{L}_{}^{2},\hat{L}_z^{}\big]=0
\end{align}
A convenient set of non-Hermitian operators, named the raising and lowering operators, can be defined in terms of \(\hat{L}_x^{}\) and \(\hat{L}_y^{}\)
\begin{align}
 \hat{L}_{+}^{}= \hat{L}_{x}^{}+i\hat{L}_{y}^{}, &&
 \hat{L}_{-}^{}= \hat{L}_{x}^{}-i\hat{L}_{y}^{}
\end{align}
with the following commutation relations involving \(\hat{L}_z^{}\)
\begin{align}
 \big[\hat{L}_{+}^{},\hat{L}_{-}^{}\big]=2\hat{L}_z^{}, &&
 \big[\hat{L}_{z}^{},\hat{L}_{+}^{}\big]=\hat{L}_{+}^{}, &&
 \big[\hat{L}_{z}^{},\hat{L}_{-}^{}\big]=-\hat{L}_{-}^{}
\end{align}

\subsection{The uncertainty principle}\label{Section:Uncertainty}

Given an Hermitian operator $\hat{L}$, representing an observable property of the system in a state $\big|\Psi\big\rangle$; the uncertainty in a measurement of that observable property can be represented as a standard deviation \cite{WEYL27}:
\begin{equation}
 \Delta \hat{L}=\sqrt{\big\langle \hat{L}_{}^2\big\rangle-\big\langle \hat{L}\big\rangle_{}^{2}}=\sqrt{\big\langle\psi\big| \hat{L}_{}^2\big|\psi\big\rangle-\big\langle\psi\big| \hat{L}\big|\psi\big\rangle_{}^{2}}
\end{equation}
\textit{Heisenberg's uncertainty principle} \cite{HEISENBERG27} states that for two hermitian operators:
\begin{align}
 && \Delta \hat{A}\Delta \hat{B}\geqslant \frac{\hbar}{2}\bigg|\Big\langle\big[\hat{A},\hat{B}\big]\Big\rangle\bigg| && \big[\hat{A},\hat{B}\big]=ic
\end{align}
where \(c\) is a constant. This can be seen with angular momentum operators applied to the Heisenberg uncertainty relation in atomic units, where \(\hbar=1\):
\begin{align}
 \Delta \hat{L}_x^{}= & \sqrt{\big\langle \hat{L}_x^{2}\big\rangle-\big\langle \hat{L}_x^{}\big\rangle_{}^{2}}\nonumber\\
 \Delta \hat{L}_y^{}= & \sqrt{\big\langle \hat{L}_y^{2}\big\rangle-\big\langle \hat{L}_y^{}\big\rangle_{}^{2}}\nonumber\\
 \Rightarrow\quad\Delta \hat{L}_x^{}\Delta \hat{L}_y^{}\geqslant & \frac{1}{2}\bigg|\Big\langle\big[\hat{L}_x^{},\hat{L}_y^{}\big]\Big\rangle\bigg| =  \frac{1}{2}\Big|\big\langle\hat{L}_z^{}\big\rangle\Big|
\end{align}

The angular momentum operator components do no commute with each other. This indicates that all components of angular momentum cannot be simultaneously defined, unless all equal zero -- which sets it apart from linear momentum\footnote{The result obtained by considering infinitesimal translations and the homogeneity of space, would give relations for the linear momentum operator.}. However, the modulus of the angular momentum, the Casimir operator, can have a definite value at the same time as each component of the angular momentum and is therefore simultaneously observable.

\subsection{Spherical polar coordinates}\label{Section:SphericalCoordinates}

\begin{figure}
\centering{\includegraphics{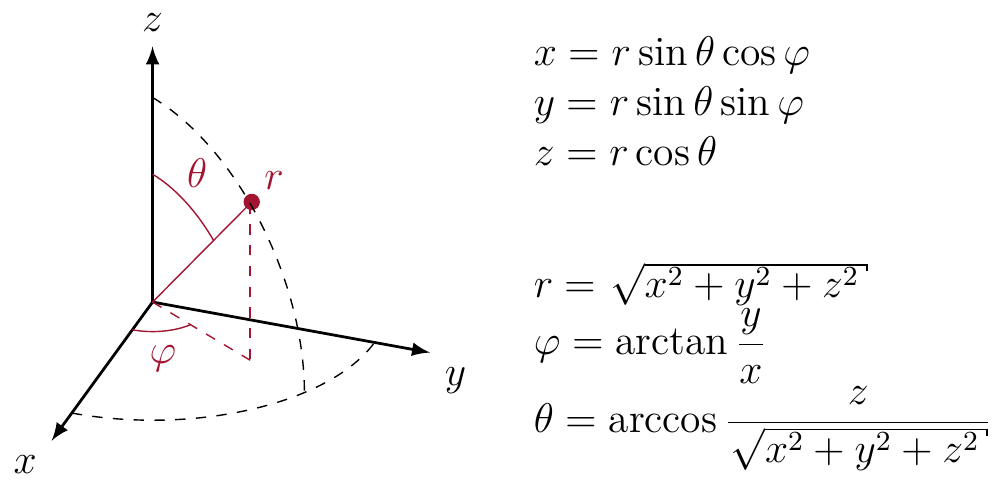}}
 \caption[Spherical coordinates]{Diagram of a spherical coordinate system in relation to a Cartesian coordinate system.}\label{Fig:SphericalCoordinates}
\end{figure}

The three Cartesian components of the angular momentum operators can be expressed in a spherical polar coordinate system, more natural than a Cartesian coordinate system when considering rotations from \cref{Section:Relativity}. Depicted in \cref{Fig:SphericalCoordinates}, their formulation is:
\begin{gather}
 \hat{L}_x^{} =  -i\bigg(-\sin{\varphi}\frac{\partial}{\partial\theta}-\cos{\varphi}\cot{\theta}\frac{\partial}{\partial\varphi}\bigg)\label{eqn:spc1}\\
 \hat{L}_y^{} =  -i\bigg(\cos{\varphi}\frac{\partial}{\partial\theta}-\sin{\varphi}\cot{\theta}\frac{\partial}{\partial\varphi}\bigg)\label{eqn:spc2}\\
 \hat{L}_z^{} =  -i\frac{\partial}{\partial\varphi}
\end{gather}
Similarly, the transform of the Casimir operator to spherical polar coordinates gives
\begin{equation}
 \hat{L}_{}^{2}=-\frac{1}{\sin^2{\theta}}\frac{\partial^2}{\partial \varphi^2}-\frac{1}{\sin{\theta}}\frac{\partial}{\partial\theta}\bigg(\sin{\theta}\frac{\partial}{\partial\theta} \bigg)
\end{equation}
The simultaneous eigenvalues of \(\hat{L}_{}^{2}\) and \(\hat{L}_z^{}\) are
\begin{align}
&& \begin{rcases}
 \hat{L}_{}^{2}\big|Y_{\!\ell}^{m_{}^{\vphantom{\prime}}}\!(\theta,\varphi)\big\rangle=\ell(\ell+1)\big|Y_{\!\ell}^{m_{}^{\vphantom{\prime}}}\!(\theta,\varphi)\big\rangle\\
 \hat{L}_z^{}\big|Y_{\!\ell}^{m_{}^{\vphantom{\prime}}}\!(\theta,\varphi)\big\rangle=m\big|Y_{\!\ell}^{m_{}^{\vphantom{\prime}}}\!(\theta,\varphi)\big\rangle
\end{rcases} &&\begin{array}{l} \ell\in\mathbb{N}\\ m=-\ell,-\ell+1,\dotsc, \ell\end{array}\label{eqn:eigspc1}
\end{align}
where the normalised, simultaneous eigenfunctions of the Casimir operator, \(\hat{L}_{}^{2}\), and the \(z-\)projection operator, \(\hat{L}_z^{}\), are the spherical harmonics
\begin{gather}
 Y_{\!\ell}^{m_{}^{\vphantom{\prime}}}\!(\theta,\varphi)=(-1)^{\!\frac{m+|m|}{2}}\!\sqrt{ \frac{2\ell+1}{4\pi}\frac{\big( \ell-\big|m\big| \big)!}{\big( \ell+\big|m\big| \big)!}}P_{\!\ell}^{m_{}^{\vphantom{\prime}}}\!\big( \cos{\theta} \big)\e^{im\varphi}\\
 P_{\!\ell}^{m_{}^{\vphantom{\prime}}}\!(\xi)=\big( 1-\xi^2 \big)^{\!\frac{|m|}{2}}\frac{\dd^{\:|m|}}{\dd \xi^{|m|}}P_{\!\ell}^{}(\xi)
\end{gather}
where \(P_{\!\ell}^{}(\xi)\) are the Legendre polynomials which can be defined by the formula
\begin{equation}
P_{\!\ell}^{}(\xi)=\frac{1}{2^\ell \ell!}\frac{{\dd}^{\:\ell}}{\dd \xi^\ell}\Big[\big( \xi^2-1 \big)^{\!\ell}\Big]
\end{equation}
\(\ell\) is the total momentum quantum number, \(m\) is the projection quantum number, \(\ell\in\mathbb{N}, m=-\ell,-\ell\!+\!1,\dotsc,\ell\).

Following \cref{eqn:spc1,eqn:spc2} the transform of raising and lowering operators to spherical polar coordinates are
\begin{align}
 \hat{L}_+^{}=&\e^{i\varphi}\big(\frac{\partial}{\partial\theta}+i\cot{\theta}\frac{\partial}{\partial\varphi}\big)\nonumber\\
 \Rightarrow \hat{L}_+^{}\big|Y_{\!\ell}^{m_{}^{\vphantom{\prime}}}\big\rangle= & \sqrt{\ell(\ell+1)-m(m+1)}\big|Y_{\!\ell}^{m+1_{}^{\vphantom{\prime}}}\big\rangle\label{eqn:eigspc2}\\
 \hat{L}_-^{}=&-\e^{i\varphi}\big(\frac{\partial}{\partial\theta}-i\cot{\theta}\frac{\partial}{\partial\varphi}\big)\nonumber\\
 \Rightarrow \hat{L}_-^{}\big|Y_{\!\ell}^{m_{}^{\vphantom{\prime}}}\big\rangle= & \sqrt{\ell(\ell+1)-m(m-1)}\big|Y_{\!\ell}^{m-1_{}^{\vphantom{\prime}}}\big\rangle\label{eqn:eigspc3}
\end{align}
The raising and lowering operators perform operations that increment and decrement the magnetic quantum number \(m\).

\subsection{Intrinsic angular momentum}\label{section:spin}

In dealing with the subject of magnetic resonance the idea of \textit{intrinsic angular momentum} (as opposed to \textit{orbital angular momentum} discussed above), termed \textit{spin}, must first be introduced \cite{UHLENBECK25,PAULI27}. The most direct evidence for this \textit{spin} is the Stern-Gerlach experiment, proposed by Otto Stern \cite{STERN21}: Silver atoms are sent in a focused beam through an non-uniform magnetic field, \(B_0^{}\) \cite{GERLACH22}. Classically, the atoms should give a continuous trace, between \(+\mu\) and \(-\mu\), when they emerge from the magnetic field. However, observation of the emerging atoms shows a number of distinct traces, rather than the expected continuous trace, spaced equally between \(+\mu\) and \(-\mu\). Conventionally, \(\mu\) is considered to be the magnetic moment of the particle. A diagram of the experiment is shown in \cref{Fig:SternGerlach}.

\begin{figure}
\centering{\includegraphics{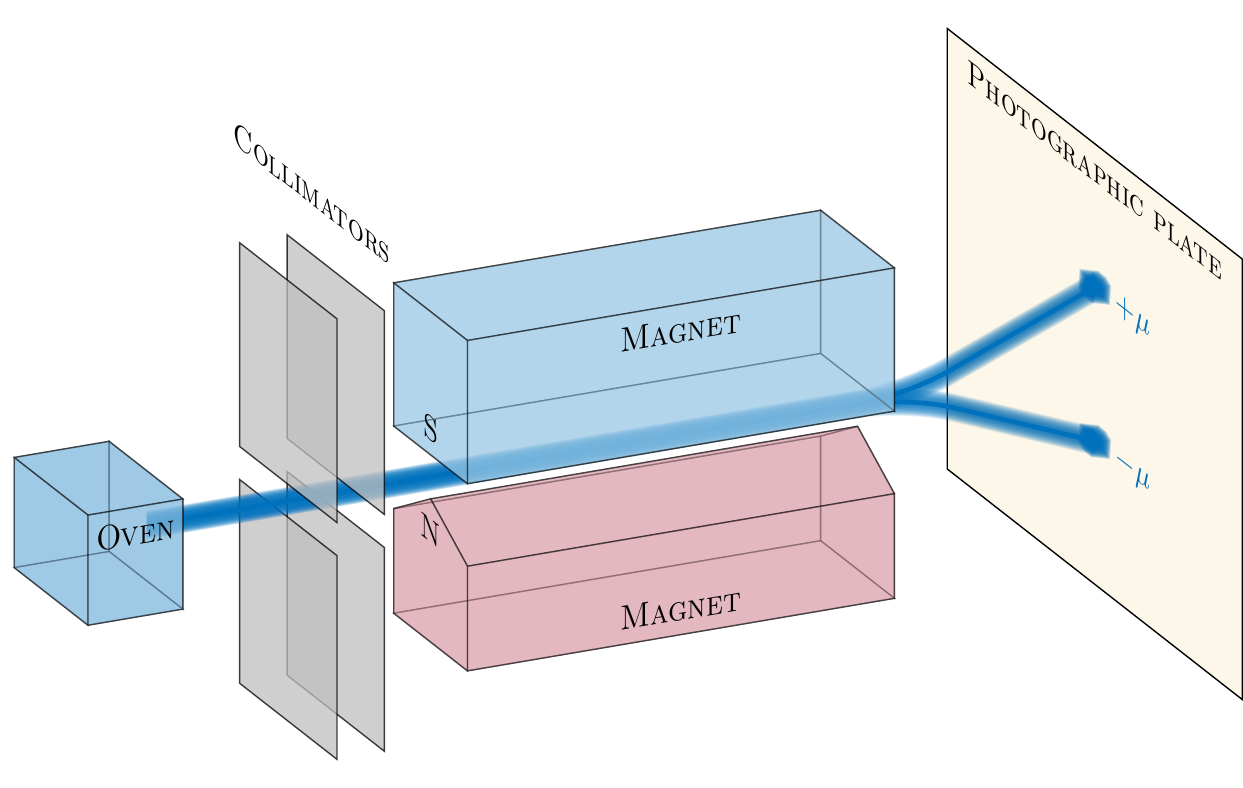}}
\caption[Stern-Gerlach experiment]{Schematic diagram of the Stern-Gerlach experiment showing a beam of silver atoms split by an inhomogeneous magnetic field. The classical picture of this experiment would show a continuous, rather than discrete, trace on the photographic plate.}\label{Fig:SternGerlach}
\end{figure}

This is attributed to the two spin states of atoms within the incident beam -- \textit{spin up} and \textit{spin down}. The ``up'' and ``down'' could also be termed \(+1\) and \(-1\), or \(\alpha\) and \(\beta\). The nomenclature is only to indicate that there is a quantised property of the silver atoms, having two states which become physically separated in the presence of an external magnetic field.

This spatial splitting of the wave functions describing the silver atoms must be included in addition to, and independent of (commuting with), the spatial coordinate description of the wave function. The discrete spin variable describing the splitting in the Stern-Gerlach can take two values\footnote{for a spin-\(\sfrac{1}{2}\) particle, which is the case for silver atoms}, \(+1\) or \(-1\). 

Spin degrees of freedom can be split from the spatial degrees of freedom, forming a \textit{spinor}:
\begin{equation}
 \big|\Psi(\vec{r},s)\big\rangle=\sum\limits_{n,k}\big|\Psi_{\!n}^{}(\vec{r}\,)\big\rangle\otimes\big|\Psi_{\!k}^{}(s)\big\rangle
\end{equation}
The spin part of the this construct can be described by a pair of orthogonal vectors projecting onto the two discrete values of the spin,
\begin{equation}
 \chi=\begin{pmatrix}c_{\sfrac{+1}{2}}^{} \\ c_{\sfrac{-1}{2}}^{}\end{pmatrix}=c_{\sfrac{+1}{2}}^{}\begin{pmatrix}1 \\ 0\end{pmatrix} + c_{\sfrac{-1}{2}}^{}\begin{pmatrix}0 \\ 1\end{pmatrix}=c_{\sfrac{+1}{2}}^{}\big|\alpha\big\rangle + c_{\sfrac{-1}{2}}^{}\big|\beta\big\rangle
\end{equation}
which must be normalised, requiring \(\chi^{\dagger}\chi=|c_{\sfrac{+1}{2}}^{}|^{2}+|c_{\sfrac{-1}{2}}^{}|^{2}=1\). In the case of a spin-\(\sfrac{1}{2}\) particle, this basis can be used for describing the two spin eigenstates of the Stern-Gerlach experiment. Without derivation, we make the assertion that the spin quantum number can have \(2s+1\) values, and for a spin-\(\sfrac{1}{2}\) particle these values are \(s=\pm\sfrac{1}{2}\). This imposes the condition that \(s=0,\frac{1}{2},1,\frac{3}{2},2,\dots\).

In analogy to orbital angular momentum, for intrinsic angular momentum a matrix representation can be formulated with raising and lowering operators to move between the \(|\alpha\rangle\) and \(|\beta\rangle\) states
\begin{align}
 \hat{S}_+^{}\big|\alpha\big\rangle=&\big|\bm{0}\big\rangle,&&
 \hat{S}_+^{}\big|\beta\big\rangle=\big|\alpha\big\rangle, &&
 \Rightarrow && 
 \hat{S}_+^{}=\begin{bmatrix}0 & 1\\ 0 & 0\end{bmatrix}\\
 \hat{S}_-^{}\big|\alpha\big\rangle=&\big|\beta\big\rangle,&&
 \hat{S}_-^{}\big|\beta\big\rangle=\big|\bm{0}\big\rangle, &&
 \Rightarrow && 
 \hat{S}_-^{}=\begin{bmatrix}0 & 0\\ 1 & 0\end{bmatrix}
\end{align}
and with analogy to orbital angular momentum, the raising and lowering operators can be formulated in terms of \(\hat{S}_x^{}\) and \(\hat{S}_y^{}\):
\begin{align}
 \hat{S}_{+}^{}=& \hat{S}_{x}^{}+i\hat{S}_{y}^{}, &&
 \hat{S}_{-}^{}= \hat{S}_{x}^{}-i\hat{S}_{y}^{}
\end{align}
The Cartesian operators can be constructed from the raising and lowering operators, and by requiring the eigenvalues of the \(z\)-component to be the allowed values of spin \(\pm\sfrac{1}{2}\), giving
\begin{align}
 \hat{S}_x^{}=&\frac{\hat{S}_+^{}+\hat{S}_-^{}}{2}, &&
 \hat{S}_y^{}=\frac{\hat{S}_+^{}-\hat{S}_-^{}}{2i}, &&
 \begin{matrix}\hat{S}_z^{}\big|\alpha\big\rangle=+\frac{1}{2}\big|\alpha\big\rangle\\\hat{S}_z^{}\big|\beta\big\rangle=-\frac{1}{2}\big|\beta\big\rangle\end{matrix}\nonumber\\
 \hat{S}_x^{}=&\frac{1}{2}\begin{bmatrix}0 & 1\\ 1 & 0\end{bmatrix}, &&
 \hat{S}_y^{}=\frac{1}{2}\begin{bmatrix}0 & -i\\ i & 0\end{bmatrix}, &&
 \hat{S}_z^{}=\frac{1}{2}\begin{bmatrix}1 & 0\\ 0 & -1\end{bmatrix}
\end{align}
These are called the Pauli matrices of a spin-\(\sfrac{1}{2}\) particle. It is easy to see that the square of each of these operators is the same and proportional to the unit matrix, \(\Unit\). Unlike orbital angular momentum, the square of spin is also proportional to the identity
\begin{equation}
 \hat{S}_{}^{2}=\hat{S}_x^{2}+\hat{S}_y^{2}+\hat{S}_z^{2}=\frac{1}{4}\Unit+\frac{1}{4}\Unit+\frac{1}{4}\Unit=\frac{3}{4}\Unit
\end{equation}

For spin greater than \(\sfrac{1}{2}\), the matrix dimension becomes larger, and similar operators can be constructed e.g. for spin-\(1\) particles:
\begin{align}
 \hat{S}_x^{}=&\frac{1}{\sqrt{2}}\begin{bmatrix}0 & 1 & 0\\ 1 & 0 & 1\\ 0 & 1 & 0\end{bmatrix}, &&
 \hat{S}_y^{}=\frac{1}{\sqrt{2}}\begin{bmatrix}0 & -i & 0\\ i & 0 & -i\\ 0 & i & 0\end{bmatrix}, &&
 \hat{S}_z^{}=\frac{1}{2}\begin{bmatrix}1 & 0 & 0\\ 0 & 0 & 0\\ 0 & 0 & -1\end{bmatrix}
\end{align}
The commutation relations follow by inspection:
\begin{align}
 && \Big[\hat{S}_x^{},\hat{S}_y^{}\Big]=i\hat{S}_z^{} &&\Big[\hat{S}_z^{},\hat{S}_x^{}\Big]=i\hat{S}_y^{} && \Big[\hat{S}_y^{},\hat{S}_z^{}\Big]=i\hat{S}_x^{} && \Big[\hat{S}_{}^2,\hat{S}_{\{x,y,z\}}^{}\Big]=0 &&
\end{align} 

\section[Irreducible representations of the rotation group]{Irreducible representations\linebreak of the rotation group}\label{Section:IrreducibleRep}

Returning to the \textit{Euclidean principle of relativity}, space is required to be isotropic and homogeneous. The consequence is that when a state vector \(\big|\Psi\big\rangle\) undergoes a rotation operation in space, \(\hat{R}\), all probabilities must be invariant. Mathematically, this means that all inner products of two rotated states remain invariant
\begin{gather}
 \big|\Psi\big\rangle\longmapsto \hat{R}\big|\Psi\big\rangle=\big|\Psi^{\prime}\big\rangle\\
 \Big|\big\langle\Psi^{\prime}\big|\Phi^{\prime}\big\rangle\Big|_{}^{2}= \Big|\big\langle\Psi\big|\Phi\big\rangle\Big|_{}^{2}
\end{gather}
for every pair of state vectors, \(\Psi\) and \(\Phi\). This mapping is called \textit{isometry} and must be reversible. A second rotation results only in a phase change \cite{MERZBACHER61}:
\begin{equation}
 \big|\Psi^{\prime\prime}\big\rangle=\e^{i\varphi}\big|\Psi^{\prime}\big\rangle\label{eqn:rotation}
\end{equation}
The transform of a sum of two vectors is equal to the sum of the transforms of two vectors. It follows that \cite{MERZBACHER61}
\begin{align}
 \big\langle\Psi_a^{\prime\prime}\big|\Psi_b^{\prime\prime}\big\rangle+\Big(\big\langle\Psi_a^{\prime\prime}\big|\Psi_b^{\prime\prime}\big\rangle\Big)^{\ast} & = \big\langle\Psi_a^{}\big|\Psi_b^{}\big\rangle+\Big(\big\langle\Psi_a^{}\big|\Psi_b^{}\big\rangle\Big)^{\ast}\label{eqn:sumvectors}\\
 \begin{split}
 \Re\big\langle\Psi_a^{\prime\prime}\big|\Psi_b^{\prime\prime}\big\rangle & = \Re\big\langle\Psi_a^{}\big|\Psi_b^{}\big\rangle\\
 \Im\big\langle\Psi_a^{\prime\prime}\big|\Psi_b^{\prime\prime}\big\rangle & = \pm\Im\big\langle\Psi_a^{}\big|\Psi_b^{}\big\rangle\end{split}\\
 \Rightarrow \Big(\lambda\big|\Psi\big\rangle\Big)^{\prime\prime} & =\lambda\big|\Psi^{\prime\prime}\big\rangle\label{eqn:condunitary}\\
  \Big(\lambda\big|\Psi\big\rangle\Big)^{\prime\prime} & =\lambda^{\ast}\big|\Psi^{\prime\prime}\big\rangle\label{eqn:antilinear}
\end{align}
where \(_{}^{\ast}\) has been used here to denote the complex conjugate. \cref{eqn:sumvectors,eqn:condunitary} are the properties of a linear operator, and \cref{eqn:condunitary} implies the transform is unitary. A unitary transformation operator \(\hat{\mathcal{U}}\) satisfies \(\hat{\mathcal{U}}_{}^{\dagger}\hat{\mathcal{U}} = \hat{\mathcal{U}}\hat{\mathcal{U}}_{}^{\dagger} = \Unit\) and gives a symmetry operation\footnote{A transformation that leaves the physics unaltered is called a \textit{symmetry operation}, and a symmetry operation applied to all eigenstates of a basis must leave the Schr\"{o}dinger equation invariant. This condition is satisfied if the system Hamiltonian, \(\HamH\), commutes with the transform: \([\HamH,\hat{\mathcal{U}}_R^{}]=0\)}, an important property of a rotation. \cref{eqn:antilinear} characterises an antilinear operator, not considered in generating rotation operations, which is important in describing a system under \textit{time reversal}.

Spin does not rotate when a particle is rotated, but anisotropic interactions (see \cref{Section:SpinHam}) do rotate. To represent these rotations, the following complex Hermitian matrices will become generators of the rotations
\begin{align}
 && \hat{J}_1^{}\triangleq\begin{bmatrix}0&0&0\\0&0&i\\0&-i&0\end{bmatrix}, &&
  \hat{J}_2^{}\triangleq\begin{bmatrix}0&0&i\\0&0&0\\-i&0&0\end{bmatrix}, &&
   \hat{J}_3^{}\triangleq\begin{bmatrix}0&i&0\\-i&0&0\\0&0&0\end{bmatrix} &&\label{eqn:ROTgenerators}
\end{align}
A rotation from a state \(|\Psi\rangle\) to \(|\Psi^{\prime}\rangle\) can be expressed as a unitary operator, \(|\Psi\rangle=\hat{\mathcal{U}}_R^{}|\Psi^{\prime}\rangle\). From the generators in \cref{eqn:ROTgenerators} and the rotation in \cref{eqn:rotation}, the unitary rotation operators representing rotations through angles \(\{\alpha,\beta,\gamma\}\) are
\begin{align}
\e^{i\hat{J}_1^{}\alpha}=\begin{bmatrix}1&0&0\\0&\cos{\alpha}&-\sin{\alpha}\\0&\sin{\alpha}&\cos{\alpha}\end{bmatrix}\nonumber \\
\e^{i\hat{J}_2^{}\beta}=\begin{bmatrix}\cos{\beta}&0&-\sin{\beta}\\0&1&0\\\sin{\beta}&0&\cos{\beta}\end{bmatrix}\label{eqn:unitaryop}\\ \e^{i\hat{J}_3^{}\gamma}=\begin{bmatrix}\cos{\gamma}&-\sin{\gamma}&0\\\sin{\gamma}&\cos{\gamma}&0\\0&0&1\end{bmatrix}\nonumber
\end{align}
\(\hat{J}\) must satisfy the general commutation relations to give a well-defined transform:
\begin{align}
 \big[\hat{J}_i^{},\hat{J}_j^{}\big]=i\varepsilon_{ijk}^{}\hat{J}_k^{}, && \varepsilon_{ijk}^{}=\begin{cases}1 & ijk\in\big\{123,231,312\big\}\\-1 & ijk\in\big\{321,213,132\big\}\\0  & \text{all else}\end{cases}
\end{align}
where \(\hat{J}\) is a rotation operator and \(i,j,k\) take the values \(1,2,3\) and represent the Cartesian components \(x,y,z\). Further to the unitary operator representing a rotation in \cref{eqn:unitaryop}, we also require the same rotation at further \(2\pi\) angles
\begin{equation}
 \hat{\mathcal{U}}_R^{}=\exp{\big[-i\hat{J}\big(\varphi+2\pi k\big)\big]}=\exp{\big[-i\hat{J}\varphi\big]}\exp{\big[-i\hat{J}2\pi k\big]}\label{eqn:unitaryop2}
\end{equation}
where \(k\) is an arbitrary integer. The additional factor in \cref{eqn:unitaryop2} has the effect of multiplying an eigenstate \(Y_{\ell}^{m}\) by \((-1)_{}^{2km}\), where \(m\) is the magnetic quantum number and the eigenvalue of an operator \(\hat{J}_z^{}\).

Any rotation \(\hat{R}\) operation in 3D can be described by three \textit{Euler angles}\footnote{other rotation conventions exist, such are quaternions or directional cosine matrices, but Euler angles are shown here for convenience.}, shown in \cref{fig:euler_angles}, which represent three consecutive rotations by angles \(\alpha,\beta,\gamma\) \cite{ARFKEN99},
\begin{equation}
 \hat{R}(\alpha,\beta,\gamma)= \begin{bmatrix}\cos{\gamma}&-\sin{\gamma}&0\\\sin{\gamma}&\cos{\gamma}&0\\0&0&1\end{bmatrix} \begin{bmatrix}\cos{\beta}&0&-\sin{\beta}\\0&1&0\\\sin{\beta}&0&\cos{\beta}\end{bmatrix} \begin{bmatrix}\cos{\alpha}&-\sin{\alpha}&0\\\sin{\alpha}&\cos{\alpha}&0\\0&0&1\end{bmatrix}\label{eqn:eulerangles}
\end{equation}

\begin{figure}
\centering{\includegraphics{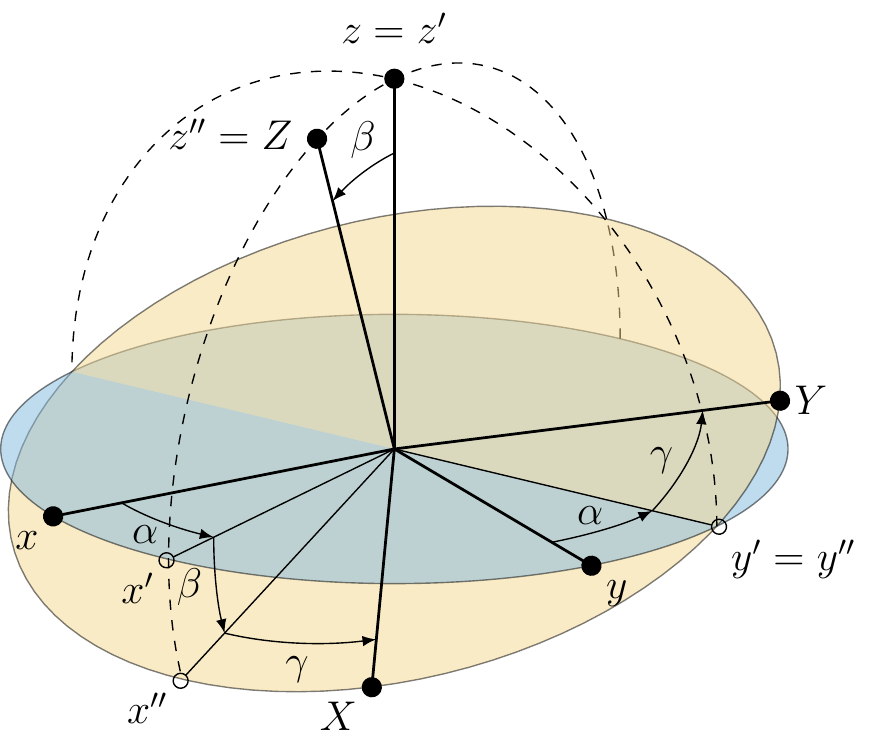}}
\caption[Euler angles]{Euler angles: This rotation convention corresponds to \circled{1} rotation of angle \(\alpha\) around the \(z\)-axis, \circled{2} rotation of angle \(\beta\) around the \(y_{}^{\prime}\)-axis, \circled{3} rotation of angle \(\gamma\) around the \(z_{}^{\prime\prime}\)-axis.}\label{fig:euler_angles}
\end{figure}

\section{Liouville-von Neumann equation}\label{Section:LvN}

Magnetic resonance is concerned with ensemble averages of spin particles. Simply describing the basis in a Hilbert space of \cref{Section:Schrodinger} is not enough, and a more descriptive, larger space must be constructed.

Given the initial state of the system, \(|\psi(0)\rangle\), the solution to this \cref{eqn:schrodinger} at a future time \(T>0\) is
\begin{equation}
 \big|\psi(T)\big\rangle = \exp_{(\text{o})}{\!\!\Bigg[\!-i\!\!\int\limits_0^{T}\!\!\HamH(t)\dd t\Bigg]}\big|\psi(0)\big\rangle\label{eqn:timedependentsolutions}
\end{equation}
where \(\exp_{(\text{o})}{\![\dots]}\) indicates a time ordered exponential \cite{DYSON49, DYSON49b,ERNST87,TANNOR07}. 

The density operator is a description of the state of a system ensemble
\begin{align}
 \hat{\rho}(t) & =\sum\limits_{i,j} p_{ij}^{}\big|\psi_i^{}(t)\big\rangle\!\big\langle\psi_{\!j}^{}(t)\big|\\
 & = \big|\psi_{}^{}(t)\big\rangle\!\big\langle\psi_{}^{}(t)\big|
\end{align}
where \(p_{ii}^{}\) are the associated probabilities for each state \(\psi_i^{}(t)\), normalised so that \(\sum\limits_ip_{ii}=1\). The diagonal elements of \(\hat{\rho}(t)\) are the probabilities of being in an eigenstate of the Hamiltonian, and the off-diagonal elements of \(\hat{\rho}(t)\) indicate a superposition of states \cite{ERNST87}.

The equation of motion for the density operator can be found from the time-dependent Schr\"{o}dinger equation:
\begin{align}
 \frac{\partial}{\partial t}\hat{\rho}(t)=&\frac{\partial}{\partial t}\Big(\big|\psi(t)\big\rangle\!\big\langle\psi(t)\big|\Big) = \bigg( \frac{\partial}{\partial t}\big|\psi(t)\big\rangle\bigg)\!\big|\psi(t)\big\rangle\!\big\langle\psi(t)\big|\!\bigg( \frac{\partial}{\partial t}\big\langle\psi(t)\big|\bigg)\nonumber\\
 =&-i\HamH(t)\big|\psi(t)\big\rangle\!\big\langle\psi(t)\big|+i\big|\psi(t)\big\rangle\!\big\langle\psi(t)\big|\HamH(t)\nonumber\\
 =&-i\Big(\HamH(t)\hat{\rho}(t)-\hat{\rho}(t)\HamH(t)\Big)\nonumber
\end{align}
Identifying the commutator, the equation of motion can be written as
\begin{equation}
 \frac{\partial}{\partial t}\hat{\rho}(t)=-i\Big[\HamH(t),\hat{\rho}(t)\Big]\label{eqn:liouvillvonneumann}
\end{equation}
This is known as the Liouville-von Neumann equation. In a similar manner to finding the solutions of \cref{eqn:schrodinger} with \cref{eqn:timedependentsolutions}, the solutions to \cref{eqn:liouvillvonneumann} are
\begin{equation}
 \hat{\rho}(T)=\hat{\mathcal{U}}(T)\hat{\rho}(0)\hat{\mathcal{U}}_{}^{-1}(T)\label{eqn:densityoperatorprop}
\end{equation}
where \(\hat{\mathcal{U}}(T)\) is the time propagator over a time interval $[0,T]$ defined by
\begin{equation}
 \hat{\mathcal{U}}(T)=\exp_{(\text{o})}\!\!\Bigg[\!-i\!\!\int\limits_0^{T}\!\!\HamH(t)\dd t\Bigg]\label{eqn:propagator}
\end{equation}

\subsection{Superoperator algebra}\label{Section:Superoperator}

The transform in \cref{eqn:densityoperatorprop} can be represented as the multiplication of three \(n\times n\) matrices in the following schematic:
\begin{equation}
 \scalemath{0.75}{\begin{bmatrix}
 \rho_{11} & \rho_{12} & \cdots & \rho_{1n} \\
 \rho_{21} & \rho_{22} & \cdots & \rho_{2n} \\
 \vdots & \vdots & \ddots\vphantom{\Bigg|} & \vdots \\
 \rho_{n1} & \rho_{n2} & \cdots & \rho_{nn}
 \end{bmatrix}}
=
 \scalemath{0.75}{\begin{bmatrix}
  & \hphantom{\text{\hspace{5em}}} &  \\
  & &  \\
  & \hat{\mathcal{U}}_{} &  \\
  & \textsc{size: }(n \times n) &  \\
  & &  
 \end{bmatrix}}
  \scalemath{0.75}{\begin{bmatrix}
 \sigma_{11} & \sigma_{12} & \cdots & \sigma_{1n} \\
 \sigma_{21} & \sigma_{22} & \cdots & \sigma_{2n} \\
 \vdots & \vdots & \ddots\vphantom{\Bigg|} & \vdots \\
 \sigma_{n1} & \sigma_{n2} & \cdots & \sigma_{nn}
 \end{bmatrix}}
  \scalemath{0.75}{\begin{bmatrix}
  & \hphantom{\text{\hspace{5em}}} &  \\
  & &  \\
  & \hat{\mathcal{U}}_{}^{-1} &  \\
  & \textsc{size: }(n \times n) &  \\
  & &  
 \end{bmatrix}}
\end{equation}
The same transform in \cref{eqn:densityoperatorprop} can also be written as a matrix-vector multiplication, numerically more appealing than the product of three matrices, by using the linear transform of density matrix vectorisation:
\begin{equation}
 \scalemath{0.75}{\begin{pmatrix}
 \rho_{11} \\
 \rho_{21} \\
 \vdots \\
 \rho_{n1} \\ \hdashline[1pt/1.5pt]
 \rho_{12} \\
 \rho_{22} \\
 \vdots \\
 \rho_{n2} \\ \hdashline[1pt/1.5pt]
 \\
 \vdots \\
 \\ \hdashline[1pt/1.5pt]
 \rho_{1n} \\
 \rho_{2n}\\
 \vdots \\
 \rho_{nn} \\
 \end{pmatrix}}
=
 \scalemath{0.75}{\begin{bmatrix}
  \hphantom{\text{\hspace{25em}}}\\
  \\
  \\
  \\
  \\
  \\
  \\
  \hhat{\mathcal{U}}_{}\\\vspace{-0.5em}
  \\
  \textsc{size: }(n^2 \times n^2) \\
  \\
  \\
  \\
  \\
  \\
  \\
 \end{bmatrix}}
 \scalemath{0.75}{\begin{pmatrix}
 \sigma_{11} \\
 \sigma_{21} \\
 \vdots \\
 \sigma_{n1} \\ \hdashline[1pt/1.5pt]
 \sigma_{12} \\
 \sigma_{22} \\
 \vdots \\
 \sigma_{n2} \\ \hdashline[1pt/1.5pt]
 \\
 \vdots \\
 \\ \hdashline[1pt/1.5pt]
 \sigma_{1n} \\
 \sigma_{2n}\\
 \vdots \\
 \sigma_{nn} \\
 \end{pmatrix}}
\end{equation}

The commutation relation in \cref{eqn:liouvillvonneumann} can be represented as a matrix in itself, called a superoperator, and transforms the Liouville-von Neumann equation into a matrix-vector multiplication
\begin{equation}
 \frac{\partial}{\partial t}\hat{\rho}=-i\HamHH\hat{\rho}\label{eqn:liouvillvonneumannsuperoperator}
\end{equation}
where the ``double-hat'' superoperator notation has been introduced, with the Liouville superoperator acting on another operator defined as a commutation \(\HamHH\hat{\rho}=[\HamH,\hat{\rho}]\), and is represented by a matrix of size \(n^2\times n^2\). Using this superoperator representation \cref{eqn:densityoperatorprop} becomes
\begin{align}
 && \big|{\rho}(T)\big\rangle=\exp_{(\text{o})}\!\!\Bigg[\!-i\!\!\int\limits_0^{T}\!\!\HamHH(t)\dd t\Bigg]\big|{\rho}(0)\big\rangle && \HamHH=\Unit\otimes\HamH -  \HamH_{}^{\dagger}\otimes\Unit + i\hhat{\mathcal{R}}\label{eqn:liouvillvonneumann_rho}
\end{align}
where \(\hhat{\mathcal{R}}\) is the relaxation superoperator, describing damping and dissipations effects, and \(\Unit\) is the identity matrix\footnote{The identity matrix, \(\Unit\), is a matrix with ones on every diagonal element and zeros everywhere else.}. The Liouville superoperator defined in this way, as a direct product, allows interacting composite systems described in a numerically efficient way.

Continuing from the spherical harmonics of \cref{Section:SphericalCoordinates}, the vector space of \cref{eqn:eigspc1,eqn:eigspc2,eqn:eigspc3} can be decomposed to \((2\ell+1)\)-dimensional subspaces, which are invariant under rotation. Under a rotation, a state in the spherical polar coordinate basis transforms as
\begin{equation}
 \hat{R}\big|Y_{\ell}^{m_{}^{\vphantom{\prime}}}\big\rangle=\sum\limits_{m^{\prime}=-\ell}^{\ell}\big|Y_{\ell}^{m_{}^{\prime}}\big\rangle\mathfrak{D}_{m_{}^{\prime}m}^{(\ell)}(\alpha,\beta,\gamma)
\end{equation}
where the matrix \(\mathfrak{D}_{m_{}^{\prime}m}^{(\ell)}(\alpha,\beta,\gamma)\) is the \textit{Wigner rotation matrix} (\cref{TAB:Wigner}) with the elements described by
\begin{equation}
 \mathfrak{D}_{m_{}^{\prime}m}^{(\ell)}(\alpha,\beta,\gamma)=\big\langle Y_{\ell}^{m_{}^{\prime}}\big|\hat{R}\big| Y_{\ell}^{m}\big\rangle
\end{equation}
shown explicitly in \cref{Chapter:AppendixA}. A rotation in one of these subspaces can be expressed by an \textit{irreducible spherical tensor operator} \(\hat{T}_{\ell m}^{}\) in a matrix representation, which is defined as \cite{SANCTUARY76}
\begin{equation}
 \hat{R}\hat{T}_{\ell m}^{}\hat{R}_{}^{-1}=\hhat{R}\hat{T}_{\ell m}^{}=\sum\limits_{m_{}^{\prime}=-\ell}^{\ell}\hat{T}_{\ell m_{}^{\prime}}\mathfrak{D}_{m_{}^{\prime}m}^{(\ell)}(\alpha,\beta,\gamma)\label{eqn:wigner_rot1}
\end{equation}
where \(\alpha,\beta,\gamma\) are Euler angles in \cref{eqn:eulerangles}. For the infinitesimal rotations used in \cref{Section:AngularMops,Section:Relativity}, \(\hat{J}\), \cref{eqn:wigner_rot1} simplifies to
\begin{equation}
 \big[\hat{J},\hat{T}_{\ell m}^{}\big]=\sum\limits_{m_{}^{\prime}=-\ell}^{\ell}\hat{T}_{\ell m_{}^{\prime}}^{}\big\langle Y_{\ell}^{m_{}^{\prime}}\big|\hat{J} \big|  Y_{\ell}^{m_{}^{}}\big\rangle
\end{equation}
giving an alternate form of \cref{eqn:eigspc1,eqn:eigspc2,eqn:eigspc3} as commutation relations for the spherical tensor operator:
\begin{gather}
 \big[\hat{J}_z^{},\hat{T}_{\ell m}^{}\big]=  m\hat{T}_{\ell m}^{}\\
 \big[\hat{J}_+^{},\hat{T}_{\ell m}^{}\big]=  \sqrt{\ell(\ell+1)-m(m+1)}\,\hat{T}_{\ell m+1}^{}\\
 \big[\hat{J}_-^{},\hat{T}_{\ell m}^{}\big]=  \sqrt{\ell(\ell+1)-m(m-1)}\,\hat{T}_{\ell m-1}^{}
\end{gather}
The single particle spherical tensor operators of rank-1 are
\begin{gather}
 \hat{T}_{0 0}^{} =  \Unit\\
 \hat{T}_{1 0}^{} =  \hat{L}_z^{}\label{eqn:T10}\\
 \hat{T}_{1 \pm1}^{} =  \mp \frac{1}{\sqrt{2}}\hat{L}_\pm^{}
\end{gather}
The full list of irreducible spherical tensor operators is given in \cref{Chapter:AppendixA} \cref{TAB:ISTO}. Within the software toolbox \textit{Spinach} \cite{HOGBEN11}, all rotations are active, meaning the system is rotated and not the coordinates.

\subsection{Composite systems}\label{Section:Composite}

The total wavefunction of two non-interacting uncorrelated subsystems is a \textit{direct product} of the subsystems\footnote{Also called the Kronecker product.}:
\begin{equation}
 \big|\psi_\textsc{1,2}^{}\big\rangle=\big|\psi_\textsc{1}^{}\big\rangle\otimes\big|\psi_\textsc{2}^{}\big\rangle
\end{equation} 
and the density matrices are
\begin{equation}
 \hat{\rho}_\textsc{1,2}^{}=\big|\psi_\textsc{1,2}^{}\big\rangle\!\big\langle\psi_\textsc{1,2}^{}\big|=\big(\big|\psi_\textsc{1}^{}\big\rangle\!\big\langle\psi_\textsc{1}^{}\big|\big)\!\otimes\!\big(\big|\psi_\textsc{2}^{}\big\rangle\!\big\langle\psi_\textsc{2}^{}\big|\big)=\hat{\rho}_\textsc{2}^{}\otimes\hat{\rho}_\textsc{2}^{}
\end{equation}
For a system of \(k\) non-interacting uncorrelated subsystems, the system is described by a chain of \(k\) direct products:
\begin{equation}
 \hat{\rho}_{1,2,\dots k}^{}=\hat{\rho}_\textsc{2}^{}\otimes\hat{\rho}_\textsc{2}^{}\otimes\cdots\otimes\hat{\rho}_\textsc{k}^{}
\end{equation}

Similarly, when the angular momentum of a composite system is represented by a direct product space, so to are the normalised simultaneous eigenvectors:
\begin{equation}
 \big|Y_{\ell_{1}^{},\ell_{2}^{}}^{m_{1}^{},m_{2}^{}}\big\rangle=\big|Y_{\ell_1^{}}^{m_{1}^{}}\big\rangle\otimes \big|Y_{\ell_2^{}}^{m_{2}^{}}\big\rangle
\end{equation}
and the state for \(n\) particles can be expressed with \(n\) direct products
\begin{equation}
 \big|Y_{\ell_{1}^{},\ell_{2}^{},\dots,\ell_{n}^{}}^{m_{1}^{},m_{2}^{},\dots,m_{n}^{}}\big\rangle=\big|Y_{\ell_1^{}}^{m_{1}^{}}\big\rangle\otimes \big|Y_{\ell_2^{}}^{m_{2}^{}}\big\rangle\otimes\cdots\otimes \big|Y_{\ell_n^{}}^{m_{n}^{}}\big\rangle
\end{equation}
which form the basis of this direct product space.

The operators for a multi-particle system are direct products, where the operator of the \(k^\text{th}\) subsystem occurs at the \(k^\text{th}\) position in the direct product chain e.g.
\begin{gather}
 \hat{S}_x^{(k)} =  \Unit\!\otimes\!\Unit\!\otimes\!\dotsm\!\otimes\!\hat{S}_x^{}\!\otimes\!\dotsm\!\otimes\!\Unit\nonumber\\
 \hat{S}_y^{(k)} =  \Unit\!\otimes\!\Unit\!\otimes\!\dotsm\!\otimes\!\hat{S}_y^{}\!\otimes\!\dotsm\!\otimes\!\Unit\\
 \hat{S}_z^{(k)} =  \Unit\!\otimes\!\Unit\!\otimes\!\dotsm\!\otimes\!\hat{S}_z^{}\!\otimes\!\dotsm\!\otimes\!\Unit\nonumber
\end{gather} 
where \(\hat{S}_{\{x,y,z\}}^{(k)}\) is the operator representing an \(x-\),\(y-\), or \(z-\) projection of spin on the \(k^\text{th}\) particle in the system of particles. 

The Hamiltonians of two non-interacting uncorrelated subsystems must not allow the subsystems to affect each other, giving a composite Hamiltonian of two subsystems as
\begin{equation}
 \HamH_{1,2}^{}=\HamH_1^{}\!\otimes\!\Unit_2^{}+\Unit_1^{}\!\otimes\!\HamH_2^{}
\end{equation} 
where \(\Unit\) is the identity operator.

In extending this description of spherical tensor operators to coupled particles, \textit{Clebsch-Gordan coefficients} can be used to couple two tensors of a given rank to another rank \cite{ARFKEN99}. The \textit{Clebsch-Gordan coefficients} can be interpreted as a unitary transformation from a linear combination of products of single particle operators to two particle operators \cite{ERNST87}. Two particle tensor operators, \(\hat{T}_{\ell m}^{(12)}\), from the single particle operators \(\hat{T}_{\ell_1^{} m_1^{}}^{(1)}\) and \(\hat{T}_{\ell_2^{} m_2^{}}^{(2)}\) defined in this way are
\begin{equation}
 \hat{T}_{\ell m}^{(12)}=\sum\limits_{m_1} \mathcal{C}_{m_1^{}m_2^{}m}^{\ell_1^{}\ell_2^{}\ell}\hat{T}_{\ell_1^{} m_1^{}}^{(1)}\hat{T}_{\ell_2^{} m-m_1^{}}^{(2)}
\end{equation}
where \(\mathcal{C}_{m_1^{}m_2^{}m}^{\ell_1^{}\ell_2^{}\ell}\) are \textit{Clebsch-Gordan coefficients}. The two particle spherical tensor operators of rank-2 are
\begin{gather}
\hat{T}_{2\pm 2}^{(2)}=\frac{1}{2}\hat{L}_{\pm}^2\label{eqn:T22}\\
\hat{T}_{2\pm 1}^{(2)}=\mp\frac{1}{2}\big(\hat{L}_z^{}\hat{L}_{\pm}^{}+\hat{L}_{\pm}^{}\hat{L}_z^{}\big)\\
\hat{T}_{20}^{(2)}=\sqrt{\frac{2}{3}}\big(\hat{L}_z^2-\frac{1}{4}\big(\hat{L}_+^{}\hat{L}_-^{}+\hat{L}_-^{}\hat{L}_+^{}\big)\big)
\end{gather}
The full list of irreducible spherical tensor operators is given in \cref{Chapter:AppendixA} \cref{TAB:SPericalTensor}.

\section{The Spin Hamiltonian}\label{Section:SpinHam}

Spin Hamiltonians are subject to three types of interactions, and can only have three generic algebraic forms \cite{ERNST87}:
\begin{description}
\item[Linear in the spin operators]\hfill\\\(\longrightarrow\) Interactions with the static magnetic field, \(\B{0}\), and the irradiated radio/microwave frequency \(\B{1}(t)\) field.
\item[Bilinear in the spin operators]\hfill\\\(\longrightarrow\) Interactions between two spins.
\item[Quadratic in the spin operators]\hfill\\\(\longrightarrow\) Indirect interactions e.g. with field gradient, interpreted as the interaction of a spin and itself.
\end{description}
In generating the total Hamiltonian of a arbitrary spin system, it is useful to partition the system into general parts corresponding to different physical processes
\begin{equation}
 \HamH=\HamZ+\HamNN+\HamNE+\HamEE+\HamC
\end{equation}
Each of the split Hamiltonian parts represents the following interactions
\begin{description}
\item[\(\HamZ=\) Zeeman interactions] \hfill\\\(\longrightarrow\) Chemical shielding tensors and g-tensors.
\item[\(\HamNN=\) Inter-Nuclear interactions] \hfill\\\(\longrightarrow\) J-couplings, quadrupolar interactions, dipolar interactions.
\item[\(\HamNE=\) Electron-Nuclear interactions] \hfill\\\(\longrightarrow\) Isotropic and anisotropic hyperfine couplings.
\item[\(\HamEE=\) Inter-Electron interactions] \hfill\\\(\longrightarrow\) Exchange interaction, zero field splitting, dipolar interactions.
\item[\(\HamC=\) Microwave/Radiofrequency irradiation] \hfill\\\(\longrightarrow\) Irradiated pulses with amplitude and phase, or control pulses with associated amplitude coefficients.
\end{description}
In describing these interactions it is useful to introduce the notation for a vector of spin operators, \(\vhat{S}\), which contains the three Cartesian operators as elements, ordered \(\big(\hat{S}_x^{}, \hat{S}_y^{}, \hat{S}_z^{}\big)\).

\subsection[Zeeman Interactions]{Zeeman Interactions}\label{Section:Zeeman}

An applied external magnetic field, \(\vec{B}_{}^{}\) induces an energy difference $\Delta E$ between the quantised magnetic moment values, with the energy is given by
\begin{equation}
 E=-\vec{\mu}\cdot\vec{B}_{}^{}
\end{equation}
where $\vec{B}_{}^{}$ is the external magnetic field. For nucleons, this is the nuclear Zeeman interaction, and for electrons, this is called the electron Zeeman interaction. Its Hamiltonian is 
\begin{equation}
 \HamH_{\mathcal{Z}}=\vhat{S}\ccdot\bm{A}\ccdot\vec{B}=\begin{pmatrix}\hat{S}_x^{} & \hat{S}_y^{} & \hat{S}_z^{}\end{pmatrix}\begin{pmatrix}a_{xx} & a_{xy} & a_{xz}\\ a_{yx} & a_{yy} & a_{yz}\\ a_{zx} & a_{zy} & a_{zz}\end{pmatrix}\begin{pmatrix}B_x^{}\\B_y^{}\\B_z^{}\end{pmatrix}\label{eqn:zeeman}
\end{equation} 
where the tensor $\bm{A}$ is the Zeeman interaction tensor. This is the coupling of the spin dipole moment to the external magnetic field. Without any further interaction, the Zeeman tensor reduces to the constant \(\frac{g\mu}{\hbar}\), where \(g\) is the \(g-\)factor and \(\mu\) is the nuclear magneton for nuclei and the Bohr magneton for electrons.

\subsubsection{Chemical shielding tensor}

The magnetic field also perturbs the electronic structure of the nucleus, with the induced field \(\bm{\delta}\vec{B}_0\). Using perturbation theory \cite{MERZBACHER61} gives the spin interaction from the contribution of \(\bm{\delta}\vec{B}_0\) as
\begin{equation}
 \HamH_{\mathcal{CS}}=\vhat{S}\ccdot\bm{\delta}\ccdot\vec{B}= \begin{pmatrix}\hat{S}_x^{} & \hat{S}_y^{} & \hat{S}_z^{}\end{pmatrix} \begin{pmatrix}\delta_{xx} & \delta_{xy} & \delta_{xz}\\ \delta_{yx} & \delta_{yy} & \delta_{yz}\\ \delta_{zx} & \delta_{zy} & \delta_{zz}\end{pmatrix} \begin{pmatrix}B_x^{}\\B_y^{}\\B_z^{}\end{pmatrix}
\end{equation} 
where $\bm{\delta}$ is known as the chemical shielding tensor.

\subsubsection{Electron g-tensor}

A similar interaction to the perturbation from chemical shielding can be derived for electrons. The energy operator for the electron dipole in an external magnetic field is
\begin{equation}
 \HamH_{\mathcal{G}}=\mu_B\vhat{L}\ccdot\bm{g}\ccdot\vec{B}=\begin{pmatrix}\hat{L}_x^{} & \hat{L}_y^{} & \hat{L}_z^{}\end{pmatrix}\begin{pmatrix}g_{xx} & g_{xy} & g_{xz}\\ g_{yx} & g_{yy} & g_{yz}\\ g_{zx} & g_{zy} & g_{zz}\end{pmatrix}\begin{pmatrix}B_x^{}\\B_y^{}\\B_z^{}\end{pmatrix}
\end{equation} 
where $\bm{g}$ is known as the \(g-\)tensor and the constant $\mu_B$ is the Bohr magneton.

\subsection[Inter-Nuclear Interactions]{Inter-Nuclear Interactions}\label{Section:InterNuclearInt}

\subsubsection{Inter-nuclear dipolar interactions}

Treating nuclear spins as point magnetic dipoles, \(\vec{\mu}\), the energy of the interaction between two nuclear dipoles, \(\vec{\mu}_1\) and \(\vec{\mu}_2\) is
\begin{equation}
 E=-\vec{\mu}\ccdot\vec{B}=-\frac{\mu_0}{4\pi}\frac{3\big(\vec{\mu}_1\ccdot\vec{r}\big)\big(\vec{r}\ccdot\vec{\mu}_2\big)-r^3\big(\vec{\mu}_1\ccdot\vec{\mu}_2\big)}{r^5}
\end{equation} 
where $\vec{r}$ is the vector that connects the dipoles, and the constant \(\mu_0\) is the permeability of free space. Hamiltonian is for this interaction between two spins, \(\vhat{S}_{}^{(1)}\) and \(\vhat{S}_{}^{(2)}\) 
\begin{equation}
 \HamH_{\mathcal{DD}}= -\frac{\mu_0}{4\pi}\frac{\gamma_1\gamma_2\hbar}{r^5} \Big( 3\big( \vhat{S}_{}^{(1)}\ccdot\vec{r} \big) \big(\vec{r}\ccdot\vhat{S}_{}^{(2)}\big) - r^2 \big(\vhat{S}_{}^{(1)}\ccdot\vhat{S}_{}^{(2)}\big) \Big)
\end{equation} 
where $\gamma_1$ and $\gamma_2$ are the magnetogyric ratios of the two spins.

\subsubsection{Spin-rotation coupling}

Collections of particles have angular momentum, just as single particles do. A magnetic induction is generated by a collection charges moving at velocity. Using \cref{eqn:zeeman} and considering the motion of the \(n^\text{th}\) charge with mass \(m_n\) and angular momentum operator \(\vhat{L}_n\) -- the spin rotation coupling interaction is
\begin{equation}
 \HamH_{\mathcal{SR}}=\sum\limits_k\gamma_k\vec{B}\ccdot\vhat{S}_k=-\frac{\mu_0}{4\pi}\sum\limits_{n,k}\frac{\gamma_kq_n}{m_nr_n^3}\Big(\vhat{L}_n\ccdot\vhat{S}_k\Big)
\end{equation} 
where $\gamma_k$ is the magnetogyric ratio of the \(k^\text{th}\) nucleus and $\vhat{S}_k$ are their spin operators. Evaluating the sum over the \(n\) moving charges, the spin-rotation Hamiltonian is
\begin{equation}
 \HamH_{\mathcal{SR}}=-\sum\limits_k\vhat{L}\ccdot\bm{A}_k\ccdot\vhat{S}_k
\end{equation} 
where $\bm{A}_k$ are spin-rotation coupling tensors.

\subsubsection{Quadrupolar coupling}

Nuclei with spin\(>\spinfrac{1}{2}\) have non-spherical charge distribution, and the direction of this \textit{nuclear quadrupole moment} is fixed, it interacts with the spin itself.
\begin{equation}
 \HamH_{\mathcal{Q}}=\frac{-eQ}{2s(2s-1)}\vhat{L}\ccdot\bm{V}\ccdot\vhat{L}=\frac{-eQ}{2s(2s-1)}\begin{pmatrix}\hat{L}_x^{} & \hat{L}_y^{} & \hat{L}_z^{}\end{pmatrix}\begin{pmatrix}v_{xx} & v_{xy} & v_{xz}\\ v_{yx} & v_{yy} & v_{yz}\\ v_{zx} & v_{zy} & v_{zz}\end{pmatrix}\begin{pmatrix}\hat{L}_x^{}\\\hat{L}_y^{}\\\hat{L}_z^{}\end{pmatrix}\label{eqn:quadrupolar}
\end{equation} 
where $eQ$ is the quadrupole moment of the nucleus, $s$ is the spin of the nucleus, $-\frac{eQ}{2s(2s-1)}\bm{V}$ is the quadrupolar coupling tensor. Quadrupolar coupling only exists for \(\spinint{1}\) nuclei and higher.

\subsubsection{J-coupling}

J-coupling is the scalar spin-spin coupling between two spins, after ignoring anisotropy, its Hamiltonian is
\begin{align}
 \HamH_{\mathcal{J}}&=\begin{pmatrix}\hat{L}_x^{} & \hat{L}_y^{} & \hat{L}_z^{}\end{pmatrix}\begin{pmatrix}a & 0 & 0\\0 & a & 0\\0 & 0 & a\end{pmatrix}\begin{pmatrix}\hat{S}_x^{} \\ \hat{S}_y^{} \\ \hat{S}_z^{}\end{pmatrix}\nonumber\\
 &=a\big(\hat{L}_x^{}\hat{S}_x^{}+\hat{L}_y^{}\hat{S}_y^{}+\hat{L}_z^{}\hat{S}_z^{}\big)=a\big(\vhat{L}\ccdot\vhat{S}\big)\label{eqn:Jcoupling}
\end{align} 
where \(a\) is the scalar coupling strength.

\subsection[Microwave and Radiofrequency terms]{Microwave \& Radiofrequency terms}\label{Section:ControlOps}

These interactions are the most important to the area of optimal control -- they are the part of the Hamiltonian that can be directly controlled -- all else is considered uncontrollable and is summed to a single \textit{drift Hamiltonian}. Commonly referred to as the time dependent \(\B{1}(t)\) field, areas of magnetic resonance manipulating nuclear spins use radiofrequency fields, and areas of magnetic resonance manipulating electrons use microwave fields.

Optimal control, set out in \cref{Chapter:GRAPE}, uses commonly control operators \(\HamH_x^{}\) and \(\HamH_y^{}\) with their amplitude coefficients becoming the variables of an optimisation algorithm. These amplitude coefficients relate directly to the pulse power in Hz, obtained from pulse calibration on the instrument.

This interaction has the same form as the Zeeman interaction in \cref{eqn:zeeman}
\begin{equation}
 \HamH=-\vhat{L}_k\ccdot\gamma_k^{}\ccdot\vec{B}_{rf}^{}(t)
\end{equation}
It is normal to polarise the radiofrequency field, of frequency \(\omega\), with a phase \(\varphi\) \cite{ERNST87}:
\begin{equation}
 \vec{B}_{rf}^{}=2\B{1}\cos\Big(\omega t\big(\vec{\e}_x\cos\varphi + \vec{\e}_y\sin\varphi\big)\Big)
\end{equation}
where \(\vec{\e}_x\) and \(\vec{\e}_y\) are the orthogonal unit vectors in the \(x\) and \(y\) directions. The above equation can be decomposed into two counter-rotating components (one of which can be neglected with a high \(\B{0}\) field), giving the Hamiltonian as \cite{ERNST87}\footnote{The factor in the sum of \(\frac{1}{2}K\) will become apparent when discussing optimal control in detail, but it is common to sum to the total number of control operators, usually \(2\times\) the number of spins -- it is kept here for consistency.}
\begin{equation}
 \HamC=-\B{1}\sum\limits_{k=1}^{\frac{1}{2}K}\gamma_k^{}\Big(\hat{L}_x^{(k)}\cos\big(\omega t+\varphi\big) + \hat{L}_y^{(k)}\sin\big(\omega t+\varphi\big)\Big)
\end{equation}
and in the time-independent rotating frame, applying a rotation transform, gives:
\begin{gather}
 \HamH_x^{(k)}=-\B{1}\gamma_k^{}\hat{L}_x^{(k)}\cos\varphi\\
 \HamH_y^{(k)}=-\B{1}\gamma_k^{}\hat{L}_y^{(k)}\sin\varphi
\end{gather}

\chapter{Optimal Control Theory} \label{Chapter:GRAPE}

\begin{chapquote}{Albert Camus, \textsc{The Fall}}
``You know what charm is: a way of getting the answer yes without having asked any clear question.''
\end{chapquote}
\renewcommand*{\CurrentPath}{./Chapter_3}


The focus of this thesis is advancing optimal control methods, specifically the \textsc{grape} method \cite{KHANEJA05}. Before this thesis progresses to present a novel optimal control method in \cref{Chapter:Hessian} and advances in efficiency of calculation within this method in \cref{Chapter:AuxMat}, the current chapter will present the theory of optimal control in the context of the \textsc{grape} method.


The state of a magnetic resonance system can be controlled using a set of electromagnetic pulses, discrete or continuous in time. Design of these pulses may prove difficult for control of complicated systems; numerical optimisation methods can be used to find a maximum overlap between the desired state and the state produced by the set of pulses, with the pulse schedule being the parameter of the objective function in the optimisation problem.


The art of running an experiment to a given accuracy with minimal expenditure of time and resources is known as \textit{optimal control theory}. A simple interpretation of optimal control theory is the application of numerical optimisation \cite{GILL81,FLETCHER87,BERTSEKAS96,BOYD04,NOCEDAL06} to control theory \cite{PONTRYAGIN64,TABAK71,GAMKRELIDZE78,DALESSANDRO07,LEWIS12}. This chapter will be divided into two sections, firstly outlining gradient based numerical optimisation with Newton-type methods, and secondly building a derivation of the optimal control method \textsc{grape}.

\section{Numerical optimisation}\label{Section:NumOptim}

Continuing from \cref{Chapter:Introduction} and the example of Newton's root finding method adapted to find the roots of an analytical gradient in \cref{Fig:NewtonIterates}, this section will formally extend that idea to build a family of generalised optimisation algorithms called Newton-type methods. An optimisation problem can be written as
\begin{align}
 && \min_{\objv{}} \Big\{\Obj{\objv{}}\Big\},&& \objv{}\in\Real^n
\end{align}
finding a local minimum, and a transformation of the above problem can find a local maximum:
\begin{align}
 &&  -\min_{\objv{}} \Big\{-\Obj{\objv{}}\Big\} = \max_{\objv{}} \Big\{\Obj{\objv{}}\Big\},&& \objv{}\in\Real^n\label{eqn:objmax}
\end{align}
The function \(\Obj{\objv{}}\) is a scalar function of \(n\) variables called the \textit{objective function}. The maximisation problem of \cref{eqn:objmax} was recast in the form of a minimisation problem with a simple transform, and it seems sensible to consider only minimisation problems for the remainder of this section. 

In this work an objective function should be a smooth function -- one which is continuously differentiate -- that can be represented by a convergent power series of terms calculated from the derivatives in the vicinity of a point i.e. a local approximation. For a function of one variable, this is
\begin{equation}
\Obj{\objv{}+\bm{h}}=\Obj{\objv{}}+\dfrac{\bm{h}}{1!}\Gradop{}\!\Obj{\objv{}}+\frac{\bm{h}^2}{2!}\Hessop{}\!\Obj{\objv{}}+\dots=\sum\limits_{r=0}^\infty\frac{\bm{h}^r}{r!}\nabla_{}^r\!\Obj{\objv{}}\label{eqn:taylorseries}
\end{equation}
where \(\bm{h}\) is a small displacement to the vector \(\objv{}\), and \(\nabla_{}^r\!\Obj{\objv{}}\) is the \(r^\text{th}\) order derivative of the function \(\Obj{\objv{}}\). This series is known as the many variable \textit{Taylor series} \cite{TAYLOR1715,MACLAURIN1742}.

The \textit{gradient vector}, \(\Grad{}\), and the \textit{Hessian matrix}, \(\Hess{}\), are introduced here. These two definitions will be used extensively throughout the remainder of this thesis and further explanation of their notation is required: \(\Gradop\) is the gradient operator defined by
\begin{align}
 && \Gradop\triangleq\big|\Gradop\big\rangle\triangleq\begin{pmatrix}
  \frac{\partial}{\partial c_1^{}} \\
  \frac{\partial}{\partial c_2^{}} \\
  \vdots \\
  \frac{\partial}{\partial c_n^{}} \\
 \end{pmatrix} && \objv{}\in\Real^n
\end{align}
which can be used to construct a Hessian operator \(\Hessop\):
\begin{align}
 \Hessop\triangleq\big|\Gradop\big\rangle\!\big\langle\Gradop\big|=\!\begin{pmatrix}
  \frac{\partial}{\partial c_1^{}}\frac{\partial}{\partial c_1^{}} & \frac{\partial}{\partial c_1^{}}\frac{\partial}{\partial c_2^{}} & \cdots & \frac{\partial}{\partial c_1^{}}\frac{\partial}{\partial c_n^{}} \\
  \frac{\partial}{\partial c_2^{}}\frac{\partial}{\partial c_1^{}} & \frac{\partial}{\partial c_2^{}}\frac{\partial}{\partial c_2^{}} & \cdots & \frac{\partial}{\partial c_2^{}}\frac{\partial}{\partial c_n^{}} \\
  \vdots & \vdots & \ddots & \vdots \\
  \frac{\partial}{\partial c_n^{}}\frac{\partial}{\partial c_1^{}} & \frac{\partial}{\partial c_n^{}}\frac{\partial}{\partial c_2^{}} & \cdots & \frac{\partial}{\partial c_n^{}}\frac{\partial}{\partial c_n^{}} \\
 \end{pmatrix}\! =\! \begin{pmatrix}
  \frac{\partial^2}{\partial c_1^{2}} & \frac{\partial^2}{\partial c_1^{}\partial c_2^{}} & \cdots & \frac{\partial^2}{\partial c_1^{}\partial c_n^{}} \\
  \frac{\partial^2}{\partial c_2^{}\partial c_1^{}} & \frac{\partial^2}{\partial c_2^{2}} & \cdots & \frac{\partial^2}{\partial c_2^{}\partial c_n^{}} \\
  \vdots & \vdots & \ddots & \vdots \\
  \frac{\partial^2}{\partial c_n^{}\partial c_1^{}}& \frac{\partial^2}{\partial c_n^{}\partial c_2^{}} & \cdots & \frac{\partial^2}{\partial c_n^{2}} \\
 \end{pmatrix}\!
\end{align}
A point in a vector space is denoted by a bold symbol, representing the coordinates of that point e.g. \(\objv{\itr}\), where the superscript in parentheses denotes the specific iteration, \(\itr\), of the optimisation. A direction is a line in the vector space \(\Real^n\), denoted with Dirac bra-ket symbols e.g. \(\big|\objv{\itr}\big\rangle\). A point that minimises or maximises a function will be denoted with an asterisk, \(\objv{\ast}\), and is called a \textit{minimiser} or \textit{maximiser} respectively\footnote{It is important to point out that it is only practicable to find a \textit{local minimiser}, \(\objv{\ast}\), which may not be a \textit{global minimiser}.}.

In describing optimisation problems, it will be useful to introduce two descriptions of the objective function as a specified point: the \textit{slope} and \textit{curvature}, which are defined at a point, \(\objv{\itr}\), as
\begin{subequations}
\begin{gather}
\textit{slope}\triangleq \big\langle\Grad{}\big|\objv{\itr}\big\rangle\nonumber\\
\textit{curvature}\triangleq \big\langle\objv{\itr}\big|\Hess{}\big|\objv{\itr}\big\rangle\nonumber
\end{gather}
\end{subequations}

\begin{figure}
\centering{\includegraphics{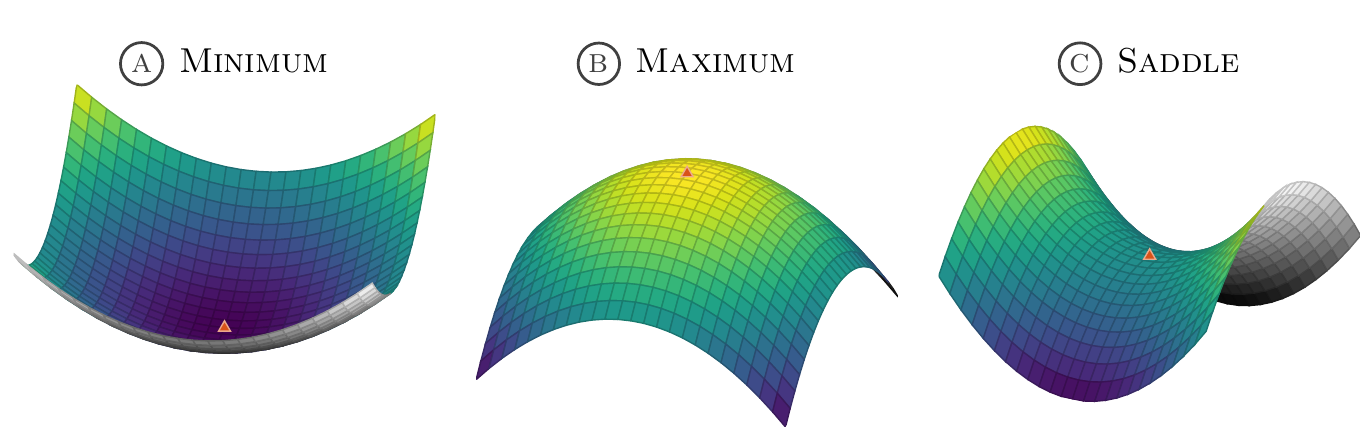}}
\caption[Types of stationary points]{Types of stationary points, where the gradient is equal to zeros. \plotlabel{a} shows the minimiser of the function \(x^2+y^2\), \plotlabel{b} shows the maximiser of the function \(-x^2-y^2\), and \plotlabel{c} shows the saddle point of the function \(x^2-y^2\). Each plot has a single point where the gradient is equal to zero.}\label{Fig:StationaryPoints}
\end{figure}

At the minimiser of a smooth function, two observations hold\footnote{for example, see \cref{Fig:NewtonIterates} \plotlabel{d}}:
\begin{enumerate}
 \item at the minimiser, the slope is zero
 \item at the minimiser, the curvature is non-negative in all directions
\end{enumerate}
The two observations of a point being a minimiser lead to the \textit{necessary conditions} for a local solution to a minimisation problem \cite{FLETCHER87}. The first, that the slope is zero at a minimiser is equivalent to
\begin{align}
 \big\langle\Grad{\ast}\big|\objv{}\big\rangle & = 0\nonumber\\
 \implies\lim_{\objv{}\to\objv{\ast}}\Big\|\Grad{\ast}\Big\|_{\infty}^{} & = 0 \label{eqn:necessary1}
\end{align}
and is called the \textit{first-order necessary condition} for a minimiser, and can be interpreted as requiring the gradient to be zero at a local minimiser. \cref{Fig:StationaryPoints} illustrates the three different types of \textit{stationary points} satisfy the \textit{first-order necessary condition}. Notice, only \cref{Fig:StationaryPoints} \plotlabel{a} shows a minimiser, and \cref{Fig:StationaryPoints} \plotlabel{c} has a point of zero gradient that is not a minimiser or a maximiser. \cref{Fig:StationaryPoints} shows by example that \cref{eqn:necessary1} alone is not enough to identify a local minimiser.

The observation concerning curvature of the objective function at the minimiser is required to identify a minimum. This observation is equivalent to
\begin{align}
 && \lim_{\objv{}\to\objv{\ast}}\big\langle\objv{}\big|\Hess{\ast}\big|\objv{}\big\rangle\geqslant 0, && \forall \objv{}\label{eqn:necessary2}
\end{align}
and is called the \textit{second-order necessary condition} for a minimiser, and can be interpreted as requiring a positive semi-definite\footnote{A positive semi-definite matrix has all eigenvalues \(\lambda\geqslant0\)} Hessian matrix in the vicinity of a local minimiser. However, a semi-definite matrix will become a problem when an inverse of the Hessian is required. Numerically, it is useful to reformulate this last condition as
\begin{align}
 && \lim_{\objv{}\to\objv{\ast}}\big\langle\objv{}\big|\Hess{\ast}\big|\objv{}\big\rangle> 0, && \forall \objv{}\neq\mathbf{0}\label{eqn:sufficient2}
\end{align}
which requires the Hessian matrix to be positive definite\footnote{A positive semi-definite matrix has all eigenvalues \(\lambda>0\)} in the vicinity of a local minimiser. \cref{eqn:necessary1} together with \cref{eqn:sufficient2} set out the \textit{sufficient conditions} for a local minimiser, and can be checked numerically. Returning again to \cref{Fig:StationaryPoints}, it is clear that \cref{Fig:StationaryPoints} \plotlabel{a} has positive curvature in all direction around the vicinity of the minimiser, \cref{Fig:StationaryPoints} \plotlabel{b} has negative curvature around the maximiser, and \cref{Fig:StationaryPoints} \plotlabel{c} as a mixture of positive and negative curvature around the saddle point. These sufficient conditions for an optimal solution will be used extensively in \cref{Chapter:Hessian}, and will be elaborated there.

\subsection{The gradient ascent method}\label{Section:GradAscent}

In searching for an optimum, the desired minimum or maximum of an objective function, a general strategy must be developed to find new points giving better objective function values. Two popular strategies are the \textit{restricted step methods}\footnote{also known as the \textit{trust region methods}.} and the \textit{line search method}. This chapter will be concerned with the \textit{line search method}, and \cref{Chapter:Hessian} will be concerned with \textit{restricted step methods}.

An example of a strategy to perform a line search at an iteration \(\itr\) is
\begin{enumerate}
 \item find a search direction \(\sdir{\itr}\)
 \item find a search length, \(\slen{}\), to minimise or maximise \(\Obj{\objv{\itr}+\slen{}\sdir{\itr}}\) with respect to \(\slen{}\).
 \item set the new iteration, \(\objv{\itr+1}=\objv{\itr}+\slen{}\sdir{\itr}\).
\end{enumerate}
Step \circled{2} is an idealised solution called an \textit{exact line search}, and is conceptually useful but not practicable in a finite number of operations. The subproblem of finding \(\slen{}\) will be outlined in \cref{Section:Line_Search}, for now it is sufficient to use the concept of an exact line search.

Line search methods introduce the search direction \(\sdir{\itr}\), at an iteration \(\itr\). One choice of search direction is that of the gradient. For minimisation, the search direction should be antiparallel to the gradient, and for maximisation the search direction should be parallel to the gradient:
\begin{subequations}
\begin{align}
 \textit{minimisation},&& \sdir{\itr} & =-\Grad{\itr} &&\label{eqn:graddescent}\\
 \textit{maximisation},&& \sdir{\itr} & =\Grad{\itr} &&\label{eqn:gradascent}
\end{align}
\end{subequations}
Together with an adequate line search method, these two methods are called the \textit{gradient descent method} for minimisation and the \textit{gradient ascent method} for maximisation. The update to the objective variable using a scaled step length found from a line search is
\begin{subequations}
\begin{align}
 \textit{minimisation}: && \objv{\itr+1}=\objv{\itr}-\slen{}\Grad{\itr}&  &&\label{eqn:grad_update_min}\\
 \textit{maximisation}: && \objv{\itr+1}=\objv{\itr}+\slen{}\Grad{\itr}& &&\label{eqn:grad_update_max}
\end{align}
\end{subequations}
These methods can also be used when no derivative information is available, although less effective, by using an estimate of the derivatives:
\begin{subequations}
\begin{gather}
 \Grad{} = \frac{\Obj{\objv{}(1 + h)}-\Obj{\objv{}}}{h}+ \BigO(h) \label{eqn:forward_diff}\\
 \Grad{} = \frac{\Obj{\objv{}(1 + h)}-\Obj{\objv{}(1 - h)}}{2h} + \BigO(h^2) \label{eqn:central_diff}
\end{gather}
\end{subequations}
where \(h=\sqrt{\eps}\) is a small perturbation to \(\objv{}\). These two approximations of the gradient are called finite difference approximations, \cref{eqn:forward_diff} is the forward differences approximation, and \cref{eqn:central_diff} is the central differences approximation. In a similar way, the second derivatives can be approximated with finite differences of the first derivatives. In addition to their use when no derivative information is available, these approximations are also useful in checking derivative calculations when derivative information is available.

\subsection{Newton's Method}\label{Section:NewtonMethod}

Considering that the linear approximation of a function, such as gradient descent/ascent methods, will fail to differentiate between an extremum and a stationary point, as in \cref{Fig:StationaryPoints}, more is required in the form of the sufficient conditions already set out in \cref{eqn:sufficient2}. A useful model in numerical optimisation is the \textit{quadratic model} of the function to be optimised. Although the optimisation methods should work on many functions, not just quadratic functions, in deriving the methods for optimisation the \textit{quadratic model} has proved fruitful. Desirable properties of this model are \cite{FLETCHER87}:
\begin{enumerate}
 \item Quadratic functions are smooth, with a well defined minimum.
 \item A continuously differentiable function is locally well approximated by a quadratic function near a minimum.
 \item More accurate than a linear approximation of a function, as in gradient descent/ascent methods, and higher derivative orders will likely be small.
 \item Can be made invariant under linear transformation of variables\footnote{An inductive proof of this property is given for Newton-like methods is given in \cite{FLETCHER87}.}.
\end{enumerate}
The aim of using a \textit{quadratic model} of a function is to find its extremum in a single iteration by analytically solving the model of the function to find its single stationary point shown in \cref{Fig:StationaryPoints}. Given an initial starting point of \(\objv{0}\), a quadratic model of the objective function at an extremum is a truncated Taylor series:
\begin{align}
\Obj{\objv{\ast}}\triangleq \Obj{\objv{0}+\snewt}\approx \Obj{\objv{0}}+\big\langle\Grad{0}\big|\snewt\big\rangle+\frac{1}{2}\big\langle\snewt\big|\Hess{0}\big|\snewt\big\rangle\label{eqn:quadmodel}
\end{align}
This equation is solved to find the vector \(\snewt{}\) required to step from the initial point \(\objv{0}\) to the extremum \(\objv{\ast}\). The solution is found when the derivative of \cref{eqn:quadmodel}, with respect to the step \(\snewt{}\), is equal to zero\footnote{\cref{eqn:quadderiv} requires the Hessian matrix to be symmetric, \(\nabla_{}^2=(\nabla_{}^2)^{\dagger}\)}:
\begin{align}
 \frac{\dd }{\dd \snewt}\Obj{\objv{0}+\snewt}= & \big\langle\Grad{0}\big| + \big\langle\snewt\big|\Hess{0}=\big\langle\mathbf{0}\big|\nonumber\\
\implies & \big|\Grad{0}\big\rangle + \Hess{0}\big|\snewt\big\rangle=\big|\mathbf{0}\big\rangle\label{eqn:quadderiv}
\end{align}
This gives two solutions: either \(\big|\snewt\big\rangle=\big|\mathbf{0}\big\rangle\) which means the starting point is an extremum, \(\objv{0}=\objv{\ast}\); or the non-trivial solution
\begin{equation}
 \big|\snewt\big\rangle=-\big[\Hess{0}\big]^{-1}\big|\Grad{0}\big\rangle\label{eqn:Newtonstep}
\end{equation}
Specifically, this model is true for the function in \cref{Fig:StationaryPoints} \plotlabel{a}, which has one minimum and the function must be minimised. To be able to find the maximum in \cref{Fig:StationaryPoints} \plotlabel{b}, a small modification must be made to \cref{eqn:quadmodel}. There are now two conditions, one is a minimisation condition that $\Obj{\objv{\ast}}\leqslant \Obj{\objv{0}}$ and the other is the maximisation condition that $\Obj{\objv{\ast}}\geqslant \Obj{\objv{0}}$:
\begin{subequations}
\begin{align}
 \textit{minimisation}: && \Obj{\objv{\ast}}-\Obj{\objv{0}} & \leqslant 0 &&\\
 \textit{maximisation}: && \Obj{\objv{\ast}}-\Obj{\objv{0}} & \geqslant 0 &&
\end{align}
\end{subequations}
Using the Newton step found in \cref{eqn:Newtonstep}, these conditions lead to the search directions
\begin{equation}
\big|\sdir{\textsc{Newton}}\big\rangle=-\big[\Hess{}\big]^{-1}\big|\Grad{}\big\rangle\label{eqn:newt_dir_min}
\end{equation}
where the direction of ascent/descent is accounted for by a positive/negative definite Hessian matrix. If a \textit{linear model} of the objective function is used, the Taylor series truncates after the gradient term in \cref{eqn:quadmodel} and the update formulae would be described by \cref{eqn:graddescent,eqn:gradascent}, the gradient descent/ascent methods using an exact line search. Following the modification of the exact line search, a Newton method using a line search method would modify the update rule to
\begin{equation}
\big|\objv{\itr+1}\big\rangle=\big|\objv{\itr}\big\rangle-\slen{}\big[\Hess{\itr}\big]^{-1}\big|\Grad{\itr}\big\rangle\label{eqn:Newton_update}
\end{equation}
where \(\slen{}\) is calculated with an appropriate line search method in \cref{Section:Line_Search}. The final case of using a Newton method to find (or avoid) the stationary point in \cref{Fig:StationaryPoints} \plotlabel{c} is an important one, but left to \cref{Chapter:Hessian}.

\subsection{Quasi-Newton Methods}\label{Section:QuasiNewton}

Although Newton's method has attractive quadratic convergence properties, it has the obvious disadvantage of needing a quadratic number of Hessian element calculations. Moreover, the formulae for the Hessian elements must be supplied. The disadvantages associated with a Hessian calculation can be avoided by using a popular class of \textit{quasi-Newton methods}. These methods are similar to Newton's method, except the Hessian is approximated by a symmetric, positive definite matrix, \(\HessA{}{}\) which is updated at every iteration with information from gradient calculations.
\begin{subequations}
\begin{align}
 \textit{minimisation}: && \big|\sdir{\textsc{qn}}\big\rangle=-\big[\HessA{}{}\big]^{-1}\big|\Grad{}\big\rangle&  &&\label{eqn:qnewt_dir_min}\\
 \textit{maximisation}: && \big|\sdir{\textsc{qn}}\big\rangle=+\big[\HessA{}{}\big]^{-1}\big|\Grad{}\big\rangle& &&\label{eqn:qnewt_dir_max}
\end{align}
\end{subequations}
A procedure to update the Hessian approximation follows
\begin{enumerate}
 \item set the current search direction \(\big|\sdir{\itr}\big\rangle=\pm\big[\HessA{\itr}{}\big]^{-1}\big|\Grad{\itr}\big\rangle\)
 \item perform a line search to find a step length, \(\slen{}\), giving \(\big|\objv{\itr+1}\big\rangle=\big|\objv{\itr}\big\rangle+\slen{}\big|\sdir{\itr}\big\rangle\)
 \item update the Hessian approximation, \(\HessA{\itr}{}\), with an appropriate method to give a new approximation \(\HessA{\itr+1}{}\)
\end{enumerate}
Maximisation and minimisation are accounted for with \(\pm\): maximisation with positive direction; minimisation with negative direction, allowing a strictly positive definite Hessian matrix.

A useful property of this procedure is that the Hessian approximation can use any gradient information obtained in the line search. Without better information, it is normal to set the initial Hessian approximation equal to the identity matrix \(\HessA{0}{}=\Unit\). An advantage of \textit{quasi-Newton methods} over other methods, including the \textit{conjugate gradient method} \cite{HESTENES52,LASDON67}, is that the quasi-Newton iteration does not need to be restarted periodically \cite{FLETCHER87}.

After repeated updating, the matrix \(\HessA{}{}\) should approximate the Hessian matrix $\Hess{}$, and one method to achieve this is to use the difference of gradients from iteration to iteration. For brevity, the change in objective variable and the change in gradient will be denoted by
\begin{subequations}
\begin{align}
 \dobjvar{\itr} & \triangleq \Big(\objv{\itr+1}-\objv{\itr}\Big) \\
 \dGrad{\itr} & \triangleq \Big(\Grad{\itr+1}-\Grad{\itr}\Big)
\end{align}
\end{subequations}
Using a truncated Taylor series of the gradient, with a quadratic model of the objective, an equation for the change in gradient can be derived:
\begin{align}
 \big|\Gradop\!\Obj{\objv{\itr}+\slen{}\sdir{\itr}}\big\rangle & =\big|\Grad{\itr}\big\rangle+\Hess{\itr}\big|\slen{}\sdir{\itr}\big\rangle+\dots\nonumber\\
 \implies\big|\dGrad{\itr}\big\rangle & =\Hess{\itr}\big|\dobjvar{\itr}\big\rangle\label{eqn:gradchange}
\end{align}
Since the Hessian approximation is calculated after a line search as the final stage of an iteration, \(\HessA{\itr+1}{}\) correctly relates the gradient change in \cref{eqn:gradchange}, giving
\begin{equation}
 \big|\dGrad{\itr}\big\rangle =\HessA{\itr+1}{}\big|\dobjvar{\itr}\big\rangle\label{eqn:secant}
\end{equation}
which is known as the \textit{secant equation} \cite{NOCEDAL06}. It maps the symmetric, positive-definite Hessian to the gradient, which is only possible if
\begin{equation}
 \big[x_{k+1}-x_k\big]^{T}\big[\nabla f_{k+1}-\nabla f_k\big]>0\label{eqn:curvaturecond}
\end{equation}
This is known as the \textit{curvature condition} and is satisfied for convex function \cite{NOCEDAL06}.

Alternatively, the matrix approximating the inverse to the Hessian can be updated, reducing the computational load\footnote{The inverse of a matrix is regarded as an expensive matrix operation that can be ill-defined \cite{GOLUB13}}.
\begin{subequations}
\begin{align}
 \textit{minimisation}: && \sdir{\textsc{qn}}=-\big[\HessAI{}{}\big]\big|\Grad{}\big\rangle&  &&\label{eqn:iqnewt_dir_min}\\
 \textit{maximisation}: && \sdir{\textsc{qn}}=+\big[\HessAI{}{}\big]\big|\Grad{}\big\rangle& &&\label{eqn:iqnewt_dir_max}
\end{align}
\end{subequations}
A similar procedure as above updates the approximation to the inverse-Hessian, \(\HessAI{\itr}{}\), except the search direction does not involve an inverse of a matrix: \(\big|\sdir{\itr}\big\rangle=\pm\big[\HessA{\itr}{}\big]^{-1}\big|\Grad{\itr}\big\rangle\). Similarly, the change in gradient is used to approximate the inverse-Hessian with
\begin{equation}
 \big|\dobjvar{\itr}\big\rangle=\big[\HessAI{\itr+1}{}\big]\big|\dGrad{\itr}\big\rangle \label{eqn:quasicond}
\end{equation}
which is known as the \textit{quasi-Newton condition} \cite{FLETCHER87}.

\subsubsection{Rank-1 Hessian update}\label{Section:quasi_rank1}

Initial attempts to find a method of updating the Hessian approximation that ensures its symmetry and satisfies \cref{eqn:secant}, resulted in a method called the \textit{symmetric rank-1} or \textsc{sr1} method \cite{BROYDEN67,DAVIDON68}. A rank-1 symmetric matrix can be constructed from the outer product of a real vector with itself, \(\big|\mathbf{u}\big\rangle\!\big\langle\mathbf{u}\big|\), giving the symmetric rank-1 update to the Hessian approximation
\begin{equation}
 \HessA{\itr+1}{\textsc{sr1}}=\HessA{\itr}{\textsc{sr1}}+a\big|\mathbf{u}\big\rangle\!\big\langle\mathbf{u}\big|\label{eqn:sr1_start}
\end{equation}
where \(a\) is an arbitrary constant and \(\big|\mathbf{u}\big\rangle\) is a vector to be determined. Substituting the last equation into the secant equation, \cref{eqn:secant}, gives
\begin{equation}
 \big|\dGrad{\itr}\big\rangle =\HessA{\itr}{\textsc{sr1}}\big|\dobjvar{\itr}\big\rangle + a\Big[\big\langle\mathbf{u}\big|\dobjvar{\itr}\big\rangle\Big]\big|\scptspc\mathbf{u}\big\rangle
\end{equation}
The vector \(\big|\mathbf{u}\big\rangle\) must be proportional to \(\big|\dGrad{\itr}\big\rangle - \HessA{\itr}{\textsc{sr1}}\big|\dobjvar{\itr}\big\rangle\), with proportionality absorbed into the arbitrary constant \(a\), giving
\begin{equation}
 \big|\dGrad{\itr}\big\rangle - \HessA{\itr}{\textsc{sr1}}\big|\dobjvar{\itr}\big\rangle = a \bigg[\big\langle\dobjvar{\itr}\big|\Big(\big|\dGrad{\itr}\big\rangle - \HessA{\itr}{\textsc{sr1}}\big|\dobjvar{\itr}\big\rangle\Big)\bigg]\Big(\big|\dGrad{\itr}\big\rangle - \HessA{\itr}{\textsc{sr1}}\big|\dobjvar{\itr}\big\rangle\Big)
\end{equation}
This is satisfied for \(a\) if \cite{NOCEDAL06}
\begin{equation}
 a=\frac{\sign{\bigg[\big\langle\dobjvar{\itr}\big|\Big(\big|\dGrad{\itr}\big\rangle - \HessA{\itr}{\textsc{sr1}}\big|\dobjvar{\itr}\big\rangle\Big)\bigg]}}{\bigg[\big\langle\dobjvar{\itr}\big|\Big(\big|\dGrad{\itr}\big\rangle - \HessA{\itr}{\textsc{sr1}}\big|\dobjvar{\itr}\big\rangle\Big)\bigg]}
\end{equation}
and gives the only symmetric rank-1 update, satisfying \cref{eqn:secant}, to be
\begin{equation}
 \HessA{\itr+1}{\textsc{sr1}}=\HessA{\itr}{\textsc{sr1}} + \frac{\big|\dGrad{\itr}-\HessA{\itr}{\textsc{sr1}}\dobjvar{\itr}\big\rangle\big\langle\dGrad{\itr}-\HessA{\itr}{\textsc{sr1}}\dobjvar{\itr}\big|}{\big\langle\dGrad{\itr}\big|\dobjvar{\itr}\big\rangle-\big\langle\dobjvar{\itr}\big|\HessA{\itr}{\textsc{sr1}}\big|\dobjvar{\itr}\big\rangle}\label{eqn:sr1}
\end{equation}

A similar derivation can be made to find a symmetric rank-1 update that satisfies \cref{eqn:quasicond}, giving an update rule for the inverse-Hessian approximation. Alternatively, a useful formula to use \cref{eqn:sr1} to find a corresponding update rule for the inverse-Hessian approximation is the \textit{Sherman-Morrison-Woodbury formula} \cite{GOLUB13}\footnote{\textit{Sherman-Morrison-Woodbury formula:} \cite{GOLUB13}
\begin{equation*}\big(\mathbf{A}+\mathbf{U}\mathbf{V}^{T}\big)^{-1}=\mathbf{A}^{-1}-\mathbf{A}^{-1}\mathbf{U}\big(\Unit+\mathbf{V}^{T}\mathbf{A}^{-1}\mathbf{U}\big)^{-1}\mathbf{V}^{T}\mathbf{A}^{-1}\end{equation*}with \(\mathbf{A}\in\Real^{n\times n}\) and both \(\mathbf{U},\mathbf{V}\in\Real^{n\times k}\) for \(1\leqslant k\leqslant n\). This shows a rank-$k$ update to a matrix gives a rank-$k$ update to its inverse.}. Using the formula for a rank-1 update amounts to the interchanges \(\HessA{}{}\longleftrightarrow\HessAI{}{}\) and \(\dGrad{}\longleftrightarrow\dobjvar{}\), giving\footnote{This is a property of a \textit{self-dual formula}.}
\begin{equation}
 \HessAI{\itr+1}{\textsc{sr1}}=\HessAI{\itr}{\textsc{sr1}} + \frac{\big|\dobjvar{\itr}-\HessAI{\itr}{\textsc{sr1}}\dGrad{\itr}\big\rangle\big\langle\dobjvar{\itr}-\HessAI{\itr}{\textsc{sr1}}\dGrad{\itr}\big|}{\big\langle\dobjvar{\itr}\big|\dGrad{\itr}\big\rangle-\big\langle\dGrad{\itr}\big|\HessAI{\itr}{\textsc{sr1}}\big|\dGrad{\itr}\big\rangle}\label{eqn:inv_sr1}
\end{equation}

\cref{eqn:sr1,eqn:inv_sr1} both ensure any update is a symmetric matrix, however, the update is not guaranteed to be positive definite \cite{NOCEDAL06,FLETCHER87}. This drawback may be alleviated by forcing a positive definite matrix with a method described in \cref{Chapter:Hessian}. However, there is a second problem with \cref{eqn:sr1,eqn:inv_sr1}, that the denominator may vanish.

\subsubsection{Rank-2 Hessian update}\label{Section:quasi_rank2}

In an attempt to avoid the problems of using \cref{eqn:sr1,eqn:inv_sr1}, a more flexible Hessian update formula can be constructed be replacing the rank-1 update of \cref{eqn:sr1_start} with a rank-2 update
\begin{equation}
 \HessA{\itr+1}{\textsc{dfp}}=\HessA{\itr}{\textsc{dfp}}+a\big|\mathbf{u}\big\rangle\!\big\langle\mathbf{u}\big|+b\big|\mathbf{v}\big\rangle\!\big\langle\mathbf{v}\big|\label{eqn:dfp_start}
\end{equation}
Using the secant equation, \cref{eqn:secant}, and substituting into this rank-2 update gives
\begin{equation}
  \big|\dGrad{\itr}\big\rangle =\HessA{\itr}{\textsc{dfp}}\big|\dobjvar{\itr}\big\rangle + a\Big[\big\langle\mathbf{u}\big|\dobjvar{\itr}\big\rangle\Big]\big|\scptspc\mathbf{u}\big\rangle + b\Big[\big\langle\mathbf{v}\big|\dobjvar{\itr}\big\rangle\Big]\big|\scptspc\mathbf{v}\big\rangle\label{eqn:rank2}
\end{equation}
There was only one symmetric rank-1 solution to \cref{eqn:secant}, there are a family of symmetric rank-2 update formulae which will be summarised at the end of this section in a single equation of one variable, \cref{eqn:broyden,eqn:inv_broyden}, for a Hessian approximation and an inverse-Hessian approximation.

To start a derivation of the general rank-2 formulae \cref{eqn:broyden,eqn:inv_broyden} a set of non-unique solutions to \cref{eqn:rank2} are trialled, while ensuring the solutions satisfy the curvature condition of \cref{eqn:curvaturecond}: \(\big|\mathbf{u}\big\rangle=\HessA{\itr}{\textsc{dfp}}\big|\dobjvar{\itr}\big\rangle\) and \(\big|\mathbf{v}\big\rangle=\big|\dGrad{\itr}\big\rangle\), giving the arbitrary constants \cite{FLETCHER87}
\begin{align}
 a=\frac{1}{\big\langle\dobjvar{\itr}\big|\HessA{\itr}{\textsc{dfp}}\big|\dobjvar{\itr}\big\rangle} && b=\frac{-1}{\big\langle\dGrad{\itr}\big|\dobjvar{\itr}\big\rangle}
\end{align}
This gives one of two special cases to a general formula to compute the approximation to the Hessian matrix is \cite{NOCEDAL06}, which in itself can be used as a rank-2 update approximating a Hessian:
\begin{multline}
 \HessA{\itr+1}{\textsc{dfp}}=\HessA{\itr}{\textsc{dfp}} + \Bigg[\Unit+\frac{\big\langle\dobjvar{\itr}\big|\HessA{\itr}{\textsc{dfp}}\big|\dobjvar{\itr}\big\rangle}{\big\langle\dGrad{\itr}\big|\dobjvar{\itr}\big\rangle}\Bigg]\frac{\big|\dGrad{\itr}\big\rangle\!\big\langle\dGrad{\itr}\big|}{\big\langle\dGrad{\itr}\big|\dobjvar{\itr}\big\rangle} \\ - \Bigg[\frac{\big|\dGrad{\itr}\big\rangle\!\big\langle\dobjvar{\itr}\big|\HessA{\itr}{\textsc{dfp}}+\HessA{\itr}{\textsc{dfp}}\big|\dobjvar{\itr}\big\rangle\!\big\langle\dGrad{\itr}\big|}{\big\langle\dGrad{\itr}\big|\dobjvar{\itr}\big\rangle}\Bigg] \label{eqn:dfp}
\end{multline}
The formula is called the \textsc{dfp} updating formula\footnote{The method is commonly named after the authors Davidon, Fletcher, and Powell. It also known as the \textit{variable metric algorithm}.} and was originally published in a technical paper \cite{DAVIDON59,DAVIDON91} and later presented using \cref{eqn:dfp_start} by Fletcher and Powell \cite{FLETCHER63}. Again, using the \textit{Sherman-Morrison-Woodbury formula} \cite{GOLUB13}, the update in \cref{eqn:dfp} can be transformed to update an inverse-Hessian approximation:
\begin{equation}
\HessAI{\itr+1}{\textsc{dfp}}=\HessAI{\itr}{\textsc{dfp}}-\Bigg[\frac{\HessAI{\itr}{\textsc{dfp}}\big|\dGrad{\itr}\big\rangle\!\big\langle\dGrad{\itr}\big|\HessAI{\itr}{\textsc{dfp}}}{\big\langle\dGrad{\itr}\big|\HessAI{\itr}{\textsc{dfp}}\big|\dGrad{\itr}\big\rangle}\Bigg]+\frac{\big|\dobjvar{\itr}\big\rangle\!\big\langle\dobjvar{\itr}\big|}{\big\langle\dGrad{\itr}\big|\dobjvar{\itr}\big\rangle}\label{eqn:inv_dfp}
\end{equation}
In comparison to the \textsc{sr1} updates in \cref{eqn:sr1,eqn:inv_sr1}, the \textsc{dfp} updates in \cref{eqn:dfp,eqn:inv_dfp} ensure a positive definite update, which can be proved with a Cholesky factorisation\footnote{if a matrix \(\mathbf{A}\) is positive definite, there exists a factorisation \(\mathbf{A}=\mathbf{L}\mathbf{L}^{\dagger}\) where $\mathbf{L}$ is a lower triangular matrix with real, positive diagonal elements. This is called a Cholesky factorisation \cite{GOLUB13}} of \(\HessAI{\itr}{\textsc{dfp}}\). There still exists a drawback with these update methods, that they need an accurate line search \cite{FLETCHER87}, which can become computationally expensive when gradient calculations are needed for a line search (see \cref{Section:Line_Search}).

Although the \textsc{sr1} formulae in \cref{eqn:sr1,eqn:inv_sr1} are \textit{self-dual}, it is obvious that the \textsc{dfp} formulae in \cref{eqn:dfp,eqn:inv_dfp} are not. An important formula, called the \textsc{bfgs} formula \cite{BROYDEN70a,BROYDEN70b,FLETCHER70,GOLDFARB70,SHANNO70}\footnote{named after the authors Broyden, Fletcher, Goldfarb, and Shanno}, can be constructed by observing \(\HessA{\itr+1}{\textsc{bfgs}}\HessAI{\itr+1}{\textsc{bfgs}}=\Unit\) and finding the \textit{dual formula} of \cref{eqn:inv_dfp} by interchanging \(\HessA{}{}\longleftrightarrow\HessAI{}{}\) and \(\dGrad{}\longleftrightarrow\dobjvar{}\):
\begin{equation}
\HessA{\itr+1}{\textsc{bfgs}}=\HessA{\itr}{\textsc{bfgs}}-\Bigg[\frac{\HessA{\itr}{\textsc{bfgs}}\big|\dobjvar{\itr}\big\rangle\!\big\langle\dobjvar{\itr}\big|\HessA{\itr}{\textsc{bfgs}}}{\big\langle\dobjvar{\itr}\big|\HessA{\itr}{\textsc{bfgs}}\big|\dobjvar{\itr}\big\rangle}\Bigg]+\frac{\big|\dGrad{\itr}\big\rangle\!\big\langle\dGrad{\itr}\big|}{\big\langle\dobjvar{\itr}\big|\dGrad{\itr}\big\rangle}\label{eqn:bfgs}
\end{equation}
A \textsc{bfgs} update for an approximation to the inverse-Hessian can be found by transforming this last formula with the \textit{Sherman-Morrison-Woodbury formula} \cite{GOLUB13}:
\begin{multline}
 \HessAI{\itr+1}{\textsc{bfgs}}=\HessAI{\itr}{\textsc{bfgs}} + \Bigg[\Unit+\frac{\big\langle\dGrad{\itr}\big|\HessAI{\itr}{\textsc{bfgs}}\big|\dGrad{\itr}\big\rangle}{\big\langle\dobjvar{\itr}\big|\dGrad{\itr}\big\rangle}\Bigg] \frac{\big|\dobjvar{\itr}\big\rangle\!\big\langle\dobjvar{\itr}\big|}{\big\langle\dobjvar{\itr}\big|\dGrad{\itr}\big\rangle} \\ - \Bigg[\frac{\big|\dobjvar{\itr}\big\rangle\!\big\langle\dGrad{\itr}\big|\HessAI{\itr}{\textsc{bfgs}}+\HessAI{\itr}{\textsc{bfgs}}\big|\dGrad{\itr}\big\rangle\!\big\langle\dobjvar{\itr}\big|}{\big\langle\dobjvar{\itr}\big|\dGrad{\itr}\big\rangle}\Bigg] \label{eqn:inv_bfgs}
\end{multline}
The \textsc{bfgs} update formulae are widely accepted to perform better than the \textsc{dfp} formulae \cite{FLETCHER87,GILL81,NOCEDAL06}, requiring a line search with comparatively fewer inner iterations.

The \textsc{dfp} and \textsc{bfgs} are both special cases for the general \textit{Broyden family} rank-2 formulae \cite{BROYDEN67}:
\begin{subequations}
\begin{align}
 \HessA{\itr+1}{\phi} & =\big(1-\phi\big)\HessA{\itr+1}{\textsc{dfp}}+\phi\HessA{\itr+1}{\textsc{bfgs}}\label{eqn:broyden}\\
 \HessAI{\itr+1}{\phi} & =\big(1-\phi\big)\HessAI{\itr+1}{\textsc{dfp}}+\phi\HessAI{\itr+1}{\textsc{bfgs}}\label{eqn:inv_broyden}
\end{align}
\end{subequations}
with \(\phi\in[0,1]\).

In considering the benefit of calculating the inverse-Hessian approximation of \cref{eqn:inv_bfgs,eqn:inv_dfp}, that there is no need to find the inverse of a matrix explicitly, there is a drawback -- as will be pointed out in \cref{Chapter:Hessian}, the inverse-Hessian approximations do not allow the implementation of \textit{trust region methods} and other conditioning methods.

\subsubsection[Limited memory quasi-Newton methods]{Limited memory quasi-Newton methods}

Large optimisation problems may require Hessian approximations which are not sparse and cannot be computed at a reasonable cost. Rank-1 and rank-2 updates only add further vector information to the to the Hessian approximation, and the end use of the Hessian information is to calculate a vector search direction. \textit{Limited memory quasi-Newton methods} \cite{NOCEDAL80} make use of a store of the \(m\) most recent vectors used to describe the Hessian approximation -- saving storage space if \(m<n\), where \(n\) is the number of objective function variables. The following review of this class of methods is from \cite{NOCEDAL06}. As the \(\ell\)-\textsc{bfgs} (limited-memory \textsc{bfgs}) method \cite{LIU89,BYRD94} is a popular limited-memory quasi-Newton method in optimal control \cite{FOUQUIERES11} (and more compactly derived), the following derivation will follow this update to the inverse Hessian approximation. Before proceeding, it will be useful to present \cref{eqn:inv_bfgs} in a product form \cite{NOCEDAL80}:
\begin{align}
&& \HessAI{\itr+1}{\textsc{bfgs}}= \mV_{\itr}^{\dagger}\HessAI{\itr}{\textsc{bfgs}}\mV_{\itr}^{} + \frac{\big|\dobjvar{}_{\itr}^{}\big\rangle\!\big\langle\dobjvar{}_{\itr}^{}\big|}{\big\langle\dGrad{}_{\itr}^{}\big|\dobjvar{}_{\itr}^{}\big\rangle}, && \mV_{\itr}^{}= \Bigg[\Unit-\frac{\big|\dGrad{}_{\itr}^{}\big\rangle\!\big\langle\dobjvar{}_{\itr}^{}\big|}{\big\langle\dGrad{}_{\itr}^{}\big|\dobjvar{}_{\itr}^{}\big\rangle}\Bigg]\label{eqn:inv_bfgs2}
\end{align}

Considering that quasi-Newton methods are updates to the Hessian, or inverse-Hessian approximation from vectors describing the change in objective function, \(\dobjvar{}\), and the change in its gradient \(\dGrad{}\) -- a Hessian approximation at an iteration \(\itr\) can be recursively calculated from a store of these vectors and the initial Hessian approximation at \(\itr=\itr-m\) i.e. \(m\) iterations before the current iteration \(\itr\).
\begin{multline}
 \HessAI{\itr}{\textsc{bfgs}}=\Bigg[ \prod\limits_{k=\itr-1}^{\itr-m}\!\!\mV_{k}^{\dagger}\Bigg] \HessAI{\itr-m}{\textsc{bfgs}} \Bigg[ \prod\limits_{k=\itr-m}^{\itr-1}\!\!\!\mV_{k}^{} \Bigg]\\
\hfill+ \frac{1}{\big\langle\dGrad{}_{\itr-m}^{}\big|\dobjvar{}_{\itr-m}^{}\big\rangle} \Bigg[ \prod\limits_{k=\itr-1}^{\itr-m+1}\!\!\!\mV_{k}^{\dagger}\Bigg] \big|\dobjvar{}_{\itr-m}^{}\big\rangle\!\big\langle\dobjvar{}_{\itr-m}^{}\big| \Bigg[ \prod\limits_{k=\itr-m+1}^{\itr-1}\!\!\!\!\!\!\mV_{k}^{}\Bigg]\\
\hfill+ \frac{1}{\big\langle\dGrad{}_{\itr-m+1}^{}\big|\dobjvar{}_{\itr-m+1}^{}\big\rangle} \Bigg[ \prod\limits_{k=\itr-1}^{\itr-m+2}\!\!\!\mV_{k}^{\dagger}\Bigg] \big|\dobjvar{}_{\itr-m+1}^{}\big\rangle\!\big\langle\dobjvar{}_{\itr-m+1}^{}\big| \Bigg[ \prod\limits_{k=\itr-m+2}^{\itr-1}\!\!\!\!\!\!\mV_{k}^{}\Bigg]\\
\hfill+\cdots+\frac{1}{\big\langle\dGrad{}_{\itr-1}^{}\big|\dobjvar{}_{\itr-1}^{}\big\rangle} \big|\dobjvar{}_{\itr-1}^{}\big\rangle\!\big\langle\dobjvar{}_{\itr-1}^{}\big|
\end{multline}
where the ordered products of \(\mV_k^{}\) and its transpose \(\mV_k^{\dagger}\) form ordered products. This formula lends itself to a recursive algorithm to calculate the inverse-Hessian approximation given a store of objective and gradient vectors. However, the inverse-Hessian approximation \(\HessAI{\itr-m}{\textsc{bfgs}}\) must still be stored, defeating the point of the algorithm. The proposition of the limited-memory methods is to use a scaling matrix \(\HessAI{0}{}\), computed at each update \cite{NOCEDAL80}:
\begin{equation}
 \HessAI{\itr}{0}=\Unit\frac{\big\langle\dobjvar{}_{\itr-1}^{}\big|\dGrad{}_{\itr-1}^{}\big\rangle}{\big\langle\dGrad{}_{\itr-1}^{}\big|\dGrad{}_{\itr-1}^{}\big\rangle}
\end{equation}
Further to a recursive algorithm to update the inverse-Hessian approximation, a short-cut can be taken to update the search directional of \cref{eqn:qnewt_dir_min,eqn:qnewt_dir_max} directly. A popular \(\ell\)-\textsc{bfgs} method factors in the gradient to a two-loop recursive algorithm \cite{NOCEDAL80,LIU89} to give the search direction\footnote{The \(\ell\)-\textsc{bfgs} two-loop recursive algorithm is the implementation in \textit{Spinach}, and the one used in all \(\ell\)-\textsc{bfgs} optimisations of this thesis}. It should be noted that the \(\ell\)-\textsc{bfgs} method is expected to perform badly when applied to ill-conditioned problems, where the Hessian proper has a wide range of eigenvalues \cite{NOCEDAL06} and the scaling matrix fails to be effective. This will be investigated in the context of optimal control in \cref{Chapter:Hessian}.

To derive similar recursive update algorithms for other quasi-Newton methods is not so neat because a similar product form of \cref{eqn:inv_bfgs2} has not been derived for other Broyden class methods \cite{NOCEDAL06}. However, an ``outer product'' representation of quasi-Newton methods is useful in deriving other limited-memory algorithms and shows a compact form for general quasi-Newton methods. For the \(\ell\)-\textsc{bfgs} Hessian approximation of \cref{eqn:bfgs}, and using the diagonal scaling matrix \(\HessA{0}{}=\big[\HessAI{0}{}\big]^{-1}\), this compact notation is \cite{BYRD94}:
\begin{equation}
 \HessA{\itr}{\textsc{bfgs}}=\HessA{0}{} - \begin{bmatrix}\HessA{0}{}\dOBJVAR{}_{\!\itr}^{} & \dGRAD{}_{\!\itr}^{} \end{bmatrix} \begin{bmatrix} \HessA{0}{}\big\langle\dOBJVAR{}_{\!\itr}^{}\big|\dOBJVAR{}_{\!\itr}^{}\big\rangle & \mathbf{L}_{\itr}^{} \\ \mathbf{L}_{\itr}^{\dagger} & -\mathbf{K}_{\itr}^{} \end{bmatrix}^{\!\!-1} \begin{bmatrix} \HessA{0}{}\dOBJVAR{}_{\!\itr}^{\dagger} \\ \dGRAD{}_{\!\itr}^{\dagger}\end{bmatrix}\label{eqn:compactLBFGS}
\end{equation}
where the stores of \(\dobjvar{}\) and \(\dGrad{}\) are \(n\times m\) matrices defined by
\begin{subequations}
\begin{gather}
 \dOBJVAR{}_{\!\itr}^{}\triangleq \begin{bmatrix}\dobjvar{}_{\itr-m}^{} & \dobjvar{}_{\itr-m+1}^{} & \cdots & \dobjvar{}_{\itr-1}^{} \end{bmatrix}\\
 \dGRAD{}_{\!\itr}^{}\triangleq \begin{bmatrix}\dGrad{}_{\itr-m}^{} & \dGrad{}_{\itr-m+1}^{} & \cdots & \dGrad{}_{\itr-1}^{} \end{bmatrix}
\end{gather}
\end{subequations}
\(\mathbf{L}_{\itr}^{}\) is an \(m\times m\) matrix defined by
\begin{align}
 && \Big(\mathbf{L}_{\itr}^{}\Big)_{\!\!p,q}\triangleq\begin{cases} \big\langle\dobjvar{}_{\itr-m-1+p}^{}\big|\dGrad{}_{\itr-m-1+q}^{}\big\rangle \\ 0 \end{cases} &&
\begin{array}{l}
 p>q\\
 p\leqslant q
 \end{array}
 \end{align}
and \(\mathbf{K}_{\itr}^{}\) is a diagonal \(m\times m\) matrix defined by
 \begin{equation}
 \mathbf{K}_{\itr}^{}\triangleq \begin{bmatrix} \big\langle\dobjvar{}_{\itr-m}^{}\big|\dGrad{}_{\itr-m}^{}\big\rangle & 0 & \cdots & 0\\ 0 & \big\langle\dobjvar{}_{\itr-m+1}^{}\big|\dGrad{}_{\itr-m+1}^{}\big\rangle & \cdots & 0\\ \vdots & \vdots & \ddots & \vdots \\ 0 & 0 & \cdots & \big\langle\dobjvar{}_{\itr-1}^{}\big|\dGrad{}_{\itr-1}^{}\big\rangle\end{bmatrix}
\end{equation}
The central matrix in \cref{eqn:compactLBFGS} is of small size, \(2m\times 2m\), when compared to the number of optimisation variables, \(n\): their factorisation requires negligible amount of computation \cite{NOCEDAL06}. This \(\ell\)-\textsc{bfgs} method is an update to the Hessian approximation, and so allows the implementation of Hessian conditioning methods.

Using the definitions above, an \(\ell\)-\textsc{sr1} algorithm can be derived as \cite{BYRD94}:
\begin{equation}
 \HessA{\itr}{\textsc{sr1}}=\HessA{0}{} + \!\Big[\dGRAD{}_{\!\itr}^{}\!- \HessA{0}{}\dOBJVAR{}_{\!\itr}^{}\Big]\!\Big[ \mathbf{K}_{\itr}^{} + \mathbf{L}_{\itr}^{} + \mathbf{L}_{\itr}^{\dagger} - \dOBJVAR{}_{\!\itr}^{\dagger}\HessA{0}{}\dOBJVAR{}_{\!\itr}^{}\Big]_{}^{\!\!-1}\Big[ \dGRAD{}_{\!\itr}^{}\ - \HessA{0}{}\dOBJVAR{}_{\!\itr}^{}\Big]^{\!\dagger}
\end{equation}
Further compact notations can be derived for all the Broyden class of quasi-Newton methods, for inverse-Hessian and Hessian approximations, allowing design of their limited-memory algorithms \cite{BYRD94}.

\subsection{Line search subproblem}\label{Section:Line_Search}

This section will present criteria for an acceptable point using the scalar quantity called the \textit{step length}, mentioned in \cref{eqn:grad_update_min,eqn:grad_update_max,eqn:Newton_update,eqn:gradchange}, deciding how far to move along the search direction.

The \textit{linear approximation} to a function in \cref{Section:GradAscent}, and the \textit{quadratic approximation} to a function in \cref{Section:NewtonMethod,Section:QuasiNewton}, usually need more than their derived search direction to find an acceptable point. Although the gradient based methods will step in the correct direction, without a step length, the methods could wildly misjudge a predicted extremum. Algorithms to find an appropriate step length are called \textit{line search methods} or \textit{inexact line search}. The update to an objective variable at the next iteration is
\begin{equation}
 \objv{\itr+1}=\objv{\itr}+\slen{}\sdir{\itr}\label{eqn:update}
\end{equation}
where the search direction is calculated with a method chosen from \cref{eqn:graddescent,eqn:gradascent,eqn:newt_dir_min,eqn:qnewt_dir_min,eqn:qnewt_dir_max,eqn:iqnewt_dir_min,eqn:iqnewt_dir_max}. The line search subproblem can be defined as a minimisation\footnote{A maximisation problem follows a similar definition\begin{equation}
\max_{\slen{}>0}\Obj{\objv{\itr}+\slen{}\sdir{\itr}}\nonumber
\end{equation}The rest of this section will follow a minimisation problem, but a maximisation problem is easily accounted for.} problem in itself:
\begin{equation}
\min_{\slen{}>0}\Obj{\objv{\itr}+\slen{}\sdir{\itr}}\label{eqn:MinProblem}
\end{equation}
The exact minimisation defined by \cref{eqn:MinProblem} is not required, and may not find this solution in a finite number of iterations, only an approximation of \(\slen{}\) is required to progress the optimisation procedure. 

In deciding if the approximation is acceptable, a number of conditions should be tested. The first of these conditions is trivial, \(\Obj{\objv{\itr+1}}<\Obj{\objv{\itr}}\), and should be accounted for with calculation of a search direction giving a descent direction:
\begin{equation}
\big\langle\sdir{\itr}\big|\Grad{\itr}\big\rangle<0
\end{equation}
However, this alone does not guarantee convergence. There must be \textit{sufficient decrease} in the slope, \(\big\langle\Grad{}\big|\objv{}\big\rangle\), of the objective along the search direction. Assuming that an iterative algorithm produces calculated step lengths, \(\slen{}\), the acceptable step length should satisfy
\begin{align}
 \parbox{5em}{\centering\textit{Armijo\\ condition:}} && \Obj{\objv{\itr}+\slen{}\sdir{\itr}}\leqslant\Obj{\objv{\itr}}+\lsconst{1}\slen{}\big\langle\Grad{\itr}\big|\sdir{\itr}\big\rangle, && \lsconst{1}\in(0,1)\label{eqn:sufficientdecrease}
\end{align}
for a pre-defined constant, \(\lsconst{1}\). This condition is known as the \textit{sufficient decrease condition} or the \textit{Armijo condition} \cite{ARMIJO66}, and the constant is chosen to be small, usually \(\lsconst{1}\sim 10^{-4}\).

The condition of \cref{eqn:sufficientdecrease} will be satisfied for small \(\slen{}\), and the line search algorithm may make insufficient progress by underestimating an acceptable step length. In addition to the condition in \cref{eqn:sufficientdecrease}, a condition known as the \textit{Goldstein condition} \cite{GOLDSTEIN65}, ensures that the step length is not underestimated:
\begin{align}
 \parbox{5em}{\centering\textit{Goldstein\\ condition:}} &&  \Obj{\objv{\itr}+\slen{}\sdir{\itr}} & \geqslant\Obj{\objv{\itr}}+(1-\lsconst{1})\slen{}\big\langle\Grad{\itr}\big|\sdir{\itr}\big\rangle, && \lsconst{1}\in(0,\sfrac{1}{2})\label{eqn:goldsteinconditions}
\end{align}
where the bounds of \(\lsconst{1}\) are now more narrow than in \cref{eqn:sufficientdecrease}.

An alternative to \cref{eqn:goldsteinconditions} in avoiding underestimation of the step length is to account for curvature. This requires the gradient to be calculated at each step length calculation but should give a much better estimation of \cref{eqn:MinProblem}. The \textit{curvature condition} \cite{WOLFE69, WOLFE71} requires \(\slen{}\) to satisfy
\begin{align}
 \parbox{5em}{\centering\textit{Curvature\\ condition:}} && \big\langle\Gradop\!\!\Obj{\objv{\itr}+\slen{}\sdir{\itr}}\big|\sdir{\itr}\big\rangle\geqslant\lsconst{2}\big\langle\Grad{\itr}\big|\sdir{\itr}\big\rangle, && \lsconst{2}\in(\lsconst{1},1)\label{eqn:curvaturecondition}
\end{align}
This ensures that the slope found during step length iteration, \(\slen{j}\), is greater than the slope at \(\slen{}=1\). The constant \(\lsconst{2}\) is usually chosen to be relatively large for quasi-Newton and Newton methods \cite{NOCEDAL06}, \(\lsconst{2}\simeq 0.9\), but should be reduced for gradient descent methods, \(\lsconst{2}\simeq 0.1\). \cref{eqn:sufficientdecrease,eqn:curvaturecondition} are known as the \textit{Wolfe conditions} \cite{WOLFE69, WOLFE71}.

The \textit{sufficient decrease condition} stipulates that the decrease in \(\Obj{\objv{}}\) must be proportional to the step length and directional derivative. We can extend the \textit{weak curvature condition} of \cref{eqn:curvaturecondition} to ensure that the step length is in the region of a local minimum, ensuring the derivative in \cref{eqn:curvaturecondition} is not too positive, by using a \textit{strong curvature condition} \cite{WOLFE69, WOLFE71}:
\begin{align}
 \parbox{6em}{\centering\textit{Strong Wolfe\\ condition:}} && \Big|\big\langle\Gradop\!\!\Obj{\objv{\itr}+\slen{}\sdir{\itr}}\big|\sdir{\itr}\big\rangle\Big|\geqslant\lsconst{2}\Big|\big\langle\Grad{\itr}\big|\sdir{\itr}\big\rangle\Big|, && \lsconst{2}\in(\lsconst{1},1)\label{eqn:strongcurvaturecondition}
\end{align}
This equation, together with \cref{eqn:sufficientdecrease}, are known collectively as the \textit{strong Wolfe conditions}. These conditions are particularly effective when used in a quasi-Newton method from \cref{Section:QuasiNewton}.

\subsubsection{Backtracking line search}

Perhaps the most simple line search method is to iteratively reduce the step length by a predefined factor until \cref{eqn:sufficientdecrease} is satisfied. This algorithm is called the \textit{backtracking line search}, and is useful when gradient calculations are accurate but expensive. The iterative reduction of \(\slen{}\) follows
\begin{align}
&& \slen{j+1}=\beta\slen{j} && \text{while} &&\Obj{\slen{j}\sdir{\itr}}\geqslant\lsconst{1}\big\langle\Grad{\itr}\big|\sdir{\itr}\big\rangle
\end{align}
where \(\beta\in(0,1)\), continuing until \cref{eqn:sufficientdecrease} is satisfied, when \(\slen{}\rightarrow\slen{j}\).The initial step length is chosen as the Newton step, \(\slen{0}=1\), for quasi-Newton and Newton methods. The initial step length can be chosen to be larger for gradient methods and sometimes quasi-Newton methods to avoid small steps at the initial iterations of an optimisation, far from an extremum\footnote{A suggestion in \cite{NOCEDAL06} is to set \(\slen{0}=\frac{1}{\big\|\Grad{\itr}\big\|_{\infty}^{}}\)}.

\subsubsection{Bracketing \& sectioning line search}\label{Section:BracketSection}

A bracketing and sectioning line search starts with \textit{bracketing phase} which contains an interval of acceptable points, $[a,b]$. Once the bracket is identified, a \textit{sectioning phase} follows dividing the bracket to give a sequence of brackets, \([a_i,b_i]\), where their length tends to zero \cite{FLETCHER87}. 

Assuming gradients are available and they are cheap to calculate\footnote{It will be outlined later in this chapter, in \cref{Section:FidelityGradient}, that gradients needed in optimal control used in this thesis are particularly cheap to calculate, following at little extra cost to calculating the required objective function value.}, the bracketing and sectioning line search can use the Wolfe conditions in \cref{eqn:sufficientdecrease,eqn:curvaturecondition,eqn:strongcurvaturecondition}.

The initial step in the bracketing phase would be choosing an initial step length, usually this is the Newton-step $\slen{0}=1$ as in the backtracking line search. Assuming a lower bound on an acceptable point, \(J_{\min}^{}\), and a line search is restricted to the interval \((0,\mu]\) where
\begin{equation}
 \mu=\frac{J_{\min}^{}-\Obj{\objv{}}}{\lsconst{1}\Grad{}}
\end{equation}
a general procedure starting from \(\Obj{\objv{}+\slen{0}\sdir{}}\) at an iteration (the optimisation iteration \(\itr\) does not change during a line search) \(\itr\), set out in \cite{FLETCHER87} is

\begin{enumerate}
 \item Set \(j=1\).
 \item Calculate \(\Obj{\objv{}+\slen{j}\sdir{}}\), where \(0<\slen{j}\leqslant\mu\).\label{b2}
 \item If \(\Obj{\objv{}+\slen{j}\sdir{}}\leqslant J_{\min}^{}\),\\\textbf{terminate} with exit message \textit{Boundary $J_{\min}^{}$ has been reached}.
 \item If \(\Obj{\objv{}+\slen{j}\sdir{}}>\Obj{\objv{}+\sdir{}} + \slen{j}\Grad{}\) or \(\Obj{\objv{}+\slen{j}\sdir{}}\geqslant \Obj{\objv{}+\slen{j-1}\sdir{}}\) go to \cref{b4a}, else go to \cref{b5}.
 \begin{enumerate}
  \item Then \(a_i\rightarrow\slen{j-1}\) and \(b_i\rightarrow\slen{j}\), \\ Bracket located -- case 1 (Wolfe conditions).\\ Continue to \textbf{Sectioning phase}.\label{b4a}
 \end{enumerate}
 \item Calculate \(\Gradop\!\!\Obj{\objv{}+\slen{j}\sdir{}}\).\label{b5}
 \item If \(\big|\Gradop\!\!\Obj{\objv{}+\slen{j}\sdir{}}\big|\leqslant-\lsconst{2}\Grad{}\), \textbf{terminate}.
 \item If \(\big|\Gradop\!\!\Obj{\objv{}+\slen{j}\sdir{}}\big|\geqslant 0\), go to \cref{b7a}, else go to \cref{b8}.
 \begin{enumerate}
  \item Then \(a_i\rightarrow\slen{j}\) and \(b_i\rightarrow\slen{j-1}\), \\ Bracket located -- case 2 (Wolfe conditions).\\ Continue to \textbf{Sectioning phase}.\label{b7a}
 \end{enumerate}
 \item If \(\mu\leqslant 2\slen{j}-\slen{j-1}\), go to \cref{b8a}, else go to \cref{b8b}. \label{b8}
 \begin{enumerate}
  \item Then \(\slen{j+1}\rightarrow\mu\).\label{b8a}
  \item Else choose \(\slen{j+1}\in\{2\slen{j}-\slen{j-1},\min{(\mu,\slen{j}+\tau_1(\slen{j}-\slen{j-1}))}\}\).\\Continue to \textbf{Interpolation phase}\label{b8b}
 \end{enumerate}
 \item \(j=j+1\), return to \cref{b2}.
\end{enumerate}

Once a bracket has been identified the sectioning phase can begin:
\begin{enumerate}
 \item Set \(k=j\).
 \item Choose \(\slen{k}\in\{a_k+\tau_2(b_k-a_k),b_k-\tau_3(b_k-a_k)\}\).\\Continue to \textbf{Interpolation phase}\label{s2}
 \item Calculate \(\Obj{\objv{}+\slen{k}\sdir{}}\).
 \item If \(\Obj{\objv{}\!+\!\slen{k}\sdir{}}>\Obj{\objv{}\!+\!\sdir{}}\!+\!\lsconst{1}\slen{k}\Grad{}\) or \(\Obj{\objv{}\!+\!\slen{k}\sdir{}}\geqslant \Obj{\objv{}\!+\!a_k\sdir{}}\), go to \cref{s4a}, else go to \cref{s5}.
 \begin{enumerate}
  \item \(a_{k+1}\rightarrow a_k\) and \(b_{k+1}\rightarrow\slen{k}\), go to \cref{s9}.\label{s4a}
 \end{enumerate}
 \item Calculate \(\Gradop\!\!\Obj{\objv{}+\slen{k}\sdir{}}\).\label{s5}
 \item If \(\big|\Gradop\!\!\Obj{\objv{}+\slen{k}\sdir{}}\big|\leqslant-\lsconst{2}\Grad{}\), \\\textbf{terminate} with an acceptable step length.
 \item \(a_{k+1}\rightarrow \slen{k}\), go to \cref{s9}.
 \item If \(b_k-a_k)\Gradop\!\!\Obj{\objv{}+\slen{k}\sdir{}}\geqslant 0\), go to \cref{s8a}, else go to \cref{s8b}.
 \begin{enumerate}
  \item \(b_{k+1}\rightarrow a_k\).\label{s8a}
  \item \(b_{k+1}\rightarrow b_k\).\label{s8b}
 \end{enumerate}
 \item \(k=k+1\), return to \cref{s2}.\label{s9}
\end{enumerate}

The constants \(\tau_1>0\) and \(0<\tau_2<\tau_3\leqslant\sfrac{1}{2}\) are set as the following in \textit{Spinach}: Bracketing expansion factor \(\tau_1=3\); Left bracket contraction factor \(\tau_2=0.1\); Right bracket contraction factor \(\tau_3=0.5\).

It is common to include an \textit{interpolation phase}, \cref{b8b,s2}, particularly within the sectioning phase, to find a point close to a local minimiser. Here, a quadratic or cubic polynomial if fitted to data from the acceptable bracket\footnote{The work in this thesis uses a cubic polynomial interpolation.}.

\subsection[Comparison of Newton-type methods]{Comparison of Newton-type methods}\label{Section:CompareNewtonTypes}

\begin{figure}
\centering{\includegraphics{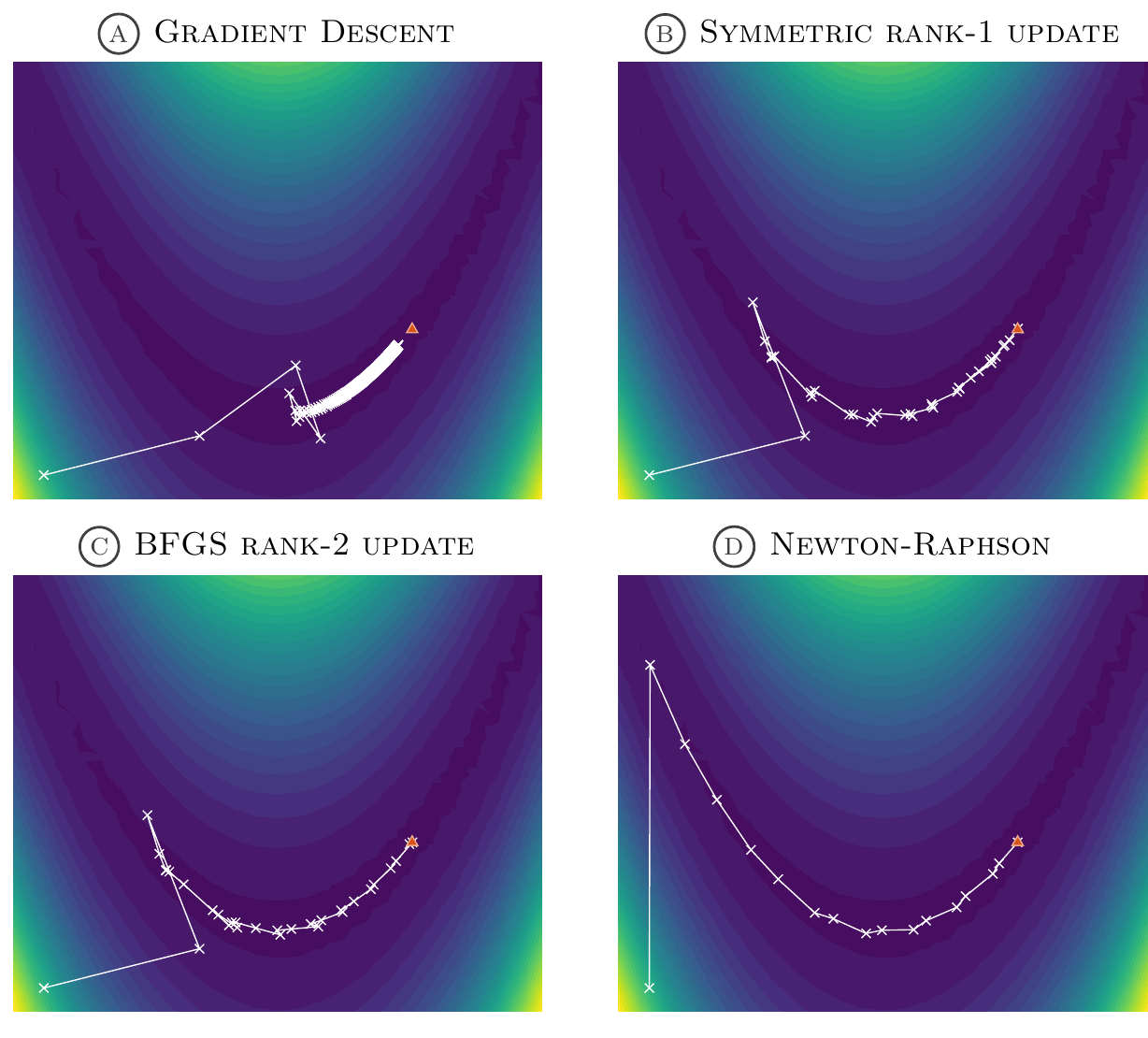}}
\caption[Comparison of optimisation methods on the Rosenbrock function]{Comparison of gradient based optimisation methods on the Rosenbrock function \cite{ROSENBROCK60}, with a backtracking line-search method. The optimisation algorithms terminate when $f(x)<5\times 10^{-2}$. \plotlabel{a} -- gradient descent method, taking $376$ iterations with $5101$ functional evaluations. \plotlabel{b} -- symmetric rank-1 update, taking $33$ iterations with $185$ functional evaluations. \plotlabel{c} -- \textsc{bfgs} rank-2 update, taking $29$ iterations with $78$ functional evaluations. \plotlabel{d} -- Newton-Raphson method, taking $16$ iterations with $43$ functional evaluations.}\label{Fig:Rosenbrock}
\end{figure}

A common test for optimisation methods is on the Rosenbrock function \((1-x)_{}^2 + 100(y-x_{}^2)_{}^2\) \cite{ROSENBROCK60}. The function has one minimum at \((1,1)\), but the difficulty in its minimisation is that the valley is very flat. Although it is easy to find the valley, once there, optimisation usually converges slowly to the point \((1,1)\). 

\cref{Fig:Rosenbrock} compares the convergence of the gradient descent method of \cref{eqn:grad_update_min}, the Newton-Raphson method of \cref{eqn:Newton_update}, and two quasi-Newton method of symmetric rank-1 update, \cref{eqn:inv_sr1}, and the \textsc{bfgs} rank-2 update in \cref{eqn:inv_bfgs}.

It is clear that the gradient descent method performs badly, even when using an analytical gradient. The convergence trajectory is seen zig-zagging in small steps on either side of the valley. It is clear that the gradient used in this way is quite ineffective. The Newton-Raphson method performs the best, with an analytic Hessian to use local curvature information in finding the best route down the valley. The quasi-Newton methods also perform quite well, starting from a gradient descent iteration, then being able to approximate the curvature within the valley to find the best route, although taking smaller steps compared to the Newton-Raphson method. Comparing the two quasi-Newton methods, the rank-1 and rank-2 updates both have a similar trajectory iteration count, however, the rank-1 update puts a greater load on the line search algorithm -- needing more than twice the amount of gradient evaluations to get similar curvature information compared to the rank-2 update.

\section[Gradient Ascent Pulse Engineering]{Gradient Ascent\linebreak Pulse Engineering}\label{Section:GRAPE}

The task of taking a quantum system from one state to another to a specified accuracy with minimal expenditure of time and energy, with the emphasis on the word \textit{minimal}, is increasingly important in physics and engineering \cite{TOSNER09,KOBZAR04,KHANEJA05,NIMBALKAR12}.

Optimal control can be thought of as an algorithm; there is a start and a stop. in the language of physics; a dynamic system with a initial state, \(\big|\init\big\rangle\), and a desired destination state, \(\big|\targ\big\rangle\). A solution should be found with a minimum of effort, an optimal amount of effort. In forming a problem for use in a numerical optimisation algorithm, the first step is to create a system model. Within this model the controllable parameters are exposed then fed into an optimisation algorithm.

\subsection{Bilinear systems}\label{Section:Bilinear}

The Hamiltonian of a practical \textsc{nmr} quantum system can be split into two parts \cite{KHANEJA01}: that which is beyond the control of the instrument and a \textit{control} part that measurement instrumentation can vary within certain limits.
\begin{equation}
\HamHH(t)=\HamHH_0^{} + i\hhat{\mathcal{R}} + \sum\limits_{k=1}^K\ctrl{k}(t)\HamHH_k^{}\label{eqn:HamSep}
\end{equation}
\(\HamHH_0^{}\) is the Hamiltonian describing the internal part of the system called the \textit{drift} or \textit{free Hamiltonian}, and \(\HamHH_k^{}\) are the \(K\) parts of the Hamiltonian that can be changed externally, called the \textit{controls}, and \(\ctrl{k}(t)\) are their time-dependent coefficients. In the case of magnetic resonance the instrumentally controllable subsystem consists of radio frequency/microwave control fields defined in \cref{Section:ControlOps}. \(\hhat{\mathcal{R}}\) is the relaxation superoperator characterising dissipation from the system, and has been separated from \(\HamHH_0^{}\) for convenience \cite{REDFIELD57,GOLDMAN01,KUPROV11}.

Systems with equations of motion that are linear both in the drift and controls \cite{AIZU63}, as in \cref{eqn:HamSep}, are known as \textit{bilinear control systems} \cite{SONTAG98,SCHULTE_HERBRUEGGEN10,ZEIER11,SCHULTE_HERBRUEGGEN12} in classical control theory. Optimal control of the bilinear system in \cref{eqn:HamSep} is the task of finding the time-dependent control amplitudes \(\ctrl{k}(t)\) -- controlling the system with their corresponding controls \(\HamHH_k^{}\), while under the influence of the time-independent drift \(\HamHH_0^{}\)\footnote{Relaxation will be neglected in many of the derivations in this thesis. For convenience, it will be considered an uncontrollable \textit{damping term}. Although this is only an approximation, the control of open systems is beyond the scope of this thesis.} -- taking the system from a defined initial state, \(\big|\init\big\rangle\), to a desired final state,  \(\big|\targ\big\rangle\), a time \(T\) later.

\subsection{Piecewise constant approximation}\label{Section:PiecewiseConstant}

\begin{figure}
\centering{\includegraphics{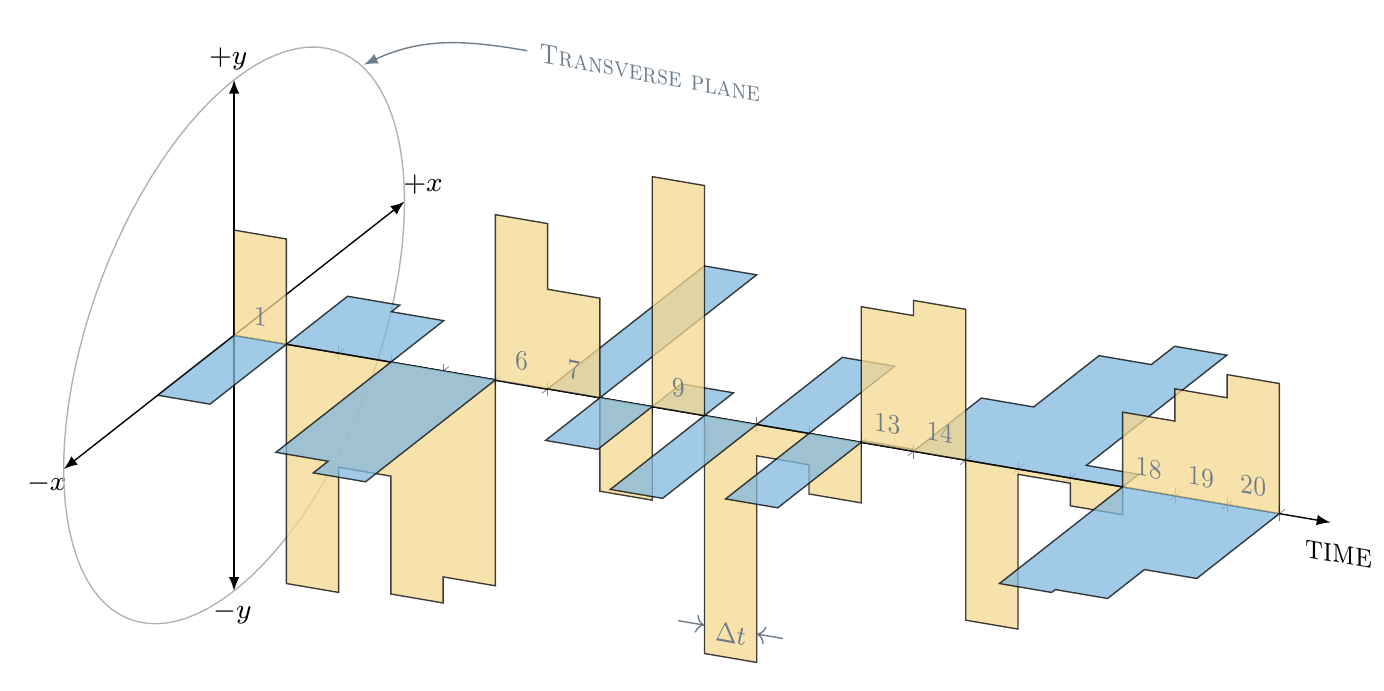}}
\caption[Piecewise-constant approximation]{Digram of the piecewise-constant approximation, for 20 equal time slices of length \(\Delta t\). Pulses are performed simultaneously in the \(x\) and \(y\) directions of the transverse magnetisation plane.}\label{fig:PiecewiseConstant}
\end{figure}

The evolution of a quantum ensemble is governed by the Liouville-von Neumann equation (\cref{eqn:liouvillvonneumann}). Using a column-wise vector representation of the density operator, \(\hat{\state}\), the Liouville-von Neumann equation has the general solution from \cref{eqn:liouvillvonneumann_rho}. With the separated Hamiltonian formalism of \cref{eqn:HamSep}, this gives
\begin{equation}
 \big|{\rho}(T)\big\rangle=\exp_{(\text{o})}{\!\!\Bigg[\!-i\!\!\int\limits_0^{T}\!\!\Big(\HamHH_0^{}+i\hhat{\mathcal{R}}+\sum\limits_{k=1}^K\ctrl{k}(t)\HamHH_k^{}\Big)\dd t\Bigg]}\big|\init\big\rangle\label{eqn:rhoT}
\end{equation}

An approximation of the integral in \cref{eqn:rhoT} can be made by assuming that \(\ctrl{k}(t)\) are piecewise-constant \cite{CONOLLY86,KHANEJA05}, depicted in \cref{fig:PiecewiseConstant}; \textit{slicing} the problem into discrete time intervals to obtain control sequences
\begin{align}
 && \ctrl{k}(t) \longrightarrow \ctrlv{k}\triangleq\begin{pmatrix}
                                        \ctrl{k}(t_1^{}) & \ctrl{k}(t_2^{}) & \cdots & \ctrl{k}(t_N^{})
                                       \end{pmatrix}, && t_1^{}<t_2^{}<\cdots<t_N^{}
\end{align}
where the individual elements \(\ctrl{k}(t_n^{})=\ctrl{k,n}\) are treated as continuous parameters. With this approximation, the evolution of the system from a state \(\state(t_0^{})\) to \(\state(t_N^{})\) is
\begin{equation}
  \big|\state(t_N^{})\big\rangle=\Bigg[\prod_{n=1}^N\exp{\bigg[-i\Big(\HamHH_0^{}+i\hhat{\mathcal{R}}+\sum\limits_{k=1}^K\ctrl{k,n}\HamHH_k^{}\Big)\Delta t\bigg]}\Bigg]\big|\state(t_0^{})\big\rangle\label{eqn:rhoTn}
\end{equation}
where the product must be \textit{time-ordered} and the time interval \(\Delta t\). Time intervals considered in the remainder of this thesis will be the same for each time slice, although this is for convenience and not a restriction of the piecewise constant approximation.

Although the physical arrangement of the control vector \(\ctrlv{}\) is as $K$ rows of control sequences vectors, the design of an optimisation algorithm is simpler with a single vector representing the many variables of its scale objective function. For this reason, the control vector is concatenated to column vector when used within the optimiser:
\begin{equation}
 \ctrlv{}=\begin{pmatrix}
  \ctrl{1,1} & \ctrl{1,2} & \cdots & \ctrl{1,N}\\
  \ctrl{2,1} & \ctrl{2,2} & \cdots & \ctrl{2,N}\\
  \vdots & \vdots & \ddots& \vdots\\
  \ctrl{K,1} & \ctrl{K,2} & \cdots & \ctrl{K,N}\\
 \end{pmatrix}
\longmapsto
 \begin{pmatrix}
  \ctrl{1,1}\\
  \ctrl{2,1}\\
  \vdots\\
  \ctrl{K,1}\\\hdashline[1pt/1.5pt]
  \ctrl{1,2}\\
  \ctrl{2,2}\\
  \vdots\\
  \ctrl{K,2}\\\hdashline[1pt/1.5pt]
  \\
  \vdots\\
  \\\hdashline[1pt/1.5pt]
  \ctrl{1,N}\\
  \ctrl{2,N}\\
  \vdots\\
  \ctrl{K,N}
 \end{pmatrix}\label{eqn:ctrl_arrange}
\end{equation}
Although the order \((k,n)\) is used here, the ordering is arbitrary and could equally be ordered \((n,k)\) as long as the vectorisation of the control amplitudes is consistent within the optimiser.

\subsection{The Pontryagin maximum principle}\label{Section:Pontryagin}

To investigate the gradient that would be needed for numerical optimisation, in \cref{Section:NumOptim}, a principle from optimal control theory needs to be imposed, called the \textit{Pontryagin maximum principle} \cite{PONTRYAGIN64,GAMKRELIDZE78}. It follows from the classical Lagrangian mechanics \cite{LANDAU60}, which introduces a \textit{Lagrangian}, \(\mathcal{L}(t,\state(t),\dot{\state}(t))\), -- a function that characterises the dynamics a system. In this case, the Lagrangian \(\mathcal{L}\) is defined by the variables of time \(t\), the state of the system \(\state(t)\), and the equation of motion of the state \(\dot{\state}(t)\). Lagrangian mechanics is the calculus of variations, and so too is optimal control.

The Pontryagin maximum principle sets an optimal control problem in a similar formulation, with the modification of replacing the equation of motion with the system controls, giving the Lagrangian as \(\mathcal{L}(t,\state(t),\ctrl{k}(t))\) and the equation of motion depends on the controls. In addition to its Lagrangian formulation, optimal control is also the application of mathematical optimisation, and its solutions are set out by the conditions of Pontryagin maximum principle.

Physical limitations of the instrumentation and constraints on control variable are part of the optimal control problem, and should be formulated within the problem \cite{TABAK71}. In the context of magnetic resonance, the optimal control problem is set out as \cite{SKINNER03,FISHER_Thesis,LAPERT10}:
\begin{align}
 \max_{\ctrl{}(t)}{\big(\mathcal{L}\big)}, && \mathcal{L}\big(\ctrl{}(t)\big)\triangleq \Fid{}\big(\state(T)\big) + \int\limits_0^T\Pen\big(t,\state(t),\ctrl{}(t)\big)\dd t &&\label{eqn:OCProblem}
\end{align}
where the scalar function \(\Fid{}\big(\state(T)\big)\) is a \textit{terminal cost} depending only on the final state of the system at a time \(T\), and the integral of the scalar function \(\Pen\big(t,\state(t),\ctrl{}(t)\big)\) is a \textit{running cost} which depends on the controls and the state of the system during the time interval \([0,T]\). The controls, \(\ctrl{}(t)\), affect the equation of motion of the system -- the Liouville-von Neumann equation of \cref{eqn:liouvillvonneumannsuperoperator}
\begin{equation}
 \frac{\dd}{\dd t}\big|\state(t)\big\rangle=-i\Bigg[\int\limits_0^{T}\!\!\HamHH\big(\ctrl{}(t)\big)\dd t\Bigg]\big|\init\big\rangle
\end{equation}
where \(\HamHH(t)\) can be separated as a bilinear system in \cref{eqn:HamSep}.

In analogy to \textit{Lagrange multipliers} of Lagrangian mechanics \cite{LANDAU60} a variable called the \textit{adjoint state} (also known as the \textit{costate}) is introduced to solve the optimal control problem. Without exploring the dry depths of Lagrangian mechanics, the \textit{adjoint state} will be introduced by example.

The formal mathematical description of the adjoint states will follow, but it is useful to show an interpretation of their physical significance with an example of state-to-state transfer in \cref{Fig:AdjointState}. The state of the system at each time slice forms a state trajectory over a time interval \([0,T]\), containing information of the basis state population at each time increment. If propagated from a defined initial state \(\big|\init\big\rangle\) with \cref{eqn:propagatorN}, the control pulses form a unique trajectory of the system under the influence of the drift.

\begin{figure}
\centering{\includegraphics{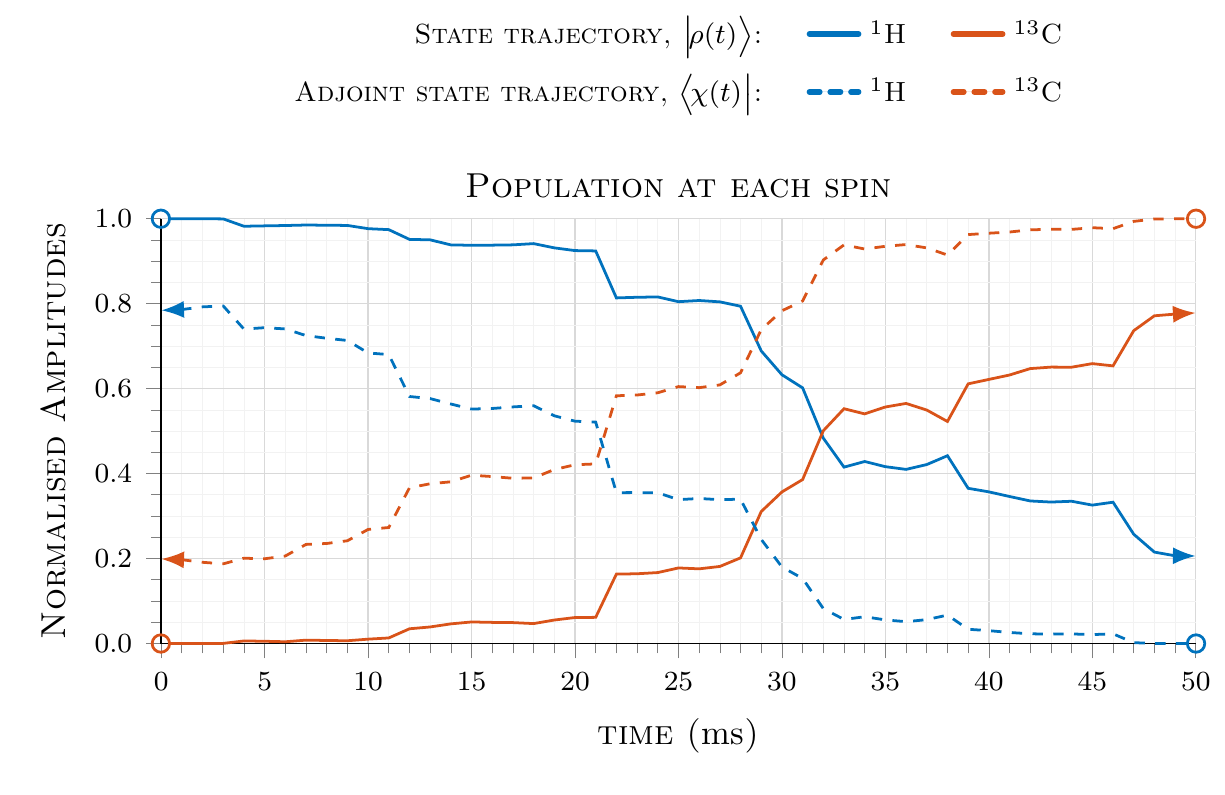}}
 \caption[A representation of the adjoint state]{An example of two state vector trajectories, and the corresponding adjoint state trajectories. The state vector propagates forward in time from the initial condition, under the influence of drift and calculated controls. The adjoint state propagates backwards in time from the desired state, under the influence of the time-reversed drift and controls. This example shows polarisation transfer from a hydrogen spin to a coupled carbon spin (\(140\)~Hz). The two trajectories show polarisation local to each of the spins. In this case, nearly 80\% of the polarisation has been transferred with calculated control pulses (not shown).}\label{Fig:AdjointState}
\end{figure}

\cref{Fig:AdjointState} shows a projection of the state of the system at each time interval onto a representation of magnetisation population local to each spin. The aim is to transfer magnetisation from one spin to the other in a coupled two-spin system with calculated control pulses. It is easy to see the population gradually transfers from one spin to the other over the time interval \([0,T]\).

The state trajectory is the trajectory formed by propagating the system from a defined initial state under the effect of control pulses. The \textit{adjoint state} also forms a trajectory, propagated under the influence of the same drift and the same controls. However, the adjoint state is a propagation backwards in time from the desired final state of the system. The \textit{adjoint state} trajectory is the control problem in reverse.

When an optimal trajectory is found, the state trajectory should end at the desired final state, and the adjoint state should start at the defined initial state of the system. This can be seen in \cref{Fig:AdjointState}, where an optimal trajectory has not yet been found; there is a population difference between the adjoint state trajectory and the state trajectory at times \(t=0\) and at times \(t=T\). The adjoint state \(\big|\Astate(T)\big\rangle=\big|\targ\big\rangle\), and the state \(\big|\state(0)\big\rangle=\big|\init\big\rangle\) form the boundary conditions for the optimal control problem.

The \textit{Pontryagin maximum principle} requires a number of conditions to be met for an optimal trajectory to be found. Just as in the area of numerical optimisation of \cref{Section:NumOptim}, necessary conditions of the first and second derivatives must be met (\cref{eqn:necessary1,eqn:necessary2}), in this case applied to the function called a \textit{pseudo-Hamiltonian}\footnote{The term \textit{pseudo-Hamiltonian} is introduced to distinguish it from Hamiltonians of quantum mechanics. The \textit{pseudo-Hamiltonian} is a canonical system and therefore it is a Hamiltonian function \cite{TABAK71,GAMKRELIDZE78}} \cite{FISHER_Thesis,LAPERT10}, derived from \cref{eqn:OCProblem} and the \textit{adjoint state} \cite{PONTRYAGIN64,GAMKRELIDZE78}:
\begin{equation}
\mathcal{H}_\textsc{p}^{}\big(\state(t),\ctrl{}(t)\big)\triangleq\big\langle\Astate(t)\big|-i\Bigg[\int\limits_0^{T}\HamHH\big(\ctrl{}(t)\big)\dd t\Bigg]\big|\state(t)\big\rangle + \Pen\big(\state(t),\ctrl{}(t)\big)
\end{equation}
The first and second order optimality conditions of \cref{eqn:necessary1,eqn:necessary2}, now become
\begin{subequations}
\begin{gather}
\frac{\partial \mathcal{H}_\textsc{p}^{}}{\partial \ctrl{}(t)} = 0\\
\frac{\partial^2 \mathcal{H}_\textsc{p}^{}}{\partial \ctrl{}(t)^2}> 0
\end{gather}
\end{subequations}
in the vicinity of a maximum. The \textit{Pontryagin maximum principle} \cite{PONTRYAGIN64} also requires consideration of the relationship between the state trajectory and the adjoint state trajectory. These give the further necessary conditions
\begin{subequations}
\begin{gather}
 \frac{\dd}{\dd t}\big|\Astate(t)\big\rangle=-\frac{\partial\mathcal{H}_\textsc{p}^{}}{\partial\,|\state(t)\rangle}\\
 \frac{\dd}{\dd t}\big|\state(t)\big\rangle=\frac{\partial\mathcal{H}_\textsc{p}^{}}{\partial\,\langle\Astate(t)|}
\end{gather}
\end{subequations}
which formally sets out the conclusions made in discussion of \cref{Fig:AdjointState} above: as the state trajectory approaches as maximum, the state trajectory as a time \(t=T\) should completely overlap with the desired target state, and that the adjoint state at a time \(t=0\) should overlap completely with the initial state of the system.

The formulation of \cref{eqn:OCProblem} admits three types of optimal control problem \cite{TABAK71,DALESSANDRO07}: \textit{The Lagrange problem} has only a running cost; \textit{The Mayer problem} has only a terminal cost; \textit{The Bolza problem} both the running cost and the terminal cost are included. The Mayer problem, where the final state of the system forms the optimisation problem, will be the main consideration of this thesis until \cref{Chapter:Penalty}, when control amplitude penalties are included in a Bolza problem.

For an unconstrained optimal control problem \(\Pen=0\) \cite{FISHER_Thesis} and the optimal control problem reduces to one of the Mayer problem. The \textit{terminal cost} \(\Fid{}\) in \cref{eqn:OCProblem} will be the focus of the next section on \textit{fidelity} measures.

\subsection{Fidelity measures}\label{Section:Fidelity}

Optimal solutions can be found numerically, by optimising \textit{fidelity} -- the overlap between the final state of the system and the desired destination state \cite{GLASER98,MADAY03,KHANEJA05}. The optimum of the fidelity functional $\Fid(c)$ with respect to the control sequences \(\ctrl{k}(t)\) in \cref{eqn:HamSep}, subjected to experimental constraints of feasible amplitudes and frequencies, is calculated with \textit{optimal control theory} \cite{PONTRYAGIN64,LEWIS12}. For Hermitian states, the fidelity function is
\begin{align}
&& \Fid_0\triangleq\big\langle  \targ \big| \state(T)\big\rangle && \Fid_0\in[-1,+1]\label{eqn:fidelity0}
\end{align}
where angular bracket denotes column-wise vectorization of the corresponding density matrix, \(\big| \state(T)\big\rangle\) is the final state, \(\big| \targ\big\rangle\) is the desired destination state. Finding the optimal value of this inner product, is that of finding its maximum.

Non-Hermitian states can be accounted for with two further definitions of the fidelity functional, either by using only the real part of the overlap in \cref{eqn:fidelity0}
\begin{align}
&& \Fid_1\triangleq\Re\big\langle  \targ \big| \state(T)\big\rangle && \Fid_1\in[-1,+1]\label{eqn:fidelity1}
\end{align}
 or by using the modulus square\footnote{The modulus square of a complex valued vector is \(\big|\mathbf{v}\big|^2_{}=\mathbf{v}\mathbf{v}^{\ast}\), where \(^{\ast}\) denotes the complex conjugate.} of the overlap in \cref{eqn:fidelity0}
\begin{align}
&& \Fid_2\triangleq\Big|\big\langle  \targ \big| \state(T)\big\rangle_{}^{\vphantom{2}}\Big|_{}^2 && \Fid_2\in[0,+1]\label{eqn:fidelity2}
\end{align}
The definition of \(\Fid_1\) allows the optimisation to span a space from that parallel to, orthogonal to, and anti-parallel to $ \targ$. The definition of \(\Fid_2\) is phase insensitive. For the fidelity to have this physical meaning, the initial state and the system, \(\big|\init\big\rangle\), and the desired target state, \(\big|\targ\big\rangle\), should be be normalised.

In addition to these state-to-state definitions of fidelity, similar measures can be designed for multi-state mapping:
\begin{equation}
 \Fid_{\textsc{ur}}=\frac{1}{q_{\max}^{}}\Re\Big(\sum\limits_{q=1}^{q_{\max}^{}}\big\langle  \targ_q^{} \big| \state_q^{}(T)\big\rangle \Big)\label{eqn:fidelityUR}
\end{equation}
where \(\targ_k^{}\) define the desired state propagated in time from corresponding initial states \(\state_k^{}(0)\) respectively. When \cref{eqn:fidelityUR} is averaged over 3 state-to-state problems, together completely characterising a rotation of the state in Cartesian coordinates, the optimal control problem becomes a universal rotation in 3D e.g.
\begin{equation}
\begin{matrix}
  \big|\state_1^{}(0)\big\rangle=+\big|L_z^{}\big\rangle &\longrightarrow &-\big|L_x^{}\big\rangle=\big|\targ_1^{}\big\rangle\\
  \big|\state_2^{}(0)\big\rangle=+\big|L_x^{}\big\rangle &\longrightarrow &-\big|L_z^{}\big\rangle=\big|\targ_2^{}\big\rangle\\
  \big|\state_3^{}(0)\big\rangle=+\big|L_y^{}\big\rangle &\longrightarrow &-\big|L_y^{}\big\rangle=\big|\targ_3^{}\big\rangle\nonumber
\end{matrix}
\end{equation}

These 3 state-to-state problems can restated with a fidelity definition to find an \textit{effective propagator} \cite{KHANEJA05} or an \textit{effective Hamiltonian} \cite{TOSNER06} over the time period of, and subject to, the control sequence. A measure of the distance between two matrices can be defined by the the desired propagation over the total time period \([0,T]\).

The fidelity definition for this desired \textit{effective propagator} is
\begin{align}
&& \Fid_3\triangleq\Re\big\langle  \hhat{\mathcal{U}}_D^{} \big| \Prop\,(T)\big\rangle && \Fid_3\in[-1,+1]\label{eqn:fidelity3}
\end{align}
where \(\hhat{\mathcal{U}}_D^{}=\exp{\big(-i\hhat{\mathcal{H}}_D^{}T\big)}\) is the desired effective propagator for the effective Hamiltonian \(\hhat{\mathcal{H}}_D^{}\). In analogy to \cref{eqn:fidelity2}, the modulus square of this overlap is
\begin{align}
&& \Fid_4\triangleq\Big|\big\langle  \hhat{\mathcal{U}}_D^{} \big| \Prop\,(T)\big\rangle_{}^{\vphantom{2}}\Big|_{}^2 && \Fid_4\in[0,+1]\label{eqn:fidelity4}
\end{align}
In \cref{eqn:fidelity3,eqn:fidelity4}, the overlap of two propagators is defined by the Frobenius inner product as the trace of the operator product:
\begin{equation}
 \big\langle  \hhat{\mathcal{U}}_D^{} \big| \Prop\,(T)\big\rangle\triangleq\Tr\big(\hhat{\mathcal{U}}_D^{\dagger}\Prop\,(T)\big)
\end{equation}
Although not used in this thesis, the definitions of fidelity in \cref{eqn:fidelity3,eqn:fidelity4} are reported to avoid local minima which \cref{eqn:fidelityUR} encounter \cite{KHANEJA05,TOSNER06}.

\subsection{Fidelity gradient}\label{Section:FidelityGradient}

To make use of Newton-type methods of \cref{Section:NumOptim}, a gradient of the fidelity functional with respect to the control amplitudes is required. The first step to achieve this is to expand the fidelity definitions in \cref{eqn:fidelity0,eqn:fidelity1,eqn:fidelity2}. As a function of control amplitudes, the fidelity definitions become
\begin{subequations}
\begin{gather}
 \Fid_0(\ctrlv{n})=\big\langle  \targ \big|\Prop_N^{}\Prop_{N-1}^{}\cdots\Prop_2^{}\Prop_1^{} \big| \state_0^{}\big\rangle\\
 \Fid_1(\ctrlv{n})=\Re\big\langle  \targ \big|\Prop_N^{}\Prop_{N-1}^{}\cdots\Prop_2^{}\Prop_1^{} \big| \state_0^{}\big\rangle\\
 \Fid_2(\ctrlv{n})=\Big|\big\langle  \targ \big|\Prop_N^{}\Prop_{N-1}^{}\cdots\Prop_2^{}\Prop_1^{} \big| \state_0^{}\big\rangle_{}^{\vphantom{2}}\Big|_{}^2
\end{gather}
\end{subequations}
where the time propagator of \cref{eqn:propagator} is now piecewise constant
\begin{equation}
 \Prop_n^{}(\ctrlv{n})=\exp{\bigg[-i\Big(\HamHH_0^{}+i\hhat{\mathcal{R}}+\sum\limits_{k=1}^K\ctrl{k,n}\HamHH_k^{}\Big)\Delta t\bigg]}\label{eqn:propagatorN}
\end{equation}
where the definition of the propagator at each time slice, \(n\), is dependent on the \(K\) control amplitudes in the vector array
\begin{equation}
 \ctrlv{n}=\begin{pmatrix}
            \ctrl{1}(t_n^{}) \\ \ctrl{2}(t_n^{}) \\ \vdots \\ \ctrl{K}(t_n^{})
           \end{pmatrix}
\end{equation}
Once the controllable part of the Hamiltonian becomes parametrised, the variation ${\delta \Fid}/{\delta\HamHH_k^{}}$ becomes a gradient \(\DFid\) (derived in \cref{Section:FirstPropDirDerivs}) in the parameter space and the process of maximising \(\Fid\) becomes an instance of a non-linear optimisation problem for a continuous function in \cref{Section:NumOptim}.

Elements of the gradient are the derivative of the fidelity functional with respect to the control amplitudes \(\ctrl{k,n}\). The only dependence of the fidelity on the control amplitudes is in the propagator at that time slice, \(\Prop_n^{}\) -- calculation of the time propagator derivatives will be set out in \cref{Section:FirstPropDirDerivs}.

Once the \(K\times N\) propagator derivatives are calculated, they can be used to give the derivative of the fidelity functional with respect to the control amplitude. From the previous section \cref{Section:Pontryagin}, the fidelity at a time slice is the overlap, the inner product, of state of the system and its adjoint state at that time slice.

The state trajectory can be calculated from one forward-time propagation, and the adjoint trajectory from one backward-time propagation. The forward propagation from \(\state_0^{}\) to \(\state_n^{}\) is
\begin{equation}
 \big|\state_n\big\rangle=\Prop_{n}^{}\cdots\Prop_2^{}\Prop_1^{} \big| \state_0^{}\big\rangle\label{eqn:fwdprop}
\end{equation}
and the backward propagation from \(\targ\) to \(\state_{n+1}^{}\) is
\begin{equation}
 \big|\Astate_n\big\rangle=\Prop_{n}^{\,\,\dagger}\cdots\Prop_{N-1}^{\,\,\dagger}\Prop_N^{\,\,\dagger} \big| \targ\big\rangle\label{eqn:bwdprop}
\end{equation}

Using \cref{eqn:fwdprop,eqn:bwdprop,eqn:OCProblem,eqn:propagatorN} in the piecewise constant approximation gives gradient elements of the fidelity functional, used for a gradient ascent method \cite{KHANEJA05} or a quasi-Newton method \cite{FOUQUIERES11}:
\begin{subequations}
\begin{gather}
 \DFid_0^{}(\ctrl{k,n})=\frac{\partial\Fid_0^{}}{\partial \ctrl{k,n}}=\Big\langle\Astate_{n+1}^{}\Big|\frac{\partial\Prop_n^{}}{\partial \ctrl{k,n}}\Big|\state_{n-1}^{}\Big\rangle\label{eqn:fidgrad0}\\
 \DFid_1^{}(\ctrl{k,n})=\frac{\partial\Fid_1^{}}{\partial \ctrl{k,n}}=\Re\Big\langle\Astate_{n+1}^{}\Big|\frac{\partial\Prop_n^{}}{\partial \ctrl{k,n}}\Big|\state_{n-1}^{}\Big\rangle\label{eqn:fidgrad1}\\
 \DFid_2^{}(\ctrl{k,n})=\frac{\partial\Fid_2^{}}{\partial \ctrl{k,n}}=2\Re\Big\langle\Astate_{n+1}^{}\Big|\frac{\partial\Prop_n^{}}{\partial \ctrl{k,n}}\Big|\state_{n-1}^{}\Big\rangle\Big\langle\state_{n-1}^{}\Big|\targ\Big\rangle\label{eqn:fidgrad2}
\end{gather}
\end{subequations}
A particular strength of the \textsc{grape} method is that the gradient of the fidelity functional has the same complexity as the fidelity functional \cite{SKINNER03,KHANEJA05,TOSNER09,GERSHENZON07,FOUQUIERES11} and reduces to finding the derivative of a propagator at each time slice and for each control. Calculation of these derivatives will be the focus of \cref{Chapter:AuxMat}.

The elements of the gradient should be arranged as a vector of derivatives with respect to control amplitudes, \(\ctrl{k,n}\), for a control \(k)\) at a time steps \(n\):
\begin{equation}
 \DFid(\ctrlv{})=\begin{pmatrix}
  \DFid(\ctrl{1,1}) & \DFid(\ctrl{1,2}) & \cdots & \DFid(\ctrl{1,N})\\
  \DFid(\ctrl{2,1}) & \DFid(\ctrl{2,2}) & \cdots & \DFid(\ctrl{2,N})\\
  \vdots & \vdots & \ddots& \vdots\\
  \DFid(\ctrl{K,1}) & \DFid(\ctrl{K,2}) & \cdots & \DFid(\ctrl{K,N})\\
 \end{pmatrix}
\longmapsto
 \begin{pmatrix}
  \DFid(\ctrl{1,1})\\
  \DFid(\ctrl{2,1})\\
  \vdots\\
  \DFid(\ctrl{K,1})\\ \hdashline[1pt/1.5pt]
  \DFid(\ctrl{1,2})\\
  \DFid(\ctrl{2,2})\\
  \vdots\\
  \DFid(\ctrl{K,2})\\\hdashline[1pt/1.5pt]
  \\
  \vdots\\
  \\\hdashline[1pt/1.5pt]
  \DFid(\ctrl{1,N})\\
  \DFid(\ctrl{2,N})\\
  \vdots\\
  \DFid(\ctrl{K,N})
 \end{pmatrix}\label{eqn:grad_arrange}
\end{equation}
and should be consistent with the arrangement of the control vector in \cref{eqn:ctrl_arrange}.

During the forward-time propagation of the state \(\state_0^{}\) with the propagators in \cref{eqn:propagatorN}, a trajectory of the system dynamics is produced. This can be stored in an array of state vectors, \(\begin{pmatrix}\state_0^{} & \state_1^{} & \cdots & \state_N^{}\end{pmatrix}\), and used for trajectory analysis of the controlled system dynamics \cite{KUPROV11}.

\subsection{Algorithm}\label{Section:GRAPEAlgo}

An effective optimisation should be preconditioned \cite{NOCEDAL06}, which means that the optimisation variables should not be too small or too large. For the \textsc{grape} method this means that the control amplitudes should be normalised, requiring a nominal power level to be separated from the control amplitudes
\begin{equation}
  \ctrlv{}\to\pwr\ctrlvn{}
\end{equation}
where \(\pwr\) is a control power level

The \textsc{grape} algorithm can be summarised in the following way:
\begin{enumerate}
 \item Define the system uncontrollable Hamiltonian \(\HamHH_0\), relaxation \(\hhat{\mathcal{R}}\), and control operators \(\HamHH_k\).
 \item Set an initial state \(\big|\init\big\rangle\), and target state \(\big|\targ\big\rangle\) of the system.
 \item Start with an initial guess\footnote{Can be a random guess} of control amplitudes, \(\ctrlvn{}\). and nominal power level \(\pwr\) with \(\max{|\ctrlvn{}}|\approx+1\) and \(|\min{\ctrlvn{}}|\approx0\).
 \item Call \textsc{grape} as the objective function and multiply the control amplitudes by the nominal power level.\label{grape:var2grape}
 \item Propagate the system forward in time from \(\big|\init\big\rangle\) with the time propagators in \cref{eqn:propagatorN} and store the states at each time slice in \cref{eqn:fwdprop}
 \item Propagate the system backward in time from \(\state_0^{}\) with the complex conjugate of the time propagators in \cref{eqn:propagatorN} and store the states at each time slice in \cref{eqn:bwdprop}
 \item Calculate gradient elements with \cref{eqn:fidgrad1} and the stored states from \cref{eqn:fwdprop,eqn:bwdprop}.
 \item Multiply the gradient vector with \(\pwr\) and send back to the optimiser.
 \item Update the control amplitudes with an update rule in \cref{eqn:update}.
 \item Test termination conditions, and an optimal solution hasn't been found, return to \cref{grape:var2grape}
\end{enumerate}

\chapter{Hessian Calculations} \label{Chapter:Hessian}

\begin{chapquote}{James Joyce, \textsc{Ulysses}}
``A man of genius makes no mistakes. His errors are volitional and are the portals of discovery.''
\end{chapquote}
\renewcommand*{\CurrentPath}{./Chapter_4}


A method of optimal control, called \textsc{grape}, was reviewed in \cref{Section:GRAPE} with application to magnetic resonance systems of quantum ensembles in \cref{Chapter:MagResTheory}. Applying rapidly converging numerical optimisation methods of \cref{Section:NumOptim} to optimal control algorithms will be the focus of this chapter. The Newton-Raphson version of \textsc{grape} (Newton-\textsc{grape}) has the potential to achieve quadratic convergence, compared to the superlinear convergence of state-of-the-art methods currently used in this area of optimal control, as outlined in \cref{Chapter:Introduction}.


Scientific instruments used in many applications of quantum theory have reached the limits of what is physically, legally or financially possible. Examples include power deposition safeguards in \textsc{mri} instruments \cite{DEMPSEY02}, sample heating thresholds in biomolecular \textsc{nmr} spectroscopy \cite{GORKOV07} and the steep dependence of the cost of superconducting magnets on the induction they generate \cite{IWASA06}. Some limits, such as the length of time a patient can be persuaded to stay inside an \textsc{mri} machine, are psychological \cite{SARJI98}, but in practice no less real. In these time-constrained situations faster convergence and greater code parallelisation, compared to other quantum control algorithms in the \textsc{grape} family, have the prospect of giving financial and medical benefits.


This chapter demonstrates that the Hessian of the \textsc{grape} fidelity functional is unusually cheap, having the same asymptotic complexity scaling as the functional itself. This leads to the possibility of using very efficient numerical optimisation techniques. In particular, the Newton-Raphson method with a regularised and well-conditioned Hessian (\cref{Section:Cholesky,Section:TRM,Section:RFO}) is shown to require fewer system trajectory evaluations than any other algorithm in the \textsc{grape} family. The methods presented in this chapter were published by Goodwin and Kuprov in \cite{GOODWIN16} and are implemented in \textit{Spinach} v2.0 and later\footnote{Available at \url{www.spindynamics.org}.} and results are presented in \cref{Section:ConvAnalysis}.

\section{Fidelity Hessian}\label{Section:FidelityHessian}

The Newton-Raphson method in \cref{Section:NewtonMethod} requires a Hessian matrix so that local curvature information can be used to improve a search direction compared to a gradient-based method. From \cref{eqn:Newtonstep}, the Newton-step search direction of an iteration \(\itr\) leads the update of the control amplitudes to maximise a fidelity function
\begin{equation}
 \ctrlv{}^{(\itr+1)}=\ctrlv{}^{(\itr)} + \slen{}\Big[\DDFid\big(\ctrlv{}^{(\itr)}\big)\Big]^{-1}\Big|\DFid\big(\ctrlv{}^{(\itr)}\big)\Big\rangle\label{eqn:Newtonupdate_grape}
\end{equation}
where the step length is \(\slen{}=1\) for a Newton-step, or with \(\slen{}\) calculated with a line search method, such as those in \cref{Section:Line_Search}.

\begin{figure}
\centering{\includegraphics{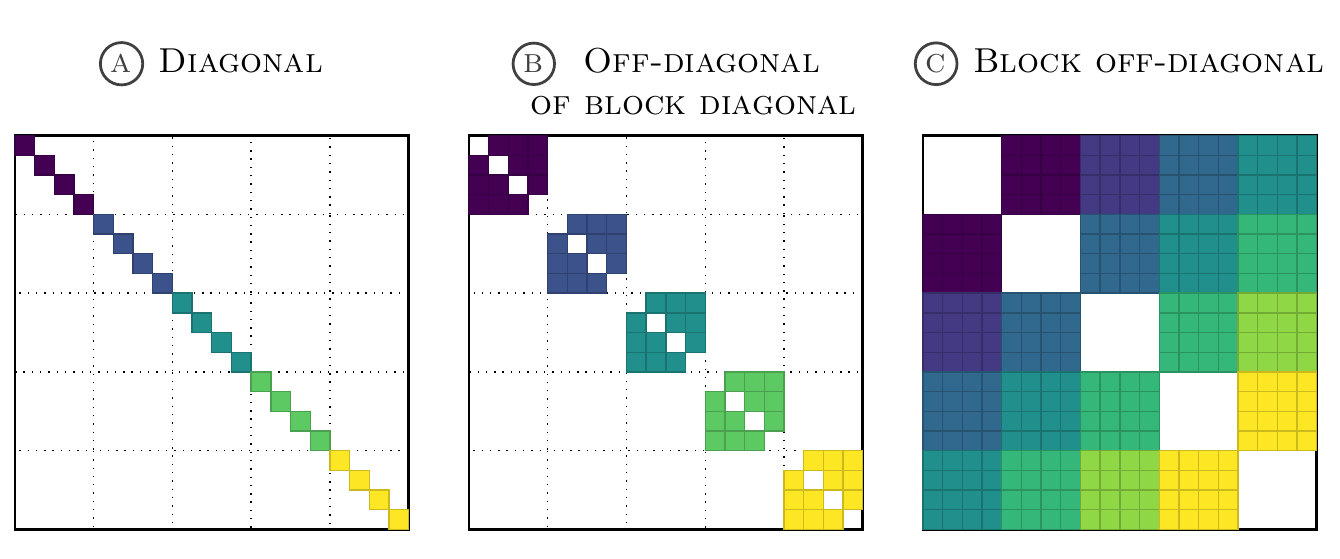}}
 \caption[Block Hessian diagram]{Diagram of the block structure of the Hessian matrix (here, with \(K=4\) and \(N=5\)). \plotlabel{a} shows the diagonal proper, with \(\big(K\times N\big)\) elements calculated with \cref{eqn:fidhess0_DIAG,eqn:fidhess1_DIAG,eqn:fidhess2_DIAG}. \plotlabel{b} shows the off-diagonal elements within the block diagonal, with \(\big(K(K-1)\times N\big)\) elements calculated with \cref{eqn:fidhess0_BLKDIAG,eqn:fidhess1_BLKDIAG,eqn:fidhess2_BLKDIAG}. \plotlabel{c} shows the block off-diagonal elements of \cref{eqn:hess_arrangeOFFblock}, with \(\big(K^2\times N(N-1)\big)\) elements calculated with \cref{eqn:fidhess0_OFFDIAG,eqn:fidhess1_OFFDIAG,eqn:fidhess2_OFFDIAG}.}\label{Fig:HessDiag}
\end{figure}

\subsection{Hessian structure}\label{Section:HessStruct}

The fidelity Hessian with more than one control is a tensor, and should be arranged as a block matrix of \(N\times N\) blocks of \(K\times K\) matrices, consistent with \cref{eqn:ctrl_arrange,eqn:grad_arrange}:
\begin{equation}
\DDFid(\ctrlv{})\triangleq
\begin{pmatrix}
 \mD_{1}^2 & \mD_{1,2} & & \cdots & & \mD_{1,N} 
\vspace{0.5em}\\
 \mD_{2,1} & \mD_{2}^2 & & \cdots & & \mD_{2,N} 
\vspace{0.5em}\\
 \vdots & \vdots & & \ddots & & \vdots 
\vspace{0.5em}\\
 \mD_{N,1} & \mD_{N,2} & & \cdots & & \mD_{N}^2
\end{pmatrix}
\end{equation}
where the matrices \(\mD_{m,n}\) are each of size \(K\times K\). The block diagonals, depicted in \cref{Fig:HessDiag} \plotlabel{a}--\plotlabel{b}, are
\begin{equation}
\mD_{n}^2\triangleq
\begin{pmatrix}
\dfrac{\partial^{\,2}\!\Fid}{\partial {\ctrl{}}_{\scriptscriptstyle 1,n}^{\,2}} &
\dfrac{\partial^{\,2}\!\Fid}{\partial {\ctrl{}}_{\scriptscriptstyle 1,n}\partial {\ctrl{}}_{\scriptscriptstyle 2,n}} &  \cdots  &
\dfrac{\partial^{\,2}\!\Fid}{\partial {\ctrl{}}_{\scriptscriptstyle 1,n}\partial {\ctrl{}}_{\scriptscriptstyle K,n}} 
\vspace{0.5em}\\
 \dfrac{\partial^{\,2}\!\Fid}{\partial {\ctrl{}}_{\scriptscriptstyle 2,n}\partial {\ctrl{}}_{\scriptscriptstyle 1,n}} &
 \dfrac{\partial^{\,2}\!\Fid}{\partial {\ctrl{}}_{\scriptscriptstyle 2,n}^{\,2}} &  \cdots & 
 \dfrac{\partial^{\,2}\!\Fid}{\partial {\ctrl{}}_{\scriptscriptstyle 2,n}\partial {\ctrl{}}_{\scriptscriptstyle K,n}} 
\vspace{0.5em}\\
 \vdots & \vdots &  \ddots  & \vdots 
\vspace{0.5em}\\
 \dfrac{\partial^{\,2}\!\Fid}{\partial {\ctrl{}}_{\scriptscriptstyle K,n}\partial {\ctrl{}}_{\scriptscriptstyle 1,n}} &
 \dfrac{\partial^{\,2}\!\Fid}{\partial {\ctrl{}}_{\scriptscriptstyle K,n}\partial {\ctrl{}}_{\scriptscriptstyle 2,n}} &  \cdots  &
 \dfrac{\partial^{\,2}\!\Fid}{\partial {\ctrl{}}_{\scriptscriptstyle K,n}^{\,2}}
 \end{pmatrix}
\label{eqn:hess_arrangeDIAGblock}
\end{equation}
and the block off-diagonals depicted in \cref{Fig:HessDiag} \plotlabel{c} are
\begin{align}
&& \mD_{m,n}\triangleq
\begin{pmatrix}
\dfrac{\partial^{\,2}\!\Fid}{\partial {\ctrl{}}_{\scriptscriptstyle 1,m}\partial {\ctrl{}}_{\scriptscriptstyle 1,n}} & 
\dfrac{\partial^{\,2}\!\Fid}{\partial {\ctrl{}}_{\scriptscriptstyle 1,m}\partial {\ctrl{}}_{\scriptscriptstyle 2,n}} &  \cdots  &
\dfrac{\partial^{\,2}\!\Fid}{\partial {\ctrl{}}_{\scriptscriptstyle 1,m}\partial {\ctrl{}}_{\scriptscriptstyle K,n}} 
\vspace{0.5em}\\
 \dfrac{\partial^{\,2}\!\Fid}{\partial {\ctrl{}}_{\scriptscriptstyle 2,m}\partial {\ctrl{}}_{\scriptscriptstyle 1,n}} &
 \dfrac{\partial^{\,2}\!\Fid}{\partial {\ctrl{}}_{\scriptscriptstyle 2,m}\partial {\ctrl{}}_{\scriptscriptstyle 2,n}} &  \cdots  &
 \dfrac{\partial^{\,2}\!\Fid}{\partial {\ctrl{}}_{\scriptscriptstyle 2,m}\partial {\ctrl{}}_{\scriptscriptstyle K,n}} 
\vspace{0.5em}\\
 \vdots & \vdots &  \ddots &  \vdots 
\vspace{0.5em}\\
 \dfrac{\partial^{\,2}\!\Fid}{\partial {\ctrl{}}_{\scriptscriptstyle K,m}\partial {\ctrl{}}_{\scriptscriptstyle 1,n}} &
 \dfrac{\partial^{\,2}\!\Fid}{\partial {\ctrl{}}_{\scriptscriptstyle K,m}\partial {\ctrl{}}_{\scriptscriptstyle 2,n}} &  \cdots  &
 \dfrac{\partial^{\,2}\!\Fid}{\partial {\ctrl{}}_{\scriptscriptstyle K,m}\partial {\ctrl{}}_{\scriptscriptstyle K,n}}
 \end{pmatrix}, && n\neq m
\label{eqn:hess_arrangeOFFblock}
\end{align}

\subsection{Hessian elements}\label{Section:HessElements}

The second derivatives of the fidelity \cite{ANAND12} have a similar form to the gradient elements in \cref{eqn:fidgrad0,eqn:fidgrad1,eqn:fidgrad2}, defined for the fidelity functional in \cref{eqn:fidelity0,eqn:fidelity1,eqn:fidelity2}. The second derivatives of the fidelity, with respect to a single time slice and a single control, are the diagonal elements of Hessian in \cref{eqn:hess_arrangeDIAGblock}, depicted in \cref{Fig:HessDiag} \plotlabel{a}, and their equations are:
\begin{subequations}
\begin{gather}
 \dfrac{\partial^{\,2}\!\Fid_0^{}}{\partial {\ctrl{}}_{\scriptscriptstyle k,n}^{\,2}}=\Big\langle\Astate_{n+1}^{}\Big|\frac{\partial^{\,2}\!\Prop_n^{}}{\partial {\ctrl{}}_{\scriptscriptstyle k,n}^{\,2}}\Big|\state_{n-1}^{}\Big\rangle\label{eqn:fidhess0_DIAG}\\
 \dfrac{\partial^{\,2}\!\Fid_1^{}}{\partial {\ctrl{}}_{\scriptscriptstyle k,n}^{\,2}}=\Re\Big\langle\Astate_{n+1}^{}\Big|\frac{\partial^{\,2}\!\Prop_n^{}}{\partial {\ctrl{}}_{\scriptscriptstyle k,n}^{\,2}}\Big|\state_{n-1}^{}\Big\rangle\label{eqn:fidhess1_DIAG}\\
 \dfrac{\partial^{\,2}\!\Fid_2^{}}{\partial {\ctrl{}}_{\scriptscriptstyle k,n}^{\,2}}=2\Re\Big\langle\Astate_{n+1}^{}\Big|\frac{\partial^{\,2}\!\Prop_n^{}}{\partial {\ctrl{}}_{\scriptscriptstyle k,n}^{\,2}}\Big|\state_{n-1}^{}\Big\rangle\label{eqn:fidhess2_DIAG}
\end{gather}
\end{subequations}
The off-diagonal elements of the block diagonal in \cref{eqn:hess_arrangeDIAGblock}, depicted in \cref{Fig:HessDiag} \plotlabel{b}, are similarly defined:
\begin{subequations}
\begin{gather}
 \dfrac{\partial^{\,2}\!\Fid_0^{}}{\partial {\ctrl{}}_{\scriptscriptstyle k,n}\partial {\ctrl{}}_{\scriptscriptstyle j,n}}=
 \Big\langle\Astate_{n+1}^{}\Big|\frac{\partial^{\,2}\!\Prop_n^{}}{\partial {\ctrl{}}_{\scriptscriptstyle k,n}\partial {\ctrl{}}_{\scriptscriptstyle j,n}}\Big|\state_{n-1}^{}\Big\rangle\label{eqn:fidhess0_BLKDIAG}\\
 \dfrac{\partial^{\,2}\!\Fid_1^{}}{\partial {\ctrl{}}_{\scriptscriptstyle k,n}\partial {\ctrl{}}_{\scriptscriptstyle j,n}}=
 \Re\Big\langle\Astate_{n+1}^{}\Big|\frac{\partial^{\,2}\!\Prop_n^{}}{\partial {\ctrl{}}_{\scriptscriptstyle k,n}\partial {\ctrl{}}_{\scriptscriptstyle j,n}}\Big|\state_{n-1}^{}\Big\rangle\label{eqn:fidhess1_BLKDIAG}\\
 \dfrac{\partial^{\,2}\!\Fid_2^{}}{\partial {\ctrl{}}_{\scriptscriptstyle k,n}\partial {\ctrl{}}_{\scriptscriptstyle j,n}}=
 2\Re\Big\langle\Astate_{n+1}^{}\Big|\frac{\partial^{\,2}\!\Prop_n^{}}{\partial {\ctrl{}}_{\scriptscriptstyle k,n}\partial {\ctrl{}}_{\scriptscriptstyle j,n}}\Big|\state_{n-1}^{}\Big\rangle\label{eqn:fidhess2_BLKDIAG}
\end{gather}
\end{subequations}
where \(j\neq k\). The states \(\big|\state_{n-1}^{}\big\rangle\) and \(\big|\Astate_{n+1}^{}\big\rangle\) are the forward and backward propagated states respectively, from \cref{eqn:fwdprop,eqn:bwdprop}, and propagators are those in \cref{eqn:propagatorN}.

The block off-diagonal elements in \cref{eqn:hess_arrangeOFFblock} have the useful relation in that they require only first derivatives of the fidelity \cite{ANAND12}, which would have been calculated previously in evaluating the fidelity gradient in \cref{eqn:fidgrad0,eqn:fidgrad1,eqn:fidgrad2}. The equations of the block off-diagonal Hessian elements in \cref{eqn:hess_arrangeOFFblock}, depicted in \cref{Fig:HessDiag} \plotlabel{a}, are
\begin{subequations}
\begin{gather}
 \dfrac{\partial^{\,2}\!\Fid_0^{}}{\partial {\ctrl{}}_{\scriptscriptstyle k,n}\partial {\ctrl{}}_{\scriptscriptstyle j,m}}=
 \Big\langle\Astate_{n+1}^{}\Big|\frac{\partial\!\Prop_n^{}}{\partial {\ctrl{}}_{\scriptscriptstyle k,n}}\Prop_{n-1}^{}\cdots\Prop_{m+1}^{}\frac{\partial\!\Prop_m^{}}{\partial {\ctrl{}}_{\scriptscriptstyle j,m}}\Big|\state_{m-1}^{}\Big\rangle\label{eqn:fidhess0_OFFDIAG}\\
 \dfrac{\partial^{\,2}\!\Fid_1^{}}{\partial {\ctrl{}}_{\scriptscriptstyle k,n}\partial {\ctrl{}}_{\scriptscriptstyle j,m}}=
 \Re\Big\langle\Astate_{n+1}^{}\Big|\frac{\partial\!\Prop_n^{}}{\partial {\ctrl{}}_{\scriptscriptstyle k,n}}\Prop_{n-1}^{}\cdots\Prop_{m+1}^{}\frac{\partial\!\Prop_m^{}}{\partial {\ctrl{}}_{\scriptscriptstyle j,m}}\Big|\state_{m-1}^{}\Big\rangle\label{eqn:fidhess1_OFFDIAG}\\
 \dfrac{\partial^{\,2}\!\Fid_2^{}}{\partial {\ctrl{}}_{\scriptscriptstyle k,n}\partial {\ctrl{}}_{\scriptscriptstyle j,m}}=
 2\Re\Big\langle\Astate_{n+1}^{}\Big|\frac{\partial\!\Prop_n^{}}{\partial {\ctrl{}}_{\scriptscriptstyle k,n}}\Prop_{n-1}^{}\cdots\Prop_{m+1}^{}\frac{\partial\!\Prop_m^{}}{\partial {\ctrl{}}_{\scriptscriptstyle j,m}}\Big|\state_{m-1}^{}\Big\rangle\label{eqn:fidhess2_OFFDIAG}
\end{gather}
\end{subequations}
where \(n>m\)\footnote{If \(n<m\) then \(n\) and \(m\) are interchanged in \cref{eqn:fidhess0_OFFDIAG,eqn:fidhess1_OFFDIAG,eqn:fidhess2_OFFDIAG}. If \(n=m+1\), there are no propagators in between the two first-order propagator derivatives. If \(n=m+2\), there is one propagator \(\Prop_{m+1}^{}=\Prop_{n-1}^{}\) in between the two first-order propagator derivatives.}.

\cref{eqn:fidhess0_DIAG,eqn:fidhess1_DIAG,eqn:fidhess2_DIAG,eqn:fidhess0_BLKDIAG,eqn:fidhess1_BLKDIAG,eqn:fidhess2_BLKDIAG,eqn:fidhess0_OFFDIAG,eqn:fidhess1_OFFDIAG,eqn:fidhess2_OFFDIAG} show the Hessian has almost the same numerical computational complexity scaling as the fidelity functional in \cref{eqn:fidelity0,eqn:fidelity1,eqn:fidelity2} \cite{ANAND12}, just as the gradient does in \cref{eqn:fidgrad0,eqn:fidgrad1,eqn:fidgrad2} \cite{KHANEJA05}. This situation is highly unusual in non-linear optimization theory -- Hessians are normally so expensive that a significant body of work exists on the subject of avoiding their calculation and recovering second derivative information in an approximate way from the gradient history \cite{FLETCHER87,NOCEDAL06,LIU89,BYRD94}. The recent \textsc{bfgs-grape} algorithm \cite{FOUQUIERES11} is an example of such approach. The fact that the Hessian is relatively cheap suggests Newton-Raphson type algorithms \cite{FLETCHER87,NOCEDAL06} with the control sequence update rule of \cref{eqn:Newtonupdate_grape}.

However, three logistical problems present themselves that must be solved before the method becomes useful in practice: 
\begin{enumerate}
\item Efficient calculation of second derivatives of the propagators in \cref{eqn:propagatorN}. This problem will be considered in detail in \cref{Chapter:AuxMat}.
\item Propagator derivative recycling between gradient elements in \cref{eqn:fidgrad0,eqn:fidgrad1,eqn:fidgrad2} and block off-diagonal Hessian elements in \cref{eqn:fidhess0_OFFDIAG,eqn:fidhess1_OFFDIAG,eqn:fidhess2_OFFDIAG}, required for efficient scaling. Outlined in \cite{GOODWIN15}, the latency of the cache storage device means that for small matrices it may be faster to recalculate the function. A caching procedure becomes beneficial for larger matrices.
\item Regularisation and conditioning of the Hessian matrix in \cref{eqn:Newtonupdate_grape} is needed to avoid the numerical difficulties associated with its inverse -- a Hessian matrix should be positive definite, from \cref{eqn:sufficient2}, and well-conditioned \cite{FLETCHER87,NOCEDAL06}. This will be the main focus of the remainder of this chapter.
\end{enumerate}

\subsection{Control operator commutativity}\label{Section:Commutaion}

The full Hessian requires \((K\times N)\) second-order propagator derivative evaluations of \cref{eqn:fidhess0_DIAG,eqn:fidhess1_DIAG,eqn:fidhess2_DIAG} and \cref{Fig:HessDiag} \plotlabel{a}. It was pointed out that the block off-diagonal second-order propagator derivatives of \cref{eqn:fidhess0_OFFDIAG,eqn:fidhess1_OFFDIAG,eqn:fidhess2_OFFDIAG} and \cref{Fig:HessDiag} \plotlabel{c} are essentially free, having already been calculated with a gradient evaluation. There is a further useful property of the Hessian matrix: that it should be real and symmetric. Potentially, this can halve the number of calculations of the type in \cref{eqn:fidhess0_BLKDIAG,eqn:fidhess1_BLKDIAG,eqn:fidhess2_BLKDIAG}.

However, it is noted in \cite{VANLOAN77,MOLER78} that the additive law fails unless there is commutativity:
\begin{align}
 && \e^{t\mA}\e^{t\mB}= \e^{t(\mA+\mB)}, &&\text{iff}\qquad \mA\mB=\mB\mA\label{eqn:addlaw}\\
\Longrightarrow && \dfrac{\partial^{\,2}\!\Fid_{}^{}}{\partial {\ctrl{}}_{\scriptscriptstyle k,n}\partial {\ctrl{}}_{\scriptscriptstyle j,n}}=\dfrac{\partial^{\,2}\!\Fid_{}^{}}{\partial {\ctrl{}}_{\scriptscriptstyle j,n}\partial {\ctrl{}}_{\scriptscriptstyle k,n}}, && \text{iff}\quad\Big[\HamH_k^{},\HamH_j^{}\Big]=0\label{eqn:symderivs}
\end{align}
The relation is only valid if the associated control operators commute with each other. To ensure the Hessian matrix is symmetric\footnote{When a Hessian is not symmetric, its eigenvalues are not real.} when condition of \cref{eqn:symderivs} is not met, the following formula must be used
\begin{align}
 && \begin{rcases}\DDFid({\ctrl{}}_{\scriptscriptstyle j,n},{\ctrl{}}_{\scriptscriptstyle k,n})\\
\DDFid({\ctrl{}}_{\scriptscriptstyle k,n},{\ctrl{}}_{\scriptscriptstyle j,n})
    \end{rcases}= \frac{1}{2}\Bigg(\dfrac{\partial^{\,2}\!\Fid_{}^{}}{\partial {\ctrl{}}_{\scriptscriptstyle k,n}\partial {\ctrl{}}_{\scriptscriptstyle j,n}}+\dfrac{\partial^{\,2}\!\Fid_{}^{}}{\partial {\ctrl{}}_{\scriptscriptstyle j,n}\partial {\ctrl{}}_{\scriptscriptstyle k,n}}\Bigg), && \text{if}\quad\Big[\HamH_k^{},\HamH_j^{}\Big]\neq0 \label{eqn:force_symmetry}
\end{align}
which requires two second-order propagator derivative evaluations.

\section{Hessian Regularisation}\label{Section:HessReg}

Maximisation with Newton-Raphson and quasi-Newton methods relies on the necessary conditions for the truncated Taylor series approximation to be valid. This was introduced using a local quadratic approximation of the objective function in \cref{eqn:quadmodel}, restated here for the fidelity functional:
\begin{equation}
\Delta\Fid=\Fid\big(\ctrlv{}^{(\itr+1)}\big)-\Fid\big(\ctrlv{}^{(\itr)}\big)\approx \Big\langle \DFid\big(\ctrlv{}^{(\itr)}\big) \Big| \ctrlv{}^{(\itr)} \Big\rangle +\frac{1}{2}\Big\langle \ctrlv{}^{(\itr)} \Big|\DDFid\big(\ctrlv{}^{(\itr)}\big)\Big| \ctrlv{}^{(\itr)} \Big\rangle\label{eqn:taylor_Fid}
\end{equation}
In addition to the gradient being zero in the vicinity of a stationary point \(\ctrlv{}^{(\ast)}\),
\begin{equation}
\lim_{\ctrlv{}\to\ctrlv{}^{(\ast)}}\Big\|\DFid\big(\ctrlv{}^{(\ast)}\big)\Big\|_{\infty}^{} = 0\label{eqn:necessary1_grape}
\end{equation}
the second-order sufficient condition of \cref{eqn:sufficient2} must also be true in the vicinity of a stationary point:
\begin{align}
&&\lim_{\ctrlv{}\to\ctrlv{}^{(\ast)}}\Big\langle \ctrlv{} \Big|\DDFid\big(\ctrlv{}^{(\ast)}\big)\Big| \ctrlv{} \Big\rangle>0, && \forall\ctrlv{}\in\Real^{(K\cdot N)}\label{eqn:necessary2_grape}
\end{align}
i.e. the Hessian must be positive definite near a stationary point \(\ctrlv{}^{(\ast)}\). This is evident from \cref{eqn:Newtonupdate_grape}, in which a negative definite Hessian would result in a step being performed down, rather than up, the corresponding gradient direction. Furthermore, an indefinite Hessian is an indication of a saddle point, shown in \cref{Fig:StationaryPoints} \plotlabel{c}, which should be avoided during maximisation of the fidelity functional. Forcing a Hessian to be a definite matrix ensures that steps are made that avoid saddle points. Forcing a positive definite Hessian, as opposed to a negative definite Hessian, ensures that a step is always made in the correct ascent/descent direction. 

A significant problem is that, far away from a stationary point, the Hessian is not expected to be positive definite \cite{PECHEN11,PECHEN12a,PECHEN12b,FOUQUIERES13,RIVIELLO14,SUN14}. Small Hessian eigenvalues are also problematic because they result in overly long steps that can be detrimental because most fidelity functionals are not actually quadratic. A significant amount of research has gone into modifying the Hessian in such a way as to avoid these undesired behaviours \cite{GOLDFELD66,GREENSTADT67,BANERJEE76,HEBDEN73,GOLDFARB80,CERJAN81,MORE82,SHEPARD82,MORE83,BANERJEE85,BAKER86}.

\subsection{Cholesky factorisation}\label{Section:Cholesky}

One fairly cheap way to work around an indefinite Hessian is to attempt \textit{Cholesky factorisation} \cite{GILL81}, which exists for any invertible positive definite matrix \cite{GOLUB13}:
\begin{align}
 \Big[\DDFid\Big] & =\mL\mL^{\text{T}} \label{eqn:cholesky1}\\
 \Rightarrow\quad \Big[\DDFid \Big]_{}^{-1} & =\Big[\mL^{-1}\Big]^{\text{T}}\mL^{-1}\label{eqn:cholesky2}
\end{align}
where \(\mL\) is a lower triangular matrix with positive diagonal elements and \(\mL^{\text{T}}\) is its transpose\footnote{The argument of the Hessian has been dropped here, and for the rest of the chapter. It should be assumed that within equations and definitions, the Hessian, gradient, and controls are all of the same iteration, \(\itr\), unless stated otherwise:\begin{gather*}
\DDFid\big(\ctrlv{}^{(\itr)}\big)\to\DDFid\\\DFid\big(\ctrlv{}^{(\itr)}\big)\to\DFid\\\ctrlv{}^{(\itr)}\to\ctrlv{}
\end{gather*}}. If a Cholesky factorisation finds solution, the Hessian is positive definite. If a Cholesky factorisation fails, the Hessian is indefinite, and an identity matrix may be used as a substitute for the Hessian, effectively reverting to a gradient ascent step for any iterations that produce an indefinite Hessian \cite{GOLDSTEIN67}.

The problem with this approach is that indefinite Hessians become more common as the dimension of the problem increases, making the maximiser spend most of the time in the gradient ascent mode and destroying any advantage of the second-order method over simple gradient ascent -- this technique is not recommended.

However, there is a further use of the Cholesky factorisation test for an indefinite Hessian: an algorithm can be designed to iteratively test the Hessian, modifying the Hessian at each stage. Using the Cholesky decomposition on a Hessian \cite{GILL81} with a multiple of the unit matrix added \cite{GOLDFELD66,HEBDEN73,MORE82,MORE83}: 
\begin{align}
&& \Big[\DDFid\Big] +\eshft \Unit=\mL\mL^{\text{T}}, && \eshft\geqslant 0
\end{align}
The choice of \(\eshft\) is made to produce a positive definite Hessian satisfying \cref{eqn:necessary2_grape}, with the trial value
\begin{align}
&& \eshft =\begin{cases}
   \Big\| \DDFid \Big\|_{\text{F}}-\min \Big[ \DDFid \Big]_{ii} \vspace{1em}\\
   \Big\| \DDFid \Big\|_{\text{F}}
\end{cases} &&
\begin{array}{l}
 \min \Big[ \DDFid \Big]_{ii}<0\vspace{1em}\\
 \min \Big[ \DDFid \Big]_{ii}\geqslant 0
 \end{array}
\end{align}
chosen in that way because the Frobenius norm of the Hessian is an upper bound on the largest absolute eigenvalue. The value of \(\eshft \) is increased iteratively until the Cholesky decomposition succeeds \cite{NOCEDAL06} and the inverse Hessian may be obtained.

A benefits of this iterative Cholesky factorisation method are that an inverse Hessian is automatically produced from \cref{eqn:cholesky2}, which is a simple task to compute for invertible triangular matrices \cite{GOLUB13}, and that the method avoids the expensive operation of calculating an eigendecomposition, which is time-consuming for large Hessian matrices.

\subsection{Trust Region Method}\label{Section:TRM}

A more sophisticated workaround is to use the eigenvalue shifting method, also called the trust region method, adapted from the Levenberg-Marquardt algorithm \cite{MARQUARDT63,GILL74}. Hessian eigenvalues may be computed explicitly \cite{GREENSTADT67}
\begin{equation}
\Big[ \DDFid \Big]=\mQ\mLbd\mQ^{-1}\label{eqn:eigdecomp}
\end{equation}
where \(\mLbd\) is a diagonal matrix containing the eigenvalues of \(\DDFid\) and \(\mQ\) is the matrix with columns of corresponding eigenvectors. A precise estimate can now be made for the largest negative eigenvalue of the Hessian, and added to the diagonal matrix of eigenvalues to reform the Hessian from this modified eigendecomposition:
\begin{align}
&& \Big[ \DDFid \Big]_{\text{reg}}\!\!\!=\mQ\big(\mLbd+ \eshft\Unit\big)\mQ^{-1}, && \eshft =\max\big(0,\trm -\mineig\big)\label{eqn:trm}
\end{align}
where \(\mineig\) is the minimum eigenvalue of the Hessian, \(\min(\mLbd_{ii})\). The arbitrary positive value of \(\trm\) is included to make the Hessian positive definite. The primary problem with this method is that, for poorly conditioned or near-singular Hessian matrices, the regularisation procedure destroys much of the curvature information and the technique effectively becomes a combination of the Newton-Raphson and gradient ascent methods. Formally, \cref{eqn:eigdecomp,eqn:trm} do solve the step direction problem, but practical testing indicates the convergence rate can fail to achieve true quadratic convergence near a maximum.

\subsection{Rational Function Optimisation}\label{Section:RFO}

A refined regularisation method, called rational function optimization \cite{BANERJEE76,SHEPARD82,BANERJEE85}, does allow a modified Hessian to be a well-conditioned, positive definite matrix. It replaces the truncated Taylor series in \cref{eqn:taylor_Fid} with a \([2/1]\) Pad\'{e} approximant \cite{PADE92,HIGHAM08}, defined in \cref{Section:PadeApprox}, approximating as a rational function: 
\begin{equation}
\Delta\Fid=\frac{\big\langle \DFid \big|\, \ctrlv{} \big\rangle + \dfrac{1}{2}\big\langle \ctrlv{}\,\big|\DDFid\big|\, \ctrlv{} \big\rangle}{1+\big\langle \ctrlv{}\, \big|\,\mS\,\big|\, \ctrlv{} \big\rangle }
\end{equation}
This preserves the derivative information of \cref{eqn:taylor_Fid}, leaving the necessary conditions unchanged, because the derivatives of \(\big[ 1+\big\langle  \ctrlv{} \big|\mS\big| \ctrlv{}\big\rangle  \big]^{-1}\) give contributions only through higher orders in \(\ctrlv{}\). The nature of the rational function means that the asymptotes of \(\Delta\Fid\) and is gradients remain finite for \(\ctrlv{}\to \pm \infty\), determined by the Hessian and the symmetric scaling matrix \(\mS\). The first-order necessary condition for \cref{eqn:necessary1_grape} gives the following eigenvalue equation:
\begin{equation}
\begin{pmatrix}
   \big[\DDFid\big] & \big|\DFid\big\rangle  \\
   \big\langle\DFid\big| & \Zero  \\
\end{pmatrix}\begin{pmatrix}
   \ctrlv{}  \\
   \Unit  \\
\end{pmatrix}=2\Delta\Fid\begin{pmatrix}
   \mS & \mathbf{0}  \\
   \mathbf{0} & \Unit  \\
\end{pmatrix}\begin{pmatrix}
   \ctrlv{}  \\
   \Unit  \\
\end{pmatrix}
\end{equation}
Choosing a uniform scaling matrix $\mS=\rfo^{-2}\Unit$ \cite{BANERJEE85}, where $0<\rfo \leqslant 1$, reduces this equation to
\begin{equation}
\begin{pmatrix}
   \rfo^2\big[\DDFid\big] & \rfo\big|\DFid\big\rangle  \\
   \rfo\big\langle\DFid\big| & \Zero  \\
\end{pmatrix}\begin{pmatrix}
   {\ctrlv{}}/\rfo\;  \\
   \Unit  \\
\end{pmatrix}=2\Delta\Fid\begin{pmatrix}
   {\ctrlv{}}/\rfo\;  \\
   \Unit  \\
\end{pmatrix} 
\end{equation}
Rational function optimisation proceeds in a similar way to eigenvalue shifting methods described above, except the shifting is applied to the augmented Hessian matrix. Defining the eigendecomposition of the augmented Hessian matrix as
\begin{equation}
  \Big[ \DDFid \Big]^\text{aug}\!\!\!=\begin{pmatrix}
   \rfo^2\big[\DDFid\big] & \rfo\big|\DFid\big\rangle  \\
   \rfo\big\langle\DFid\big| & \Zero  \\
\end{pmatrix}=\mQ\mLbd\mQ^{-1}\label{eqn:rfo1}
\end{equation}
a similar shifting formula to \cref{eqn:trm} can be used to ensure a positive definite augmented Hessian matrix
\begin{align}
 && \Big[ \DDFid \Big]_\text{reg}^\text{aug}\!\!\!=\frac{1}{\rfo^2}\mQ\big(\mLbd+\eshft \Unit\big)\mQ^{-1}, && \eshft =\max \big(0,-\mineig\big)\label{eqn:rfo2}
\end{align}
The top left corner block of the regularised augmented Hessian is then used for the Newton-Raphson step \cite{BANERJEE76,SHEPARD82,BANERJEE85} in \cref{eqn:Newtonupdate_grape}. 

\subsection{Hessian Conditioning}\label{Section:conditioning}

Motivated by the \textit{Wolfe conditions} of \cref{eqn:sufficientdecrease,eqn:curvaturecondition,eqn:strongcurvaturecondition} \cite{WOLFE69, WOLFE71}, it may be necessary to place restrictions on the search direction by excluding directions near-orthogonal to the steepest ascent vector. This condition, called the \textit{angle criterion}, states that for any angle \(\theta^{(\itr)}\) between the search direction, \(\sdir{s}\), and the gradient vector should be bounded away from orthogonality. Using the derivation set out by Fletcher \cite{FLETCHER87}, the critical angle is should be
\begin{align}
 && \theta^{(\itr)}\leqslant \frac{\pi}{2}-\mu && \forall \itr\label{eqn:angle_criterion1}
\end{align}
with \(\mu>0\) and independent of \(\itr\), and the angle \(\theta^{(\itr)}\) is calculated as
\begin{align}
 \cos\big(\theta^{(\itr)}\big)=\frac{\Big\langle\DFid\big(\ctrlv{}^{(\itr)}\big)\Big|\ctrlv{}^{(\itr)}\Big\rangle}{\big\|\DFid\big(\ctrlv{}^{(\itr)}\big)\big\|_2^{}\cdot\big\|\ctrlv{}^{(\itr)}\big\|_2^{}}\label{eqn:eqn:angle_criterion2}
\end{align}
For Newton-type methods with a continuous gradient, using inexact line searches of \cref{eqn:sufficientdecrease,eqn:curvaturecondition,eqn:strongcurvaturecondition}, maximisation is bounded from above. This leads to the conclusion that, as the change in fidelity tends to zero, \(\Delta\Fid\to 0\), then so does the slope \(\big\langle\DFid\big|\ctrlv{}\big\rangle\to 0\). Two cases remain: either \(\DFid{}^{(\itr)}\to0\) or \(\Fid{}^{(\itr)}\to\infty\) \cite{FLETCHER87}.

The condition number defined here as the ratio of its largest Hessian eigenvalue, \(\max(\mLbd_{ii})\), to the smallest Hessian eigenvalue, \(\min(\mLbd_{ii})\):
\begin{equation}
 \condno=\frac{\maxeig}{\mineig}
\end{equation}
The sufficient condition for ascent is that the Hessian is positive definite. This gives a sufficient condition for the angle criterion of \cref{eqn:angle_criterion1,eqn:eqn:angle_criterion2}, requiring the condition number of the Hessian matrix also to be bounded above. With the inequalities
\begin{gather}
\big\|\DDFid|\DFid\rangle\big\|_2\leqslant \maxeig\\
\big\langle\DFid\big|\DDFid\big|\DFid\big\rangle\geqslant \mineig\big\langle\DFid\big|\DFid\big\rangle
\end{gather}
the sufficient condition for the angle criterion is
\begin{equation}
 \theta^{(\itr)}\leqslant \frac{\pi}{2}-\frac{1}{\condno^{(\itr)}}
\end{equation}
With this condition satisfied, the Armijo condition in \cref{eqn:sufficientdecrease} is also satisfied. The methods that follow will ensure that the condition number of the Hessian matrix is bounded so the above conditions are met.

The condition number of the Hessian should be bound; one way to achieve this is to allow the scaling constant \(\rfo\) in \cref{eqn:rfo1,eqn:rfo2} to vary according to the condition of the Hessian. The value of \(\rfo\) is reduced until the condition number becomes acceptable (nearer to \(\rfo=1\), when the Hessian is well-conditioned), for example:
\begin{align}
&& \rfo_{r+1}=\phi \rfo_{r} && \text{while} &&\condno>\frac{1}{\sqrt[3]{\eps}}\label{eqn:cond_reduce}
\end{align}
where \(\eps\) is machine precision and \(\rfo_0=1\). The cube-root of machine precision reflects the allowable condition number for an acceptable cubic interpolation within the sectioning phase of a line search in \cref{Section:BracketSection}. The factor \(0<\phi <1\) is used to iteratively decrease the condition number of the Hessian -- this is the method used to condition the Hessian in the examples presented below. It should be noted that a value \(\phi >1\) may be used to increase the condition number of the Hessian when it is very small and the inequality of \cref{eqn:cond_reduce} is reversed.

Quantum mechanics is rich in situations that produce very small values of Hessian and gradient elements. Choosing \(\rfo_0\) to be the value that would shift the smallest eigenvalues to be above \(1\), giving a similar effect to choosing $\trm =1$ in \cref{eqn:trm}, accounts for optimisation problems that are inherently ill-conditioned in this way. When the choice
\begin{equation}
\rfo_0=\frac{1}{\sqrt{\big| \mineig \big|}}
\end{equation}
is made, the augmented Hessian in \cref{eqn:rfo1,eqn:rfo2} becomes
\begin{gather}
  \Big[ \DDFid \Big]^\text{aug}\!\!\!=\begin{pmatrix}
   \frac{1}{\lambda_{\min }}\big[\DDFid\big] & \frac{1}{\sqrt{\lambda_{\min }}}\big|\DFid\big\rangle  \\
   \frac{1}{\sqrt{\lambda_{\min }}}\big\langle\DFid\big| & \Zero  \\
\end{pmatrix}=\mQ\mLbd\mQ^{-1}\label{eqn:rfoscale1}\\
\begin{array}{cccr}
 & \Big[ \DDFid \Big]_\text{reg}^\text{aug}\!\!\!=\lambda_{\min }\mQ\big(\mLbd+\eshft \Unit\big)\mQ^{-1}, & & \eshft =\max \big(0,-\mineig\big)\end{array}\label{eqn:rfoscale2}
\end{gather}
Practical testing indicates that this combination of using initial scaling, then accepting large condition numbers for the Hessian, allows the Newton-Raphson method to avoid getting stuck at inflection points. The condition number would grow to a large value around the inflection point and then shrink back when the point has been avoided. Due to the tendency of the Hessian to increase condition number as it approaches the maximiser, machine precision could eventually become a limit to this type of conditioning. To avoid a slowdown at the final stages of the optimisation, an upper bound is placed on the \(\rfo\) parameter in \cref{eqn:cond_reduce} -- this guarantees that the terminal convergence is always quadratic.

In practice, at each optimisation step the function code attempts to compute the Cholesky decomposition of \cref{eqn:cholesky1}. If that is successful then no regularisation is needed, otherwise the function proceeds to regularise with the methods described above. Once the ascent direction and the initial step length are obtained from \cref{eqn:Newtonupdate_grape} it is advantageous to perform a line search procedure.

\begin{figure}
\centering{\includegraphics{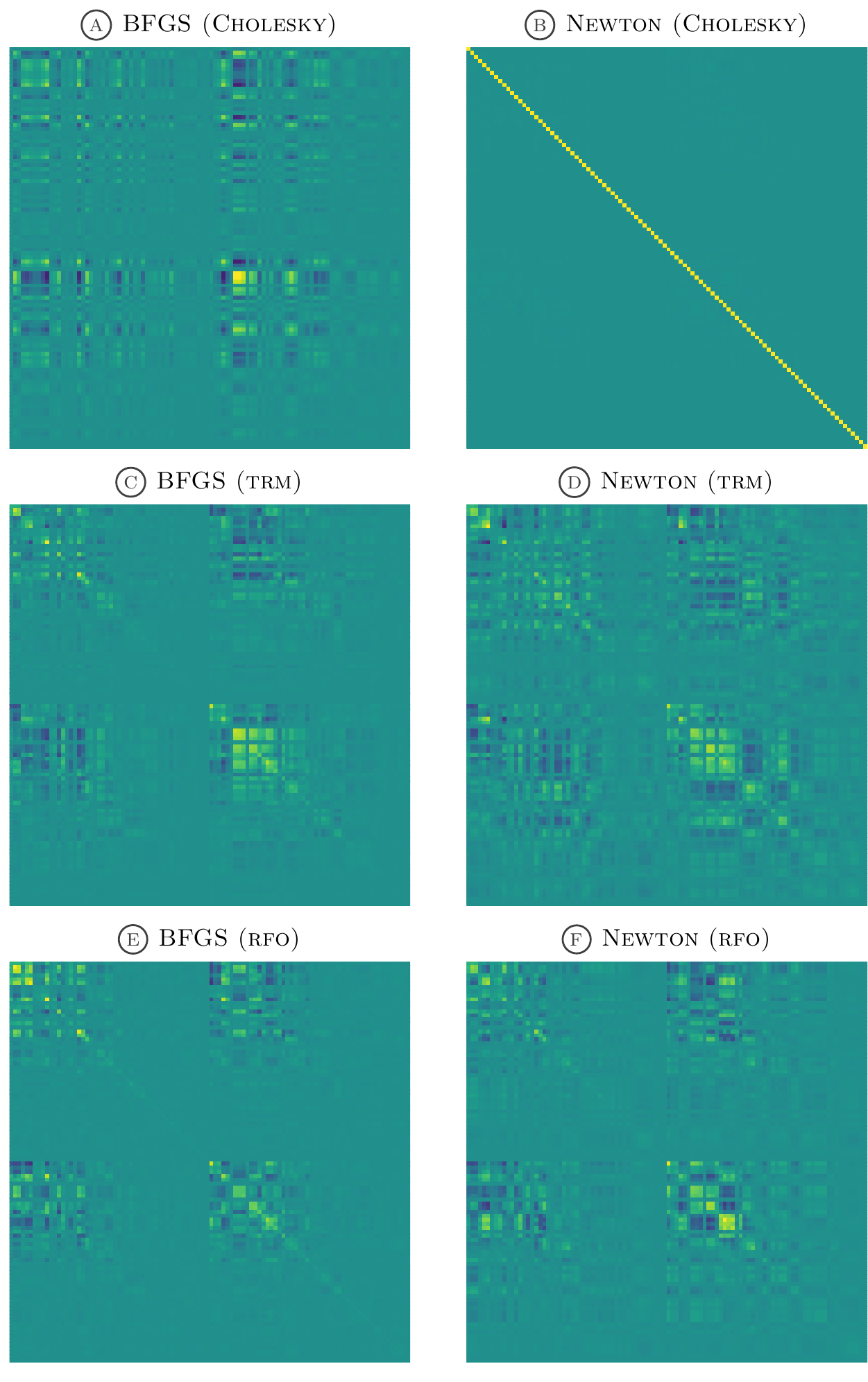}}
 \caption[Regularised and conditioned Hessian matrices]{Converged Hessian matrices, arranged \(N\times K\) (opposite to \cref{Fig:HessDiag} to clearly see the time evolution of control channels, \(\HamH_x^{(\text{H})}\) and \(\HamH_y^{(\text{H})}\), produced by one of the random seeds in \cref{Fig:HCF_CHOL_compare,Fig:HCF_compare}. \plotlabel{b} Newton method almost give a multiple of \(\Unit\), \plotlabel{a} has structure -- indicating an indefinite matrix from the Newton method and a positive definite matrix from the Quasi-Newton method (the \textsc{bfgs} method has a single iteration of the Cholesky regularisation).}\label{Fig:HessianPics}
\end{figure}

\section{Convergence analysis}\label{Section:ConvAnalysis}

In comparing the performance of this Newton-Raphson optimal control, the following optimisation problems will be used\footnote{The templates for these optimal control simulations are included in \textit{Spinach}--v2.0, downloadable from \url{www.spindynamics.org}}: 
\begin{itemize}
 \item State-to-state transfer over a scalar coupled hydrofluorocarbon fragment spin system, as described in \cite{GOODWIN16}.
 \item State-to-state transfer in a quadrupolar \(_{}^{14}\)N spin, transferring the population between the \(T_{1,0}^{}\) and \(T_{2,2}^{}\) states of \cref{eqn:T10,eqn:T22}.
 \item Phase-modulated universal rotation pulses with a sine function amplitude profile, and relaxation. The system is a chain of \(101\) non-interacting \(_{}^{13}\)C spins spread over \(50\)~kHz.
\end{itemize}
For fair comparison, all optimisations are run a number of times from different initial guesses\footnote{Random numbers were seeded with \texttt{rng(32)} at the start of each convergence analysis, and the same starting guesses were used for different methods of optimisation.}, allowing identification of the optimisation convergence characteristics.

\subsection[Scalar coupled three-spin state-to-state population transfer]{Scalar coupled three-spin state-to-state\linebreak population transfer}\label{Section:HCF}

\begin{figure}
\centering{\includegraphics{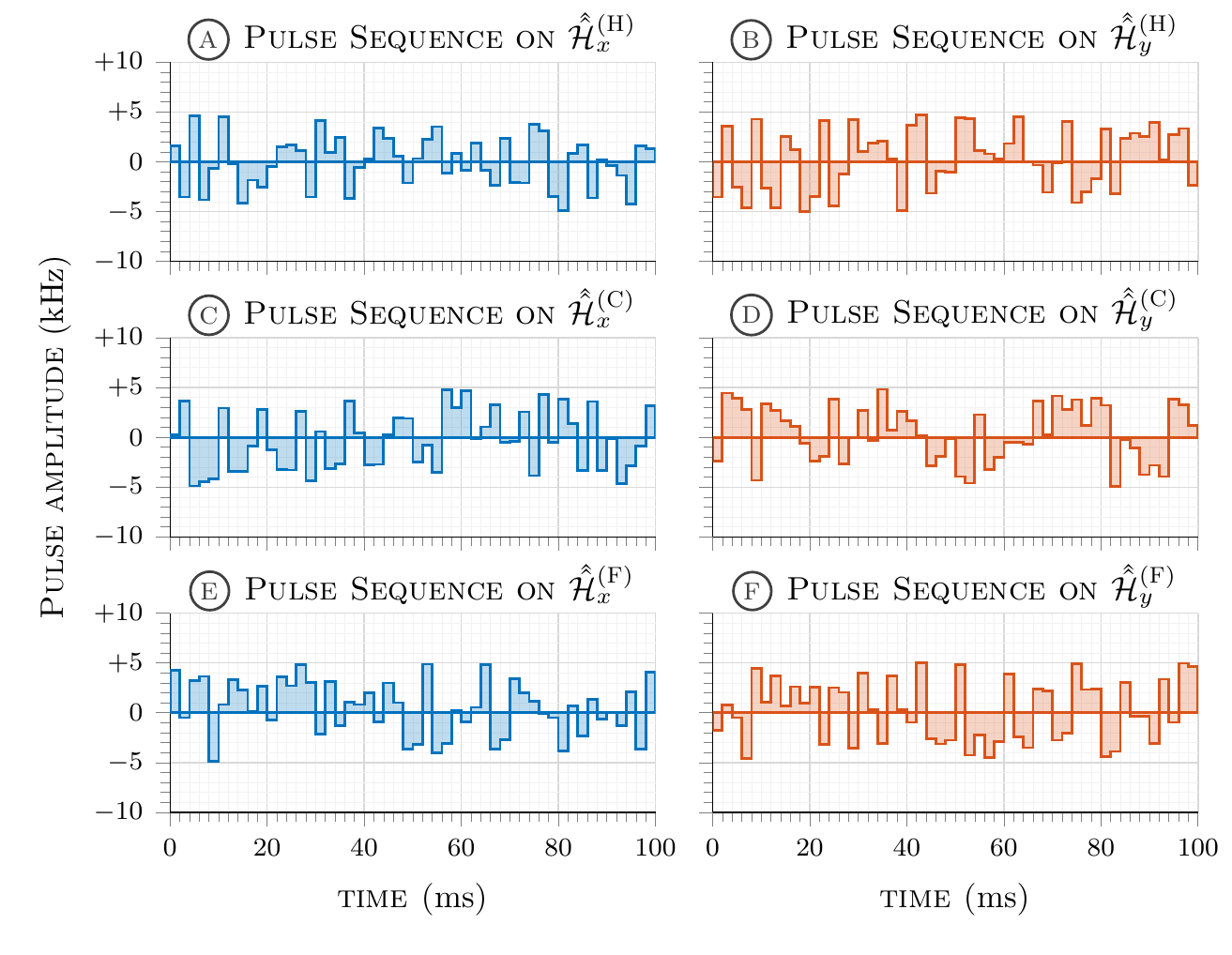}}
 \caption[Control pulses for HCF state transfer]{The six channel control pulse sequences producing the trajectories in \cref{Fig:HCF_traj}. The pulses are a result of the Newton-\textsc{grape} optimisation, which reached maximum fidelity, for a state-to-state transfer in a scalar coupled hydrofluorocarbon fragment, transferring the population between the \(\hat{L}_{z}^\text{(H)}\) and \(\hat{L}_{z}^\text{(F)}\) states}\label{Fig:HCF_controls}
\end{figure}

This system consists of a hydrofluorocarbon fragment, as described in \cite{GOODWIN16}, with spins \(_{}^{1}\)H -- \(_{}^{13}\)C -- \(_{}^{19}\)F coupled in a linear chain, in a \(9.4\)~tesla magnetic field. The scalar coupling, J-coupling, is \(140\)~Hz between \(_{}^{1}\)H -- \(_{}^{13}\)C and \(-160\)~Hz between \(_{}^{13}\)C -- \(_{}^{19}\)F. Using \cref{eqn:Jcoupling}, this gives the drift Hamiltonian of the system as
\begin{align}
 \HamHH_0 &= 280\pi\big(\vhat{L}_{}^\text{(H)}\ccdot\vhat{L}_{}^\text{(C)}\big)- 320\pi\big(\vhat{L}_{}^\text{(C)}\ccdot\vhat{L}_{}^\text{(F)}\big)\nonumber\\
 &= 280\pi\big(\hat{L}_{z}^\text{(H)}\hat{L}_{z}^\text{(C)}\big) - 320\pi\big(\hat{L}_{z}^\text{(C)}\hat{L}_{z}^\text{(F)}\big)
\end{align}
to be used in \cref{eqn:HamSep}.

\begin{figure}
\centering{\includegraphics{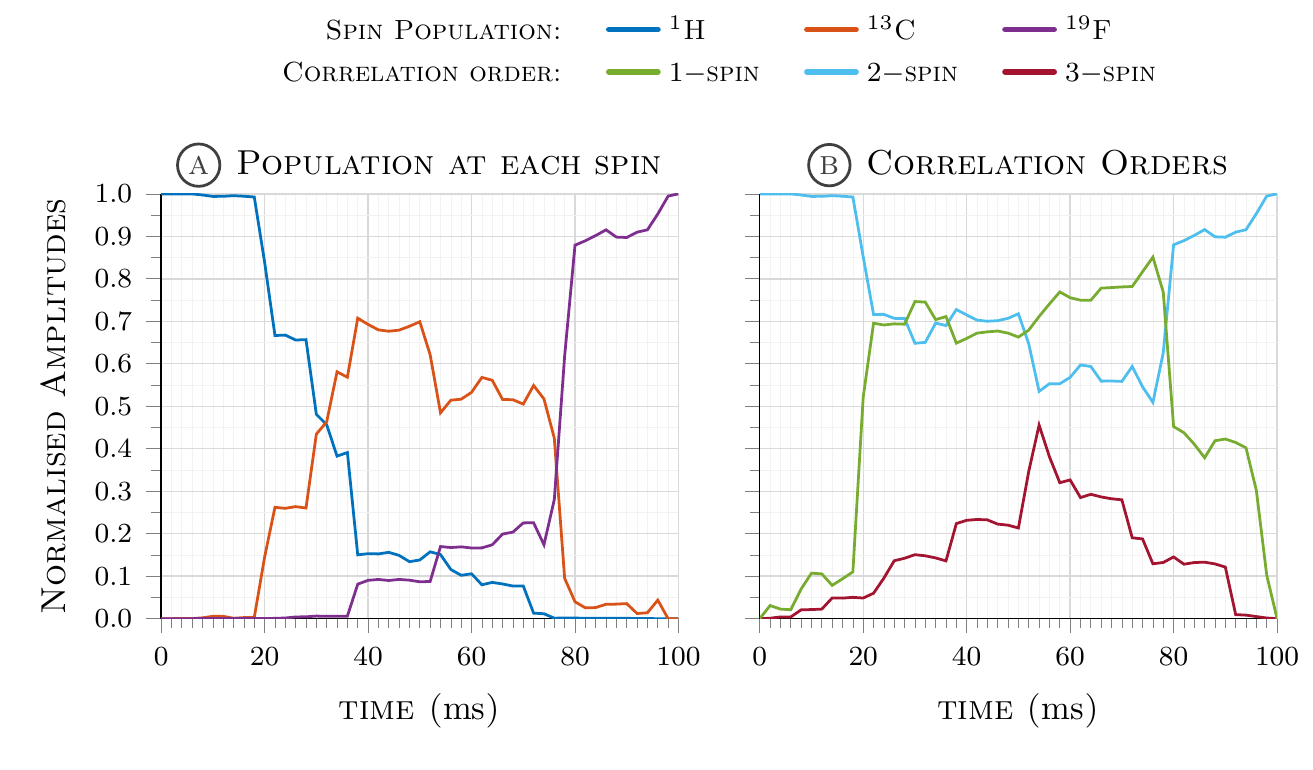}}
 \caption[Trajectory analysis of HCF state transfer]{Trajectories analysis for a state-to-state transfer, in a scalar coupled hydrofluorocarbon fragment, transferring the population between the \(\hat{L}_{z}^\text{(H)}\) and \(\hat{L}_{z}^\text{(F)}\) states. \plotlabel{a} Local spin population. \plotlabel{b} Spin correlation order amplitude. The pulse sequence producing these trajectories has \(6\) control channels and is shown in \cref{Fig:HCF_controls}.}\label{Fig:HCF_traj}
\end{figure}

The initial state is magnetisation population in the \(\hat{L}_{z}^\text{(H)}\) state in Liouville space, and the desired target is the \(\hat{L}_{z}^\text{(F)}\) state, with both states normalised. 

The total control pulse duration is \(100\)~ms and the sequence is discretised to \(50\) equal time slices. The six control operators are \(\HamHH_x^\text{(H)}\), \(\HamHH_y^\text{(H)}\), \(\HamHH_x^\text{(C)}\), \(\HamHH_y^\text{(C)}\), \(\HamHH_x^\text{(F)}\), and \(\HamHH_y^\text{(F)}\). Each of the control power levels allowed is simulated with a nominal power level of \(10\)~kHz. The maximum fidelity achievable for this system is \(\Fid_1=1\). The results of one of these, fully converged, optimisations is depicted in \cref{Fig:HCF_traj}, showing the evolution of the correlation orders and polarisation at each spin over the total pulse time. For completeness, the six control pulse sequences and are shown in \cref{Fig:HCF_controls}.

\begin{figure}
\centering{\includegraphics{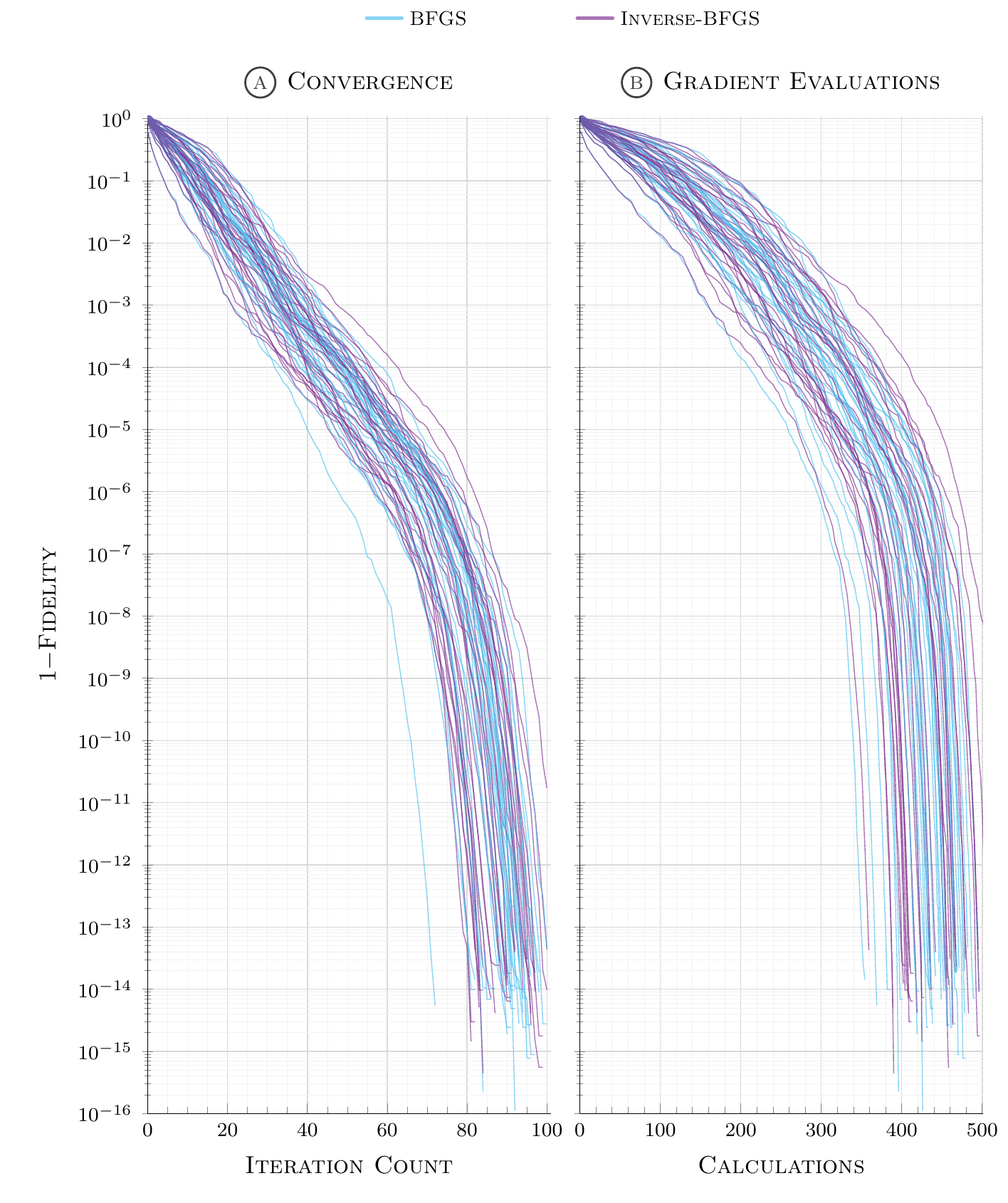}}
 \caption[BFGS method updating the Hessian and the inverse Hessian]{Comparison of \textsc{bfgs} updating the inverse Hessian, and \textsc{bfgs} updating the Hessian approximation. Simulation was run multiple times, from different initial guesses of control pulses. \plotlabel{a} Showing iterate count. \plotlabel{b} Showing gradient calculation count.}\label{Fig:HCF_BFGS_compare}
\end{figure}

Three regularisation methods are compared for Newton-\textsc{grape} and \textsc{bfgs}-\textsc{grape}:
\begin{enumerate}
 \item Comparison of the \textsc{bfgs} and inverse-\textsc{bfgs} method.
 \item Newton-\textsc{grape} and \textsc{bfgs}-\textsc{grape} methods, using iterative Cholesky factorisation as a regularisation method. At least one iteration of the Cholesky factorisation was forced for the \textsc{bfgs} method.
 \item Newton-\textsc{grape} and \textsc{bfgs}-\textsc{grape} methods, using \textsc{trm} as a regularisation method. The minimum eigenvalue of the Hessian is shifted with \(\trm=1\) in \cref{eqn:trm}. At least one iteration of \textsc{trm} was forced for the \textsc{bfgs} method.
 \item Newton-\textsc{grape} and \textsc{bfgs}-\textsc{grape} methods, using \textsc{rfo} as a regularisation and conditioning method. The iterative conditioning bounds the Hessian to \(\condno<1\times 10^4\) with \(\phi=0.9\) in \cref{eqn:cond_reduce}. At least one iteration of \textsc{rfo} was forced for the \textsc{bfgs} method.
\end{enumerate}

Newton-Raphson methods that need Hessian regularisation need a framework of comparison to current leading \cite{FOUQUIERES11} quasi-Newton methods. However, the \(\ell\)-\textsc{bfgs} method does not store an explicit Hessian matrix, only calculating the resulting search direction vector. For the use of testing regularisation methods, standard \textsc{bfgs} methods will be used -- which should give very similar convergence since the \(\ell\)-\textsc{bfgs} and \textsc{bfgs} methods are both rank-2 updates \cite{NOCEDAL06}.

The \textsc{dfp} and \textsc{bfgs} methods perform a rank-2 update, commonly on the inverse Hessian using \cref{eqn:inv_dfp,eqn:bfgs}, and with a defined positive definite inverse Hessian approximation. However, regularisation methods of \cref{eqn:trm,eqn:rfo1,eqn:rfo2} may have other beneficial effects on the Hessian matrix; making the Hessian sufficiently positive definite and well-conditioned. A fair comparison of regularised Newton-Raphson methods should be with quasi-Newton methods which update the Hessian proper with \cref{eqn:bfgs} with the same, forced, regularisation and conditioning as the Newton-Raphson Hessian. Concentrating on the arguably superior rank-2 update of the \textsc{bfgs} method, \cref{Fig:HCF_BFGS_compare} shows a comparison between the two \textsc{bfgs} update formulas in \cref{eqn:inv_bfgs,eqn:bfgs}.

The plots of \cref{Fig:HCF_BFGS_compare} clearly show there is no appreciable difference of performance or computational cost in the \textsc{bfgs} Hessian update and the \textsc{bfgs} inverse-Hessian update methods. This justifies the use of a \textsc{bfgs} Hessian update of \cref{eqn:bfgs} as a benchmark of comparison using regularised and conditioned Hessian matrices. \cref{Fig:HCF_BFGS_compare} should be used as a benchmark for the current best \textsc{grape} method with \(\ell\)-\textsc{bfgs} optimisation \cite{FOUQUIERES11}.

\subsubsection{Correcting an indefinite Hessian with Cholesky factorisation}\label{Section:IndefiniteHess}

Initial testing of the Newton-Raphson method failed to proceed further than a single iterate, failing to find a search direction based from an indefinite Hessian matrix. Further to this, forcing a step along the direction of the Newton-step shows the direction is not an ascent direction and an optimisation cannot proceed based on violation of the Wolfe conditions.

An iterative Cholesky factorisation, of \cref{eqn:cholesky1}, can be used to correct the non-ascent direction from an indefinite Hessian -- forcing the positive definite property, allowing a Newton-Raphson method to proceed. Results of this optimisation are shown compared with the equivalent \textsc{bfgs} method, where at least one Cholesky factorisation is forced, in \cref{Fig:HCF_CHOL_compare}. Both methods use a bracketing and sectioning line search with cubic interpolation.

\begin{figure}
\centering{\includegraphics{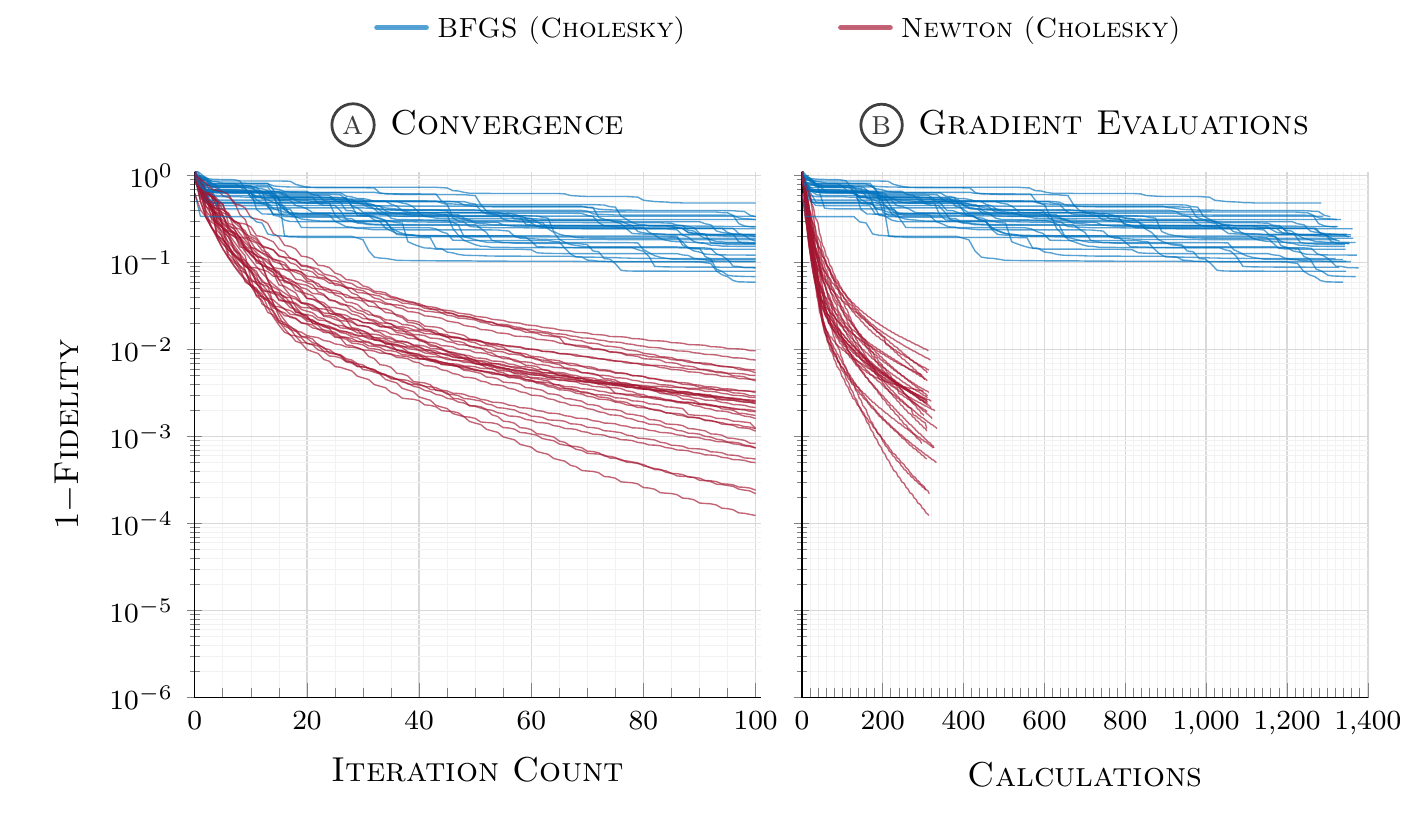}}
 \caption[Newton and BFGS methods with Cholesky factorisation]{Comparison of optimisation methods for state-to-state transfer in a scalar coupled hydrofluorocarbon fragment, transferring the population between the \(\hat{L}_{z}^\text{(H)}\) and \(\hat{L}_{z}^\text{(F)}\) states (pulse and trajectory shown in \cref{Fig:HCF_traj}): Newton-\textsc{grape} and \textsc{bfgs}-\textsc{grape} methods are regularised with a Cholesky factorisation (forced in the case of \textsc{bfgs}-\textsc{grape}). Both methods use a bracketing and sectioning line search. \plotlabel{a} Comparing the iterations count. \plotlabel{b} Comparing the gradient/Hessian calculation count.}\label{Fig:HCF_CHOL_compare}
\end{figure}

The convergence characteristics in \cref{Fig:HCF_CHOL_compare} are not promising: linear convergence at best. They show that an iterative Cholesky factorisation does allow the Newton-Raphson method to proceed, finding an ascent direction, correcting an indefinite Hessian. This also shows that regularisation is needed for the Newton-\textsc{grape} method.

However, the convergence for the Newton-Raphson method is not quadratic, reverting to the linear convergence of a gradient ascent method. Furthermore, there is a heavy load on the line search, which is indicated by the large amount of gradient calculations in \cref{Fig:HCF_CHOL_compare} \plotlabel{b}.

Comparison of the Newton method to the \textsc{bfgs} method, both with a Cholesky factorisation show that this method of regularisation quickly destroys any useful information from the Hessian and the Hessian approximation -- both methods effectively rely on the line search to make progress, with the second order information only showing a vague ascent direction. 

The iterative Cholesky factorisation reduces the Newton and \textsc{bfgs} methods to a gradient ascent method, which is clearly not adequate for a Newton-Raphson method in \textsc{grape}.

Although a Cholesky factorisation is not ideal for regularising a Hessian, it is a cheap way to recognise an indefinite matrix. Within a finished Newton-Raphson method product, the check of its positive definite property should be done with an attempted Cholesky factorisation: if it succeeds then no regularisation is required, if not then the Newton-Raphson method should proceed to a more sophisticated regularisation method.

\subsubsection{Correcting an ill-conditioned Hessian with RFO \& TRM}\label{Section:IllcondHess}

\begin{figure}
\centering{\includegraphics{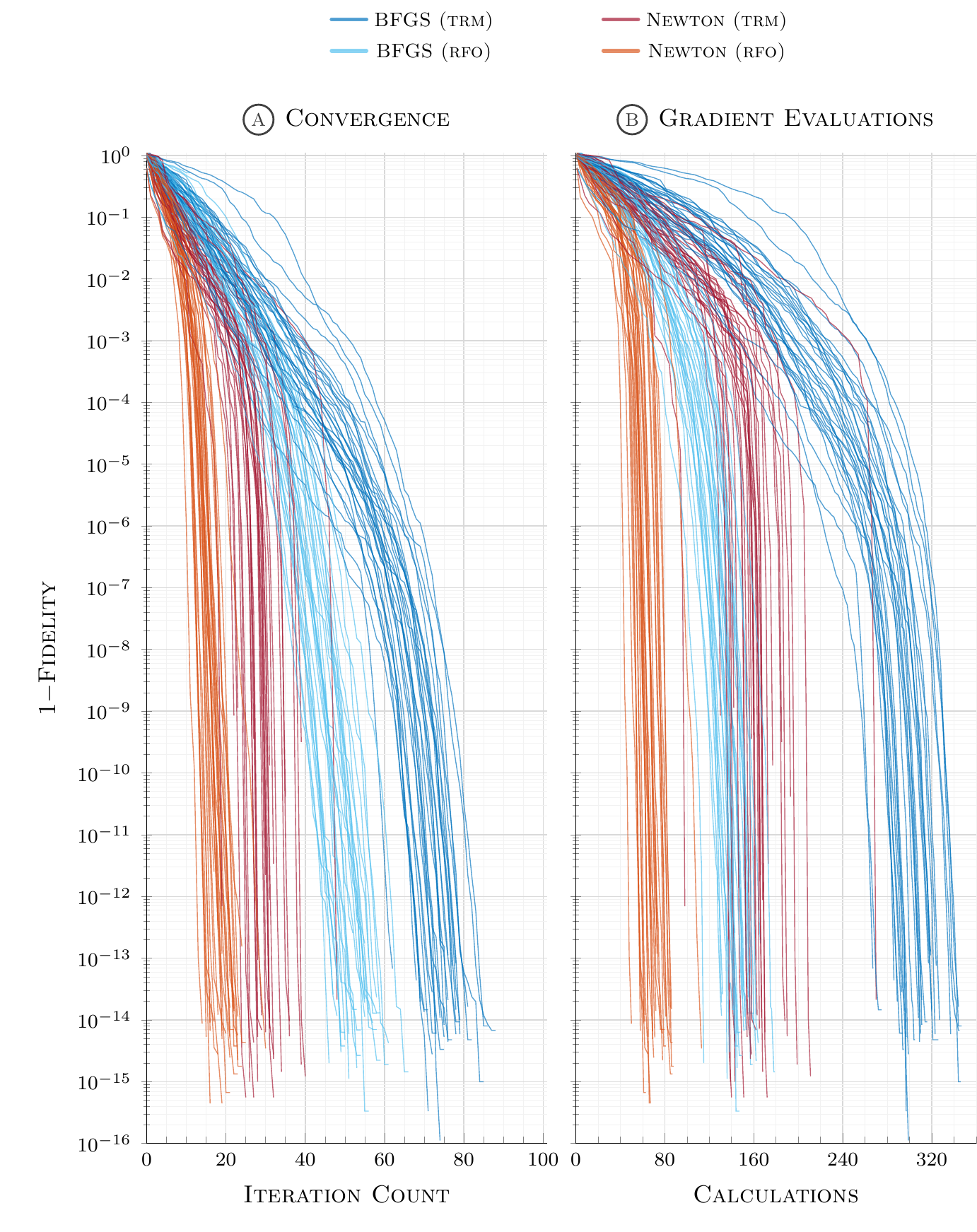}}
 \caption[Newton and BFGS methods with regularisation and conditioning]{Comparison of optimisation methods for state-to-state transfer in a scalar coupled hydrofluorocarbon fragment, transferring the population between the \(\hat{L}_{z}^\text{(H)}\) and \(\hat{L}_{z}^\text{(F)}\) states (pulse and trajectory shown in \cref{Fig:HCF_traj}): All methods use a bracketing and sectioning line search. \plotlabel{a} Comparing the iterations count. \plotlabel{b} Comparing the gradient/Hessian calculation count.}\label{Fig:HCF_compare}
\end{figure}

As indicated in \cref{Section:conditioning}, if the condition number of the Hessian is not bounded, a search direction can get too close to the vector orthogonal to the gradient. One method that can control the condition number is the \textsc{trm} of \cref{eqn:trm}. Here, a multiple of the identity matrix is added to the Hessian eigendecomposition, shifting all eigenvalues to a defined level above zero. More specifically, making a shift to \(\mineig=1\) ensures that small eigenvalues do not inflate the condition number.

Results in \cref{Fig:HCF_compare} indicate, in addition to correcting for a non-ascent direction and allowing a Newton-Raphson method to proceed, that a quadratic convergence phase does occur in the vicinity of the maximiser. Comparison of a \textsc{trm} conditioned \textsc{bfgs} method in \cref{Fig:HCF_compare} to that without \textsc{trm} in \cref{Fig:HCF_BFGS_compare} show a slight improvement in convergence, with \(~70\%\) reduction in optimisation iterations and a large saving in computational cost of \(~65\%\) reduction in gradient calculations. This indicates that a better search direction is produced with \textsc{trm} than without it.

The conditioned Newton-Raphson method cannot be compared fairly with an unconditioned Newton-Raphson method, and it must be inferred that \textsc{trm} has a similar effect in finding a good search direction. When comparing the Newton-\textsc{trm} with \textsc{bfgs}-\textsc{trm} in \cref{Fig:HCF_compare}, the benefits of calculating an explicit Hessian are evident, showing its expected quadratic convergence: Newton-\textsc{trm} approaches its quadratic phase in about half the number of iterations and at about half the calculation cost of \textsc{bfgs}-\textsc{trm}. However, there is a difference in the character of the quadratic phase of the two methods, the Newton-Raphson method follows a much steeper quadratic convergence, and the \textsc{bfgs} method follows a quadratic convergence with a similar profile to that of \textsc{bfgs} without forced regularisation and conditioning in \cref{Fig:HCF_BFGS_compare}.

\begin{figure}
\centering{\includegraphics{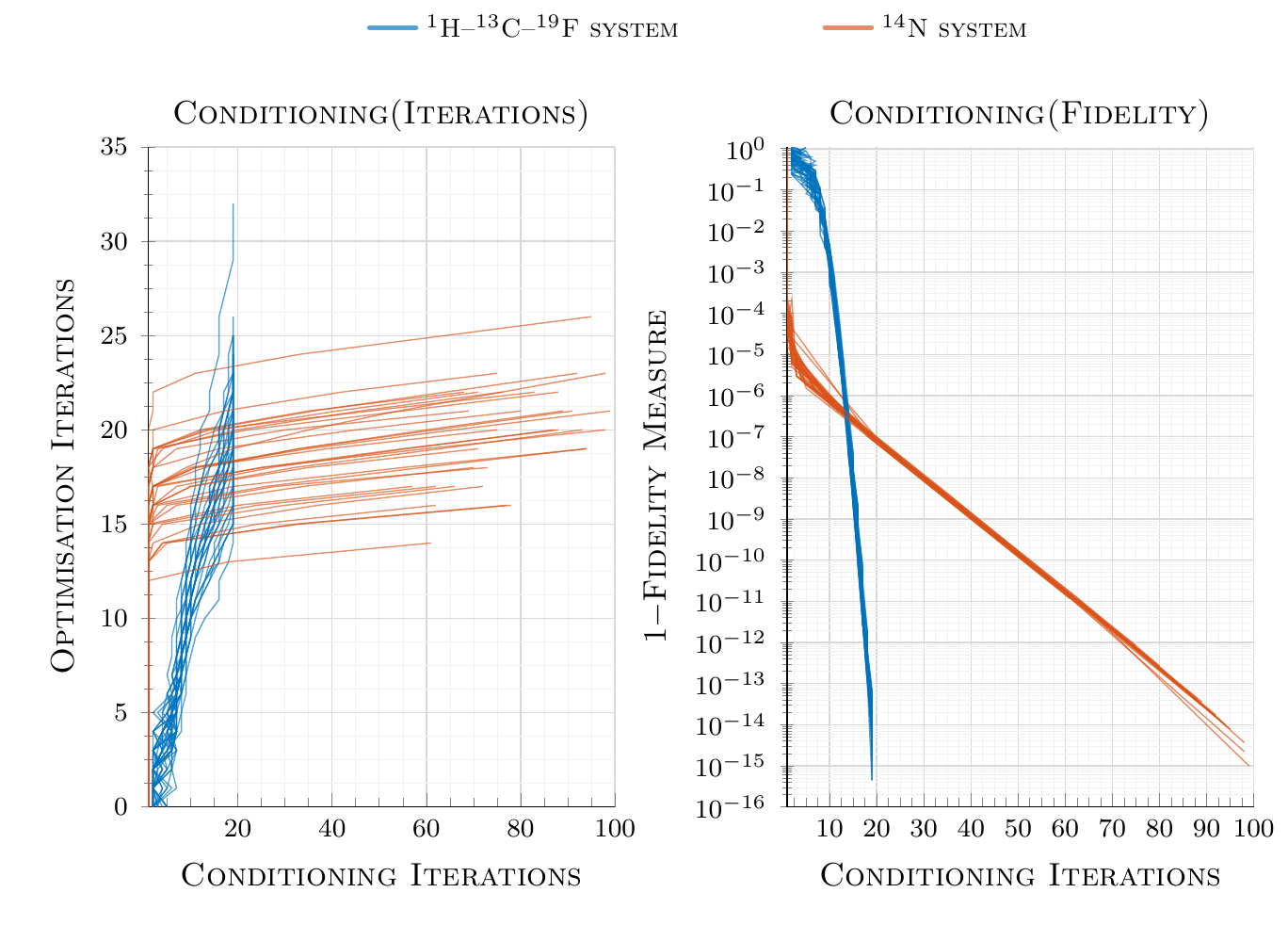}}
 \caption[Hessian conditioning iterations]{Hessian conditioning iterations for Newton-Raphson with \textsc{rfo}. Showing how  the number of conditioning iterations increases for both optimisation problems of \cref{Fig:14N_compare} and \cref{Fig:HCF_compare}; \plotlabel{a} as the optimisation iterations increase, and \plotlabel{b} as the fidelity decreases.} \label{Fig:14N_conditioning}
\end{figure}

The iteratively conditioned \textsc{rfo} method of \cref{eqn:rfoscale1,eqn:rfoscale2} including scaling of the Hessian with its corresponding gradient, is the final method compared in \cref{Fig:HCF_compare}. The Hessian condition number is bound below $10^4$ (with $\phi=0.9$ in \cref{eqn:cond_reduce})\footnote{Testing find a condition number bound of $\condno<10^4$ particularly effective. The theoretical bound of \(\condno<\sqrt[3]{\eps}\approx 1.6\times 10^{5}\) performs slightly worse in comparison, reaching quadratic convergence a couple of iterations later than with $\condno<10^4$.}. The need to increase the number of conditioning iterations as the maximum is approached can clearly be seen in \cref{Fig:14N_conditioning}, indicating that the Hessian becomes more ill-conditioned the closer it gets to a maximum. The \textsc{rfo} methods both outperform their \textsc{trm} counterparts in convergence and computational cost. It should also be noted that the \textsc{bfgs} \textsc{rfo} has less computational cost compared to the Newton-Raphson \textsc{trm}. This is because the self-scaling provided by the \textsc{rfo} greatly reduces the strain on the line search algorithm. Essentially, \textsc{rfo} scales the Hessian and gradient, which is usually performed within the line search giving a smaller step size for a correspondingly large condition number of the Hessian. 

The Newton-Raphson method with \textsc{rfo} regularisation and iterative conditioning has \(~85\%\) reduction in the computational cost of the current best \textsc{bfgs} method in \cite{FOUQUIERES11} (when comparing to \cref{Fig:HCF_BFGS_compare}), and has a predictable convergence (evident from the small variance from the average convergence trajectories), with a reduction of \(~80\%\) in optimisation iterations, compared to the \textsc{bfgs} method. Further information on the Hessian structure is depicted in \cref{Fig:HessianPics}, indicating \textsc{rfo} regularisation preserves more of the Hessian structure compared to other regularisation methods.

\begin{figure}
\centering{\includegraphics{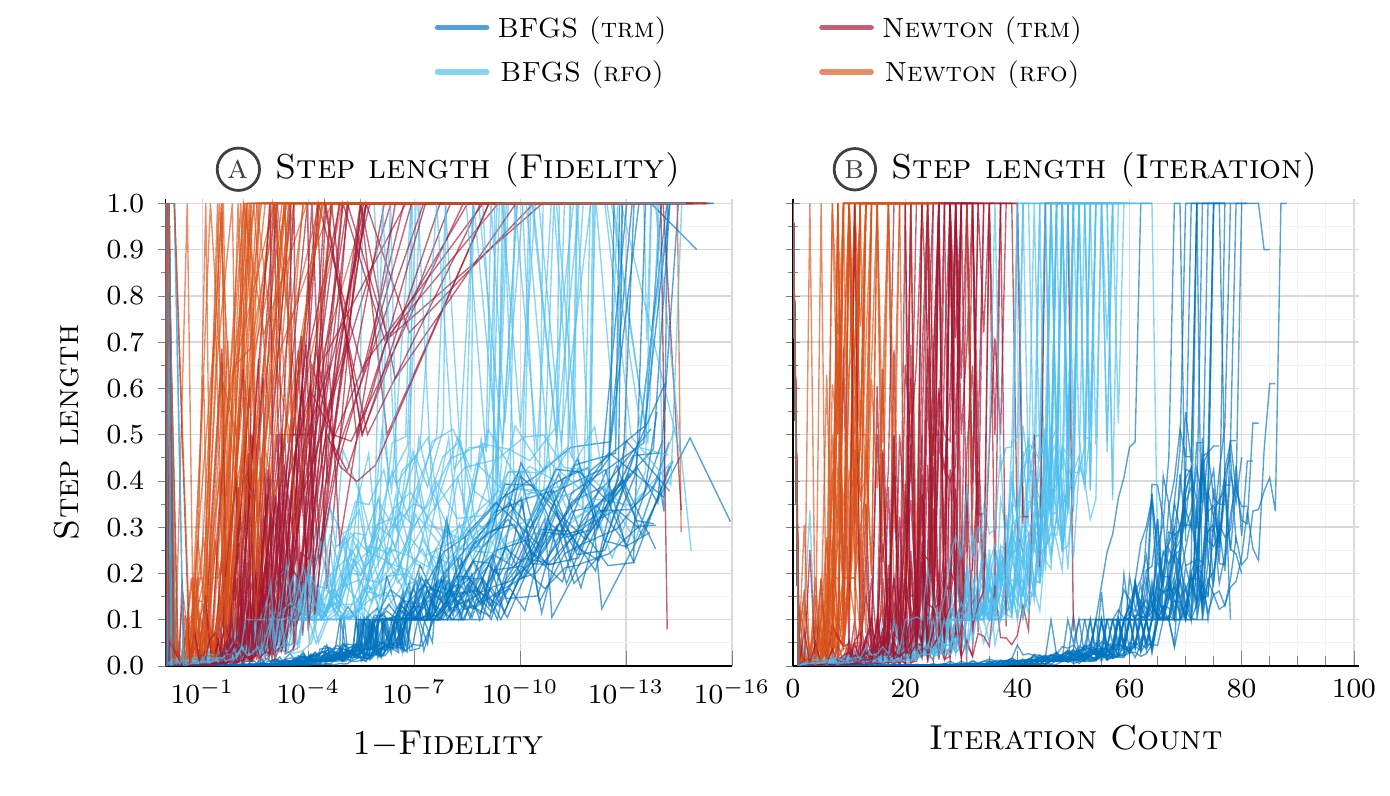}}
 \caption[HCF optimisation step lengths]{Step lengths for optimisations is \cref{Fig:HCF_compare}, \plotlabel{a} as a function of \(1-\Fid_1\), and \plotlabel{b} as a function of iteration count.} \label{Fig:HCF_steplength}
\end{figure}

Looking deeper into the line search, step lengths at each iterate show how the Newton-Raphson method with \textsc{rfo} has the best convergence characteristics of regularisation methods tested. Step lengths at each iterate for those simulations in \cref{Fig:HCF_compare} are shown in \cref{Fig:HCF_steplength}. \textsc{bfgs} and inverse-\textsc{bfgs} methods without regularisation have the same spread of step-lengths at each iterate, just as they have the same convergence characteristics in \cref{Fig:HCF_BFGS_compare} (not shown here). The general trend in all step length calculations of \cref{Fig:HCF_steplength} is that they start small and converge to \(\slen{}=1\) near the maximiser.

Quadratic convergence is reached very quickly in \(~10\) iterations for the Newton-\textsc{grape} method with \textsc{rfo}. This is evident from \cref{Fig:HCF_steplength} \plotlabel{b}, where the step length \(\slen{}=1\) and remains there until a optimum is reached in (to the accuracy of \(\eps\))\footnote{The only time there is true quadratic convergence is when the step length has converged to \(1\) and remains \(1\) for the rest of the simulation. This is characteristic of the Newton-Raphson method and is a good indication that Hessian calculations and regularisation calculations have no errors - and the method is indeed a Newton-Raphson method proper.}.

\textsc{bfgs} methods with regularisation have a similar shape in step length evolution in \cref{Fig:HCF_steplength} to \textsc{bfgs} method without regularisation. This improvement is seen at the start of the step-length evolution, approximately the first ten iterations, where step-lengths are larger than those of the \textsc{bfgs} methods without regularisation. The regularised \textsc{bfgs} methods then follow a similar shape to non-regularised methods, having a similar rate of change per iterate. The conclusion here is that regularisation does indeed improve the \textsc{bfgs} method, and this is through better initial step lengths calculations.

It should be noted that when a Hessian calculation is made, there is no need for a separate gradient calculation (see \cref{Chapter:AuxMat}). The only gradient calculations needed for the Newton-\textsc{grape} method are those during the line search, where no Hessian is calculated. There is redundancy when the Newton-Raphson method reaches quadratic convergence, when the step length is the Newton-step; a gradient calculation must be made at the start of a line search to check that no line search is needed. This accounts for the larger than expected calculation count in \cref{Fig:HCF_compare} \plotlabel{b}. If this redundancy can be removed, with a method that can switch the line search off when it detects quadratic convergence -- the gradient count for the Newton-\textsc{grape} method with \textsc{rfo} would have between 10-20 fewer gradient calculations\footnote{There is no such switching method for the \textsc{bfgs} method because the Hessian update occurs after the line search, whereas the Newton-Raphson method calculates a Hessian before a line search.}.

\subsection[Quadrupolar spin state-to-state population transfer]{Quadrupolar spin state-to-state\linebreak population transfer}\label{Section:test2}

\begin{figure}
\centering{\includegraphics{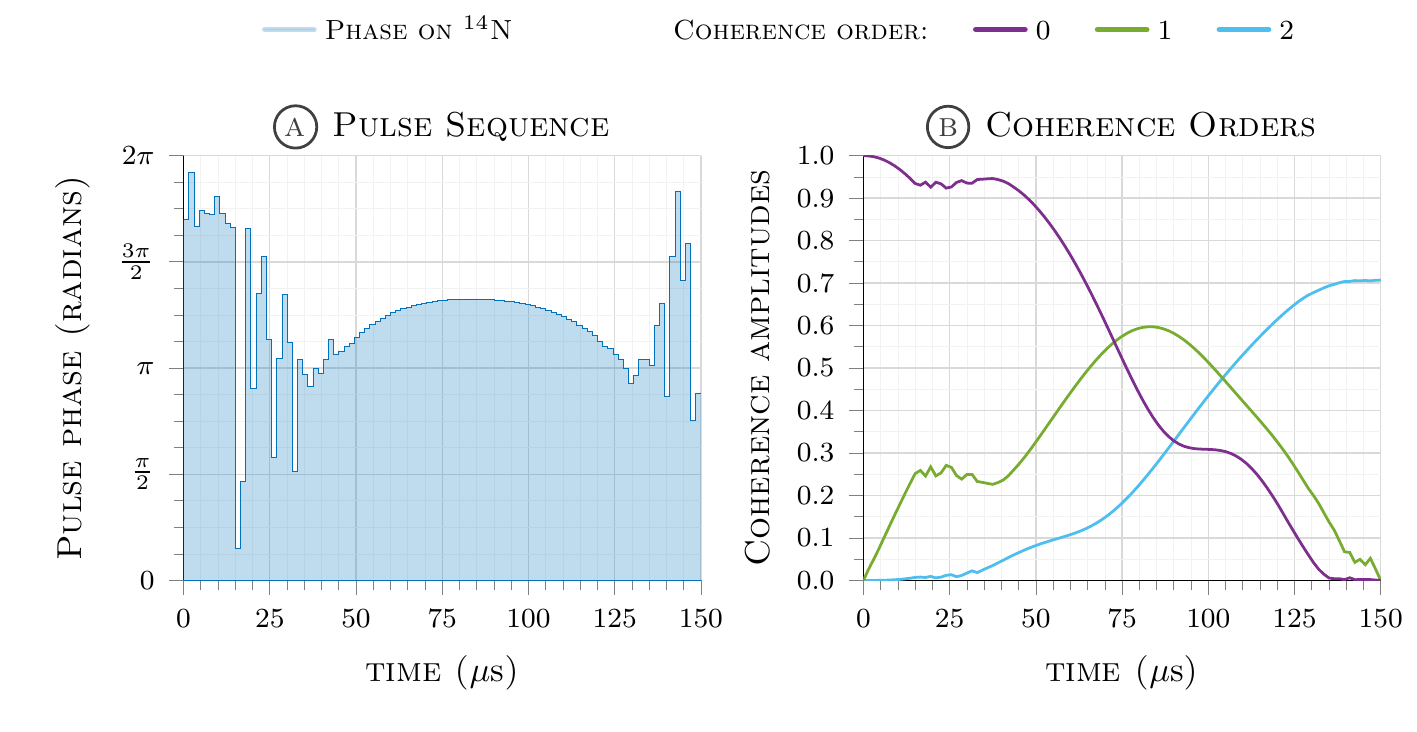}}
 \caption[Trajectory analysis of 14N with Newton-step]{Results of a Newton-Raphson optimisation for a state-to-state transfer in a quadrupolar \(_{}^{14}\)N spin, transferring the population between the \(T_{1,0}^{}\) and \(T_{2,2}^{}\) states (\cref{Chapter:AppendixA}). The convergence characteristics for this optimisation \cref{Fig:14N_compare}. \plotlabel{a} The phase modulated pulse sequence at the end of the optimisation. The amplitude profile is kept at a constant level of \(4\sqrt{2}\times 10^4\)~Hz. \plotlabel{b}
 Trajectory of the coherence order evolution during the control pulse sequence.}\label{Fig:14N_traj}
\end{figure}

This system consists of a single \(\spinint{1}\) \(_{}^{14}\)N in a \(14.1\)~tesla magnetic field, with coupling tensor of \cref{eqn:quadrupolar} in Hz
\begin{equation}
 \bm{V}=2\pi\begin{pmatrix}
  1 & 0 & 0\\
  0 & 2 & 0\\
  0 & 0 & -3
 \end{pmatrix}\cdot 10^4
\end{equation}
specifying the quadrupolar interaction, giving the drift Hamiltonian of the system as
\begin{equation}
 \HamHH_0=\vhat{L}\ccdot\bm{V}\ccdot\vhat{L}
\end{equation}
to be used in \cref{eqn:HamSep}.

The initial state is the irreducible spherical tensor operator \(\hat{T}_{1,0}^{}\) state of \cref{eqn:T10} in Liouville space, and the desired target state is the \(\hat{T}_{2,2}^{}\) state of \cref{eqn:T22}, with both states normalised.

The total control pulse duration is \(150\)~\(\mu\)s and the sequence is discretised to \(100\) equal time slices. Control operators are \(\HamHH_x^{(\text{N})}\) and \(\HamHH_y^{(\text{N})}\) operators, although the optimisation variables are the pulse sequence phase. The transform for this is
\begin{align}
 && \ctrlv{A}=\sqrt{\ctrlv{}_x^{\,2}+\ctrlv{}_y^{\,2}} && \ctrlv{\varphi}=\atan2(\ctrlv{y},\ctrlv{x}) &&\label{eqn:controlphase}
\end{align}
During the optimisation, the control phase, \(\ctrlv{\varphi}\) is modified and the control amplitude, \(\ctrlv{A}\) is kept at the constant level of \(\sqrt{2}\times 40\)~kHz.

The results of one of these, fully converged, optimisations is depicted in \cref{Fig:14N_traj}, showing the final phase modulated pulse sequence and the evolution of the coherence order over the total pulse time. The final fidelity reaches the maximum bound of \(\Fid_1=\frac{1}{\sqrt{2}}\) after 150~\(\mu\)s of pulsing.

\begin{figure}
\centering{\includegraphics{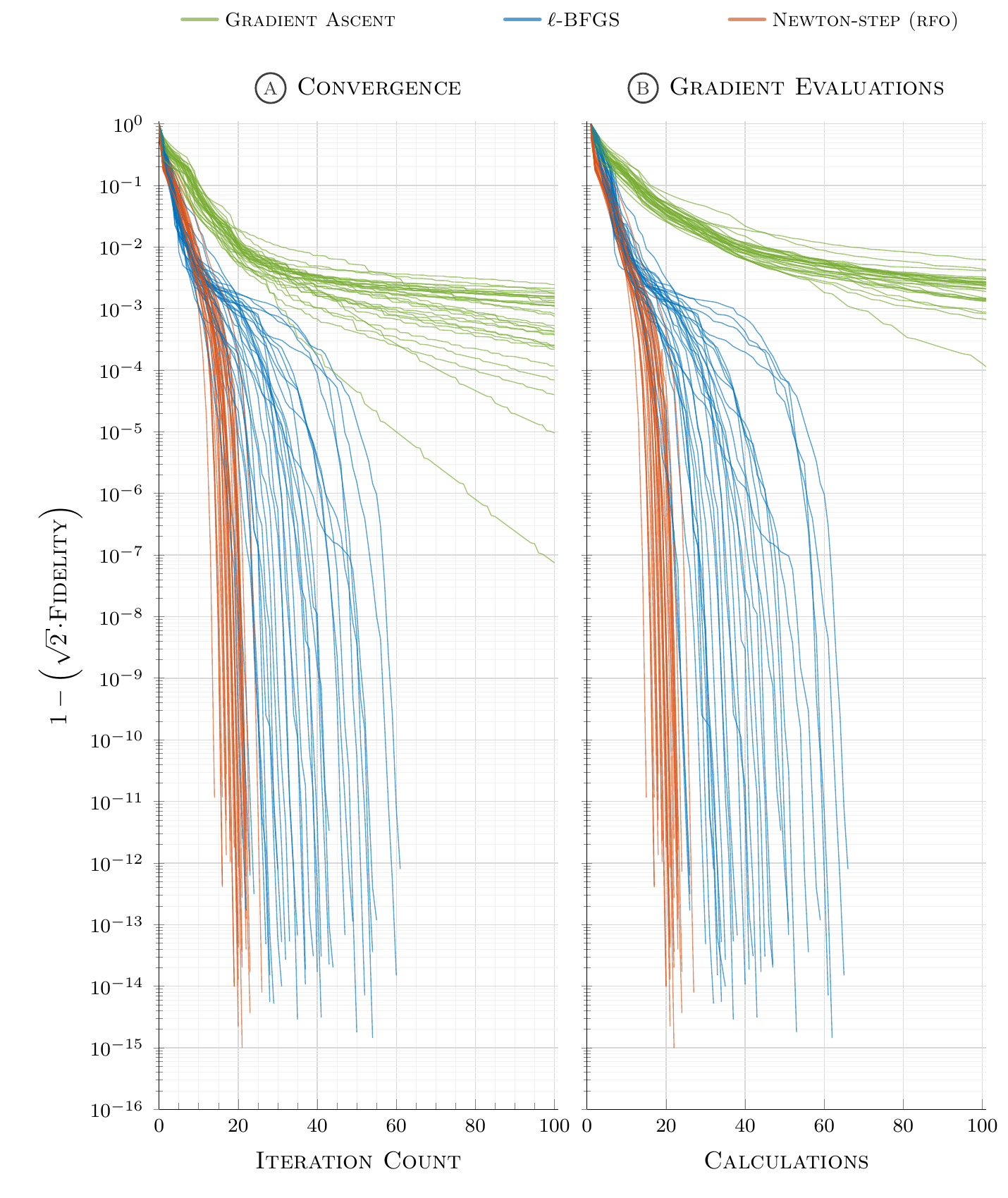}}
 \caption[Convergence of Newton-GRAPE, without a line search.]{Comparison of optimisation methods for state-to-state transfer in a quadrupolar \(_{}^{14}\)N spin, transferring the population between the \(T_{1,0}^{}\) and \(T_{2,2}^{}\) states (pulse and trajectory shown in \cref{Fig:14N_traj}): Newton-\textsc{grape} method using its Newton-step (i.e. without a line search); \(\ell\)-\textsc{bfgs}-\textsc{grape} using a bracketing and sectioning line search; the gradient ascent method, also using a bracketing and sectioning line search. \plotlabel{a} Comparing the iterations count. \plotlabel{b} Comparing the gradient/Hessian calculation count.}\label{Fig:14N_compare}
\end{figure}

Three optimisation methods were compared:
\begin{enumerate}
 \item Newton-\textsc{grape} method using its Newton-step i.e. without a line search. Hessian regularisation is performed with  \textsc{rfo}, and iterative conditioning bounds the Hessian to \(\condno<1\times 10^4\) with \(\phi=0.9\) in \cref{eqn:cond_reduce}.
 \item \(\ell\)-\textsc{bfgs}-\textsc{grape} using a bracketing and sectioning line search, with cubic interpolation during the sectioning phase.
 \item \textsc{grape} method (gradient ascent) using a bracketing and sectioning line search, with cubic interpolation during the sectioning phase.
\end{enumerate}

\cref{Fig:14N_compare} compares the convergence of the three optimisation methods, plotting \(1-\sqrt{2}\cdot\Fid_1\) on a log scale to clearly see the convergence to the maximum fidelity value of \(\Fid_1=\frac{1}{\sqrt{2}}\).

The Newton-\textsc{grape} performs well for this optimisation -- without the need for a line search, the method reaches quadratic convergence after a small number of iterations and full quadratic convergence occurs in a small range of iterations, \(~15-20\). Admittedly, this is an easy optimisation with only a single spin, and the \(\ell\)-\textsc{bfgs}-\textsc{grape} method also performs well considering it only uses gradient calculations. Furthermore, the \(\ell\)-\textsc{bfgs}-\textsc{grape} method rarely makes use of its available line search, with its rank-2 update giving a Newton-like step length at most iterations. The rank-2 update uses a gradient history to give its search direction, and it should be expected that the algorithm gets better as it has a longer history of gradients -- after around \(20\) iterations in \cref{Fig:14N_compare} \plotlabel{a}.

The benefit of using a Newton-step for the Newton-\textsc{grape} method, needing far fewer gradient evaluations, becomes a small benefit considering that the \(\ell\)-\textsc{bfgs}-\textsc{grape} method rarely uses the available line search. Furthermore, running the same Newton optimisations with any line search method (not shown here) results in the same -- the method rarely uses a step length other than \(1\). The effect is to double the number of gradient evaluations without any increase in the convergence rate.

There are a few important uses of this test system. The first is that it shows the calculated Hessian is the correct one, removing any doubt that it contains numerical errors. The second is that it is a useful test to show how the condition number inflates compared to the optimisation of the previous section \cref{Section:HCF}. \cref{Fig:14N_conditioning} compares the number of conditioning iterations to those of \cref{Section:HCF} -- showing a very different profile of how the condition number increases as the optimisation progresses. The HCF system gradually increases its condition number, starting from the first iteration, but the \(_{}^{14}\)N system does not need any conditioning until it reaches a certain fidelity, at which point the condition number increases exponentially with the fidelity increase.

\subsection{Universal rotation}\label{Section:test3}

This system consists of a \(101\) non-interacting \(\spinint{1}\) \(_{}^{13}\)C spins, spread over \(50\)~kHz, in a \(14.1\)~tesla magnetic field. Relaxation is included in the form of the \(T_1/T_2\) approximation, keeping only the diagonal terms of the relaxation superoperator. Longitudinal states, corresponding to \(\hhat{L}_z\), relax with a rate of \(10\)~Hz, and the transverse states, corresponding to \(\hhat{L}_{\pm}\), relax at a rate of \(20\)~Hz.

\begin{figure}
\centering{\includegraphics{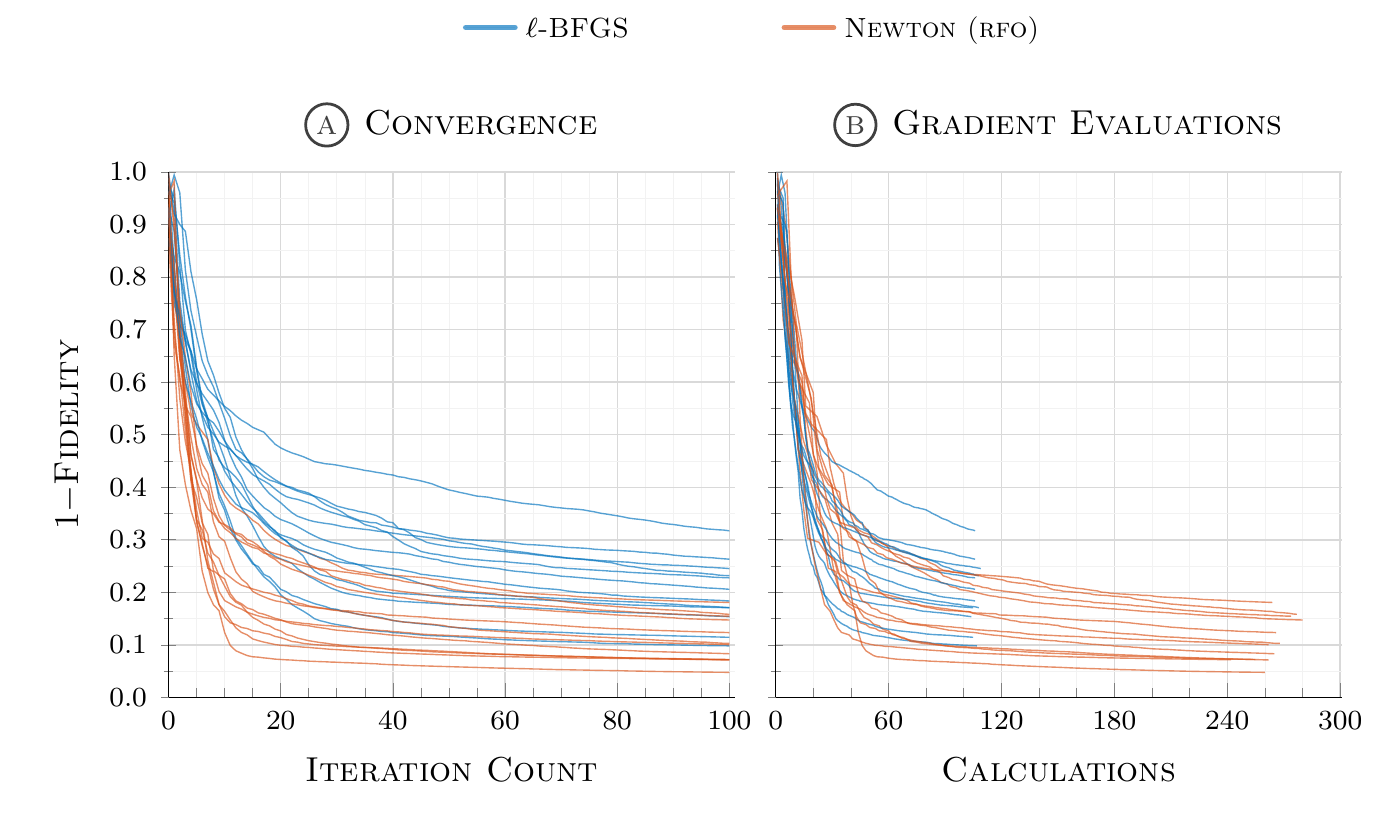}}
 \caption[Convergence of Newton-GRAPE method for a universal rotation.]{Comparison of optimisation methods for a universal rotation in 3D, on a \(101\) non-interacting \(\spinint{1}\) \(_{}^{13}\)C spins, spread over \(50\)~kHz. The Hamiltonian includes relaxation with the \(T_1/T_2\) approximation. Two methods are compared: Newton-\textsc{grape} method using a bracketing and sectioning line search; \(\ell\)-\textsc{bfgs}-\textsc{grape} using a bracketing and sectioning line search. \plotlabel{a} Comparing the iterations count. \plotlabel{b} Comparing the gradient/Hessian calculation count.}\label{Fig:UR_compare}
\end{figure}

The initial and final states of the system are set as a universal rotation in 3D, mentioned in \cref{Section:Fidelity}:
\begin{equation}
\begin{matrix}
  \big|\state_1^{}(0)\big\rangle=+\big|L_z^{}\big\rangle &\longrightarrow &-\big|L_x^{}\big\rangle=\big|\targ_1^{}\big\rangle\\
  \big|\state_2^{}(0)\big\rangle=+\big|L_x^{}\big\rangle &\longrightarrow &-\big|L_z^{}\big\rangle=\big|\targ_2^{}\big\rangle\\
  \big|\state_3^{}(0)\big\rangle=+\big|L_y^{}\big\rangle &\longrightarrow &-\big|L_y^{}\big\rangle=\big|\targ_3^{}\big\rangle\nonumber
\end{matrix}
\end{equation}
The total control pulse duration is \(5\)~ms and the sequence is discretised to \(400\) equal time slices. Control operators are \(\HamHH_x^{(\text{C})}\) and \(\HamHH_y^{(\text{C})}\) operators, although the optimisation variables are the pulse sequence phase in \cref{eqn:controlphase}. During the optimisation, the control phase, \(\ctrlv{\varphi}\) is modified and the control amplitude, \(\ctrlv{A}\) is kept constant as a shaped amplitude profile: \(\sfrac{1}{2}(1+\sin(\theta))\), \(\forall\,\theta\in[0,\pi]\). Pulses are designed to be robust to \(\B{1}\) miscalibration by simulating over a range of power levels between \(3.75\)~kHz and \(6.25\)~kHz. This is achieved by simulating 5 optimisations over the given range, then averaging their fidelity, gradient and Hessian to be fed back to the optimisation algorithm.

As can be seen in \cref{Fig:UR_compare}, quadratic convergence is not reached. This may be attributed to a solution not existing for this universal rotation, or that the Newton-\textsc{grape} method may not perform well when finding robust solutions. Evidence leans to the latter, where the \(\ell\)-\textsc{bfgs}-\textsc{grape} reaches a similar convergence after an initial phase of fast convergence. Finding a solution robust over the given range, in addition to finding a universal rotation pulse that works at each power level in the range, seem too much for either algorithm to handle effectively. However, the Newton-\textsc{grape} method did show a marginally quicker initial convergence, although the number of gradient/Hessian calculations was similar. The tests on this type of optimisation problem indicate further work is required to find a more effective algorithm.

\chapter{Propagator Directional Derivatives} \label{Chapter:AuxMat}

\begin{chapquote}{Alexandre Dumas, \textsc{The Count of Monte Cristo}}
``Learning does not make one learned: there are those who have knowledge and those who have understanding. The first requires memory and the second philosophy.''
\end{chapquote}
\renewcommand*{\CurrentPath}{./Chapter_5}

In a general context, magnetic resonance simulation frequently requires chained exponential integrals of the form
\begin{equation}
 \int\limits_0^t\!\dd t_1^{}\!\int\limits_0^{t_1^{}}\!\dd t_2^{} \cdots\!\!\int\limits_0^{t_{r-2}^{}}\!\!\dd t_{r-1}^{}\Big\{\e^{(t-t_1^{})\mA_1^{}}\mB_1^{}\e^{(t_1^{}-t_2^{})\mA_2^{}}\mB_2^{}\cdots\mB_{r-1}^{}\e^{t_r^{}\mA_r^{}}\Big\}\label{eqn:chainedintegral}
\end{equation}
where \(\mA_r^{}\) and \(\mB_r^{}\) are square matrices. Examples include perturbative relaxation theories \cite{REDFIELD57,KUBO62,GOLDMAN01}, reaction yield expressions in radical pair dynamics \cite{HABERKORN76,TIMMEL98,JONES10}, average Hamiltonian theory \cite{HAEBERLEN68}, and pulsed field gradient propagators in nuclear magnetic resonance \cite{EDWARDS12}. Their common feature is the complexity of evaluation: expensive matrix factorisations are usually required \cite{GOLDMAN01,ABRAGAM61,ERNST87}. 

The exponentiation of a matrix \cite{LAGUERRE67} is far from a trivial operation. The operation was recognised as an important tool for solving differential equations \cite{FRAZER38} -- optimal control methods rely heavily on it. The literature review in \cref{Chapter:Introduction} identified the bottleneck in computation during the \textsc{grape} method as calculation of a matrix exponential \cite{GRADL06,SCHULTE_HERBRUEGGEN09,TOSNER09,AUCKENTHALER10,FOUQUIERES11}. An outline of the \textsc{grape} method was presented in \cref{Chapter:GRAPE}, with the stages of forward/backward propagation \cref{eqn:propagatorN,eqn:fwdprop,eqn:bwdprop}, and calculation of propagator derivatives \cref{eqn:propagatorN,eqn:fidgrad0,eqn:fidgrad1,eqn:fidgrad2}, requiring exponentiation of a matrix. The second-order derivatives of propagators used to construct a Hessian matrix in \cref{Chapter:Hessian} also require the exponentiation of a matrix \cref{eqn:fidhess0_DIAG,eqn:fidhess1_DIAG,eqn:fidhess2_DIAG,eqn:fidhess0_BLKDIAG,eqn:fidhess1_BLKDIAG,eqn:fidhess2_BLKDIAG}.

This chapter will introduce the matrix exponential in \cref{Section:LitEXP} and outline some of the ``\textit{nineteen dubious way to compute the exponential of a matrix}'' to compute this, from the well cited work of Cleve Moler\footnote{It is no coincidence that Cleve Moler was the inventor of Matlab and at the forefront of research into numerical computation and matrix computations. He was also the doctoral advisor of Charles Van Loan.} and Charles Van Loan \cite{MOLER78,MOLER03}. Much of the work used in this chapter can be found in textbooks on computations of matrix functions \cite{GOLUB13,HIGHAM08}. 

The derivatives required for \textsc{grape} are directional derivatives of the matrix exponential and these are defined in \cref{Section:DirDerivEXP,Section:2ndDirDeriv}. With specific application to optimal control \cite{KHANEJA05,TOSNER09,FOUQUIERES11}, chained exponential integrals will be used to find to analytic derivatives required by \textsc{grape}, with a novel technique presented \cref{Section:AuxiliaryMatrix} and algorithmic subtleties presented in \cref{Section:KrylovAlgo} with a measure of how parallel the algorithm is in \cref{Section:Parallel} using propagator recycling in \cref{Section:PropRecycle}.

\section{The Matrix Exponential}\label{Section:LitEXP}

The matrix exponential operation is not a trivial one \cite{VANLOAN77} and approximations must be used to numerically compute the matrix exponential within Matlab \cite{HIGHAM05,AL-MOHY09}. Much of the material in this section is taken from derivations in \cite{GOLUB13,HIGHAM08}. An indispensable resource, by Moler and van Loan \cite{MOLER78,MOLER03} outlines an extensive review, showing ``\textit{nineteen dubious way to compute the exponential of a matrix}''.

\subsection{Taylor series approximation}\label{Section:TaylorExp}

The Taylor series \cite{TAYLOR1715} has already been used to form the quadratic model used as the basis of Newton-type methods of numerical optimisation in \cref{eqn:taylorseries}. The Taylor series is also the first call in approximating the function of matrix, \(f(\mM)\), and is obtained with the direct application of its scalar counterpart:
\begin{align}
 f(\mM)&=f(z_0^{})+\frac{\nabla f(z_0^{})}{1!}\big(\mM-z_0^{}\Unit\big) + \frac{\nabla^{2}f(z_0^{})}{2!}\big(\mM-z_0^{}\Unit\big)_{}^2+\cdots\nonumber\\&=\sum\limits_{r=0}^{\infty}\frac{\nabla^{r}f(z_0^{})}{r!}\big(\mM-z_0^{}\Unit\big)_{}^r
\end{align}
where \(\nabla^{r}f(z_0)\) is the \(r^\text{th}\) derivative of \(f\) at \(z_0^{}\). More specifically, the MacLaurin series \cite{MACLAURIN1742}, when \(z_0^{}=0\) above, of the scalar exponential function translates directly to the matrix exponential function of \(\e^{t\mM}\) \cite{LAGUERRE67}:
\begin{align}
 &&\e^{t\mM}=\Unit+\frac{t\mM}{1!}+\frac{t_{}^2\mM_{}^2}{2!}+\frac{t_{}^3\mM_{}^3}{3!}+\cdots=\sum\limits_{r=0}^{\infty}\frac{t_{}^r\mM_{}^r}{r!},&& \mM\in\Complex^{(n\times n)}\label{eqn:taylorexp}
\end{align}
In computing the exponential of a matrix, the Taylor series above should be summed to an appropriate number of terms, until adding another term with floating point arithmetic does not alter the numbers stored in the computer. This method is known to give bad results, even in the scalar case, when the series is truncated too soon -- when the approximation of what ``does not alter the numbers'' is a large increment compared to \(\eps\) \cite{MOLER78}. The error bound on this truncation is considered in more depth in \cite{LIOU66,MATHIAS93,HIGHAM08}.

The attraction of the Taylor series approximation, when truncated appropriately, is that it only involves matrix multiplications. Furthermore, the computation of the powers of a matrix \(\mM_{}^r\) in \cref{eqn:taylorexp} is made more efficient by expressing the matrix power in a binary expansion \cite{GOLUB13} e.g. \(\mM_{}^{13}\) requires 5 matrix multiplications:
\begin{equation*} 
\mM_{}^{13}=\Big(\big(\mM^2\big)_{}^2\Big)^2\big(\mM^2\big)^2\mM 
\end{equation*}

The \textit{Spinach} \cite{HOGBEN11} implementation of the matrix exponential uses a Taylor series approximation with scaling and squaring (see \cref{Section:scalingsquaring} below), designed to preserve sparsity, essential for large matrix dimensions -- \textsc{nmr} systems are in the millions, common in \textsc{nmr} \cite{EDWARDS14}). 

Spin Hamiltonians are guaranteed to be sparse in the Pauli basis \cite{DUMONT97,VESHTORT06,EDWARDS12} and their exponential propagators are also sparse when \(\|\HamB\Delta t\|<1\), if care is taken to eliminate insignificant elements after each matrix multiplication in the scaled and squared Taylor series procedure.

\subsection{Pad\'{e} approximation}\label{Section:PadeApprox}

A rational function approximation of a matrix is directly translated from that of a scalar function to give
\begin{equation}
 R_{pq}^{}(\mM)=\big[D_{pq}^{}(\mM)\big]^{-1}N_{pq}(\mM)\label{eqn:rationalmatrixfunction}
\end{equation}
where \(N_{pq}^{}\) and \(D_{pq}^{}\) are polynomials in \(\mM\) of at most degree-\(p\) of \(N_{pq}^{}\) and \(q\) of \(D_{pq}^{}\). The numerator polynomial is
\begin{equation}
 N_{pq}(\mM)=\sum\limits_{k=0}^p\frac{(p+q-k)!p!}{(p+q)!k!(p-k)!}\mM_{}^k\label{eqn:RationalNumerator}
\end{equation}
and the denominator polynomial is
\begin{equation}
 D_{pq}(\mM)=\sum\limits_{k=0}^q\frac{(p+q-k)!q!}{(p+q)!k!(q-k)!}(-\mM)_{}^k\label{eqn:RationalDenominator}
\end{equation}
When \(q=0\), the rational function approximation in \cref{eqn:rationalmatrixfunction} is the Taylor series truncated to order-\(p\).

If \(R_{pq}^{}(\mM)\in\Real^{(p\times q)}\) and \(D_{pq}^{}(\Zero)=\Unit\), the rational function approximation in \cref{eqn:rationalmatrixfunction} is the \([p/q]\) Pad\'{e} approximant \cite{PADE92}. The exact expression for the Pad\'{e} approximation of the matrix exponential is \cite{MOLER78,MOLER03}
\begin{equation}
 \e^{t\mM}=R_{pq}^{}(t\mM)+\frac{(-1)_{}^q}{(p+q)!}(t\mM_{})^{p+q+1}\big[D_{pq}^{}(t\mM)\big]^{-1} \!\!\int\limits_0^1\!u_{}^p(1-u)_{}^q\e^{t\mM(1-u)}\dd u\label{eqn:PadeApproxExp}
\end{equation}
Diagonal approximants, where \(p=q\), are more accurate than when \(p\neq q\) and are preferred in practice because \(R_{rr}^{}(\mM)\), with \(r=\max(p,q)\), has the same computational cost as when \(p\neq q\) \cite{HIGHAM08}.

As it stands in \cref{eqn:PadeApproxExp}, the Pad\'{e} approximant is only a good approximation of the matrix exponential near the origin, when \(\|\mM\|\) is small \cite{MOLER78,MOLER03,GOLUB13} (although the Chebyshev rational approximation can perform well in some circumstance \cref{Section:Chebychev}) -- requiring the \textit{scaling and squaring} technique in \cref{Section:scalingsquaring} to overcome the problem.

It should be noted that calculation of the ``perilous inverse'' matrix \(\big[D_{pq}^{}(\mM)\big]^{-1}\) is required, which can be an undesirable calculation \cite{VANLOAN78}.

\subsection{Chebyshev rational approximation}\label{Section:Chebychev}

If \(\mM\) is a Hermitian matrix, then approximating its exponential amounts to approximating the exponential function on the spectrum of eigenvalues of \(\mM\) \cite{HIGHAM08}. The Chebyshev rational approximation\footnote{The Chebyshev rational approximation is also known at the \textit{Best L\(_{\infty}\) approximation}, or the \textit{minimax} approximation \cite{HIGHAM08}.}, like the Pad\'{e} approximant of \cref{Section:PadeApprox} above, is based on the rational function approximation of \cref{eqn:rationalmatrixfunction}.
\begin{equation}
 C_{qq}^{}(\mM)=\big[D_{qq}^{}(\mM)\big]^{-1}N_{qq}(\mM)
\end{equation}
where only the diagonal approximants are considered. The polynomials, \(D_{qq}^{}(\mM)\) and \(N_{qq}(\mM)\), of order-\(q\) are those defined in \cref{eqn:RationalNumerator,eqn:RationalDenominator}. A minimiser for $\| C_{qq}^{}(\mM)-\e^{\mM}\|$ can be found from coefficients determined in \cite{CODY69,CARPENTER84}, where \(\mM\) is Hermitian with negative eigenvalues. This is a useful method for sparse matrices. Unlike the Pad\'{e} approximation of the matrix exponential, which is only valid near the origin, the Chebyshev rational approximation is valid for the whole negative real axis (or the whole positive real axis for \(\e^{-\mM}\)).

A success of this approximation is, when considering \(\e^{\mM\Delta t}\) with a time step \(\Delta t\), that the Chebyshev approximation allows much larger time steps compared to the Pad\'{e} approximant \cite{GALLOPOULOS92}. However, Chebyshev series diverge with non-Hermitian matrices \cite{NDONG10}, where the error bounds for the scalar problem do not translate to matrices \cite{HIGHAM08} -- the method is expected to fail when considering dissipative dynamics.

\subsection{Scaling and squaring}\label{Section:scalingsquaring}

The scaling and squaring technique \cite{LAWSON67,WARD77} exploits the relationship
\begin{align}
 && \e^{t\mM}=\Big(\e^{\frac{t\mM}{s}}\Big)^{s}, && \mM\in\Complex^{(n\times n)}
\end{align}
This can be used to control the round-off errors associated with the Taylor and Pad\'{e} approximants: as \(t\|\mM\|\) or the spread of eigenvalues increases \cite{MOLER78,MOLER03}. The choice of \(s\) to be the smallest power of \(2\) for which \(\sfrac{\|\mM\|}{s}\leqslant 1\) \cite{MOLER78,VANLOAN78}:
\begin{equation}
 \e^{t\mM}=\Big(\e^{\frac{t\mM}{2_{}^m}}\Big)^{2^m}
\end{equation}
Further efficiencies and error tolerance have been developed, the Pad\'{e} approximants with scaling and squaring is the method of matrix exponential used in both Matlab and Mathematica \cite{HIGHAM05,AL-MOHY09}. An attraction of its use with the Taylor series is that only approximate scaling is required \cite{NDONG10,HUISINGA99}.

The next section \cref{Section:directionalderivatives} will use a block triangular form \cite{PARLETT76} of the matrix \(\mM\), for the calculation of directional derivatives:
\begin{equation}
 \mM=\begin{pmatrix} \mA & \mB\\ \Zero & \mD\end{pmatrix}
\end{equation}
The method of Pad\'{e} approximants with scaling and squaring shows weakness when exponentiating block triangular matrices when \(\|\mB\|\gg\max\big(\|\mA\|,\|\mD\|\big)\) \cite{HIGHAM08}: the result is that the diagonal blocks are over-scaled \cite{KENNY98,DIECI00}.

%
%

\section{Directional derivatives}\label{Section:directionalderivatives}

Derivatives of functions of matrices are called \textit{directional derivatives}, or \textit{G\^{a}teaux derivatives}. which are defined in \cite{HIGHAM08} as
\begin{equation}
 D_{\mB}^{}(f(\mA))\triangleq\lim_{h\to0}\frac{f(\mA+h\mB) - f(\mA)}{h}=\frac{\dd}{\dd h}\bigg|_{h=0}\!\!\!f(\mA+h\mB) \label{eqn:directionalderivative}
\end{equation}
This has a similar form to the finite difference equation in \cref{eqn:forward_diff} and the Taylor series truncated to first order in \cref{eqn:taylorseries}, and should be considered the formal definition of the derivative of the function of a matrix\footnote{A directional derivative has a stronger notion of differentiability when it is a \textit{Frech\'{e}t derivative}. This is true if \(\exists D_{\mB}^{}(f(\mA))\) is a linear function of \(\mB\) and continuous in \(\mA\) \cite{HIGHAM08}. Assuming these qualifications are met, which is the case in this thesis, the terms directional derivative and Frech\'{e}t derivative are equivalent. The remainder of this thesis will use the more descriptive term \textit{directional derivative}.}. The interpretation of \(D_{\mB}^{}(f(\mA))\) is: \textit{the derivative of the function \(f\) at \(\mA\) in the direction \(\mB\)} \cite{NAJFELD95,HIGHAM08}.

\subsection{Directional derivatives of the matrix exponential}\label{Section:DirDerivEXP}

The directional derivative of the matrix exponential, \(\e^{t\mA}\), is then
\begin{equation}
 D_{\mB}^{}(\e^{t\mA})=\lim_{h\to0}\frac{\e^{t(\mA+h\mB)} - \e^{t\mA}}{h}=\frac{\dd}{\dd h}\bigg|_{h=0}\!\!\!\e^{t(\mA+h\mB)} \label{eqn:dirderivexp}
\end{equation}
which can also be expanded as a Taylor series \cite{HIGHAM08}:
\begin{align}
 D_{\mB}^{}(\e^{t\mA}) & =\e^{t(\mA+\mB)}-\e^{t\mA}\nonumber\\
 & =\frac{t}{1!}\mB + \frac{t_{}^2}{2!}\big(\mA\mB+\mB\mA\!\big) + \frac{t_{}^3}{3!}\big(\mA_{}^{\!2}\mB+\mA\mB\mA+\mB\mA_{}^{\!2}\!\big) +\cdots\label{eqn:taylordirderivexp}
\end{align}
This relates to the power directional derivatives \cite{NAJFELD95} by
\begin{align}
 D_{\mB}^{}(\e^{t\mA}) & = \sum\limits_{r=0}^{\infty}\frac{t_{}^r}{r!}\frac{\dd^{\,r}}{\dd t^r}D_{\mB}^{}(\e^{t\mA})\bigg|_{t=0}\nonumber\\
 & = \sum\limits_{r=1}^{\infty}\frac{t_{}^r}{r!}D_{\mB}^{}(\e^{\mA_{}^{r}})
\end{align}
which leads to the \textit{Magnus expansions} \cite{MAGNUS54,HAEBERLEN68}.

An alternative, and more useful representation than \cref{eqn:dirderivexp,eqn:taylordirderivexp} \cite{NAJFELD95} is the integral representation \cite{KARPLUS48,FEYNMAN51}:
\begin{equation}
D_{\mB}^{}(\e^{t\mA})=\!\!\int\limits_0^t\!\e^{(t-s)\mA\!}\mB\e^{s\mA}\dd s\label{eqn:integraldirderivexp}
\end{equation}
This will become a useful representation in the following sections. It is useful to note that rearranging this formula \cite{NAJFELD95} leads to a commutator series \cite{HAUSDORFF06,MAGNUS54}
\begin{align}
 D_{\mB}^{}(\e^{t\mA}) & =\Bigg[\int\limits_0^t\!\e^{s\mA}\mB\e^{-s\mA}\dd s\Bigg]\e^{t\mA}\nonumber\\
 & = \bigg\{\mB,\!\!\int\limits_0^t\!\e^{s\mA}\dd s\bigg\}\e^{t\mA}= \bigg\{\mB,\frac{\e^{t\mA}-\Unit}{\mA} \bigg\}\e^{t\mA}\nonumber\\
 & = \Bigg[\sum\limits_{r=0}^{\infty}\frac{t_{}^{r+1}}{(r+1)!}\big\{\mB,\mA_{}^{\!r} \big\}\Bigg]\e^{t\mA}
 \label{eqn:integraldirderivexpCommutator}
\end{align}
where the notation for the commutator power series relates to the commutator by \(\big\{\mB,\mA_{}^{\!r} \big\}=\Big[\big\{\mB,\mA_{}^{\!r-1} \big\},\mA\Big]\).

\subsection{Block triangular auxiliary matrix exponential}\label{Section:AuxiliaryMatrix}

A method for computing these chained exponential integrals was proposed by Van Loan in 1978 \cite{VANLOAN78}. While investigating the sensitivity of the matrix exponential \cite{VANLOAN77}, he noted that the integrals in question are solutions to linear block matrix differential equations and suggested that block matrix exponentials are used to compute them. In the simplest case of a single integral \cite{KALLSTROM73}, a block triangular auxiliary matrix \cite{PARLETT76} is exponentiated:
\begin{subequations}
\begin{empheq}[left={\mM=\begin{pmatrix} \mA & \mB\\ \Zero  & \mA \end{pmatrix}\Longrightarrow\empheqlbrace\,}]{align}
 \e^{\mM} & =\begin{pmatrix} \e^{\mA} & \int\limits_0^1 \!\e^{(1-s)\mA}\mB\e^{s\mA}\dd s\\ \Zero  & \e^{\mA} \end{pmatrix}\\
 \e^{t\mM} & =\begin{pmatrix} \e^{t\mA} & \int\limits_0^t \!\e^{(t-s)\mA}\mB\e^{s\mA}\dd s\\ \Zero  & \e^{t\mA} \end{pmatrix}\label{eqn:vanloan1}
\end{empheq}
\end{subequations}
where the above-diagonal block is easily identified as directional derivative in the integral representation of \cref{eqn:integraldirderivexp} \cite{KARPLUS48,FEYNMAN51}. Now \cref{eqn:vanloan1} can be recast in a form to show explicitly the first directional derivative of the matrix exponential:
\begin{align}
 \mM=\begin{pmatrix} \mA & \mB\\ \Zero  & \mA \end{pmatrix} && \Longrightarrow && \e^{t\mM} =\begin{pmatrix} \e^{t\mA} & D_{\mB}^{}(\e^{t\mA})\\ \Zero  & \e^{t\mA} \end{pmatrix}
\end{align}
Further to \cref{eqn:integraldirderivexp}, the \(r^\text{th}\) directional derivative can be defined recursively \cite{NAJFELD95} by 
\begin{equation}
D_{\mB}^{r}(\e^{t\mA})=r\!\!\int\limits_0^t\!\e^{\mA(t-s)}\mB D_{\mB}^{r-1}(\e^{s\mA})\dd s
\end{equation}

It is useful to see the pattern when using larger block-bidiagonal, semi-circulant auxiliary matrices. Higher order directional derivatives of the matrix exponential can be calculated by constructing and exponentiating sparse block-bidiagonal semi-circulant matrices \cite{NAJFELD95}:
\begin{gather}
\mM =\begin{pmatrix}
   \mA & \mB & \Zero  & \Zero  & \Zero   \\
   \Zero  & \mA & \mB & \Zero  & \Zero   \\
   \Zero  & \Zero  & \mA & \ddots  & \Zero   \\
   \Zero  & \Zero  & \Zero  & \ddots  & \mB  \\
   \Zero  & \Zero  & \Zero  & \Zero  & \mA  \\
\end{pmatrix} \nonumber\\
\Longrightarrow\e^{t\mM}=\begin{pmatrix}
   \frac{1}{0!}D_{\mB}^{0} & \frac{1}{1!}D_{\mB}^{1}\; & \frac{1}{2!}D_{\mB}^{2}\; & \cdots  & \frac{1}{r!}D_{\mB}^{r}\;  \\
   \Zero  & \frac{1}{0!}D_{\mB}^{0} & \frac{1}{1!}D_{\mB}^{1}\; & \cdots  & \frac{1}{(r-1)!}D_{\mB}^{r-1}\;  \\
   \Zero  & \Zero  & \frac{1}{0!}D_{\mB}^{0} & \ddots  & \vdots   \\
   \Zero  & \Zero  & \Zero  & \ddots  & \frac{1}{1!}D_{\mB}^{1}\;  \\
   \Zero  & \Zero  & \Zero  & \Zero  & \frac{1}{0!}D_{\mB}^{0}  \\
\end{pmatrix}\label{eqn:NAJFELD}
\end{gather}
This is set out in a general way in \cite{NAJFELD95}, and for this specific case of the first directional derivative in \cite{FLOETHER12}; derivation of the latter and for that of the second directional derivative is set out in the next section \cref{Section:PropDerivs}.

Van Loan's method was subsequently refined by Carbonell~et.~al. \cite{CARBONELL08}, deriving a convenient expression using the exponential of a block-bidiagonal semi-circulant auxiliary matrix:
\begin{gather}
\mM=\begin{pmatrix}
   {{\mA}_{1}} & {{\mB}_{1}} & \Zero  & \Zero  & \Zero   \\
   \Zero  & {{\mA}_{2}} & {{\mB}_{2}} & \Zero  & \Zero   \\
   \Zero  & \Zero  & {{\mA}_{3}} & \ddots  & \Zero   \\
   \Zero  & \Zero  & \Zero  & \ddots  & {{\mB}_{r-1}}  \\
   \Zero  & \Zero  & \Zero  & \Zero  & {{\mA}_{r}}  \\
\end{pmatrix}\nonumber\\
\Longrightarrow\e^{\mM\!t}=\begin{pmatrix}
   {{\mD}_{11}} & {{\mD}_{12}} & {{\mD}_{13}} & \cdots  & {{\mD}_{1r}}  \\
   \Zero  & {{\mD}_{22}} & {{\mD}_{23}} & \cdots  & {{\mD}_{2r}}  \\
   \Zero  & \Zero  & {{\mD}_{33}} & \ddots  & \vdots   \\
   \Zero  & \Zero  & \Zero  & \ddots  & {{\mD}_{r-1,r}}  \\
   \Zero  & \Zero  & \Zero  & \Zero  & {{\mD}_{rr}}  \\
\end{pmatrix}\label{eqn:carbonell}
\end{gather}
The integrals in question populate the rows of the resulting block matrix, having the chained integral form of \cref{eqn:chainedintegral}, e.g.
\begin{equation}
\mD_{1r}^{}=  \!\!\int\limits_0^t\!\dd t_1^{}\!\int\limits_0^{t_1^{}}\!\dd t_2^{} \cdots\!\!\int\limits_0^{t_{r-2}^{}}\!\!\dd t_{r-1}^{}\Big\{\e^{(t-t_1^{})\mA_1^{}}\mB_1^{}\e^{(t_1^{}-t_2^{})\mA_2^{}}\mB_2^{}\cdots\mB_{r-1}^{}\e^{t_r^{}\mA_r^{}}\Big\}\label{eqn:carbonell2}
\end{equation}

\section[Directional derivatives of the Propagator]{Directional derivatives\linebreak of the Propagator}\label{Section:PropDerivs}

Fidelity functional derivatives of optimal control \cite{TOSNER09,FOUQUIERES11} are a form of chained exponential integrals which have expensive matrix factorization in their evaluation \cite{GOLDMAN01,ERNST87,ABRAGAM61}. A solution to this problem is the observation that matrix exponentiation does not require factorisations and preserves the spin operator sparsity \cite{KUPROV11,HOGBEN11} and the use of an auxiliary matrix technique \cite{VANLOAN78,NAJFELD95,CARBONELL08,FLOETHER12}.

The \textsc{grape} method was shown in \cref{Chapter:GRAPE,Chapter:Hessian} to discretise time and solve \cref{eqn:rhoTn} using thin slice propagators. From the computational efficiency point of view, the central problem is therefore the calculation of first derivatives of the fidelity functional in \cref{eqn:fidgrad0,eqn:fidgrad1,eqn:fidgrad2} and mixed second derivatives in \cref{eqn:fidhess0_OFFDIAG,eqn:fidhess1_OFFDIAG,eqn:fidhess2_OFFDIAG}, e.g.
\begin{subequations}
\begin{gather}
  \dfrac{\partial\Fid_0^{}}{\partial \,\ctrl{k,n}}=\Big\langle\Astate_{n+1}^{}\Big|D_{\HamKB}^{}\!\big(\PropB_{n}^{}\big)\Big|\state_{n-1}^{}\Big\rangle\label{eqn:Jprop1}\\
 \dfrac{\partial^{\,2}\!\Fid_0^{}}{\partial\ctrl{k,n}\partial\ctrl{j,m}}=
 \Big\langle\Astate_{n+1}^{}\Big|D_{\HamKB}^{}\!\big(\PropB_{n}^{}\big)\PropB_{n-1}^{}\cdots\PropB_{m+1}^{}D_{\HamJB}^{}\!\big(\PropB_{m}^{}\big)\Big|\state_{m-1}^{}\Big\rangle\label{eqn:Jprop2}
\end{gather}
\end{subequations}
where time slice propagators (assuming a fixed time grid step $\Delta t$) are defined in \cref{eqn:propagatorN} as
\begin{equation}
 \PropB_n^{}=\exp{\Big[\!-i\HamB_n\Delta t\Big]}=\exp{\bigg[\!-i\Big(\HamB_0^{}+\sum\limits_{k=1}^K\ctrl{k,n}\HamKB\Big)\Delta t\bigg]}\label{eqn:propagatorN_2}
\end{equation}
where the relaxation term is removed and the double-hat superoperator notation is replaced by a matrix representation bold-font for notational brevity.

In addition to the first derivative of the propagators, the second derivatives are required for the construction of the Hessian block diagonal. After separating the cases where efficiencies can be made with communing control operators, discussed in \cref{Section:Commutaion}, the remaining calculations from \cref{eqn:fidhess0_DIAG,eqn:fidhess1_DIAG,eqn:fidhess2_DIAG,eqn:fidhess0_BLKDIAG,eqn:fidhess1_BLKDIAG,eqn:fidhess2_BLKDIAG} are
\begin{subequations}
\begin{gather}
 \dfrac{\partial^{\,2}\!\Fid_0^{}}{\partial {\ctrl{}}_{\scriptscriptstyle k,n}^{\,2}}=\Big\langle\Astate_{n+1}^{}\Big|D_{\HamKB}^{2}\!\big(\PropB_{n}^{}\!\big)\Big|\state_{n-1}^{}\Big\rangle\label{eqn:J2prop1}\\
 \dfrac{\partial^{\,2}\!\Fid_0^{}}{\partial\ctrl{k,n}\partial\ctrl{j,n}}=
 \Big\langle\Astate_{n+1}^{}\Big|D_{\HamKB\HamJB}^{2}\!\big(\PropB_{n}^{}\!\big)\Big|\state_{n-1}^{}\Big\rangle\label{eqn:J2prop2}
\end{gather}
\end{subequations}
The primary task therefore is to calculate first and second directional derivatives of the slice propagators.

\subsection[First propagator directional derivatives]{First directional derivatives}\label{Section:FirstPropDirDerivs}

The original derivation of the first propagator directional derivative from \cite{KHANEJA05} follows by considering a small perturbation to the control amplitude \(\delta \ctrl{k,n}\) on the control operator to derive the change in the propagator, \(\delta\PropB_n^{}\), then to form a derivative from \(\delta\PropB_n^{}/\delta \ctrl{k,n}\). Using \cref{eqn:directionalderivative,eqn:propagatorN_2} this can be written as
\begin{equation}
\frac{\delta\PropB_{n}^{}}{\delta \ctrl{n}}=D_{\HamKB}^{}\!\big(\PropB_{n}^{}\!\big)=\Bigg(\lim_{\big\{{\displaystyle\ctrl{\scriptscriptstyle k,n}}\!{\scriptstyle\to\,0}\big\}}\!\!\frac{\exp\big[-i\HamB_n\Delta t\big]- \exp\big[-i\HamB_0\Delta t\big]}{\ctrl{k,n}}\Bigg)\label{eqn:controlchange}
\end{equation}
The procedure in \cite{KHANEJA05} then finds solution by truncating \cref{eqn:integraldirderivexpCommutator} to first order in \(\Delta t\), giving
\begin{align}
 && D_{\HamKB}^{}\!\big(\PropB_{n}^{}\!\big)=\Delta t\Big[\HamKB,\PropB_{n}^{}\Big], && \Delta t\ll\Big\|\HamB_0^{}+\sum\limits_{k=1}^K\ctrl{k,n}\HamB_k^{}\Big\|^{-1}
\end{align}
An alternative representation of \cref{eqn:controlchange} is the integral form of \cref{eqn:integraldirderivexp}:
\begin{equation}
 D_{\HamKB}^{}\!\big(\PropB_{n}^{}\!\big)=\!\!\!\!\int\limits_0^{-i\Delta t}\!\e^{(-i\Delta t-\tau)\HamB_n}\HamKB\e^{\tau\HamB_n}\dd \tau
\end{equation}

Using \cref{eqn:vanloan1} and setting $\mA\leftarrow\HamB_n$, $\mB\leftarrow\HamKB$, and \(t\leftarrow-i\Delta t\), gives the block triangular auxiliary matrix exponential as \cite{FLOETHER12}
\begin{align}
 \mL_2=\begin{pmatrix} \HamB_n & \HamKB\\ \Zero  & \HamB_n \end{pmatrix} && \Longrightarrow && \e^{-i\mL_2\Delta t} =\begin{pmatrix} \PropB_{n}^{} & D_{\HamKB}^{}\!\big(\PropB_{n}^{}\!\big) \\ \Zero  & \PropB_{n}^{} \end{pmatrix}\label{eqn:PropDeriv2by2}
\end{align}
where the above-diagonal is the first directional derivative required for gradient element calculation in \cref{eqn:Jprop1,eqn:Jprop2}.

\subsection{Second directional derivatives}\label{Section:2ndDirDeriv}

The integral representation of the second directional derivatives form a chained integral as in \cref{eqn:chainedintegral}:
\begin{subequations}
\begin{gather}
 D_{\HamKB}^{2}\!\big(\PropB_{n}^{}\!\big)=\!\!\!\!\int\limits_0^{-i\Delta t}\int\limits_0^{\tau}\e^{(-i\Delta t-\tau)\HamB_n}\HamKB\e^{(\tau-\upsilon)\HamB_n}\HamKB\e^{\upsilon\HamB_n} \dd\upsilon\dd\tau\\
 D_{\HamKB\HamJB}^{2}\!\big(\PropB_{n}^{}\!\big)=\!\!\!\!\int\limits_0^{-i\Delta t}\int\limits_0^{\tau}\e^{(-i\Delta t-\tau)\HamB_n}\HamKB\e^{(\tau-\upsilon)\HamB_n}\HamJB\e^{\upsilon\HamB_n}\dd\upsilon\dd\tau
\end{gather}
\end{subequations}
Just as in the previous calculation of first directional derivatives in \cref{eqn:PropDeriv2by2}, \cref{eqn:NAJFELD} shows a similar auxiliary matrix technique can be used to calculate higher order directional derivatives. Calculation of the second directional derivatives requires a $3\times 3$ bidiagonal block matrix auxiliary matrix from \cref{eqn:NAJFELD}, 
\begin{equation}
 \mL_3=\begin{pmatrix} \HamB_n & \HamKB & \Zero\\ \Zero  & \HamB_n & \HamJB\\ \Zero & \Zero & \HamB_n \end{pmatrix}
\end{equation}
where the upper-diagonal blocks contain the two control operators needed for the mixed second directional derivatives. Following from \cref{eqn:carbonell,eqn:carbonell2}, the exponential of the auxiliary matrix \(\mL_3\) is
\begin{subequations}
\begin{empheq}[left={\Longrightarrow\e^{-i\mL_3\Delta t} =\empheqlbrace\,}]{align}
 & \begin{pmatrix} \PropB_{n}^{} & D_{\HamKB}^{}\!\big(\PropB_{n}^{}\!\big) & \frac{1}{2}D_{\HamKB}^{2}\!\big(\PropB_{n}^{}\!\big) \\ \Zero  & \PropB_{n}^{} & D_{\HamKB}^{}\!\big(\PropB_{n}^{}\!\big)\\ \Zero & \Zero & \PropB_{n}^{}\end{pmatrix}\!, && j= k\\
& \begin{pmatrix} \PropB_{n}^{} & D_{\HamKB}^{}\!\big(\PropB_{n}^{}\!\big) & \frac{1}{2}D_{\HamKB\HamJB}^{2}\!\big(\PropB_{n}^{}\!\big) \\ \Zero  & \PropB_{n}^{} & D_{\HamJB}^{}\!\big(\PropB_{n}^{}\!\big)\\ \Zero & \Zero & \PropB_{n}^{}\end{pmatrix}\!,&& j\neq k\label{eqn:Prop2Deriv3by3}
\end{empheq}
\end{subequations}
where the upper-right blocks are the second directional derivatives required for Hessian element calculation in \cref{eqn:J2prop1,eqn:J2prop2}.

\section{Krylov subspace techniques}\label{Section:KrylovAlgo}

There is an important efficiency to be made when considering the propagator derivatives of in that only the resulting vector \(\big|\frac{\partial_{}^2\state_{n+1}}{\partial\ctrl{k,n}\partial\ctrl{j,n}}\big\rangle\) is required in \cref{eqn:Jprop1,eqn:J2prop1,eqn:J2prop2}. The solutions to these equations are of the form \(f(\mM)b\) -- the action of \(f(\mM)\) on the vector \(b\) is required \cite{VANDERVORST87}. This type of problem is readily solved with \textit{Krylov subspace methods}, such as the \textit{Arnoldi process} \cite{ARNOLDI51} or the \textit{Lanczos process} \cite{LANCZOS50}, avoiding explicit calculation of the matrix exponential \cite{SAAD92,HOCHBRUCK97,HIGHAM08,GOLUB13}. 

The Arnoldi process is implemented in \textit{Spinach} to efficiently calculate \cite{KUPROV07,KUPROV08} the exponential of a Liouvillian multiplied by a state vector. The effect is that forward and backward propagation can be calculated in this way. Furthermore, the effect of first and second derivatives on the forward propagated state vector can be calculated with their auxiliary matrices:

\begin{subequations}
\begin{gather}
   \textsc{Forward propagation:}\nonumber\\
  \Big[ \exp \begin{pmatrix} -i\Delta t\HamB_n \end{pmatrix}\!\Big]\, \Big|\state_n\Big\rangle=\Big|\state_{n+1}\Big\rangle \label{eqn:fwdkrylov1x1}\\
   \textsc{Backward propagation:}\nonumber\\
  \Big[ \exp \begin{pmatrix} +i\Delta t\HamB_n \end{pmatrix}\!\Big]\, \Big|\Astate_n\Big\rangle=\Big|\Astate_{n-1}\Big\rangle \label{eqn:bwdkrylov1x1}\\
   \textsc{Propagated first directional derivative:}\nonumber\\
   \left[\exp\left(\!\!-i\Delta t\begin{pmatrix} \HamB_n & \HamKB \\ \Zero  & \HamB_n\end{pmatrix}_{\vphantom{|}}^{\vphantom{|}}\!\!\right)\!\right] \begin{bmatrix}\Zero\\ \Big|\state_n\Big\rangle\end{bmatrix}= \begin{bmatrix} \Big|\frac{\partial\state_{n+1}}{\partial\ctrl{k,n}}\Big\rangle\\\Zero\end{bmatrix} \label{eqn:krylov2x2}\\
   \textsc{Propagated second directional derivative:}\nonumber\\
   \left[\exp\left(\!-i\Delta t\begin{pmatrix} \HamB_n & \HamKB & \Zero\\ \Zero  & \HamB_n & \HamJB\\ \Zero & \Zero & \HamB_n \end{pmatrix}_{\vphantom{|}}^{\vphantom{|}}\!\right)\right] \begin{bmatrix} \Zero\\\Zero\\ \Big|\state_n\Big\rangle\end{bmatrix}= \begin{bmatrix} \frac{1}{2}\Big|\frac{\partial_{}^2\state_{n+1}}{\partial\ctrl{k,n}\partial\ctrl{j,n}}\Big\rangle\\\Zero\\ \Zero\end{bmatrix} \label{eqn:krylov3x3}
\end{gather}
\end{subequations}

\section{Propagator recycling}\label{Section:PropRecycle}

Two computational efficiencies should be taken into account when calculating the full Hessian matrix of \cref{Fig:HessDiag}. Firstly, the first propagator direction derivatives do not need to be calculated with \cref{eqn:PropDeriv2by2} when a Hessian calculation is made: these directional derivatives are already contained in the above-diagonal of \cref{eqn:Prop2Deriv3by3}. Secondly, Krylov subspace techniques can be used to calculate the effect of the propagator directional derivative on the state \(\big|\state_n\big\rangle\) in some of \cref{eqn:Jprop1,eqn:Jprop2,eqn:J2prop1,eqn:J2prop2}. The following lists outline the type of calculation recommended to construct fidelity derivatives. There are two cases: 
\begin{enumerate}
 \item A fidelity and gradient vector calculation for a gradient ascent method, quasi-Newton method, or a line search method.
 \item The simultaneous calculation of the fidelity, gradient vector and Hessian matrix for a Newton step calculation.
\end{enumerate}

\subsection{Fidelity and gradient calculation}

\begin{enumerate}
 \item \textsc{Forward and backward propagation}\hfill\\
  Number of matrix exponential calculations \(=2\times K\times N\).
 \begin{enumerate}
  \item Calculate and store forward propagated states with a Krylov subspace technique in \cref{eqn:fwdkrylov1x1}.\label{FWDstatesG}
  \item Calculate and store backward propagated states with a Krylov subspace technique in \cref{eqn:bwdkrylov1x1}.\label{BWDstatesG}
  \item Calculate fidelity as the overlap of the forward propagated state of \cref{FWDstatesG} at the time step \(n=N\), and the desired target state using one of \cref{eqn:fidelity0,eqn:fidelity1,eqn:fidelity2}.
 \end{enumerate}
 \item \textsc{Gradient elements}\hfill\\
  Number of matrix exponential calculations \(=K\times N\).
 \begin{enumerate}
  \item Use an auxiliary matrix in \cref{eqn:PropDeriv2by2} and forward propagated state from \cref{FWDstatesG} with a Krylov subspace technique in \cref{eqn:krylov2x2} and store the resulting vector.\label{gradstates}
  \item Calculate the gradient elements as the overlap of the propagated gradient from \cref{gradstates} and the backward propagated state from \cref{BWDstatesG} using one of \cref{eqn:fidgrad0,eqn:fidgrad1,eqn:fidgrad2}.
 \end{enumerate}
\end{enumerate}

\subsection{Fidelity, gradient and Hessian calculation}

\begin{enumerate}
 \item \textsc{Forward and backward propagation}\hfill\\
  Number of matrix exponential calculations \(=2\times K\times N\).
 \begin{enumerate}
  \item Calculate and store forward propagated states with a Krylov subspace technique in \cref{eqn:fwdkrylov1x1}.\label{FWDstates}
  \item Calculate and store backward propagated states with a Krylov subspace technique in \cref{eqn:bwdkrylov1x1}.\label{BWDstates}
  \item Calculate fidelity as the overlap of the forward propagated state of \cref{FWDstates} at the time step \(n=N\), and the desired target state using one of \cref{eqn:fidelity0,eqn:fidelity1,eqn:fidelity2}.
 \end{enumerate}
 \item \textsc{Diagonal Hessian elements} -- \cref{Fig:HessDiag} \plotlabel{a}\hfill\\
  Number of matrix exponential calculations \(=K\times N\).
 \begin{enumerate}
  \item Calculate explicit propagators with \cref{eqn:Prop2Deriv3by3}, exponentiated using a sparsity preserving Taylor series of \cref{Section:TaylorExp} with scaling and squaring of \cref{Section:scalingsquaring}.
  \item Store the propagator from block \((1,1)\) \label{storedprops}.
  \item Store first directional derivatives from block \((1,2)\) \label{stored1props}.
  \item Store second directional derivatives from block \((1,3)\).
  \item Calculate diagonal Hessian elements with \cref{eqn:J2prop1} using the stored states from \cref{FWDstates,BWDstates}.
 \end{enumerate}
 \item \textsc{Off-diagonal elements of the Hessian block diagonal} -- \cref{Fig:HessDiag} \plotlabel{b}\hfill\\
  Needed when \(K>1\).\hfill\\
  Control operator commutativity must be considered from \cref{Section:Commutaion}.\hfill\\
  \((K-1)!\times N<\) Number of matrix exponential calculations \(<K(K-1)\times N\).\hfill
 \begin{enumerate}
  \item Use an auxiliary matrix in \cref{eqn:Prop2Deriv3by3} with a Krylov subspace technique as in \cref{eqn:krylov3x3} and store the resulting vector.
  \item Calculate off-diagonal elements of the Hessian block diagonal with the backward propagated states from \cref{BWDstates} subject commutativity conditions in \cref{eqn:addlaw,eqn:symderivs,eqn:force_symmetry}.
 \end{enumerate}
 \item \textsc{Block off-diagonal Hessian elements} -- \cref{Fig:HessDiag} \plotlabel{c}\hfill\\
 Requires only matrix-matrix and matrix-vector multiplications.
 \begin{enumerate}
  \item Retrieve the two required first directional derivatives from \cref{stored1props}.
  \item Calculate off-diagonal Hessian elements with \cref{eqn:Jprop2}, using the stored states from \cref{FWDstates,BWDstates} and the stored propagators in \cref{storedprops}.
  \end{enumerate}
 \item \textsc{Gradient elements}\hfill\\
 Requires only matrix-matrix and matrix-vector multiplications.
 \begin{enumerate}
  \item Use the first directional derivative from \cref{stored1props} and forward propagated state from \cref{FWDstates} and store the resulting vector.\label{gradstates2}
  \item Calculate the gradient elements as the overlap of the propagated gradient from \cref{gradstates2} and the backward propagated state from \cref{BWDstates} using one of \cref{eqn:fidgrad0,eqn:fidgrad1,eqn:fidgrad2}.
 \end{enumerate}
\end{enumerate}

\subsection[Parallelisation]{Parallelisation}\label{Section:Parallel}

\begin{figure}
\centering{\includegraphics{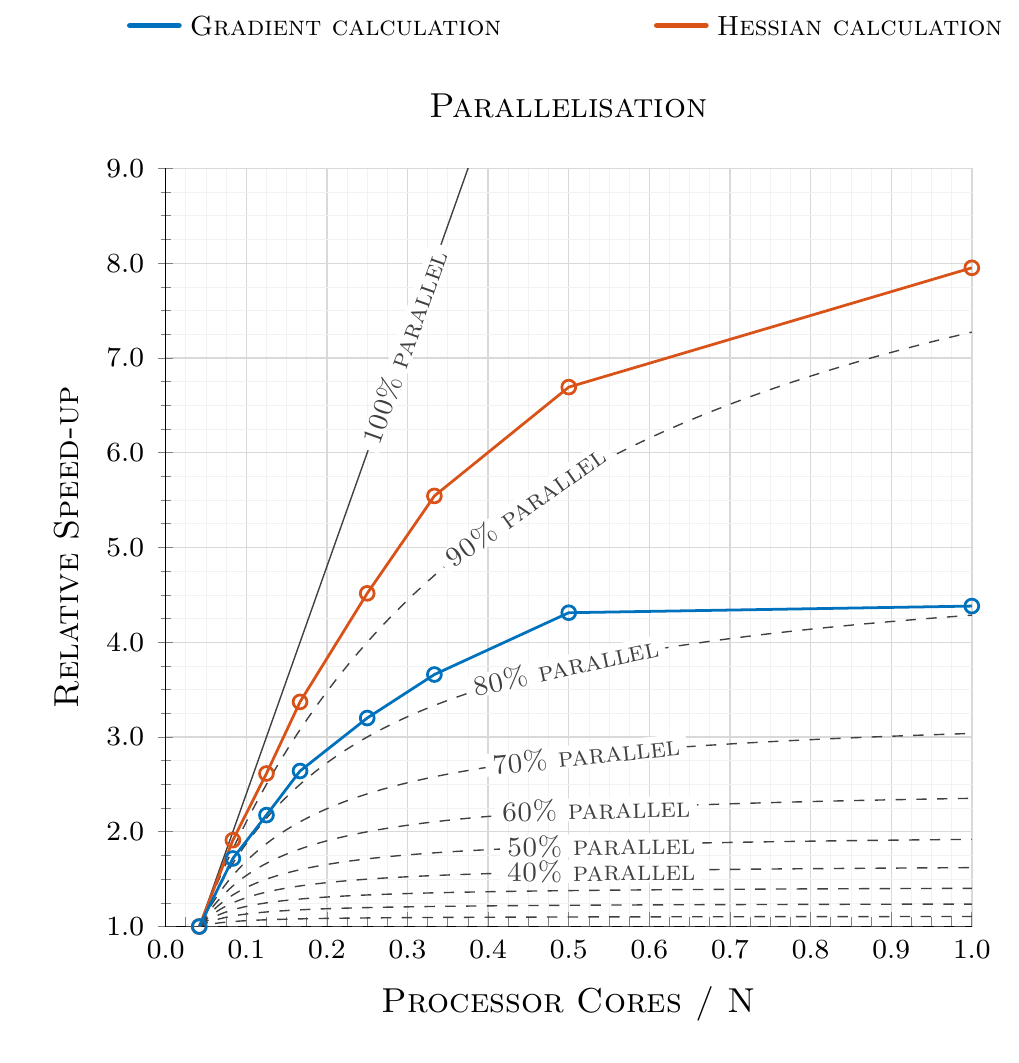}}
 \caption[Parallelism of gradient and Hessian calculations]{Amdahl's law \cite{AMDAHL67} parallelisation efficiency analysis for the Hessian calculation compared to the gradient calculation within \textit{Spinach} implementation of \textsc{grape} \cite{HOGBEN11,FOUQUIERES11}. The optimal control problem involves \(N=24\) time slices and \(K=6\) control channels, yielding a fidelity functional gradient vector with 144 elements and a \(144\times 144\) Hessian matrix.}\label{Fig:parallel}
\end{figure}

The numerical implementations of gradient and Hessian calculations should be parallelised with respect to the number of time slices in \cref{eqn:fidhess2_DIAG}. Gradient calculation uses \(2\times 2\) augmented exponentials \cref{eqn:krylov2x2} and Hessian calculation uses the \(3\times 3\) augmented exponentials shown in \cref{eqn:Prop2Deriv3by3}. The Hessian function has a further parallel loop when calculating \(n\neq m\) blocks in Equation \cref{eqn:fidhess2_OFFDIAG}. At that stage, the first derivatives have already been calculated when solving \cref{eqn:Prop2Deriv3by3}; they are recycled.

Parallelisation efficiency analysis is given in \cref{Fig:parallel}. The scaling depends on the number of time slices in the control sequence and the parallelisation is efficient all the way to the number of \textsc{cpu} cores being half the number of time slices (over which the parallel loop is running). Given the same computing resources, a Hessian calculation takes approximately five-ten times longer than a gradient calculation. Because propagator derivatives are recycled during Hessian calculation, significant efficiency gains may be made by optimising their storage and indexing, which is outlined in \cite{GOODWIN16}.

Using Amdahl's law, the parallelism of code from \cref{Fig:HCF_compare} was computed to be just less than 70\% parallel. This indicates that a maximum of about \(3\times\) speed increase to the code can be achieved using about 16 cores for parallel computation, although using only 8 cores will give a speed increase of over \(2.5\times\).

\chapter{Constrained Optimal Control} \label{Chapter:Penalty}

\begin{chapquote}{Albert Camus, \textsc{The Rebel}}
In the end, man is not entirely guilty -- he did not start history. Nor is he wholly innocent -- he continues it.
\end{chapquote}
\renewcommand*{\CurrentPath}{./Chapter_6}

The Pontryagin maximum principle of \cref{Section:Pontryagin} set out a control problem with a \textit{terminal cost} and a \textit{running cost}. The \textit{terminal cost} was investigated fully as fidelity measures in \cref{eqn:fidelity0,eqn:fidelity1,eqn:fidelity2}. This chapter will set out a number of \textit{running costs} in the form of penalty functions \(\mathcal{K}\).

The idea of a penalty function is one borne from constrained optimisation \cite{FLETCHER87,BOYD04,NOCEDAL06}. The constraint may be in the form of an inequality in addition to the objective function e.g. in the context of control operators; the control amplitudes may have a maximum allowed value corresponding to a maximum amount of power the control pulse may have -- in this case the constraint would be an experimental constraint.

The traditional area of constrained optimisation is to define an allowable region of the objective function; then only to allow the objective function evaluations that fall within this region. In the penalty functions that follow in this chapter, the idea of a penalty function from constrained optimisation is relaxed to allow objective function evaluations outside the constrained area, but at a cost.

This chapter outlines penalty functions as strategies to push solutions to the objective function into a desired region of its space. The scalar function to maximise is (using the fidelity measure of \cref{eqn:fidelity1});
\begin{equation}
\Fid_1\triangleq\Re\big\langle  \targ \big| \state(T)\big\rangle - \mathcal{K}\big(\state(t),\ctrlv{}(t)\big)
\end{equation}
where \(\mathcal{K}\) is a penalty functional enforcing the experimental constraints on the optimisation e.g. a maximum amplitude of control pulses. The task to include penalties is now to design the penalty measure, and then to find its associated gradient and Hessian.

The simplest method, but maybe affecting convergence the most, is to change the variables so that controls more strictly defined in \cref{Section:ChangeVar}, or more elegantly defined in \cref{Section:PolarTrans}. Penalties to pulse power are set out in \cref{Section:PwrPen}, with the simplest favouring lower power penalties in \cref{Section:NormSqr} and a more calibrated penalty that only penalises excursions outside a desired bound \cref{Section:SpilloutPen}. An interesting penalty function is set out in \cref{Section:Smooth}, designed to find smooth solutions by penalising rapid changes in controls over their time period.

This smoothing penalty is investigated in \cref{Section:M2S} with the problem of transferring magnetisation to a singlet state. Although the analytical solution is the optimal solution, it is not robust to \(\B{1}\) miscalibration or \(\B{0}\) resonance offsets. These type of optimal pulses are reproduced from \cite{LAUSTSEN14} in \cref{Section:RobustM2S}, then the optimisation problem is extended to attempt to find similar, but also smooth optimal, robust solutions in \cref{Section:SmoothRobustM2S}.

\section{Change of variables}\label{Section:ChangeVar}

There exists a situation in which all solutions \textit{must} be bounded i.e. a waveform must not exceed power levels which hardware cannot handle. To constrain all possible solutions to the feasible region \(R\) \cite{BOYD04}, and to guarantee a minimiser resides within the region \(R$, more is required than a method of weighted penalties. To constrain all solutions to the feasible region, a transform can be made to the objective function, where the transform maps the objective function to a similar one which only contains feasible points.

The variables forming the shape can be designed to form a repeating, mirrored structure; allowing all real numbers to be evaluated by an objective function in a continuous manner. The transform of a waveform, \(x\), to include these constraints is
\begin{equation}
 \tilde{\bm{c}}= \Bigg|\objv{}- 4\bigg\lfloor \frac{\objv{}-1}{4} \bigg\rfloor -3\Bigg|-1\label{eqn:changevar}
\end{equation}
where \(\objv{}\) is the unbounded waveform and \(\objv{}-n\lfloor \objv{}/n\rfloor\) is the remainder from dividing \(\objv{}\) by \(n\).

\subsection{Transforms between Cartesian \& Polar Coordinates}\label{Section:PolarTrans}

It is normal to represent control pulses as operators of Cartesian space i.e. \(\HamHH_x^{}\) and \(\HamHH_y^{}\), However, it is common for spectrometer hardware to implement control pulses in polar coordinate space, with amplitude and phase. Furthermore, the most simple way to enforce an equality constraint on either the phase or amplitude is to allow only one of these as the optimisation variables. This was implemented in \cref{Section:test2,Section:test3}, where only the phase was allowed to vary, ensuring that the amplitude of the pulse stayed at a constant level.

To use this type of optimisation, the control pulses, the fidelity gradient, and the fidelity Hessian must be transformed between a Cartesian representation and a spherical polar coordinate system. The definitions of these transformed control pulses are
\begin{align}
 r=&\sqrt{x^2 +y^2}\\
 {\varphi}=&\atan2{\bigg(\frac{y}{x}\bigg)}
\end{align}
with \(\atan2\) defined as the four quadrant inverse tangent, and \((x,y)\) are the components of the control pulse corresponding to \(\HamHH_x^{}\) and \(\HamHH_y^{}\). From this, the inverse transform is
\begin{align}
 x=& r \cos{{\varphi}}\\
 y=& r \sin{{\varphi}}
\end{align}
These definitions are from basic Pythagorean theorem. However, the gradient and Hessian transformations are not so obvious, and it is useful to explicitly show their transformations here.

The gradient transform from polar to Cartesian coordinates is
\begin{align}
 \frac{d}{dx}=& \cos{{\varphi}}\frac{d}{dr} - \frac{1}{r}\sin{{\varphi}}\frac{d}{dr}\\
 \frac{d}{dy}=& \sin{{\varphi}}\frac{d}{dr} + \frac{1}{r}\cos{{\varphi}}\frac{d}{dr}
\end{align}
and the inverse of these transforms are
\begin{align}
 \frac{d}{dr}=& \sin{\varphi}\frac{d}{dy} + \cos{\varphi}\frac{d}{dx}\\
 \frac{d}{d\varphi}=& x\frac{d}{dy} - y\frac{d}{dx}
\end{align}

The transform of the Cartesian Hessian element to a polar Hessian element is
\begin{equation}
\frac{d^2}{{dr}^2}=\cos^2{(\varphi)}\cdot\frac{d^2}{{dx}^2}+2\cos{(\varphi)}\cdot\sin{(\varphi)}\frac{d^2}{{dx}{dy}}+\sin^2{(\varphi)}\cdot \frac{d^2}{{dy}^2}
\end{equation}
\begin{multline}
\frac{d^2}{{d\varphi}^2}=  -r\cdot\left(\cos{(\varphi)}\cdot\frac{d}{dx}+\sin{(\varphi)}\cdot\frac{d}{dy}\right)\\
+r\cdot r\cdot \left(\cos^2{(\varphi)}\cdot\frac{d^2}{{dy}^2}-2\cos{(\varphi)}\cdot\sin{(\varphi)}\frac{d^2}{{dx}{dy}}+\sin^2{(\varphi)}\cdot \frac{d^2}{{dx}^2}\right)
\end{multline}
\begin{multline}
\frac{d^2}{{dr}{d\varphi}}=\cos{(\varphi)}\cdot \frac{d}{dy}-\sin{(\varphi)}\cdot\frac{d}{dx}\\
+r\cdot\cos{(\varphi)}\cdot\sin{(\varphi)}\frac{d^2}{{{dy}^2}}+r\cdot\cos^2{(\varphi)}\cdot\frac{d^2}{{dx}{dy}}\\
-r\cdot\sin^2{(\varphi)}\cdot\frac{d^2}{{dx}{dy}}-r\cdot\cos{(\varphi)}\cdot\sin{(\varphi)}\frac{d^2}{{{dx}^2}}
\end{multline}

The transform of the polar Hessian elements back to Cartesian Hessian elements is
\begin{multline}
\frac{d^2}{{dx}^2}=+2x\cdot y\cdot \frac{d}{d\varphi}\cdot \frac{1}{r\cdot ^4}+y\cdot y\cdot \frac{d^2}{{d\varphi}^2}\cdot \frac{1}{r\cdot ^4}+\frac{d}{dr}\cdot \frac{1}{r}\\
-x\cdot x\cdot \frac{d}{dr}\cdot \frac{1}{r\cdot ^3}-2x\cdot y\cdot \frac{d^2}{{dr}{d\varphi}}\cdot \frac{1}{r\cdot ^3}+x\cdot x\cdot \frac{d^2}{{dr}^2}\cdot \frac{1}{r\cdot ^2}
\end{multline}
\begin{multline}
\frac{d^2}{{dy}^2}=-2x\cdot y\cdot \frac{d}{d\varphi}\cdot \frac{1}{r\cdot ^4}+x\cdot x\cdot \frac{d^2}{{d\varphi}^2}\cdot \frac{1}{r\cdot ^4}+\frac{d}{dr}\cdot \frac{1}{r}\\
-y\cdot y\cdot \frac{d}{dr}\cdot \frac{1}{r\cdot ^3}+2x\cdot y\cdot \frac{d^2}{{dr}{d\varphi}}\cdot \frac{1}{r\cdot ^3}+y\cdot y\cdot \frac{d^2}{{dr}^2}\cdot \frac{1}{r\cdot ^2}
\end{multline}
\begin{multline}
\frac{d^2}{{dx}{dy}}=-\frac{d}{d\varphi}\cdot \frac{1}{r\cdot ^2}+2y\cdot y\cdot \frac{d}{d\varphi}\cdot \frac{1}{r\cdot ^4}-x\cdot y\cdot \frac{d^2}{{d\varphi}^2}\cdot \frac{1}{r\cdot ^4}-x\cdot y\cdot \frac{d}{dr}\cdot \frac{1}{r\cdot ^3}\\
+x\cdot x\cdot \frac{d^2}{{dr}{d\varphi}}\cdot \frac{1}{r\cdot ^3}-y\cdot y\cdot \frac{d^2}{{dr}{d\varphi}}\cdot \frac{1}{r\cdot ^3}+x\cdot y\cdot \frac{d^2}{{dr}^2}\cdot \frac{1}{r\cdot ^2}
\end{multline}
These transforms for the Hessian elements are basic mathematics, but at the same time they are laborious algebraic differentiation using the chain rule. They are presented here to as convenient relations to enable optimisation of phase or amplitude pulses using the Newton-\textsc{grape} method. Amplitude penalties can be implemented with these transforms and the power penalties set out below.

\section{Power penalties}\label{Section:PwrPen}

Magnetic resonance pulses may not have an infinite power, they must be bounded by that which the magnetic resonance experiment can handle. A simple strategy is to penalise the vector of control pulses based on the amplitude of each pulse in the pulse set.

\subsection{Norm-square penalty}\label{Section:NormSqr}

\begin{figure}
\centering{\includegraphics{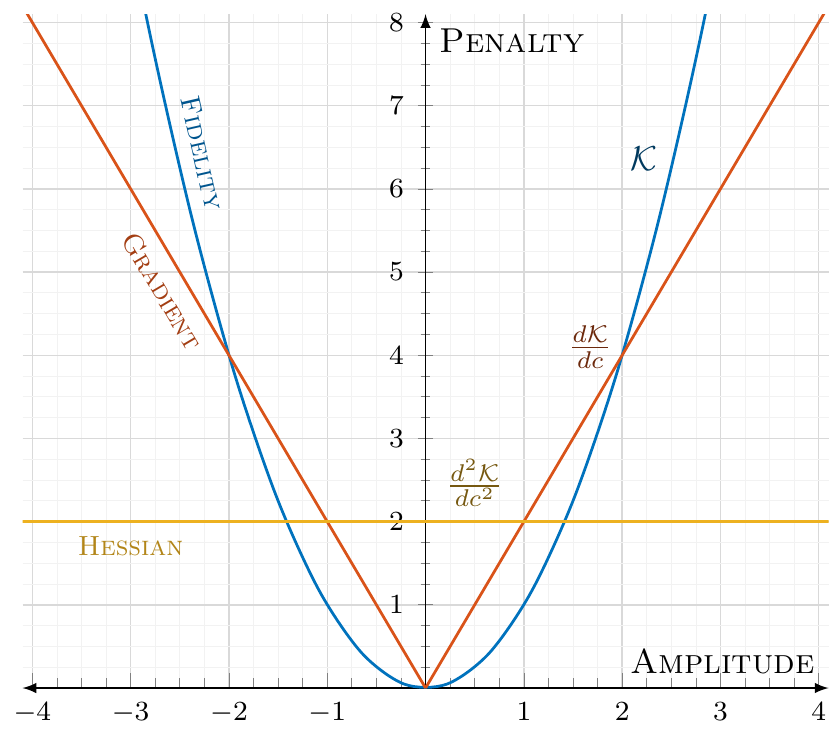}}
\caption{Norm square penalty and its derivatives}\label{norm-squared_penalty}
\end{figure}

First consider a normalised set of control pulses; normalised by a bound set by the experiment being a nominal ``maximum amplitude level'' that a control channel can experimentally implement. This maximum amplitude level, measured in Hz, can be arbitrarily set by an experienced experimenter or set by strict experimental constraints\footnote{a reasonable estimate for a non-specific \textsc{nmr} spectrometer can set this bound as \(10\)~kHz; giving a normalisation constant of \(10\times2\pi\times10^3$}.

Assuming that the vector of control pulses is normalised with a maximum power amplitude, a penalty functional can be based on the square of the control pulse vector \(\ctrlv{}\). This will give a very small penalty for low power control pulses, compared to the bound, and will give very high penalties for those above the bound. A form for this penalty is
\begin{align}
 &&\mathcal{K}=\frac{1}{N}\sum_k{c_k}^2 && \Longrightarrow &&\mathcal{K}=\frac{1}{N}\left\|\ctrlv{}\right\|^2 &&\label{norm-square}
\end{align}
where the scalar product of the vector with itself, \(\ctrlv{}\cdot\ctrlv{}\), implies the square of the euclidean norm of the control vector, \(\left\|\ctrlv{}\right\|^2\), Here, \(c_k\) denotes an element of the vector \(\ctrlv{}\) -- the penalty is simply the sum of the squares of each element of the control pulse vector. The normalisation factor for this penalty, \(N\), is the length of each control channel i.e. the number of discrete time slices. Essentially, the penalty \(\mathcal{K}\) is the average of the square of each control pulse, for each control channel.

Mentioned previously, for the use of a penalty in a gradient based optimisation scheme, we must also penalise the derivative of the penalty. Additionally, if the optimisation method calculates an explicit Hessian i.e. the Newton-Raphson method, then we must also penalise this Hessian. These derivatives of the penalty functional are with respect to the control pulse vector, just as the derivative of the fidelity functional is performed with respect to the same control pulse vector.
\begin{align}
 &&\frac{d}{dc_k}\mathcal{K}=\frac{1}{N}\sum_k2c_k && \Longrightarrow && \nabla \mathcal{K}=\frac{2}{N}\ctrlv{}&&\\
 &&\frac{d^2}{d{c_k}^2}\mathcal{K}=\frac{2}{N} && \Longrightarrow && \nabla^2 \mathcal{K}=\frac{2}{N}\Unit &&
\end{align}
The Hessian penalty is only ever a diagonal matrix because mixed derivatives of the penalty functional are zero. A graphical representation of this norm-squared penalty and its derivatives is shown in \cref{norm-squared_penalty}.

\subsection{Spillout-norm-square penalty}\label{Section:SpilloutPen}

\begin{figure}
\centering{\includegraphics{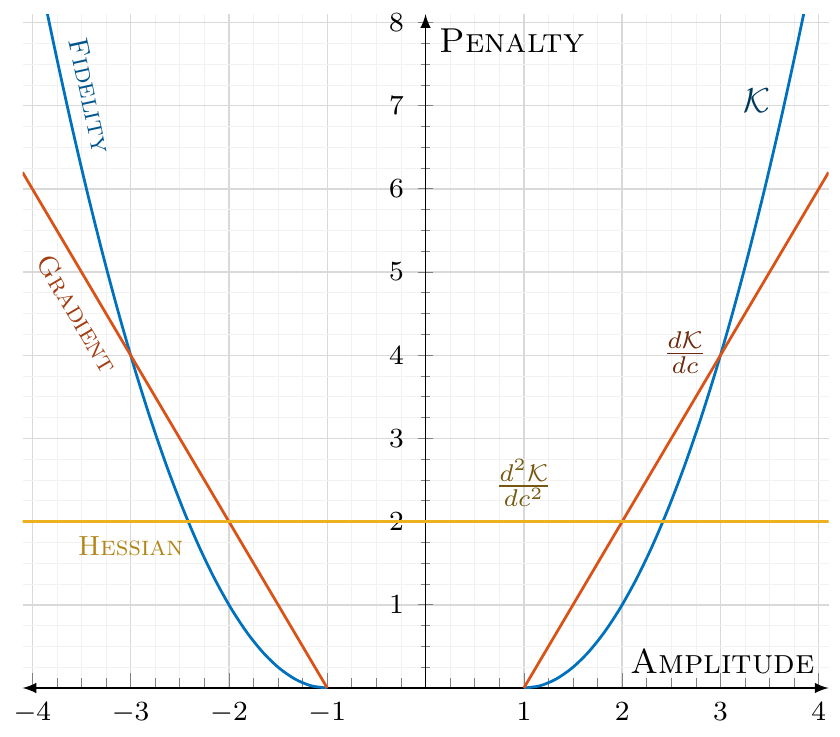}}
\caption{Spillout norm square penalty and its derivatives}\label{Spillout_norm-squared_penalty}
\end{figure}

The problem with the norm-squared penalty of \cref{norm-square} is that it penalises the objective function even when within the set bounds. This is undesirable because an optimisation will never truly converge unless there is zero amplitude on the control pulses - clearly an unrealistic solution. One way to remedy this is to introduce a level of zero penalty when within the maximum amplitude interval \([-1,+1]\), The graphical representation of this modified norm-squared penalty, penalising only that which spills out of the amplitude bounds, is shown in \cref{Spillout_norm-squared_penalty}.

The aim of this spillout-norm-squared penalty is to only penalise the control amplitudes that exceed the set bounds of the normalised control pulses, eventually constraining solutions to the set bounds. This method also allows a quadratic convergence, assuming a solution does exist within the set bounds. The form of this fidelity functional and the corresponding derivatives are:
\begin{subequations}
\begin{gather}
 \mathcal{K}=\begin{cases}\dfrac{1}{N}\displaystyle\sum\limits_k{\Big(\big|c_k\big|-1\Big)}^2, & \big|c_k\big|>1\\0, & \big|c_k\big|\leqslant 1\end{cases} \quad \Longrightarrow \quad 
 \mathcal{K}=\begin{cases}\dfrac{1}{N}\big\|\ctrlv{}-1\big\|^2, & \big|c_k\big|>1\\\Zero, & \big|c_k\big|\leqslant 1\end{cases}\\
 \frac{d}{dc_k}\mathcal{K}=\begin{cases}\dfrac{1}{N}\displaystyle\sum\limits_k{\Big(\big|c_k\big|-1\Big)}, & \big|c_k\big|>1\\0, & \big|c_k\big|\leqslant 1\end{cases} \quad \Longrightarrow \quad 
 \nabla \mathcal{K}=\begin{cases}\dfrac{2}{N}\big(\ctrlv{}-1\big), & \big|c_k\big|>1\\\Zero, & \big|c_k\big|\leqslant 1\end{cases}\\
 \frac{d^2}{d{c_k}^2}\mathcal{K}=\begin{cases}\dfrac{2}{N}, & \big|c_k\big|>1\\0, & \big|c_k\big|\leqslant 1\end{cases} \quad \Longrightarrow \quad
 \nabla^2 \mathcal{K}=\begin{cases}\dfrac{2}{N}\Unit, & \big|c_k\big|>1\\\Zero, & \big|c_k\big|\leqslant 1\end{cases}
\end{gather}
\end{subequations}
This penalty type has a discontinuity in the Hessian penalty, which is not expected to cause too much problem, but if this is the case a modification could be to use the a cubic penalty, shown in \cref{Spillout_norm-cubed_penalty}.

\begin{figure}
\centering{\includegraphics{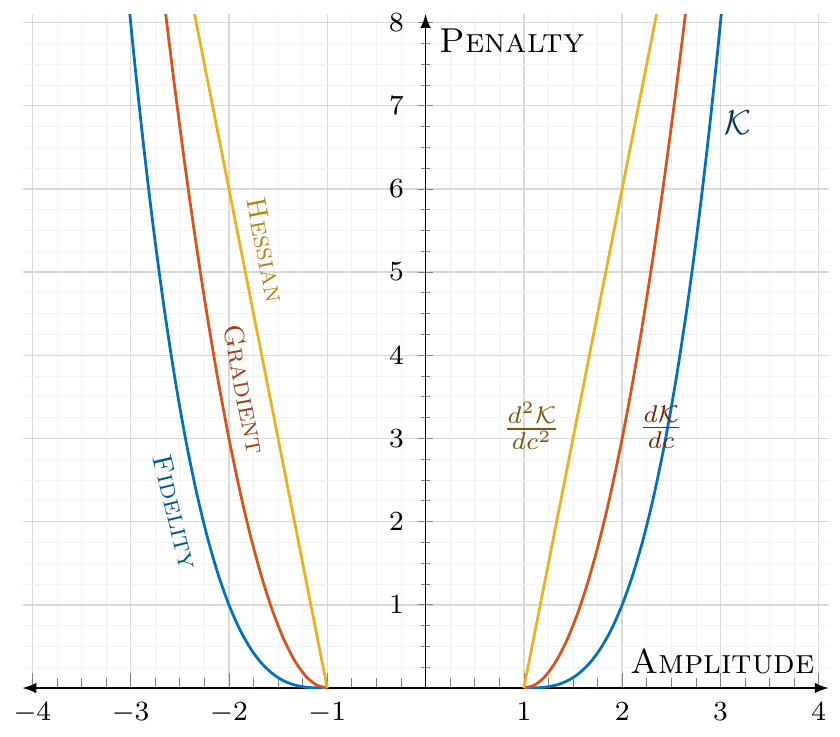}}
\caption{Spillout norm cube penalty and its derivatives}\label{Spillout_norm-cubed_penalty}
\end{figure}

\section{Smoothed Controls}\label{Section:Smooth}

A smoothing penalty is designed to penalise non-smooth pulse shape, which would be all except flat pulses. However, the metric to penalise will do so more to rapidly varying pulse shapes -- those with a derivatives that have large changes near the vicinity of the point at which the derivative is evaluated.

Called \textit{smoothing regularisation} in \cite{BOYD04}, a penalty of the following form is useful:
\begin{align}
 &&\mathcal{K}=\frac{1}{N}\big\|\hat{\mathbf{\Delta}}\ctrlv{}\big\|^2 &&\label{deriv-norm-square}
\end{align}
where \(\hat{\mathbf{\Delta}}\) is an approximate differentiation, or second order differentiation operator of a suitable type and order, or any other appropriate transformation matrix. This penalty is based on \cref{norm-square} by replacing \(\ctrlv{}\) with \(\hat{\mathbf{\Delta}}\ctrlv{}\) and penalising the ``high frequency'' roughness instead of the high power.

The first and second derivatives of the penalty in \cref{deriv-norm-square} are
\begin{align}
 &&\nabla \mathcal{K}=\frac{2}{N}\big(\hat{\mathbf{\Delta}}^{\dagger}\hat{\mathbf{\Delta}}\ctrlv{}\big)&&\\
 &&\nabla^2 \mathcal{K}=\frac{2}{N}\big(\hat{\mathbf{\Delta}}^{\dagger}\hat{\mathbf{\Delta}}\otimes\Unit\big) &&
\end{align}
where the identity matrix is of size \(K\).

\subsection{Magnetisation-to-singlet transfer}\label{Section:M2S}

\newcommand{\wffile}{\datadir/M2S/slic_wf.dat}
\newcommand{\datafile}{\datadir/M2S/rbstmat_realslic.dat}
\newcommand{\figwidth}{0.5\columnwidth}

\begin{figure}
\centering{\includegraphics{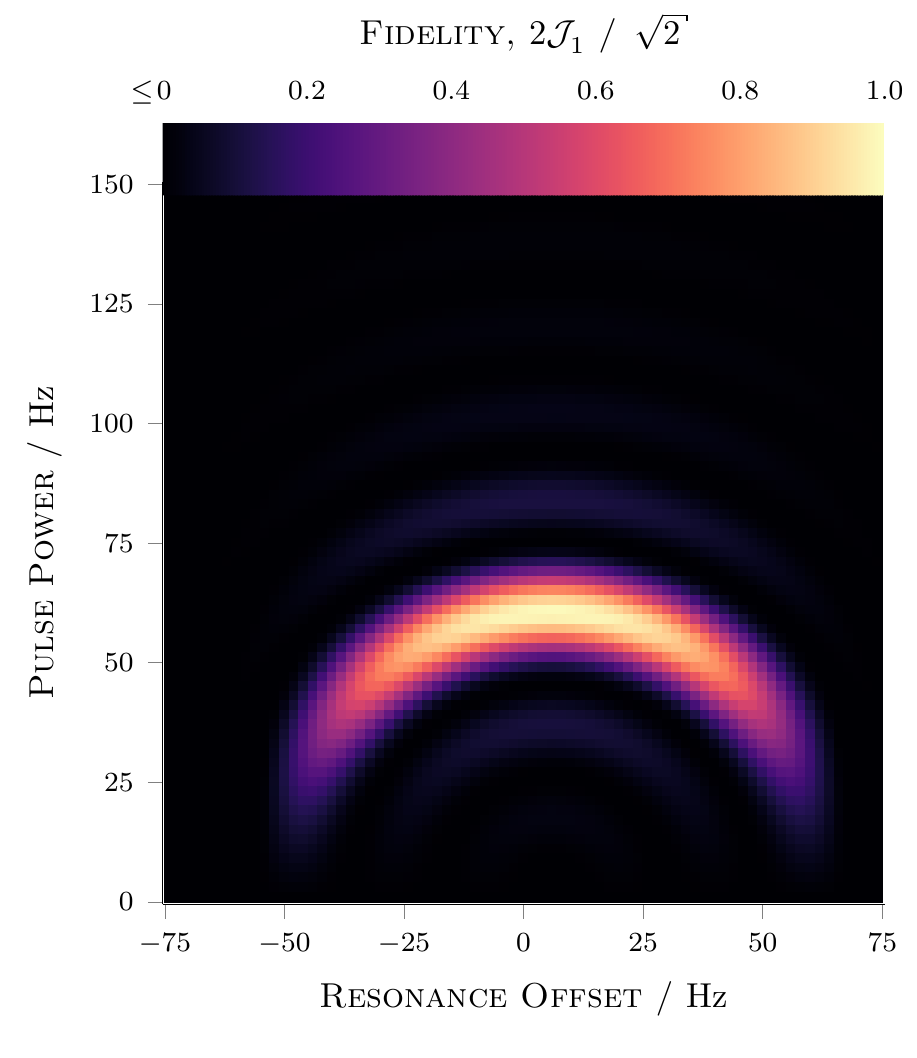}}
 \caption[The robustness of the SLIC pulse]{A plot of fidelity indicating the robustness of the \textsc{slic} pulse used as a direct comparison to optimal control pulses later in this section. The robustness is measured as a function of the nominal power level of the pulse amplitude in Hz on the \(y\)-axis, and the \(\B{0}\) resonance offsets in Hz on the \(x\)-axis.} \label{Fig:slic_robustness}
\end{figure}

The aim of examples presented here is to transfer polarisation to a singlet state, following the lead of the magnetisation-to-singlet (\textsc{m2s}) \cite{LEVITT16} and \textsc{slic} pulses \cite{DEVIENCE13,THEIS14}, and improving this method with the use of optimal control pulses. Although robust optimal control pulses have already achieved this \cite{LAUSTSEN14}, the use of a quadratic smoothing penalty with a Newton-\textsc{grape} method has not yet been realised. 

Singlet states for a pair of mutually coupled spin-\(\sfrac{1}{2}\) nuclei when the scalar coupling \(\gg\) difference in chemical shift resonances. The maximum fidelity of a system in a singlet state is 
\begin{subequations}
\begin{gather}
  \max\big[\Fid_1\big]=\max\bigg[\operatorname{Re}\big\langle {\text{singlet state}}\big|\rho(t) \big\rangle\bigg]=\frac{1}{\sqrt{2}}\nonumber\\
  \max\big[\Fid_2\big]=\max\bigg[\Big|\big\langle {\text{singlet state}}\big|\rho(t) \big\rangle\Big|^2\bigg]=\frac{1}{2}\nonumber
\end{gather}
\end{subequations}
The \textsc{slic} pulse is essentially a hard pulse of an exact amplitude on either the \(x\) or \(y\) channel. The system consists of two \(^{13}C\) spins with a chemical shift difference of \(0.1\)~ppm and a J-coupling of \(60\)~Hz. The magnetic field is set at \(11.7434\)~tesla and the total pulse duration was calculated to be \(56.25\)~ms. The aim of the \textsc{slic} pulse is to take longitudinal polarization \(\rightarrow\) 2-spin singlet state. Given these exact parameters, a Newton optimisation will find the \textsc{slic} pulse as the optimum.

The robustness of the \textsc{slic} pulses is shown in \cref{Fig:slic_robustness}. \textit{Good} pulses should be tolerant to power miscalibration and resonance offsets and large fidelities should be realised over a range of power levels and offsets. This is not really the case with the \textsc{slic} pulse -- showing a band of robustness, but only a small area of high fidelity.

\subsection{Robust magnetisation-to-singlet transfer}\label{Section:RobustM2S}

\renewcommand{\wffile}{\datadir/M2S/Laustsen_wf.dat}
\renewcommand{\datafile}{\datadir/M2S/rbstmat_reallaustsen.dat}
\renewcommand{\figwidth}{0.666\columnwidth}
\begin{figure}
\centering{\includegraphics{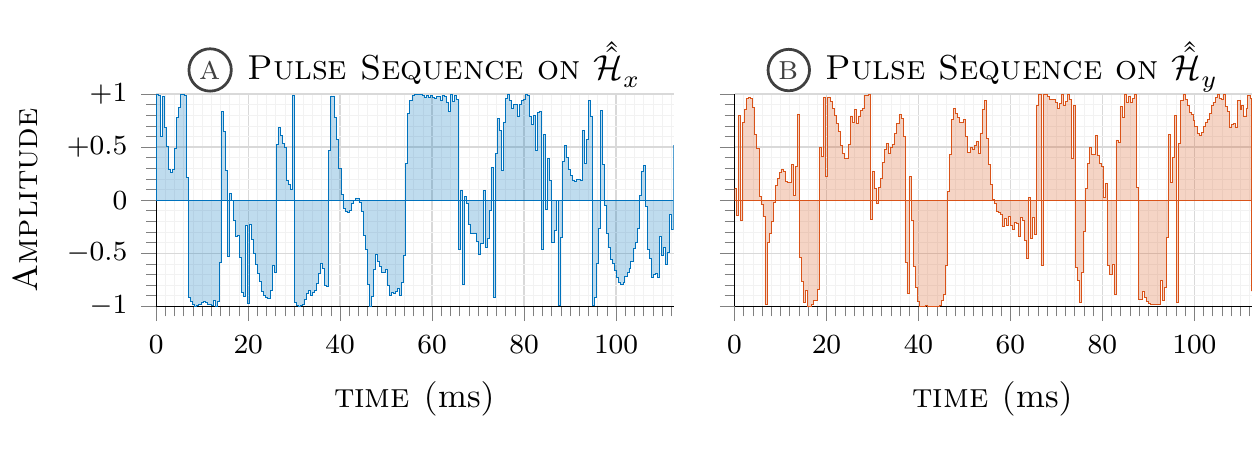}}
 \caption[Pulse shape from an optimised M2S transfer]{\(\HamHH_x^{}\) and \(\HamHH_y^{}\) components of an optimised, robust \(\ell\)-\textsc{bfgs}-\textsc{grape} pulse for a magnetisation-to-singlet transfer.} \label{Fig:laustsen_wf}
\end{figure}

\renewcommand{\wffile}{\datadir/M2S/SmoothNewton_wf.dat}
\renewcommand{\datafile}{\datadir/M2S/rbstmat_realNewtonSmooth.dat}
\renewcommand{\figwidth}{0.666\columnwidth}
\begin{figure}
\centering{\includegraphics{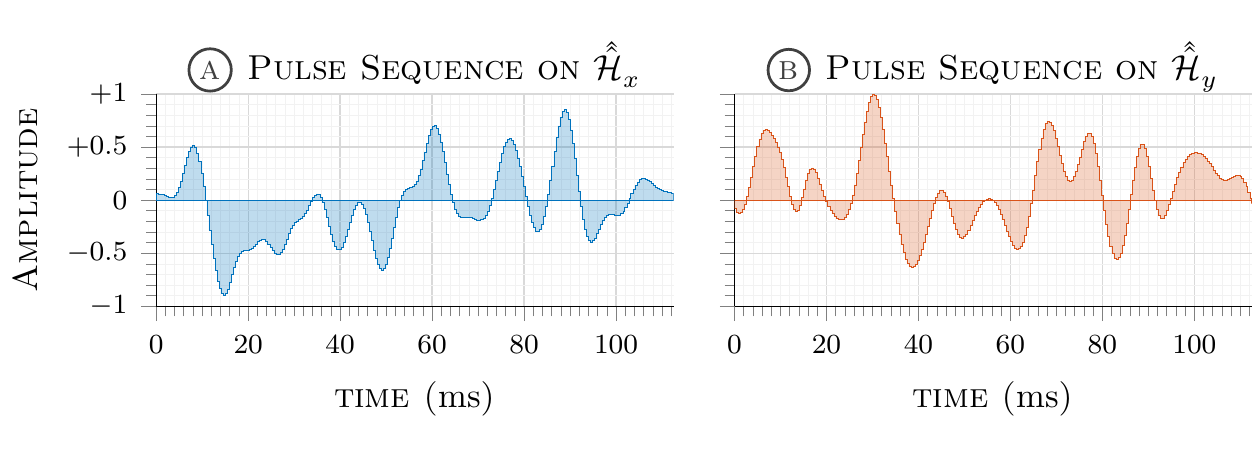}}
\caption[Smoothed pulse shape from an optimised M2S transfer]{\(\HamHH_x^{}\) and \(\HamHH_y^{}\) components of an optimised, robust, smoothed \(\ell\)-\textsc{bfgs}-\textsc{grape} pulse for a magnetisation-to-singlet transfer.} \label{Fig:smoothed_wf}
\end{figure}

Optimal control offers high fidelity state-to-state transfer and gives a shaped pulse which can be designed to be robust to \(\B{0}\) resonance offsets and \(\B{1}\) power miscalibration \cite{LAUSTSEN14}. Simulation of this robustness involves running the \textsc{grape} algorithm over a grid of nominal power levels and offsets\footnote{The resonance offset is simulated with an extra \(\HamHH_z^{}\) control at the offset power level in Hz.} and averaged fidelities and gradients are then used by the \(\ell\)-\textsc{bfgs} optimisation method.

This was done in \cite{LAUSTSEN14}, and is repeated here for comparison. To ensure there are no excursions from the power envelope, the optimisation is performed at a constant amplitude and variable phase, although the pulses shown in \cref{Fig:laustsen_wf} have been transformed back to a Cartesian representation.

\renewcommand{\wffile}{\datadir/M2S/Laustsen_wf.dat}
\renewcommand{\datafile}{\datadir/M2S/rbstmat_reallaustsen.dat}
\renewcommand{\figwidth}{0.666\columnwidth}
\begin{figure}
\centering{\includegraphics{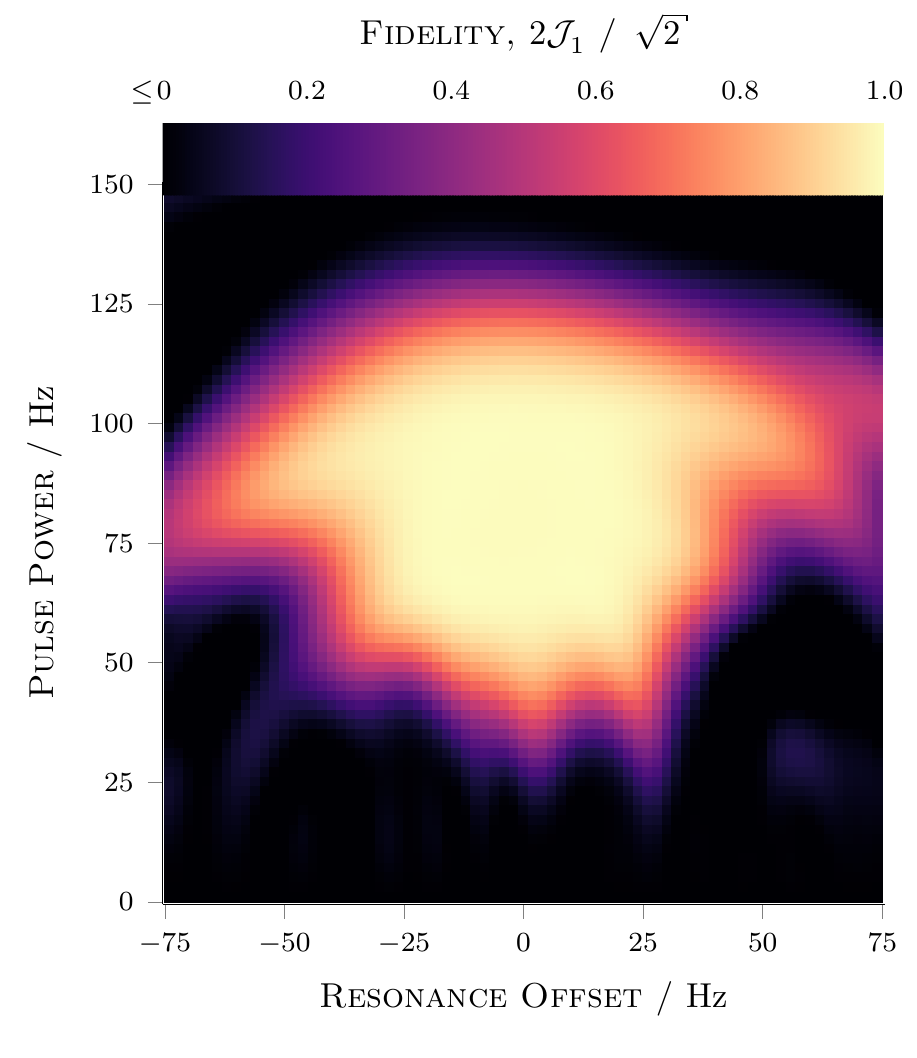}}
 \caption[The robustness of the optimised M2S transfer]{A plot of fidelity indicating the robustness of the optimised \textsc{m2s} pulses from \cref{Fig:laustsen_wf}. The robustness is measured as a function of the nominal power level of the pulse amplitude in Hz on the \(y\)-axis, and the \(\B{0}\) resonance offsets in Hz on the \(x\)-axis.} \label{Fig:laustsen_robustness}
\end{figure}

The robustness of these pulses is shown in \cref{Fig:laustsen_robustness}, and it is quite clear that these pulses are far more robust to range of \(\B{1}\) power and \(\B{0}\) offsets than the \textsc{slic} pulse of \cref{Fig:slic_robustness}. Not only is the pulse shape robust, it has near maximum fidelity over the robust range. A note should be made that the pulse duration is twice that of the \textsc{slic} pulse -- testing shows that too short a duration too critical for optimisation -- the \textsc{slic} pulse is the optimum for its calculated duration.

\subsection{Smooth, robust magnetisation-to-singlet transfer}\label{Section:SmoothRobustM2S}

\renewcommand{\wffile}{\datadir/M2S/SmoothNewton_wf.dat}
\renewcommand{\datafile}{\datadir/M2S/rbstmat_realNewtonSmooth.dat}
\renewcommand{\figwidth}{0.666\columnwidth}
\begin{figure}
\centering{\includegraphics{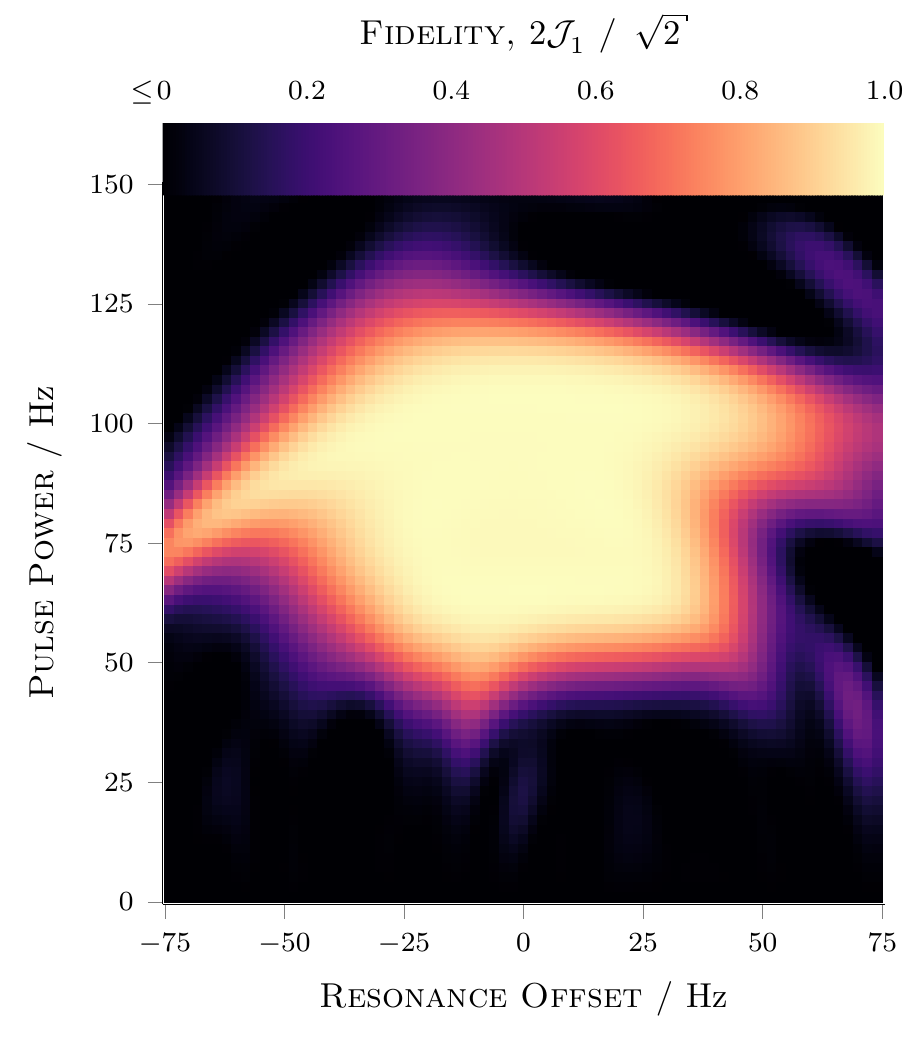}}
 \caption[The robustness of the smoothed, optimised M2S transfer]{A plot of fidelity indicating the robustness of the smoothed optimised \textsc{m2s} pulses from \cref{Fig:smoothed_wf}. The robustness is measured as a function of the nominal power level of the pulse amplitude in Hz on the \(y\)-axis, and the \(\B{0}\) resonance offsets in Hz on the \(x\)-axis.} \label{Fig:smoothed_robustness}
\end{figure}

With the assumption that non-smooth pulses will suffer amplifier non-linearity, smoothed pulse shapes should perform better experimentally.

As was set out in \cref{Chapter:Hessian,Chapter:AuxMat}, optimisation with a Newton-\textsc{grape} method converges to a local maximum. The optimisation strategy is to first gain a quick solution to the state-transfer problem with a \(\ell\)-\textsc{bfgs}-\textsc{grape} method \cite{FOUQUIERES11}, robust to \(\B{0}\) inhomogeneity, then to feed this solution into a quadratically convergent Newton-\textsc{grape} method with a quadratic smoothing penalty of \cref{deriv-norm-square}.

The pulses produced by this two-step optimisation are shown in \cref{Fig:smoothed_wf} and are evidently smooth, much more smooth than in \cref{Fig:laustsen_wf}. As in \cref{Fig:laustsen_wf}, the pulse duration is twice that of the \textsc{slic} pulse for the same reason.

The robustness of the pulse shape from \cref{Fig:smoothed_wf} is shown in \cref{Fig:smoothed_robustness}, showing very similar robustness to that of \cref{Fig:laustsen_robustness}. It is expected that these pulses perform much better in experiments than both the \textsc{slic} pulse and the non-smoothed robust optimal pulses of \cref{Fig:laustsen_robustness}.

The point of using the Newton-\textsc{grape} method is that similar smoothed pulses are not converged to in the \(\ell\)-\textsc{bfgs}-\textsc{grape} method -- the penalty competes with the fidelity measure during a line search and a pulse is found that is neither robust, as designed, nor smooth: finding a solution that is a trade-off between the two. The Newton-\textsc{grape} method seems to be able to cope with this trade-off, identifying a place where the maximum fidelity and the minimum penalty coexist. One reason for this is that the Newton method uses regularisation and conditioning of \cref{Section:conditioning}, and is able to avoid saddle-points.

\chapter[Solid State Nuclear Magnetic Resonance]{Solid State\linebreak Nuclear Magnetic Resonance} \label{Chapter:ssNMR}
\begin{chapquote}{Frank Herbert, \textsc{Dune}}
``I must not fear. Fear is the mind-killer. Fear is the little-death that brings total obliteration. I will face my fear. I will permit it to pass over me and through me. And when it has gone past I will turn the inner eye to see its path. Where the fear has gone there will be nothing. Only I will remain.''
\end{chapquote}
\renewcommand*{\CurrentPath}{./Chapter_7}

Solid-state-\textsc{nmr} can be thought of as an ensemble of systems at different orientations, rotating relative to a laboratory frame. The rotation gives a periodically time-dependent Hamiltonian. A Lebedev quadrature \cite{LEBEDEV75,LEBEDEV76} can be used to characterise a surface integral over a 3D sphere, with weighted values at particular points on the surface (\cref{Section:Powder}). This surface integral can be used to characterise an ensemble in the rotating frame in a Floquet space, \cite{FLOQUET1883,SHIRLEY65}.

This chapter is divided into two sections. The first section is a comparison of the convergence of the Newton-\textsc{grape} and \(\ell\)-\textsc{bfgs}-\textsc{grape} methods, simulating the optimal control problem over a Floquet space describing the crystalline orientations of a solid-state-\textsc{nmr} sample (\cref{Section:Floquet}). The success of the Newton-\textsc{grape} method comes with a large expense and is set out in \cref{Section:OCPowder}. The second outlines the description of the solid-state-\textsc{nmr} sample by a Fokker-Planck formalism from \cite{EDWARDS13,KUPROV16} (\cref{Section:FokkerPlanck}), for a magic-angle-spinning experiment \cite{ODELL11b} \cref{Section:OCMAS} -- using optimal control to excite the overtone transition in \cref{Section:OCHMQC}.

\section{Floquet Theory}\label{Section:Floquet}

The different orientation rotating relative to a laboratory frame gives periodically time-dependent coefficients of the Schr\"{o}dinger equation \cite{ERNST87,LEVANTE95}:
\begin{equation}
 \frac{d}{dt}\hat{\mathcal{U}}(t)=-i\HamHH(t)\hat{\mathcal{U}}(t)\label{EQN:Floquet}
\end{equation}
suggesting the time evolution operator as a Fourier series expansion of the Hamiltonian \cite{LESKES10}:
\begin{equation}
 \HamHH(t)=\sum_nH_n^{}e^{in\omega t}\label{EQN:FourierExp}
\end{equation}
where \(H_n^{}\) are Hermitian matrices of periodic time-dependent functions. The solution to \cref{EQN:Floquet} is given in the following form \cite{FLOQUET1883,SHIRLEY65}:
\begin{align}
 && \hat{\mathcal{U}}(t)= \hat{\mathcal{P}}(t)e^{-iQ t}\hat{\mathcal{P}}^{-1}(t), && \hat{\mathcal{P}}(t)=\sum_n\hat{\mathcal{P}}_ne^{in\omega t} &&
\end{align}
This formulation allows a similar representation to be used when expanding the density matrix \cite{VINOGRADOV05} from \cref{eqn:densityoperatorprop}
\begin{equation}
 \state(t)=\hat{\mathcal{U}}(t)\state(0)\hat{\mathcal{U}}^{-1}(t)=\sum_n\state_n(t)e^{in\omega t}
\end{equation}
It is easy to see that \(\HamHH(t+\tau)=\HamHH(t)e^{in\omega \tau}\), defining the periodic nature of the system. The diagonal elements of the diagonal matrix, \(Q\), are called characteristic exponents \cite{SHIRLEY65}. Subtracting multiples of \(\omega\) from the diagonal of \(Q\) can be compensated by multiplying \(\hat{\mathcal{P}}(t)\) by \(e^{ik\omega t}\).

The direct product of the basis states \(\big\{|\rho_i\rangle\big\}\) with an infinite set of Fourier states \(\big\{|n\rangle\big\}\), resulting from the Fourier expansion in \cref{EQN:FourierExp}, give the full basis set of states:
\begin{equation}
 \Big\{\big|\rho_i,n\big\rangle\Big\}= \Big\{\big|\rho_i\big\rangle\Big\}\otimes \Big\{\big|n\big\rangle\Big\}
\end{equation}
In Fourier space, the Hamiltonian and density matrix can be written as \cite{VINOGRADOV05}
\begin{subequations}
\begin{gather}
 \HamHH_\textsc{r}(t)=\sum_nH_n\hat{F}_ne^{in\omega t}=e^{i\omega \hat{N} t}\bigg\{\sum_nH_n\hat{F}_n\bigg\}e^{-i\omega \hat{N} t}\\
 \state_\textsc{r}(t)=\sum_n\state_n(t)\hat{F}_ne^{in\omega t}=e^{i\omega \hat{N} t}\bigg\{\sum_n\state_n(t)\hat{F}_n\bigg\}e^{-i\omega \hat{N} t}
\end{gather}
\end{subequations}
where \(\hat{F}_n\) are ladder operators in the Fourier space; \(\langle v+n|\hat{F}_n|v\rangle=\Unit\), and \(\hat{N}\) is the number operator; \(\langle v|\hat{N}|v^{\prime}\rangle=v\delta_{vv^{\prime}}\Unit\), and their commutation is \(\big[\hat{N},\hat{F}_n\big]=n\hat{F}_n\). The Liouville-von Neumann equation \cref{eqn:liouvillvonneumannsuperoperator} in Fourier space is
\begin{equation}
 \frac{d}{dt}\state_\textsc{r}(t)=-i\Big[\HamHH_\textsc{r}(t),\state_\textsc{r}(t)\Big]
\end{equation}
Using this Fourier representation of the Liouville-von Neumann equation, the interaction Hamiltonian becomes time independent \cite{VINOGRADOV05}. This time independent Hamiltonian in Floquet space, \(\HamHH_\textsc{f}\) is
\begin{equation}
 \HamHH_\textsc{q}=\sum_nH_n\hat{F}_n+\omega \hat{N}
\end{equation}
giving the Floquet representation of the Liouville-von Neumann equation. This defines the evolution of the density operator;
\begin{equation}
 \frac{d}{dt}\state_\textsc{q}(t)=-i\Big[\HamHH_\textsc{q},\state_\textsc{q}(t)\Big]
\end{equation}
The Floquet density matrix can be evaluated with propagation over a time \(t\) using
\begin{align}
 && \state_\textsc{q}(t)= U_\textsc{q}(t)\state_\textsc{q}(0)U_\textsc{q}^{-1}(t), &&  U_\textsc{q}(t)= e^{-i\hat{H}_\textsc{q}t} &&
\end{align}
Floquet theory has been successfully applied to describe solid-state-\textsc{nmr} \cite{SCHMIDT92,BAIN01,SCHOLZ10}. It should be noted that Kuprov pointed out in \cite{GOODWIN15} that the auxiliary matrix exponential formalism in \cref{eqn:NAJFELD} can be used to find solutions to rotating frame transformations without the usual method of nested commutators.

\subsection{Powder average}\label{Section:Powder}

Various methods exist to discretise a sphere emulating the crystalline orientations of a powder average \cite{BAK97,PONTI99,EDEN03}; a grid uniformly distributed over a unit sphere used in solid-state-\textsc{nmr} simulations:
\begin{enumerate}
 \item A spiral circling the \(z\)-axis of a unit sphere \cite{MOMBOURQUETTE92}.
 \item Finite triangular elements partitioning a unit sphere \cite{WANG95} (\textsc{sophe}).
 \item Simulating many \textit{subspectra}, then averaged to give the powder pattern \cite{VARNER96} (\textsc{sums}).
 \item Iteratively moving particles on the surface of a sphere under the influence of a repulsive force from other particles on the surface of a sphere \cite{BAK97} (\textsc{repulsion}).
 \item The use of Gaussian spherical quadrature \cite{LEBEDEV75,LEBEDEV76} to weight the distribution over the surface of the sphere \cite{EDEN98}
\end{enumerate}
The following work uses the Lebedev quadrature to weight a distribution over the spherical grid \cite{EDEN98,EDEN03}; the Euler angles and weights are shown in \cref{Tab:Lebdev_ab_5}.
\begin{table}
\begin{center}
\begin{tabular}{ccc}
\hline\hline
 $\alpha$  & $\beta$   & $w$ \\
 \hline
  0.000000 &  1.570796 & 0.066667 \\
  3.141593 &  1.570796 & 0.066667 \\
  1.570796 &  1.570796 & 0.066667 \\
 -1.570796 &  1.570796 & 0.066667 \\
  1.570796 &  0.000000 & 0.066667 \\
  1.570796 &  3.141593 & 0.066667 \\
  0.785398 &  0.955317 & 0.075000 \\
  0.785398 &  2.186276 & 0.075000 \\
 -0.785398 &  0.955317 & 0.075000 \\
 -0.785398 &  2.186276 & 0.075000 \\
  2.356194 &  0.955317 & 0.075000 \\
  2.356194 &  2.186276 & 0.075000 \\
 -2.356194 &  0.955317 & 0.075000 \\
 -2.356194 &  2.186276 & 0.075000 \\ 
\hline\hline
\end{tabular}
\caption[Orientations and weights of a Lebedev grid]{Orientations, \(\alpha\) and \(\beta\) (radians), and weights \(w\) of the Lebedev grid quadrature.}\label{Tab:Lebdev_ab_5}
\end{center}
\end{table}

\subsection[Optimal control of a static powder average]{Optimal control of a\linebreak static powder average}\label{Section:OCPowder}

With the Newton-\textsc{grape} developed in \cref{Chapter:Hessian,Chapter:AuxMat}, it seems a good place to start to test the method on a large system. As set out above, an optimisation problem involving an average over large grid will take many Hessian matrix evaluations, magnifying computational efficiency.

The system set out to test this is similar to the one in \cref{Section:test2}, except that the system is averaged over crystalline orientations. Optimal control of the powder average has a fidelity, gradients, and Hessians weighted over the crystalline orientations -- rank 17 Lebedev grid with 110 points.

\begin{figure}
\centering{\includegraphics{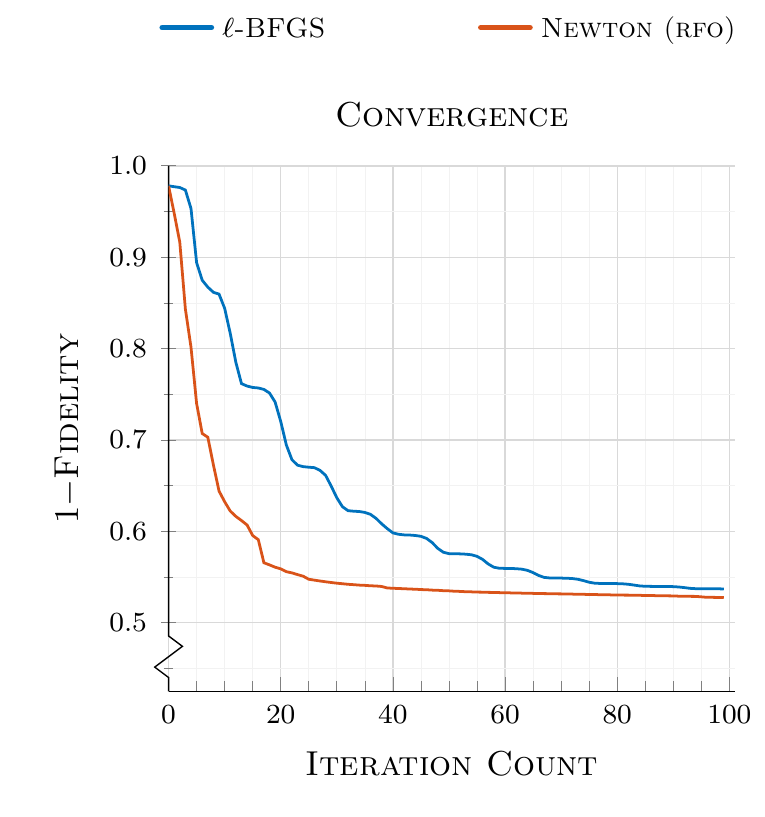}}
 \caption[Convergence of Newton-GRAPE method for glycine state transfer]{Comparison of optimisation methods for a \(^{14}\)N quadrupolar overtone signal in static glycine powder. Two methods are compared: Newton-\textsc{grape} method using a bracketing and sectioning line search; \(\ell\)-\textsc{bfgs}-\textsc{grape} using a bracketing and sectioning line search.}\label{Overtone_Newtonbfgs}
\end{figure}

The optimal control problem is to excite \(^{14}\)N from a state \(\hat{T}_{1,0}\) to \(\hat{T}_{2,2}\) (\cref{Chapter:AppendixA}) through the nuclear quadrupolar interaction with appropriate control operators. The optimisation if performed over 400 time points for total pulse duration of 40~\(\mu\)s at a power level of 40~kHz.

At a particularly short total pulse duration, the optimal solution is not at a maximum fidelity \(=1\). The comparison of results for the convergence of \(\ell\)-\textsc{bfgs}-\textsc{grape} and Newton-\textsc{grape} is shown in \cref{Overtone_Newtonbfgs}\footnote{One convergence profile is shown -- but convergence over the averaged grid gives very similar convergence from random starting pulses. It should be noted that the averaging has dictated a smooth convergence profile.}. Newton-\textsc{grape} is quicker in the initial stages of the optimisation, as is expected compared to a quasi-Newton method, and has a similar convergence characteristic to the universal rotation pulses over a power miscalibration grid in \cref{Section:test3}. The terminal fidelity is also slightly better than that of the quasi-Newton method, again similar to the results of \cref{Section:test3}. 

However, Newton-\textsc{grape} took much longer (wall clock time) to reach a fidelity only a few percent better than the quasi-Newton approach -- a couple of hours for \(\ell\)-\textsc{bfgs}-\textsc{grape} to reach 100 iterations, and half a day for Newton-\textsc{grape} to reach 100 iterations\footnote{Both methods used a 32 core computer parallelised calculations over the time steps.}.

\subsection[Optimal control under magic angle spinning]{Optimal control under\linebreak magic angle spinning}\label{Section:OCMAS}

Using Floquet space to describe the states of the solid-state-\textsc{nmr} system with a powder average, then to spin the system at the magic angle\footnote{the magic angle is \(\theta_\text{magic}=\arccos{\tfrac{1}{\sqrt{3}}}=54.7^{\circ}\)}, is the test of optimal control set out in this section.

The powder average should be modelled by an appropriate Lebedev grid describing the crystalline orientations in the power. The \textit{local fidelity} and the \textit{local gradient} of the fidelity at each orientation of the grid to produce a global value of the fidelity and its gradient -- the global measure is a weighted average over the crystalline grid. The local optimal control problem is performed with the \(\ell\)-\textsc{bfgs}-\textsc{grape} method.

Using the parameters set out in \cite{JARVIS13} for the overtone excitation in a glycine molecule, the optimal control problem here is a cross-polarisation experiment between \(^{13})C\) and the quadrupolar \(^{14}\)N. A spinning rate of frequency \(f=25\times 10^3\)~Hz is used to simulate the magic angle spinning experiment of \cite{JARVIS13}, and the Fourier rank at which the Floquet matrix is truncated is 5 and the crystalline grid is averaged over a Lebedev grid rank 5 \((\alpha,\beta,\gamma)\) relative to the rotor frame (as in \cref{Tab:Lebdev_ab_5}).

The initial state and final states of the system are the direct products with the identity matrix of size \((2\times \text{max-rank})+1\), where \(\text{max-rank}=5\) in this case. This gives the optimal control boundaries as
\begin{subequations}
\begin{gather}
  \state(0)=\left(\Unit\otimes L_z^\text{(C)}\right)\\
  \targ{}=\left(\Unit\otimes L_z^\text{(N)}\right)
\end{gather}
\end{subequations}
where both states should be normalised. Control operators are in the Floquet formalism irradiating the resonant frequencies of \(^{13}\)C and \(^{14}\)N, again as a direct product with the identity matrix:
\begin{align}
&& \HamHH_1=\Unit\otimes \HamHH_x^\text{(C)}, && \HamHH_2=\Unit\otimes \HamHH_y^\text{(C)}, && \HamHH_3=\Unit\otimes \HamHH_x^\text{(N)}, && \HamHH_4=\Unit\otimes \HamHH_y^\text{(N)} &&
\end{align}
with the nominal power of 10~kHz. Time steps are integer multiples of the characteristic time from the spinning rate, 5000 in this case, and the total pulse duration is 20~ms.

The \(\ell\)-\textsc{bfgs}-\textsc{grape} method is used; following from the long wall clock time of the Newton-\textsc{grape}, the limited memory quasi-Newton method is more appropriate considering the large number of optimisation variables, \(4\times 5000\). The optimisation was run for 256 iterations, although the optimisation had not fully converged at this point, a good set of results were obtained without waiting for super-linear convergence to find small gains.

\begin{figure}
\centering{\includegraphics{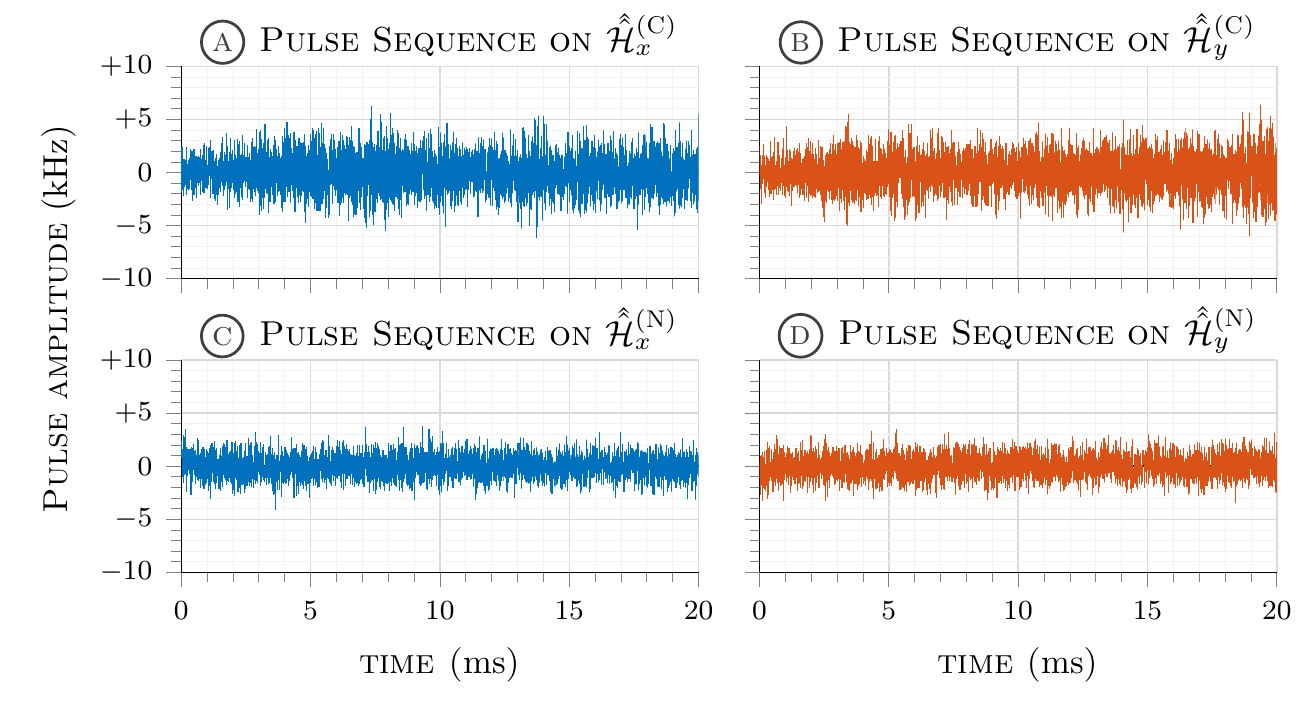}}
 \caption[Control pulses for glycine state transfer]{The four channel control pulse sequences producing the trajectories in \cref{MAS_sim}. The pulses are a result of the \(\ell\)-\textsc{bfgs}-\textsc{grape} optimisation, which reached high fidelity, state-to-state transfer from \(^{13}\)C to \(^{14}\)N of glycine in a solid-state-\textsc{nmr} simulation.}\label{Fig:CN_gly_floquet_controls}
\end{figure}

The control pulses produced at the end of the 256 iterations are shown in \cref{Fig:CN_gly_floquet_controls} -- looking like noise, apart from the \(^{14}\)N pulses generally needing less power.

\begin{figure}
\centering{\includegraphics{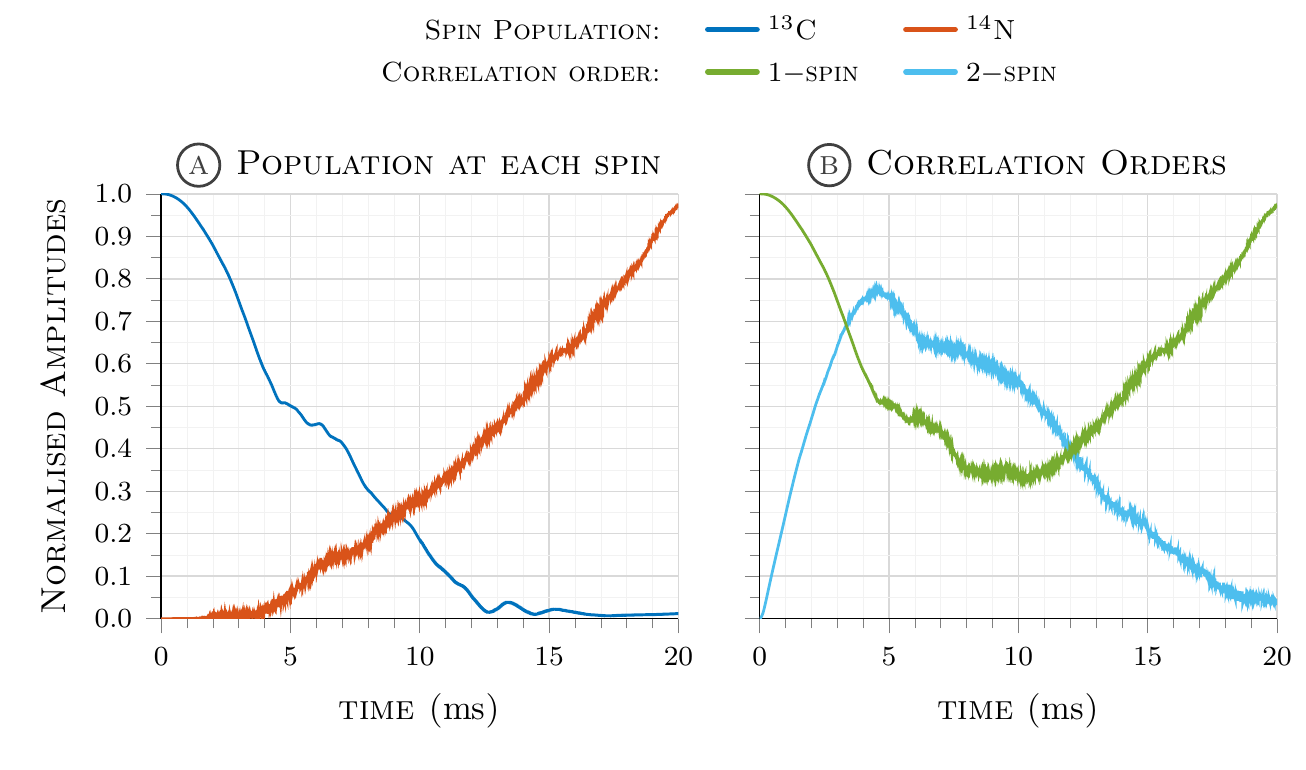}}
\caption[Trajectory analysis of glycine state transfer]{Trajectories analysis for a state-to-state transfer from \(^{13}\)C to \(^{14}\)N of glycine in a solid-state-\textsc{nmr} simulation. Averaged over the crystalline orientations using a Lebedev quadrature. 10~kHz power level of control pulses and 14.1 MHz magnet. \plotlabel{a} Local spin population. \plotlabel{b} Spin correlation order amplitude.}\label{MAS_sim}
\end{figure}

The trajectory analysis \cite{KUPROV13} of the pulses shows more useful information. As intended, the local population of the magnetisation is transferred from \(^{13}\)C to \(^{14}\)N in \cref{MAS_sim} \plotlabel{a}, with a smooth transfer except for a small amplitude, high frequency oscillation on the \(^{14}\)N trajectory.

The correlation orders in \cref{MAS_sim} \plotlabel{b} clearly show a 1-spin order giving way to the 2-spin order, to mediate the \(^{13}\)C and \(^{14}\)N coupling, then after about half of the pulse duration, the 2-spin order is transferred back to 1-spin order. There is an initial ramp in the 2-spin order and then a linear decay to zero at time \(=T\). Again, the correlation orders develop a small amplitude, high frequency oscillation in the trajectories, this time after a maximum 2-spin order has been reached.

\section{The Fokker-Planck formalism}\label{Section:FokkerPlanck}

Kuprov \cite{KUPROV16} set out a review of the Fokker-Planck \cite{FOKKER14,PLANCK17} formalism used in magnetic resonance simulations, finding the very general formulation that should be preferred to the Liouville-von Neumann equation. For this reason, the optimal control methods set out in this thesis should be constructed to be compatible with the Fokker-Planck equation of motion. As pointed out in \cite{KUPROV16}, there is a close relation between the Fokker-Planck equation and Floquet theory of the previous section \cite{LEVANTE95} -- the latter is a special case of the former.

\subsection{Optimal Control in HMQC-type experiment}\label{Section:OCHMQC}

\begin{figure}
\centering{\includegraphics{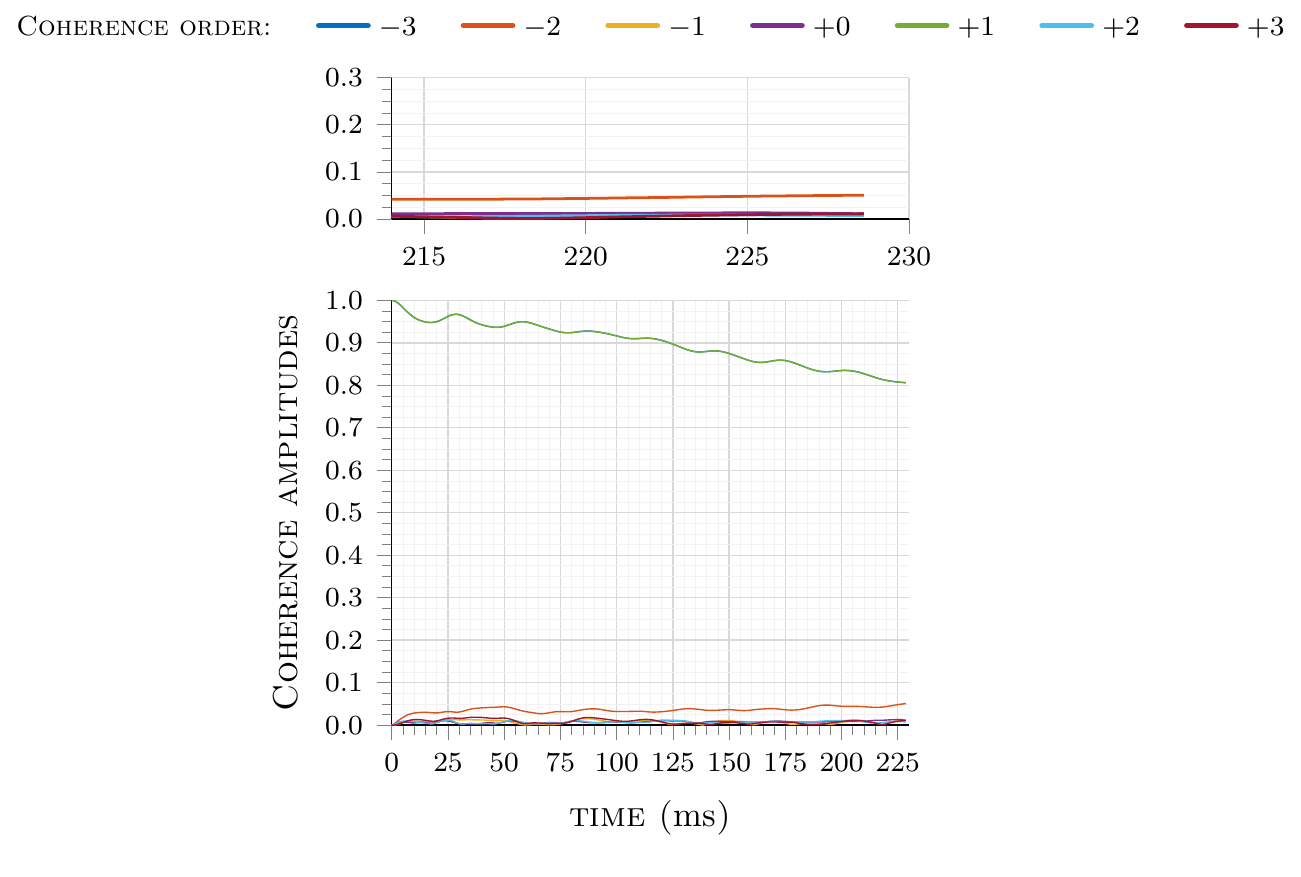}}
 \caption[Trajectory analysis of HMQC-type experiment with hard pulse]{Trajectory analysis an \textsc{hmqc}-type experiment to indirectly excite the overtone transition in glycine with hard pulses. Coherence orders are represented as the effect of a constant amplitude and zero phase. Magnified section of the plot shows higher coherence orders populated near the end of the control pulse trajectory (\textit{upper})}\label{Fig:trajanHard}
\end{figure}

The testing set out in this section is an \textsc{hmqc}-type experiment (Hetronuclear Multiple Quantum Coherence) to indirectly excite the overtone transition in glycine \cite{HAIES15} (results from hard pulses alone shown in \cref{Fig:trajanHard}). \textsc{Grape} calculations of optimal control in the Fokker-Planck formalism are used to find efficient pulses for an excitation of the \(+2\) coherence order of \(^{14}\)N in the \textsc{hmqc}-type experiment. Optimal pulses are designed to be resilient to radiofrequency pulse miscalibration of \(\pm2.5\%\). The total pulse length of the optimal set is investigated with respect to numerical convergence, maximum fidelity after 200 iterations, and coherence order trajectories.

The aim is to find optimal control pulses that spy on \(^{14}\)N through echoes on a connected \(^{1}\)H by exciting high coherence orders of \(^{14}\)N. The optimal control pulses form part of a larger pulse sequence, shown in \cref{OCpulseset}, where the second section of the optimal control pulses is the time reversed version of the first section.

\begin{figure}
\centering{\includegraphics{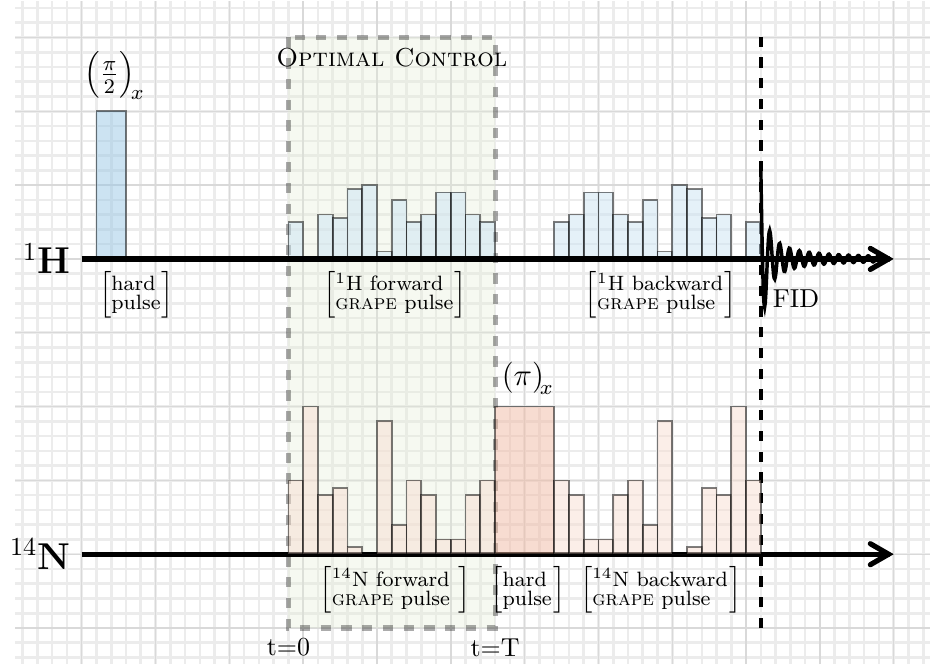}}
\caption[Optimal control in an HMQC-type experiment]{Pulse set including optimal control pulse. Pulses vary in phase only, and the amplitude should be considered constant. The second section of optimal pulses is the time-reversed version of the first section.}\label{OCpulseset}
\end{figure}

The spinning rate of the system is 35~kHz at the magic angle, in a 14.1~tesla magnet. Relaxation is included in the form of a diagonal approximation with a rate of 300~Hz, and the pulse duration is varied with different integer values of the rotor period. Optimal control of the initial state to a target state is performed as a sum of states picking out the \(+2\) coherence excitation on the \(^{14}\)N. The initial to target states are set out as
\begin{align}
  && L_+^\text{(H)} && \xrightarrow{\hphantom{aaa}\textsc{ grape }\hphantom{aaa}} && L_+^\text{(H)}I_+^\text{(N)} + L_-^\text{(H)}I_+^\text{(N)} &&\nonumber
\end{align}
Control operators are defined in the Cartesian representation, allowing pulses on both \(^{14}\)N and \(^{1}\)H to work in the rotor frame at each of the weighted crystalline orientations in the rank-5 powder average:
\begin{subequations}
\begin{gather}
 \HamHH_x^\text{(N)}=\Unit\otimes\left[\cos(\theta)\HamHH_z^\text{(N)}+\sin(\theta)\left(\frac{\HamHH_+^\text{(N)} + \HamHH_-^\text{(N)}}{2}\right)\right]\\
 \HamHH_y^\text{(N)}=\Unit\otimes\left[\cos(\theta)\HamHH_z^\text{(N)}+\sin(\theta)\left(\frac{\HamHH_+^\text{(N)} - \HamHH_-^\text{(N)}}{2i}\right)\right]\\
 \HamHH_x^\text{(H)}=\Unit\otimes\left[\cos(\theta)\HamHH_z^\text{(H)}+\sin(\theta)\left(\frac{\HamHH_+^\text{(H)} + \HamHH_-^\text{(H)}}{2}\right)\right]\\
 \HamHH_y^\text{(H)}=\Unit\otimes\left[\cos(\theta)\HamHH_z^\text{(H)}+\sin(\theta)\left(\frac{\HamHH_+^\text{(H)} - \HamHH_-^\text{(H)}}{2i}\right)\right]
\end{gather}
\end{subequations}
where \(\theta\) is the magic angle of the rotor spinning axis and \(\Unit\) is the identity operator of size \(d\), being the size of the spatial dimension of the Fokker-Planck formalism.

\begin{center}
\begin{table}
\begin{small}
{\begin{tabular}{c c c c | c }
\hline\hline
Amplitude (polar) & total time & pulses & $\nabla$ calc. & \multicolumn{1}{c}{$\% \text{ state transfer }$}\\
 (kHz) & ($\mu$s)& (\#) & (s) & initial state $=\left\{L_+^\text{(H)}\right\}$ \\
\hline
 \multicolumn{2}{c}{spin rate = 35~kHz} & & & \multicolumn{1}{c}{targets $=\left\{L_+^\text{(H)}L_+^\text{(N)} , L_-^\text{(H)}L_+^\text{(N)}\right\}$} \\ 
 $\sqrt{2}\times$ 50 & 57.143 & 200 & 47 & \bf{7.89 \%} \\ 
 $\sqrt{2}\times$ 50 & 114.286 & 400 & 92 & \bf{15.86 \%} \\ 
 $\sqrt{2}\times$ 50 & 171.429 & 600 & 140 & \bf{18.67 \%} \\ 
 $\sqrt{2}\times$ 50 & 228.571 & 800 & 189 & \bf{21.57 \%} \\ 
 \hline\hline
\end{tabular}}
\end{small}
\caption[Performance of four test optimisations: 2, 4, 6, and 8 rotor periods]{Simulation performance of each of four test optimisations: 2, 4, 6, and 8 rotor periods, each with 100 time slices per rotor period. Note: results are not converged and a greater fidelity would be achieved if the optimisation were allowed to run for more than \(200\) iterations, particularly for the longer pulse sets (see \cref{Fig:converge}) }\label{Tab:fidelity}
\end{table}
\end{center}

\begin{figure}
\centering{\includegraphics{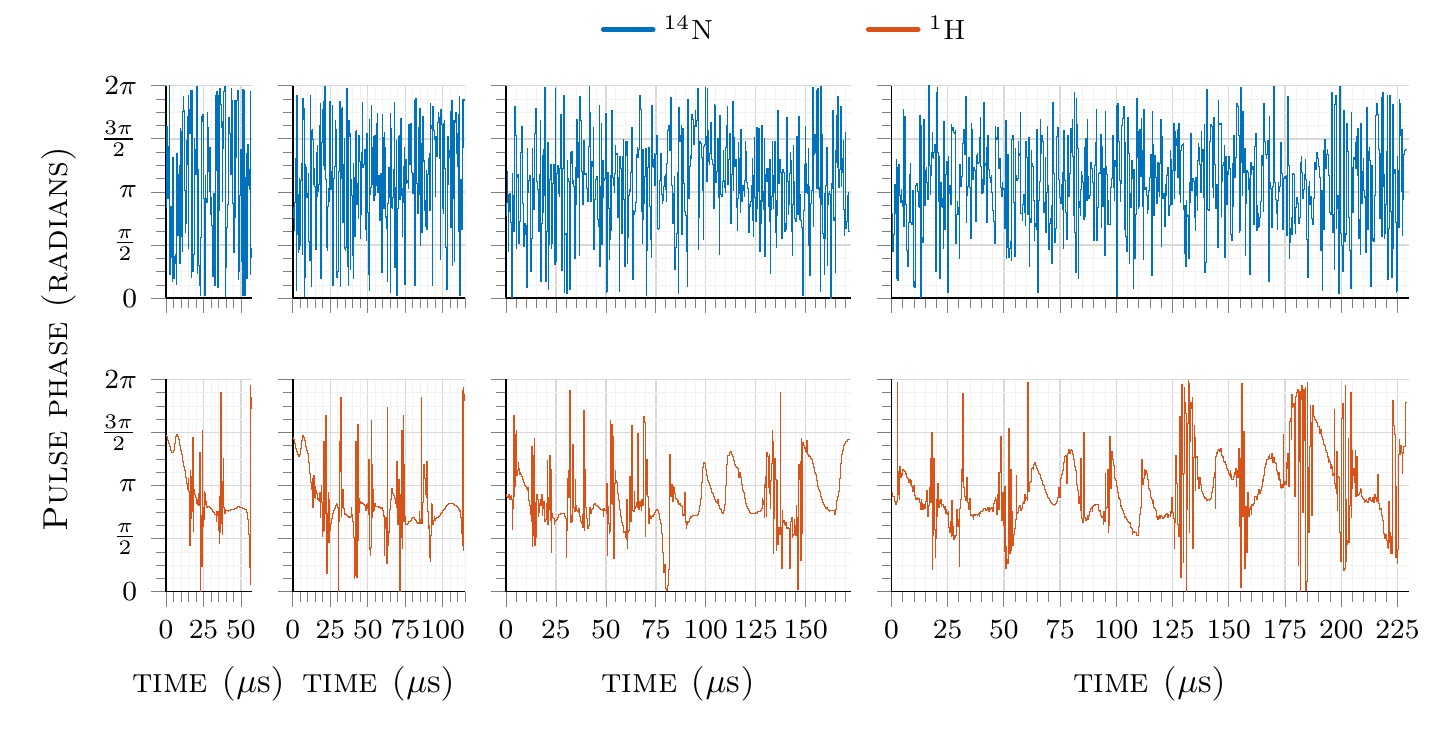}}
 \caption[Phase profile of GRAPE pulses for 2, 4, 6, and 8 rotor periods]{Phase profile resulting from the \(\ell\)-\textsc{bfgs}-\textsc{grape} method with 2, 4, 6, and 8 rotor period for total pulse duration (\textit{left to right}). The amplitude profile is constant in all cases, \(A=\sqrt{2}\cdot 50\)~kHz, and the phase pulses on \(_{}^{14}\)N and \(_{}^{1}\)H (\textit{upper and middle}) should be tolerant to a \(\pm 2.5\%\) miscalibration on the pulse amplitudes (when translated to a Cartesian representation). The total density of magnetisation at each spin is shown as an effect of the pulses (\textit{lower}). }\label{Fig:trajan1}
\end{figure}

\begin{figure}
\centering{\includegraphics{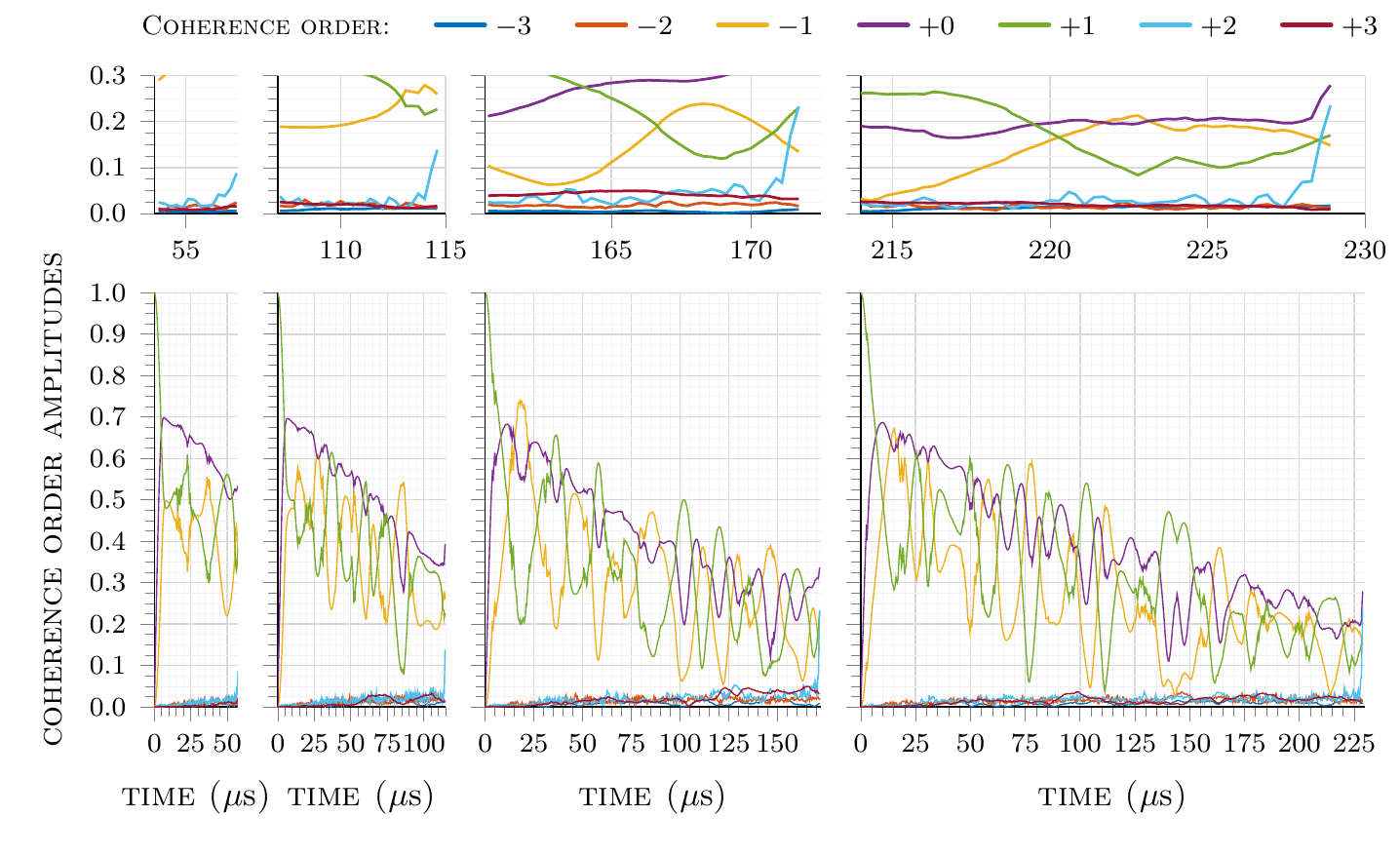}}
 \caption[Trajectory analysis of GRAPE pulses for 2, 4, 6, and 8 rotor periods]{Trajectory analysis of an l-{\sc bfgs} optimisation with 2, 4, 6, and 8 rotor periods for the total pulse duration (\textit{left to right}). Coherence orders are represented as the effect of phase profiles shown in \cref{Fig:trajan1}. Magnified section of the plot shows higher coherence orders populated near the end of the control pulse trajectory (\textit{upper}). NOTE: Trajectories are the last trajectory calculated i.e. that at \(+2.5\%\) pulse miscalibration and for the target state \(L_-^\text{(H)}L_+^\text{(N)}\).}\label{Fig:trajan2}
\end{figure}

Normalised initial states and target states are similarly defined via the Kronecker product:
\begin{subequations}
\begin{gather}
 \state(0)=\frac{1}{\sqrt{d}}\left(\Unit\otimes L_+^\text{(H)}\right)\\
 \state_1(T)=\frac{1}{\sqrt{d}}\left(\Unit\otimes L_+^\text{(H)}L_+^\text{(N)}\right)\\
 \state_2(T)=\frac{1}{\sqrt{d}}\left(\Unit\otimes L_-^\text{(H)}L_+^\text{(N)}\right)
\end{gather}
\end{subequations}
where \(\Unit\) a unity vector of length \(d\), \(\state(0)\) is the initial state at time \(t=0\), \(\state_1(T)\) and \(\state_2(T)\) are the two target states at the time \(t=T\). The \(\ell\)-\textsc{bfgs}-\textsc{grape} method is used, and performs two fidelity and gradient calculations from the initial state to the two final states, then to average the two fidelities and two gradients. The full experimental state transfer problem can be envisaged as the pulse set being part of the whole experimental pulse sequence in \cref{OCpulseset}.

Four optimisation tests were performed with the only difference being the total pulse time allowed for the control pulses. The time step remained equal in all cases, having \(100\) time steps per rotor period (\(35\)~kHz spinning rate). The resulting fidelity from each test is shown in the \cref{Tab:fidelity} (final column).

The polar amplitude was kept to a constant, \(\sqrt{2}\cdot 50\)~kHz \footnote{corresponding to \(\pm 50\)~kHz maximum/minimum pulse amplitude in the Cartesian representation}, and only the phase profiles of pulses on each spin were varied. The phase profiles resulting from the optimisation are shown in \cref{Fig:trajan1}. In addition, the coherence pathways for these pulses are shown in \cref{Fig:trajan2} -- this is the main indication of the success of the simulations in comparison to the hard pulses of \cref{Fig:trajanHard}.

The four tests were run to \(200\) iterations, rather than a converged set of solutions. \cref{Fig:converge} shows how far each of the test optimisations converged. In addition to showing the final fidelity reached at \(200\) iterations, the norm of the gradient, \(\|\nabla\|_2^{}\), is shown at each iteration -- the norm of the gradient is a measure of how optimal the solution is \footnote{a maximum or minimum proper would have \(\nabla=0\) and hence \(\|\nabla\|_2^{}=0\) -- the main point is the smaller this number, the more optimal the solution}.

\begin{figure}
\centering{\includegraphics{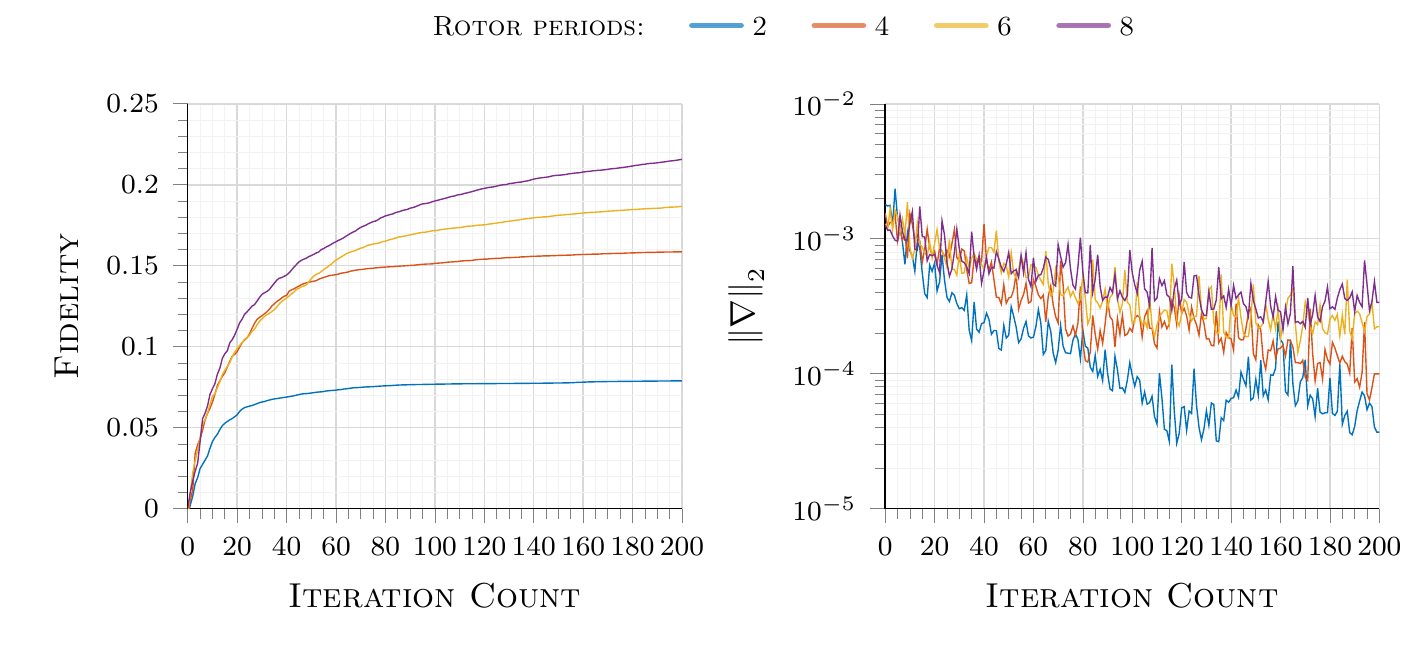}}
 \caption[Convergence analysis with 2, 4, 6, and 8 rotor periods]{Convergence characteristics. Note that the optimisations have not yet achieved convergence to a maximum (apart from the optimisation with 2 rotor periods).}\label{Fig:converge}
\end{figure}

\chapter{Feedback Control} \label{Chapter:AWG}

\begin{chapquote}{Hermann Hesse, \textsc{Steppenwolf}}
``Most men will not swim before they are able to. Is that not witty? Naturally, they won't swim! They are born for the solid earth, not for the water. And naturally they won't think. They are made for life, not for thought. Yes, and he who thinks, what's more, he who makes thought his business, he may go far in it, but he has bartered the solid earth for the water all the same, and one day he will drown.''
\end{chapquote}
\renewcommand*{\CurrentPath}{./Chapter_8}

\renewcommand{\figwidth}{0.35\columnwidth}

A significant current problem in high-field electron spin resonance (\textsc{esr}) spectroscopy is the difficulty of achieving uniform and quantitative signal excitation using microwave pulses \cite{SCHWEIGER01,BRUSTOLON09}. The greatest instrumentally available microwave power, in terms of the electron spin nutation frequency, is about 50~MHz; the shortest realistic \(\sfrac{\pi}{2}\) pulse is therefore \(5\)~ns long \cite{CRUICKSHANK09}, corresponding to the excitation bandwidth of \(200\)~MHz -- enough to affect a significant portion of many  solid state \textsc{esr} signals, but insufficient to excite such signals uniformly and quantitatively. The consequences of partial excitation of the spectrum include useful orientation selection effects \cite{SCHIEMANN09,POLYHACH07,BODE08}, but also reduced sensitivity and diminished modulation depth in two-electron dipolar spectroscopy \cite{THURNAUR80,SALIKOV92,TAN94}.

The time resolution of the best available microwave pulse shaping equipment is of the order of \(30\)~ps \footnote{Keysight M8196A \(92\)~GSa/s Arbitrary Waveform Generator}. This work uses Bruker SpinJet \textsc{awg} (Arbitrary Waveform Generator) with \(0.625\)~ns time resolution \cref{Section:AWGScheme} -- in combination with non-resonant cavities it enables generation of shaped pulses with the bandwidth of about 1~GHz \cite{CRUICKSHANK09} and allows many broadband excitation schemes originally developed for nuclear magnetic resonance (\textsc{nmr}) spectroscopy \cite{WIMPERIS94} to be used with only minor modifications \cite{SPINDLER13,DOLL13,DOLL14,SCHOPS15,JESCHKE15}. Numerically designed ``optimal control'' microwave pulses \cite{KHANEJA05,FOUQUIERES11,GOODWIN16} are also possible \cite{SPINDLER12,KAUFMANN13}, but a complication specific to \textsc{esr} is that the waveforms received by the sample are very different from those sent by the \textsc{awg} -- the response function of the \textsc{esr} instrument cannot be ignored \cite{SPINDLER12}.

One way around this is to construct a transfer matrix or a response function that connects, under the linear response approximation, the ideal pulse emitted by the computer to the real pulse seen by the sample. The transfer matrix may be measured either by adding an antenna to the resonator  \cite{SPINDLER12}, or by using a sample with a narrow \textsc{esr} line to pick up the intensity of each spectral component \cite{KAUFMANN13}. Quasi-linear responses, such as the phase variation across the excitation bandwidth in nutation frequency experiments, can be described with additional transfer matrices \cite{DOLL13}.

The transfer matrix approach is not perfect -- different samples alter the dynamical properties of the resonator in different ways, as does the antenna -- but the linear and quasi-linear models work well in practice. The standard procedure is to take the desired pulse, reverse-distort it through the transfer matrices, send the result out of the waveform generator and hope that a good rendering of the intended pulse shape arrives at the sample point. It usually does \cite{SPINDLER12,KAUFMANN13}, but the logistical overhead of measuring the transfer matrix is significant. Accurate measurement of the instrument response function in the ways described above is time-consuming (hours), and \textsc{esr} resonators, particularly at high frequencies, tend to have strongly sample-dependent response functions.

In this chapter, a different microwave pulse shape refinement strategy is explored, which does not require explicit knowledge of the transfer matrix \cref{Section:FeedbackControlOptim}. It relies instead on the possibility of repeating an \textsc{esr} experiment hundreds of times per second, and recognises the fact, discussed in detail below, that microwave pulses in \textsc{esr} need very few discretisation points due to the significant width of the instrument response function. Further details of this work are in \cref{Chapter:AppendixXepr}, including software code and experimental results.

Numerically optimised microwave pulses are used to increase excitation efficiency and modulation depth in electron spin resonance experiments performed on a spectrometer equipped with an arbitrary waveform generator (\cref{Section:AWGHardwareSoftware}). The optimisation procedure is sample-specific and reminiscent of the magnet shimming process used in the early days of nuclear magnetic resonance -- an objective function (for example, echo integral in a spin echo experiment) is defined and optimised numerically as a function of the pulse waveform vector using noise-resilient gradient-free methods (\cref{Section:AWGOptimisationMethod}). The resulting shaped microwave pulses achieve higher excitation bandwidth and better echo modulation depth than hard pulses. Although the method is theoretically less sophisticated than quantum optimal control techniques, the rapid electron relaxation means that the optimisation takes only a few seconds, and the knowledge of either the instrument response function or spin system ensemble parameters is not required. This makes the procedure fast, convenient, and easy to use. Improvements in broadband excitation efficiency, spin echo intensity \cite{HAHN50} in \cref{Section:AWGEcho} and modulation depth are demonstrated with optimal conditions of \textsc{oop-eseem} \cite{TIMMEL98} in \cref{Section:AWGOOPESEEM}, at the instrument time cost not exceeding the time it used to take to auto-shim an \textsc{nmr} magnet.

The method is known as ``feedback control'' \cite{SKOGESTAD07,FRANKLIN94,DOYLE13}; in its electron spin resonance adaptation it is similar in principle to the well known (in the \textsc{nmr} circles) task of maximising a deuterium lock signal during the magnet shimming process \cite{DEROME13} -- a target variable is chosen and maximised, using a noise-resilient optimisation algorithm, with respect to the variables of interest. In the \textsc{esr} case, these variables are amplitudes of the microwave field at each waveform discretisation point.

\section{Feedback Control Optimisation}\label{Section:FeedbackControlOptim}

\begin{figure}
\centering{\includegraphics{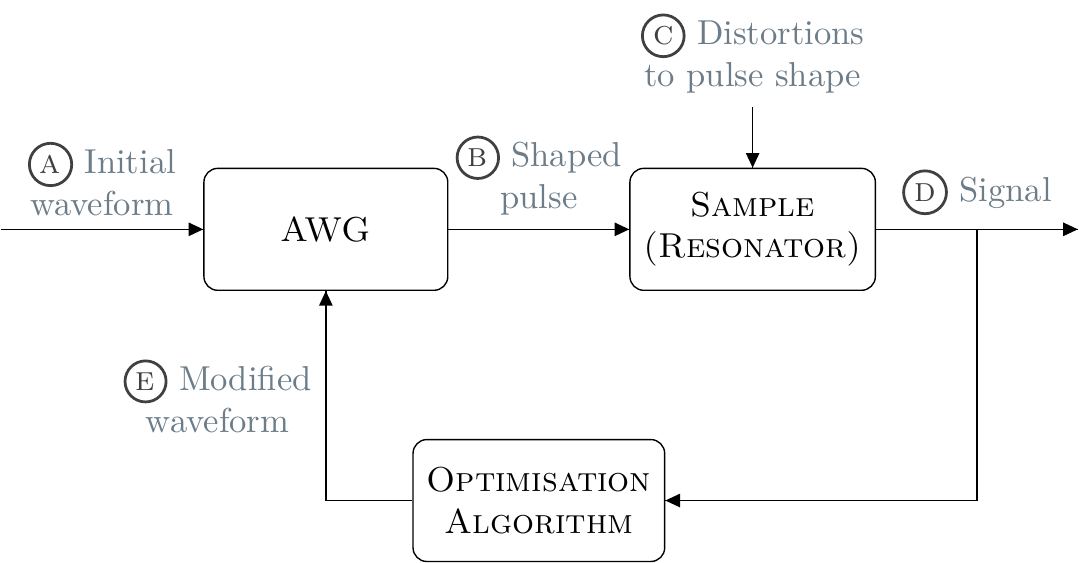}}
\caption[Closed-loop feedback control]{Block diagram of \textsc{awg} closed-loop feedback control. \plotlabel{a} An initial, random, pulse set sent to the \textsc{awg}. \plotlabel{b} Pulse-set is processed and interpolated, then sent to the sample (resonator). \plotlabel{c} The pulse-set arriving at the sample experiences shape distortions from system noise and hardware limitations. \plotlabel{d} The effect of pulse-set induced excitations on the sample is measured as a signal. \plotlabel{e} The optimisation algorithm calculates a new, modified set of pulses that are the sent back to the \textsc{awg}. \label{FIG-FeedbackControl}}
\end{figure}

One simple way to iteratively improve the result of experimental measurements is to use feedback control \cite{SKOGESTAD07,FRANKLIN94,DOYLE13}. The idea is to modify inputs to the experiment according to the result of an experimental measurement: controlling the experiment ``on the fly''. The Bruker SpinJet arbitrary waveform generator (\textsc{awg}) has the ability to shape pulses of the \textsc{esr} experiment -- giving a flexible control method. However, distortions to the pulse shape are expected in two forms -- constant distortions that are the same for repeated measurements, such as shape pulse interpolation, and those associated with non-linear effects such as noise from electronic components and sample specific miscalibration. \cref{FIG-FeedbackControl} sets out a block diagram of a closed-loop feedback control using an \textsc{awg}.

\subsection{Hardware and Software}\label{Section:AWGHardwareSoftware}

The Bruker SpinJet \textsc{awg} is a two channel arbitrary waveform generator, allowing the input of in-phase and quadrature components of shaped pulses to a spectrometer (\cref{BlockDiagSchematics}). The Bruker SpinJet \textsc{awg} used in this work has a time resolution of \(0.625\)~ns, \(14\)-bit amplitude resolution, \(1.6\)~GS/s sampling rate, and \(\pm 400\)~MHz bandwidth around the carrier frequency. The Bruker EleXSys E580 \textsc{esr} spectrometer has a \(2\)~ns time base and, in combination with the \textsc{awg}, resolves the time resolution mismatch by downsampling the pulse waveform onto a \(1\)~ns increment time grid.

The software used in this work was written in-house, and has a flow of communication between \textit{Spinach} \cite{HOGBEN11} and \textsc{Xepr} Python libraries, shown in \cref{FIG-SoftwareFlow}. The master process runs in \textit{Matlab} and calls \textsc{Xepr} Python functions as necessary to control the instrument. Experimental data is written by \textsc{Xepr} into plain text files and subsequently parsed by \textit{Matlab}. Optimisation restart capability is implemented using an \textsc{md5} hash table of the previously submitted experimental settings and outcomes \cite{GOODWIN16} -- an interrupted optimisation can therefore retrace its steps quickly without re-running previously executed experiments.

\begin{figure}
\centering{\includegraphics{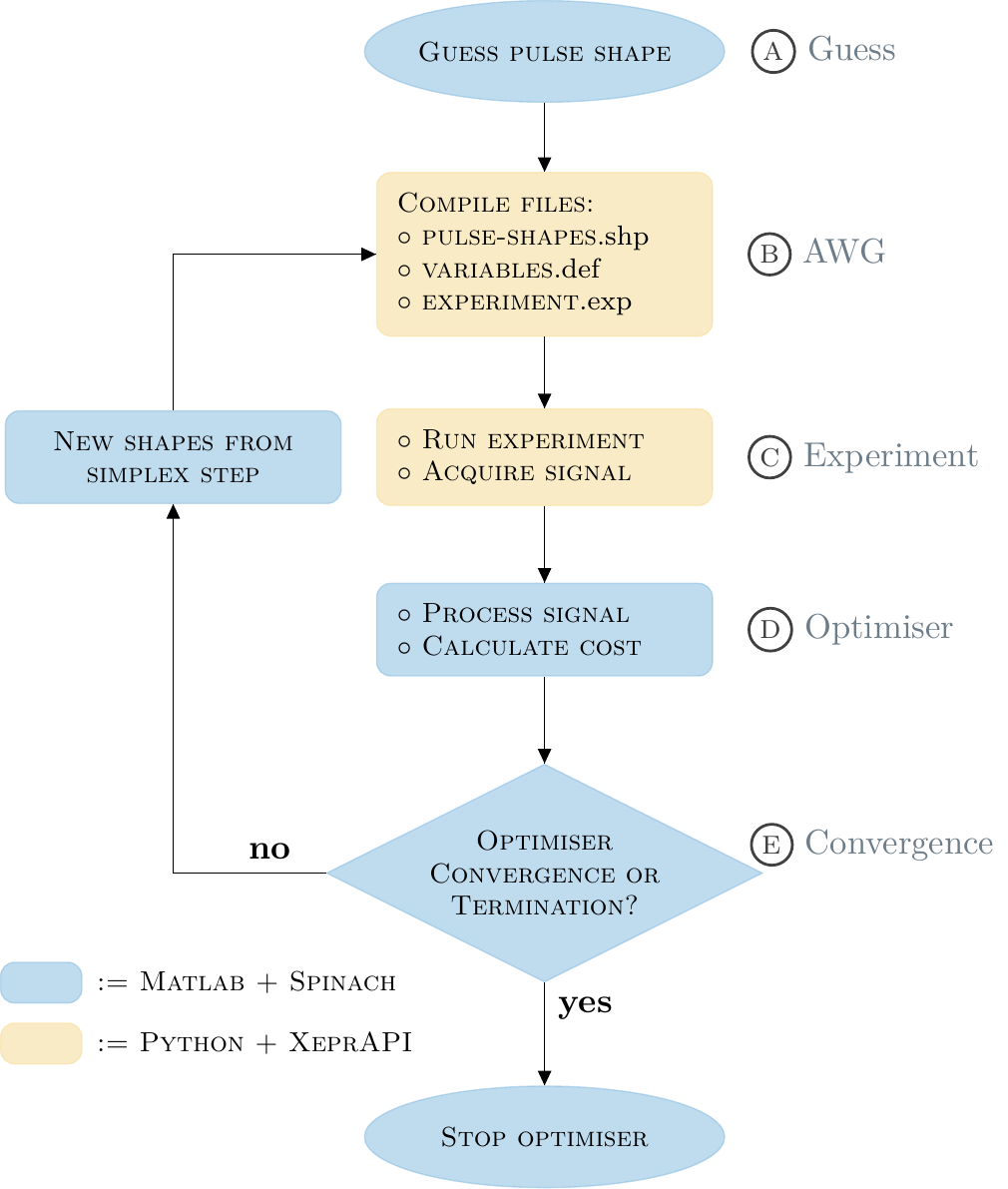}}
\caption[Software flow]{Software flow diagram. \plotlabel{a} Set initial pulses. A good guess of an ideal pulse set based on prior knowledge of the sample. \plotlabel{b} 1) Send this pulse-set to the \textsc{awg} using PulseSPEL: Load, show, and compile shape file. 2) Modify PulseSPEL variables and compile. 3) Validate and compile PulseSPEL program. \plotlabel{c} Run experiment and wait for experiment completion, then read signal data. \plotlabel{d} Interpret data in optimiser. \plotlabel{e} If the optimisation algorithm hasn't converged, calculate a new, modified set of pulses that are the sent back to the \textsc{awg}. Python code is detailed in \cref{Section:xeprScripts}.}\label{FIG-SoftwareFlow}
\end{figure}

The simplex optimiser allows both Nelder-Mead \cite{NELDER65,LAGARIAS98} and multidirectional search \cite{TORCZON89,DENNIS91} methods, with the option of evaluating independent sets of objective variables in parallel \cite{HIGHAM93}.

Many shapes can be stored within the \textsc{awg} in the form of a plain-text shape-file, limited to \(262,144\) bytes of physical disk size. A limit of \(115\) consecutive shape-file compile events must be enforced to avoid overload of the \textsc{awg}/console shared memory. Considering this last point, and that there may be many independent shapes involved in performing an optimisation iteration -- it will become advantageous to perform a 2D experiment with these pre-calculated, shapes.

Two primary sources of waveform distortions may be identified: the static ones, introduced by the instrument electronics (pulse shape interpolation, hardware response function, \textit{etc.}), and those associated with transient effects, \textit{e.g.} electrical noise and sample-specific magnetic susceptibility effects. The presence and the transient nature of these distortions makes common \textsc{esr} objective functions (signal integral, modulation depth, \textit{etc.}) impossible to differentiate numerically and necessitates the use of gradient-free noise-resilient optimisation methods.

\subsection{Optimisation Method}\label{Section:AWGOptimisationMethod}

\begin{figure}
\centering{\includegraphics{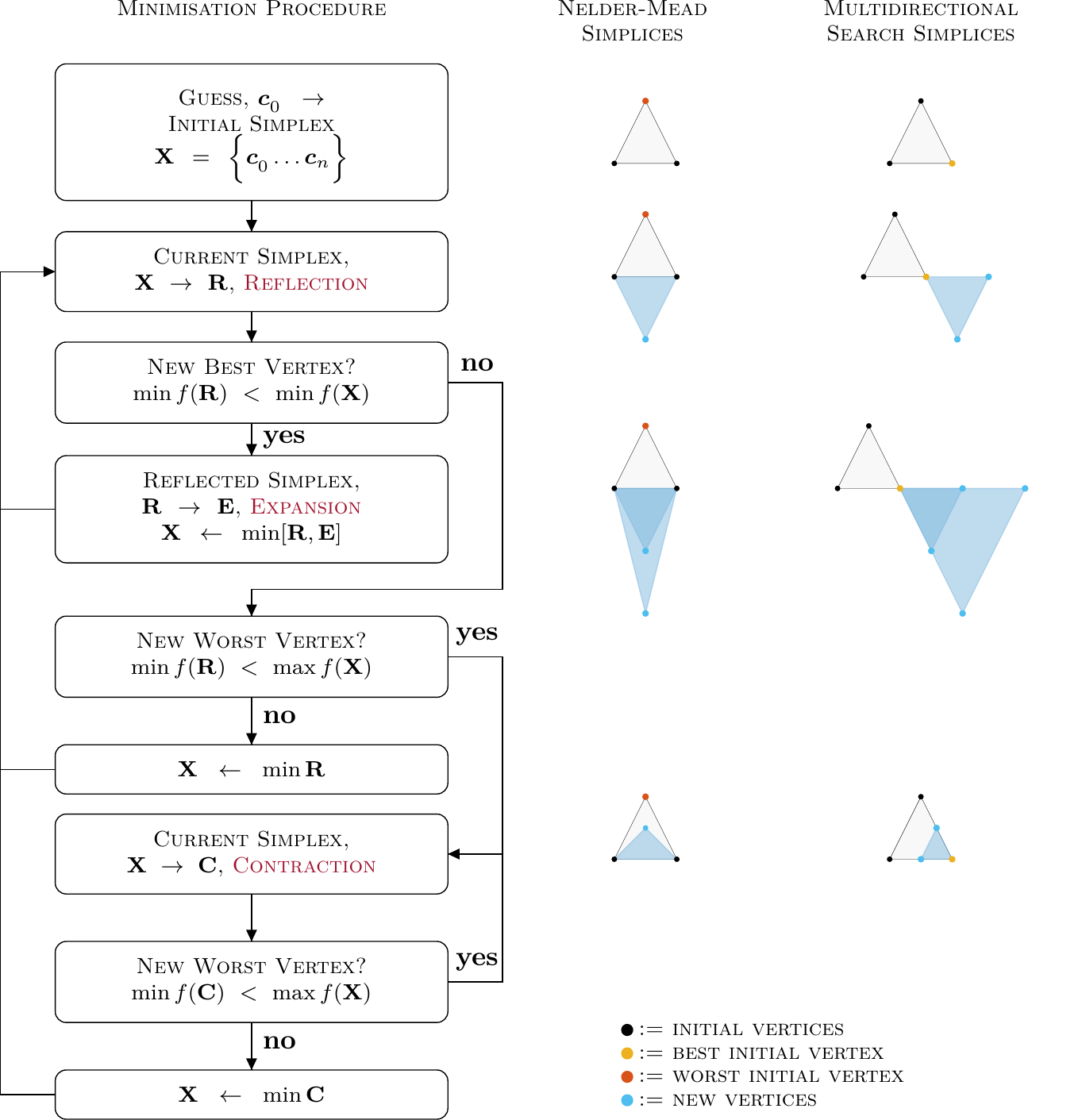}}
\caption[Simplex algorithms]{Different simplex steps for the Nelder-Mead and multidirectional search algorithms, each simplex has \(n+1\) vertices where \(n\) is the number of variables. Faces of the initial simplex are scaled to \(\big|\objv{0}\big|_{\infty}^{}\), and form an equilateral simplex. Expansion and contraction factors set to \(\beta=2\) and \(\gamma=0.5\) respectively. The best/worst vertex and the centroid \(\objv{c}=\sum_0^n \objv{j}\) of the simplex are used when reflecting, expanding, and contracting.\label{FIG-SimplexAlgorithms}}
\end{figure}

An optimisation problem usually attempts to find extrema of an objective function \cite{GILL81,FLETCHER87,NOCEDAL06} -- in financial terms these extrema could be the minimum of a cost function, or a maximum of a profit function. This function should map a set of variables to a scalar number designed as a convenient metric characterising the ``performance'' of a solution formed from those variables. 

The intensity of an echo makes a good echo experiment \cite{SCHWEIGER01,BRUSTOLON09} and the integral of the in-phase part of the echo signal makes a convenient metric to maximise. This is defined as the 1-norm of the signal plot vector \cite{DOYLE13}:
\begin{equation}
\| \vec{s}\, \|_1^{}:=\int\limits_{0}^{T}\big| s(t)\big| \mathrm{d}t
 \label{EQ-Signal_1-Norm}
\end{equation}
Any optimisation should be properly scaled \cite{FLETCHER87}, and scaling for these experiments simply involves normalising the objective metric with the signal produced by a reference experiment; in this case, an experiment with conventional hard pulses. This gives the metric of the quality of an echo as
\begin{equation}
 \Obj{\objv{}}=\frac{\| \vec{s}\, \|_1^{}}{\| \vec{r}\, \|_1^{}}
\end{equation}
where \(\Obj{\objv{}}\) is the objective measure to maximise, \(\vec{s}\) is the signal measured within an objective function and \(\vec{r}\) is the reference signal measured from an experiment with hard pulses. \(\Obj{\objv{}}>1\) is defined as a good echo, better than that from hard pulses alone, and \(0<\Obj{\objv{}}<1\) is defined as a bad echo, worse than that from hard pulses.

There is no obvious gradient with respect to the control parameters when considering an experimental measurement. Gradient-free optimisation strategies \cite{SWANN72,GILL81,FLETCHER87,NOCEDAL06,BRENT13} are the only choice for this real-time optimisation. A simple Nelder-Mead algorithm \cite{NELDER65,LAGARIAS98} (a member of a family of direct search methods termed simplex methods\cite{SWANN72}, polytope methods \cite{GILL81} and ad-hoc methods \cite{FLETCHER87}) has the benefit of a relatively small number of experimental evaluations for a single iteration compared to other popular algorithms e.g. genetic algorithms \cite{JUDSON92,BRIF10} or simulated annealing \cite{METROPOLIS53}. However, convergence is not guaranteed and is linear at best \cite{DENNIS87}. A desirable benefit of the Nelder-Mead algorithm is its tolerance to random noise (on a smooth function) \cite{GILMORE95,BORTZ98}. In addition; a \textit{restart} step has been shown to alleviate stagnation of simplex algorithms \cite{KELLEY99}.

The shape file characterising the optimisation variables has the constraint that it can only contain values in the range \(\big[ -1,+1\big]\) and these constraints can complicate the formulation considered in the optimisation problem. However, gradients are not considered in for the optimisation problem stated in this work, and a transform can easily classify the bounded pulse-shape variables \cite{FLETCHER87}: a repeating, mirrored transform can allow all real numbers to be evaluated by the objective function in a continuous manner. This waveform constraint transform can be classified as in \cref{eqn:changevar}:
\begin{equation}
 \tilde{\bm{c}}= \Bigg|\objv{}- 4\bigg\lfloor \frac{\objv{}-1}{4} \bigg\rfloor -3\Bigg|-1\nonumber
\end{equation}
where \(\objv{}\) is the unbounded waveform and \(\objv{}-n\lfloor \objv{}/n\rfloor\) is the remainder from dividing \(\objv{}\) by \(n\).

A modified Nelder-Mead simplex algorithm, the multidirectional search simplex method \cite{TORCZON89,DENNIS91,HIGHAM93}, has an efficient characteristic in that it can be used as a partially parallel simplex method, essentially creating a search grid of simplices. The added computational cost is that many more function evaluations must be made per simplex, depicted in \cref{FIG-SimplexAlgorithms}. This multidirectional search algorithm has a comparable performance to finite-difference gradient algorithms and is designed to be tolerant to problems with many local minima \cite{GILMORE95}. A further useful property of the multidirectional search algorithm is that is has guaranteed convergence \cite{TORCZON91}, similar to gradient following algorithms. The common algorithmic steps of the Nelder-Mead and multidirectional search algorithms are shown in \cref{FIG-SimplexAlgorithms}.

To improve stability of the optimisation process, the pulse shape variables are stored in a database using an \textsc{md5} hash function during the optimisation iterations  \cite{GOODWIN16} -- in the event of the optimisation algorithm halting mid-algorithm due to unexpected hardware incidents, the optimisation can effectively be resumed from the same stage after hardware problems have been resolved.

\section{Experimental Application}\label{Section:FBresults}

\subsection{Feedback Optimised Hahn Echo}\label{Section:AWGEcho}

\begin{figure}
\centering{\includegraphics{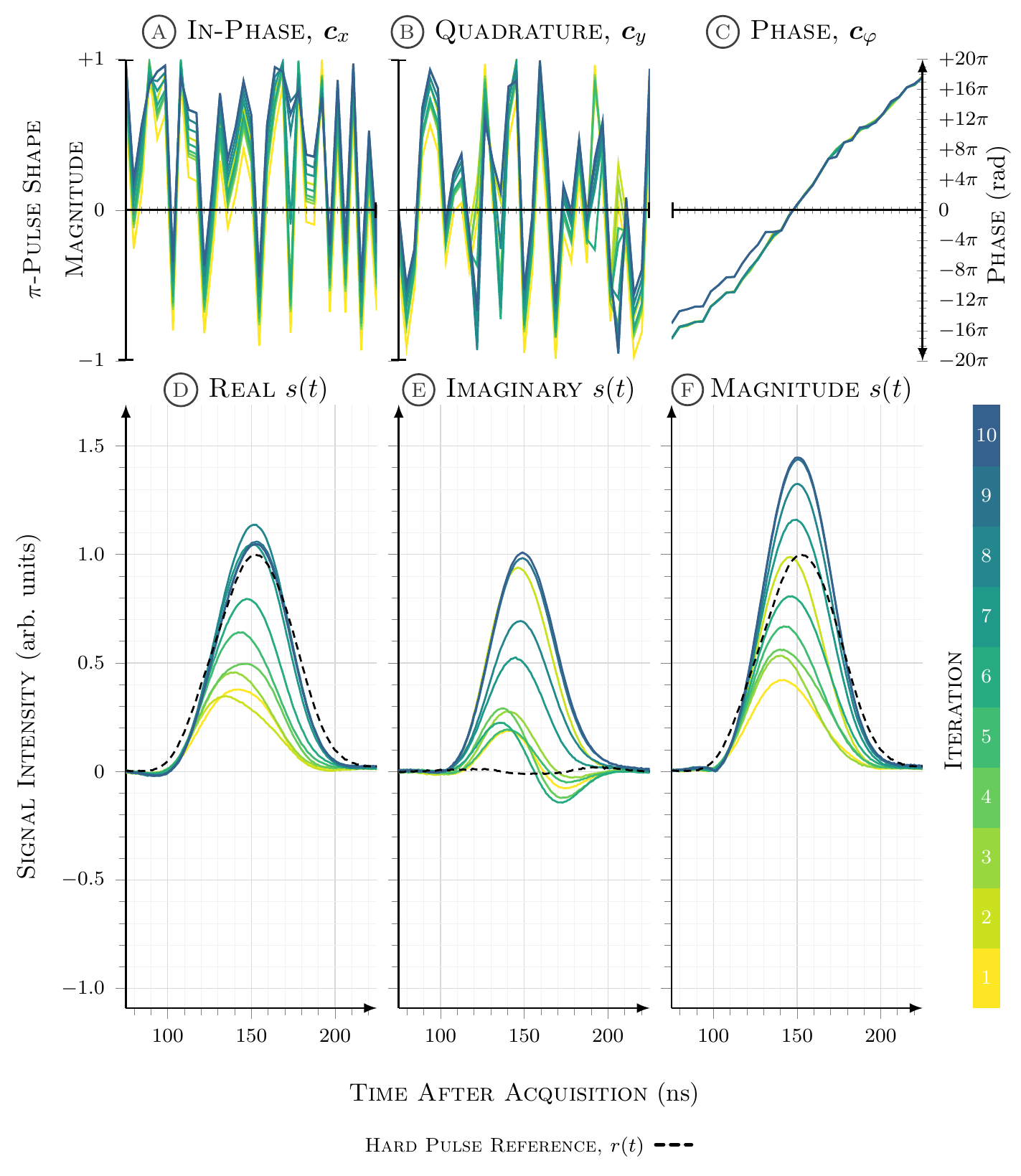}}
\caption[Echo results with 33 points]{\plotlabel{a} In-phase, \plotlabel{b} quadrature, and \plotlabel{c} phase parts of the pulse shape at each iteration. The spin echo experiment performed on a nitroxide radical and the optimisation variables are \(33\) discrete shape points of a \(32\)~ns \(\pi\)-pulse. \plotlabel{d} Real, \plotlabel{e} imaginary, and \plotlabel{f} magnitude of the measured signal at each iteration.\label{FIG-33ptEcho}}
\end{figure}

A two-pulse Hahn echo \cite{HAHN50,ROWAN65} is a good test for this closed-loop feedback control, using the 1-norm of the echo signal as a figure of merit, as in \cref{EQ-Signal_1-Norm}. An optimised \(32\)~ns \(\pi\)-pulse is the objective of the test, using discretised in-phase and quadrature components of a shaped waveform. In emulating the time resolution of the spectrometer, a shape with \(33\) discrete points should be used as the finest grid; one point at the start of every time point plus one point at the end of the last time point.

The optimisation starts at a random guess within the allowed bounds, and proceeds to search for shapes that produce better echoes. This first test of the feedback control methods uses the Nelder-Mead simplex algorithm, with termination conditions of a simplex size less than \(0.05\), or more than \(100\) consecutive contraction steps without finding a better solution (indicative of stagnation of the algorithm). The echoes produced at each of the optimisation iterations are shown in \cref{FIG-33ptEcho}.

The results of the feedback control optimisation are not encouraging for these pulse shapes \cref{AWGResults}. The most obvious failure is that the integral of the in-phase part of the echo does not improve on that from hard pulses. The optimisation is considered converged to a local maximum, evident in \cref{TAB-ExperimentalData1,TAB-ExperimentalData2}, the latter part of the optimisation makes many functional evaluations with very little gain. This is typical of a simplex optimisation which tries many contraction steps before an improved solution can be found (see \cref{FIG-SimplexAlgorithms}). There are two other characteristics of the optimisation:
\begin{enumerate}
 \item The quadrature part of the spin echo indicates an echo without the correct phase calibration.
 \item The pulse shape does not change by an appreciable amount during the optimisation.
\end{enumerate}
The first of these characteristics is not so much of a concern: a phase calibration can be performed after the optimisation and can only give a larger in-phase part of the echo integral than that shown in \cref{FIG-33ptEcho} \plotlabel{a}. The second characteristic may have two possible reasons: either the variables for the optimisation are not normalised for an effective optimisation, making too small steps at each iteration, or there are just too many variables for the optimiser to effectively handle. The Nelder-Mead algorithm is known to be effective only for low dimensional problems \cite{LAGARIAS98} -- there is no gradient to direct the algorithm and it can only be a sophisticated ad-hoc search \cite{FLETCHER87}. Directed by the limits of the Nelder-Mead algorithm, further tests of lower dimension are required.

Two further tests are performed, one with approximately \(\sfrac{2}{3}\) of the number of time points and one with approximately \(\sfrac{1}{3}\) of the time points. In both cases the total pulse duration remains the same. The first of these is a \(21\) point shape and the results at each iteration of the Nelder-Mead algorithm are shown in \cref{FIG-21ptEcho}.

\begin{figure}
\centering{\includegraphics{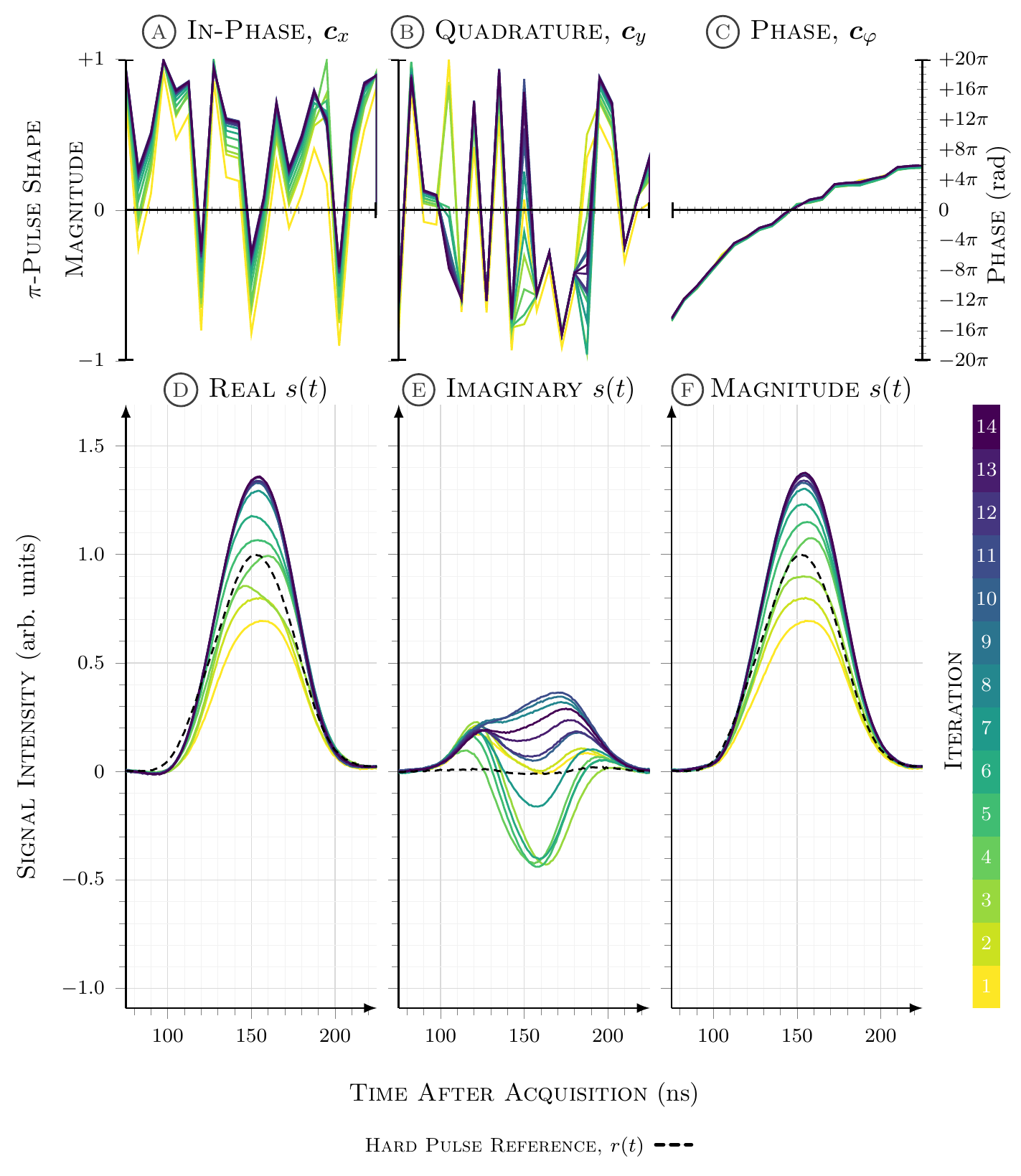}}
\caption[Echo results with 21 points]{\plotlabel{a} In-phase, \plotlabel{b} quadrature, and \plotlabel{c} phase parts of the pulse shape at each iteration. The spin echo experiment performed on a nitroxide radical and the optimisation variables are \(21\) discrete shape points of a \(32\)~ns \(\pi\)~pulse. \plotlabel{d} Real, \plotlabel{e} imaginary, and \plotlabel{f} magnitude of the measured signal at each iteration.\label{FIG-21ptEcho}}
\end{figure}

Results are more encouraging for a pulse shape using \(21\) discrete points \cref{AWGResults} -- achieving an echo integral 27\% better than that of hard pulses (see \cref{TAB-ExperimentalData1}). However, this optimisation makes many function evaluations giving a total optimisation time of more than \(20\)~minutes (much longer than an experienced spectroscopist would take to find a good echo signal). The majority of the function evaluations are from inner contraction steps of the simplex algorithm -- making smaller and smaller simplices with little gain. In fact, this optimisation terminates because the relative simplex size was less than the defined numerical tolerance. This optimisation can be considered stagnated within the noise level of the signal.

\begin{figure}
\centering{\includegraphics{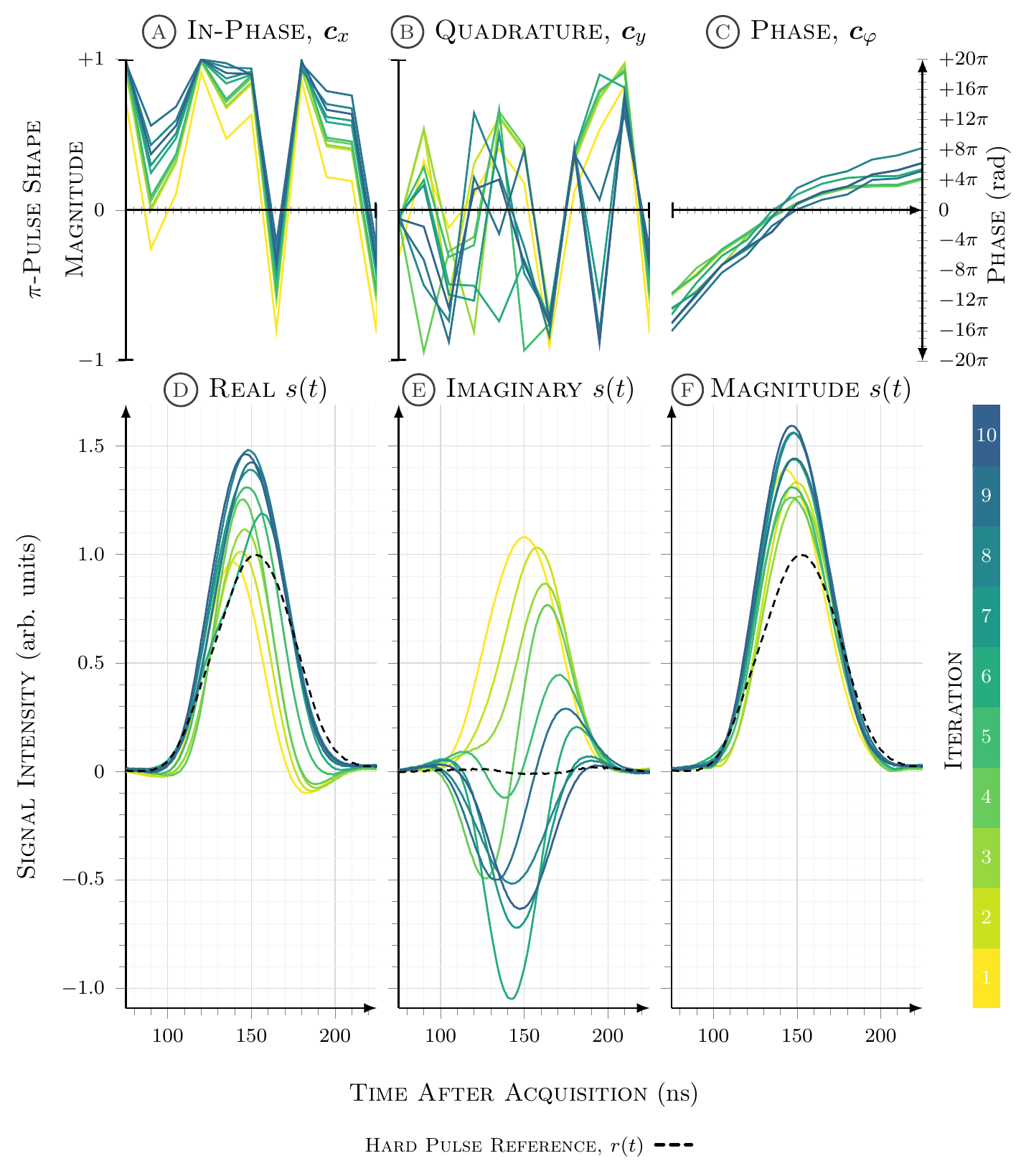}}
\caption[Echo results with 11 points]{\plotlabel{a} In-phase, \plotlabel{b} quadrature, and \plotlabel{c} phase parts of the pulse shape at each iteration. The spin echo experiment performed on a nitroxide radical and the optimisation variables are \(11\) discrete shape points of a \(32\)~ns \(\pi\)~pulse. \plotlabel{d} Real, \plotlabel{e} imaginary, and \plotlabel{f} magnitude of the measured signal at each iteration.\label{FIG-11ptEcho}}
\end{figure}

The final feedback controlled spin echo experiment shows a further insight into a hidden sophistication of the optimisation algorithm, when used within its limits. Again reducing the amount of dimensions of the optimisation problem, to approximately $1/3$ of the time points, a shape with \(11\) discrete points is used as the variables to the Nelder-Mead algorithm.

These results, shown in \cref{FIG-11ptEcho} and \cref{TAB-ExperimentalData1}, produce a good echo integral in a reasonable amount of time (\(6\)-\(7\)~minutes). They are considered converged for the same reason as the \(22\) point pulse shape -- indicating that this number a variables is low enough for the Nelder-Mead algorithm to function effectively.

A final note, which leads indirectly to the next section of results, is that the variation of the quadrature part of the echo in \cref{FIG-11ptEcho} covers all phase from positive to negative offset, finishing at a zero phase offset. Investigating further -- it is evident from the large variation in the quadrature channel of the pulse shape that the feedback control optimisation is trying to compensate for the ineffective initial phase calibration (calibrated to that for hard pulses without a quadrature element). This is also evident to a lesser degree from the \(22\) point pulse shape in \cref{FIG-21ptEcho}.

In effect, these feedback control optimisations are giving better echo signals by being able to utilise the quadrature pulse channel at the same time as the in-phase channel, and the only real optimisation is to find the correct signal phase offset to give the largest echo integral. An important conclusion is that the development of the waveform shaping equipment in \textsc{esr} spectroscopy must at this point focus on improving the instrument response function -- there is little to gain by seeking faster arbitrary waveform generators until the width of the convolution kernel is reduced. The convolution of the \(11\) point optimum pulse from \cref{FIG-11ptEcho} with the experimentally measured response function for our EleXSys II E580 spectrometer is shown in \cref{FIG-TransShapes}.

\begin{figure}
\centering{\includegraphics{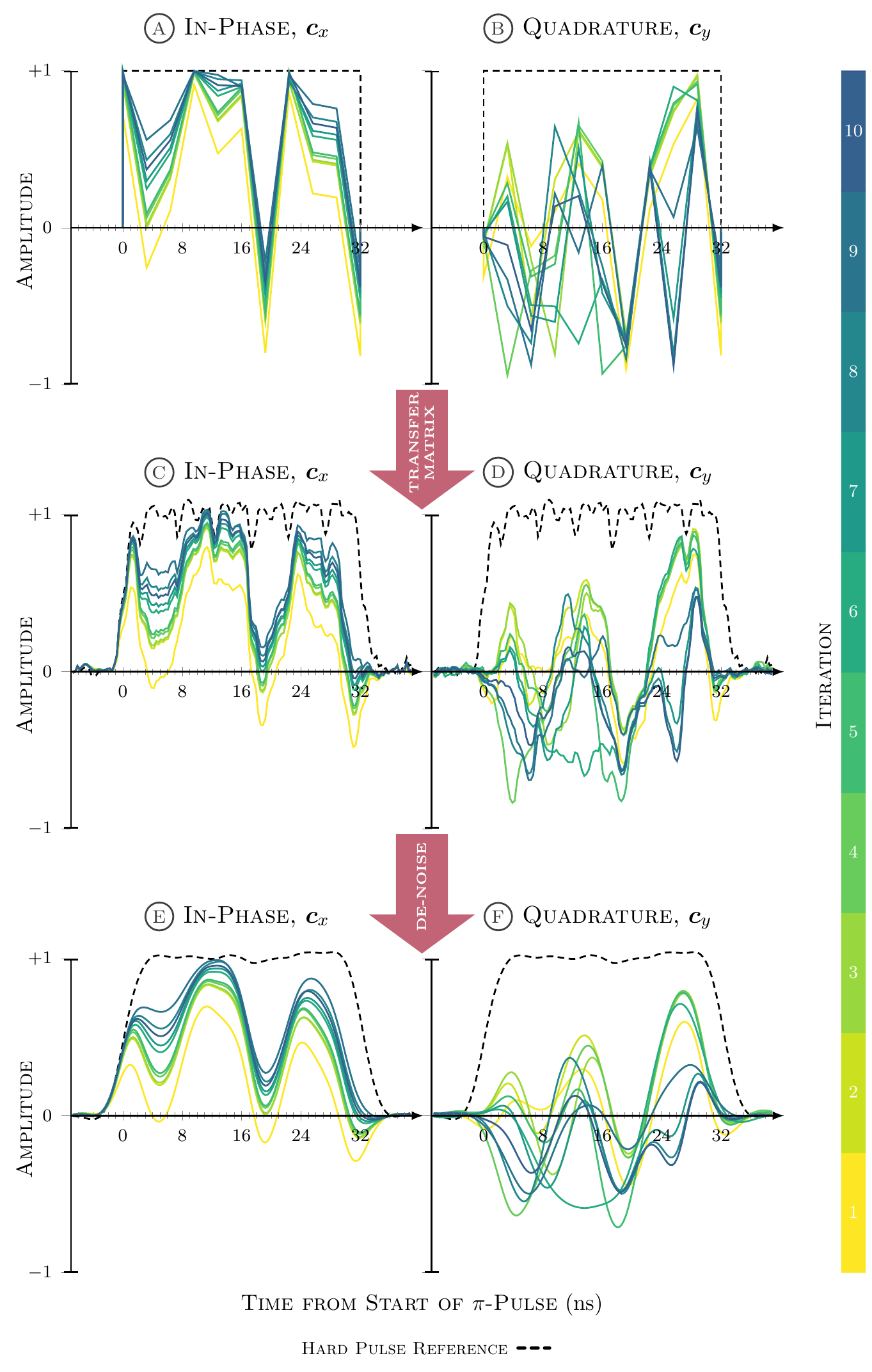}}
\caption[Transformed 11 point shapes]{In-phase and quadrature parts of pulse-shapes from \cref{FIG-11ptEcho} \plotlabel{a}-\plotlabel{b}, transformed using a \(2\)~ns resolution transfer matrix in \plotlabel{c}-\plotlabel{d}, followed by de-noising with a regularised least squares filter in \plotlabel{e}-\plotlabel{f}.\label{FIG-TransShapes}}
\end{figure}

\subsection{Feedback Optimised OOP-ESEEM}\label{Section:AWGOOPESEEM}

\begin{figure}
\centering{\includegraphics{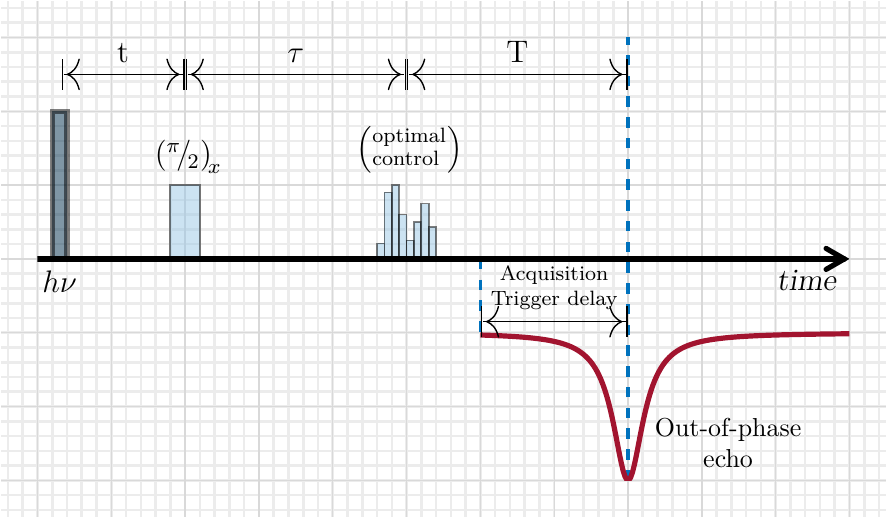}}
\caption[OOP-ESEEM pulse sequence diagram]{Out-of-phase ESEEM pulse sequence diagram, showing feedback control pulse (adapted from \cite{BRUSTOLON09})}\label{OOP-ESEEM_pulse_seq}
\end{figure}

\begin{figure}
\centering{\includegraphics{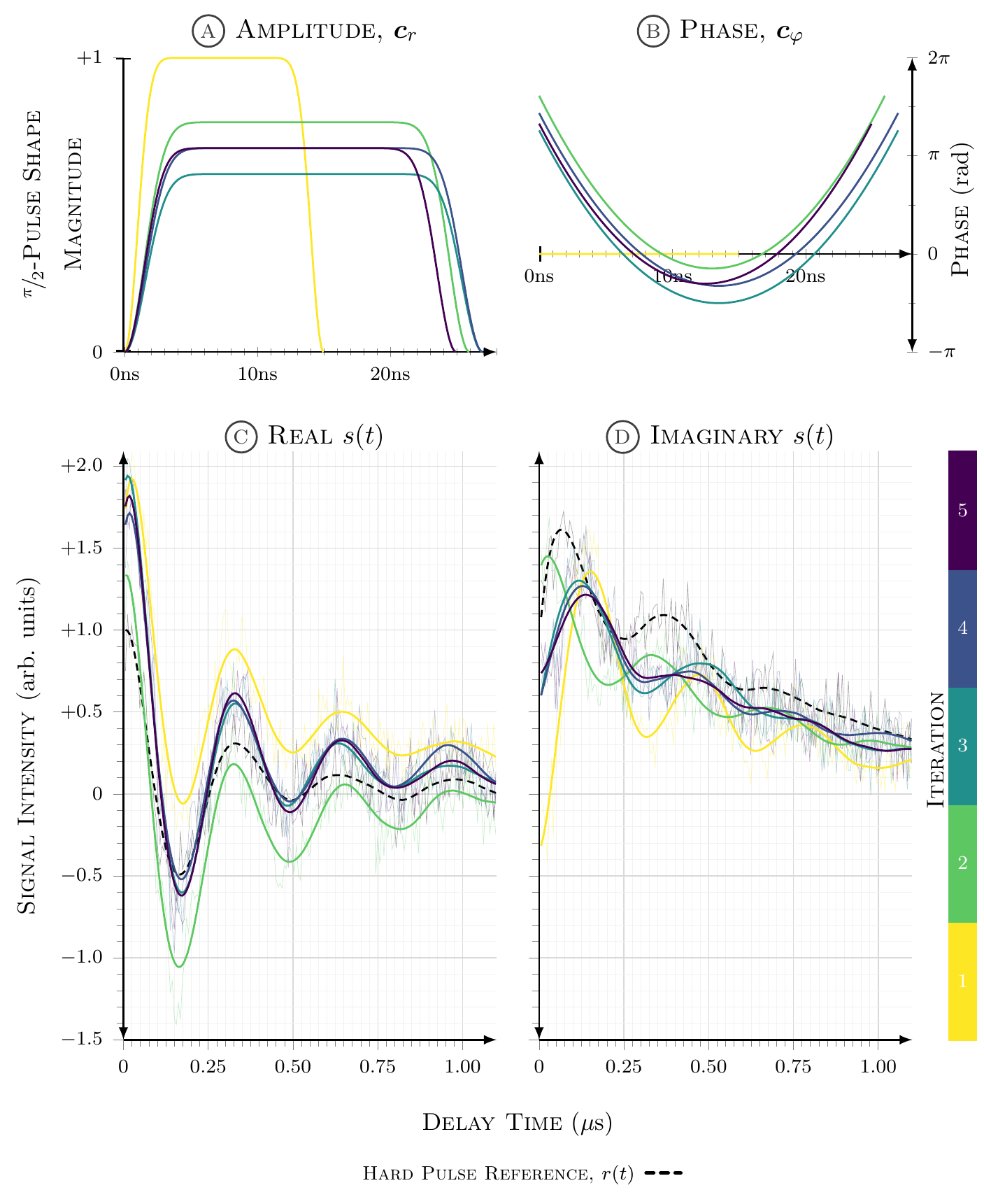}}
\caption[OOP-ESEEM from feedback-control]{\plotlabel{a} Amplitude, and \plotlabel{b} phase parts of the \textsc{wurst} shape at each iteration. \plotlabel{c} Real, and \plotlabel{d} imaginary parts of the measured \textsc{oop-eseem} signal at each iteration, normalised to the peak intensity of the real part of the reference signal from hard pulses. Opaque lines show the raw data and bold lines show data smoothed with a multi-pass Savitzky-Golay filter \cite{SAVITZY64} (\(1^{\text{st}}\) pass: \(25\)-point cubic; \(2^{\text{nd}}\) pass: \(75\)-point quintic). \label{FIG-OopeseemMD}}
\end{figure}

As an application of optimised spin echoes the \textsc{oop-eseem} experiment, shown in \cref{OOP-ESEEM_pulse_seq}, would need to optimise two echoes, one being a maximum echo at a delay time in the expected maximum modulation, and the other a minimum at the first minimum of the modulated signal. The objective metric to optimise would reflect that of a full \textsc{oop-eseem} experiment, but only at these two delays:
\begin{equation}
 \Obj{\objv{}}=\frac{\| \vec{s}_{d_1}\\|_1^{(g)}}{\| \vec{r}_{d_1} \|_1^{(g)}} - \frac{\| \vec{s}_{d_2} \|_1^{(g)}}{\| \vec{r}_{d_2} \|_1^{(g)}}
\end{equation}
where \(\vec{s}_{d_1}\) is the echo with a delay \(d_1\), corresponding to the first modulation maximum, \(\vec{s}_{d_2}\) is the echo with a delay \(d_2\), corresponding to the first modulation minimum, and similarly for the hard pulse reference \(\vec{r}\). The first maximum is placed at \(16\)~ns and the second at \(144\)~ns (approximate values are found from an \textsc{oop-eseem} experiment with hard pulses, shown in \cref{FIG-OopeseemMD}). The superscript \((g)\) on the 1-norm indicates that the integral is only performed over the integrator gate width (as would be done for a full \textsc{oop-eseem} experiment, set to \(32\)~ns for these experiments). The time stepping of the \textsc{oop-eseem} experiments is set to a coarse \(6\)~ns, and only \(22\) shots-per-point/transient averages are used. In these feedback controlled optimisations, the initial \(\sfrac{\pi}{2}\)-pulse is optimised and the \(\pi\)-pulse is left as a hard pulse. The reason for investigating the initial pulse character is that it has previously been investigated \cite{TIMMEL98} and found to be important for \textsc{oop-eseem} experiments.

Considering the observation that results from the spin echo optimisation are finding an appropriate signal phase within their quadrature pulse channel, and that the in-phase pulse channel does not change appreciably from its initial shape; this should be defined explicitly as part of the optimisation variables. 

Chirped pulses \cite{DOLL13,DOLL14,KHANEJA17} and \textsc{wurst} pulse \cite{SPINDLER12,SPINDLER13} have an easily controllable phase and will be used as the basis of pulse shapes in this section of work. A \textsc{wurst}-type chirped pulse can be characterised by its amplitude and phase as
\begin{gather}
\objv{r}= 2\pi\sqrt{\frac{\omega_\text{bw}}{T}}\left(1-\left|\sin^a{\left(\frac{\pi t}{T}\right)}\right|\right)\nonumber\\
\objv{\varphi}= \pi\frac{\omega_\text{bw}}{T} t^2 \label{EQN:wurst}
\end{gather}
where \(T\) is the total pulse duration, \(\omega_\text{bw}\) is the pulse excitation bandwidth, \(a\) is the sine power used for the amplitude envelope (set to 25\% in these experiments), and \(t\) is a time grid from \(-\sfrac{T}{2}\) to \(+\sfrac{T}{2}\). The variables of these shapes used for the objective function are the pulse excitation bandwidth, \(\omega_\text{bw}\), the total pulse duration, \(T\), the phase offset of the pulse shape, \({\objv{\varphi}}_0\), and the nominal pulse amplitude of the waveform shape, \({\objv{r}}_{\max}\) (defined as a percentage of the maximum pulse amplitude in the spectrometer configuration). These shapes are characterised by few variables, which should enable simplex methods to perform effectively after appropriate scaling.

A first test of this optimisation using the Nelder-Mead algorithm proved unstable -- the discontinuity of the objective variables forced the optimiser into local maxima, either contracting onto the pulse shape, the pulse length or the phase offset. This should be expected with discontinuous variables in the Nelder-Mead algorithm \cite{FLETCHER87}.

Now that the dimensionality of the problem has reduced substantially by using chirped pulses, the multidirectional search algorithm can now become less expensive. Furthermore, the pattern-search type nature of the algorithm allows it to deal effectively with the discontinuous variables of this optimisation. The results of using this optimisation method are shown in \cref{FIG-OopeseemMD}.

The use of a phase offset with chirped-\textsc{wurst} pulses as variables of a multidirectional search optimisation can be considered quite successful in increasing the modulation depth of the \textsc{oop-eseem} experiment compared to a similar experiment with hard pulses alone. The number of iterations is very low and utilising a partially parallel version of the simplex method reduces the total time taken to less than \(2\)~minutes \cref{AWGResults}.

The optimisation can be considered converged because the simplex size reduces to less than the defined tolerance, which is a feature of the multidirectional search convergence \cite{TORCZON91}. These solutions should not be considered unique: the optimisation was started from a chirped pulse with a very small bandwidth, almost a hard pulse. When the optimisation is started from other random starts, it will find other solutions that give a good modulation depth -- these optimisation methods are not global searches.

It should be noted that the results shown in \cref{FIG-OopeseemMD} are explicit \textsc{oop-eseem} experiments performed at each improved solution. This is for display in this thesis rather than a need of the optimisation -- this only needs two of the $256$ echo experiments needed for an \textsc{oop-eseem} experiment. This extra time to perform the \textsc{oop-eseem} experiment is not included in the timing of \cref{TAB-ExperimentalData2}. Furthermore, only $22$ transient averages were used in the echo signal -  because this method takes a very short amount of time, a much larger number of transient averages could be used, removing much of the noise in the raw data of \cref{FIG-OopeseemMD} at very little cost compared to increasing the shots-per-point by a similar amount in the \textsc{oop-eseem} experiment.

Using a simple feedback control with a gradient-free optimisation algorithm of a shaped pulse can improve the signal intensity of a spin echo \textsc{esr} experiment. The simplicity, almost na\"{i}vety, of this feedback controlled optimisation is that experiment and sample specific pulse distortions need not be classified, which could be time-consuming and require rare expertise. Furthermore, this blind method of finding broadband pulses does not need to be time-consuming and can become a useful tool, similar to the convenience of an auto-shim.

This work may be considered an extra tool in the \textsc{esr} optimisation toolbox. Code to produce examples within this manuscript are a combination of Matlab functions of the \textit{Spinach} libraries \cite{HOGBEN11} to parse data, optimise an objective function, and invoke Python scripts to communicate with the \textsc{awg} using the \textsc{Xepr}API libraries. The function of Matlab in this context is a triviality, and code is not shown here. However, using Python scripts to communicate with the \textsc{awg} in ways set out in this manuscript is not well documented. These communications scripts are bundled with \textit{Spinach}. A user may implement these methods with relative ease and very little specialist knowledge of the hardware electronics or optimisation methods. Furthermore, this feedback control can be used in conjunction with transfer function methods as a fine-tuning of the numerical method \cite{HELLGREN13,EGGER14}.

\chapter{Summary}\label{Chapter:Conclusions}

\begin{chapquote}{Miguel de Cervantes Saavedra, \textsc{Don Quixote}}
Finally, from so little sleeping and so much reading, his brain dried up and he went completely out of his mind.
\end{chapquote}

A Newton-\textsc{grape} method was developed using efficient calculation of analytical second directional derivatives. The method was developed to scale with the same complexity as methods that use only first directional derivatives. Algorithms to ensure a well-conditioned and positive definite matrix of second directional derivatives are used so the sufficient conditions of gradient-based numerical optimisation are met.

\begin{appendices}
\renewcommand\appendixname{}
\chapter[Irreducible Spherical Tensor Operators]{Irreducible Spherical\linebreak Tensor Operators} \label{Chapter:AppendixA}
\renewcommand*{\CurrentPath}{./Appendix_A}

\begin{table}[ht]
\begin{center}
\bgroup
\def\arraystretch{2.5}
\small{
\begin{tabular}{c|c c c c c }
\hline\hline
  & \(\!d_{m,2}^{(2)}\!\) & \(d_{m,1}^{(2)}\) & \(d_{m,0}^{(2)}\) & \(d_{m,-1}^{(2)}\) & \(d_{m,-2}^{(2)}\) \\\hline
$\!d_{2,n}^{(2)}\!\) & \(\frac{(1+\cos{\beta})^2}{4}\) & \(-\frac{(1+\cos{\beta})\sin{\beta}}{2}\) & \(\sqrt{\frac{3}{8}}\sin^2{\beta}\) & \(-\frac{(1-\cos{\beta})\sin{\beta}}{2}\) & \(\frac{(1-\cos{\beta})^2}{4}\) \\
$\!d_{1,n}^{(2)}\!\) & \(\frac{(1+\cos{\beta})\sin{\beta}}{2}\) & \(\frac{\cos{\beta}-1}{2}+\cos^2{\beta}\) & \(-\sqrt{\frac{3}{8}}\sin{2\beta}\) & \(\frac{\cos{\beta}+1}{2}-\cos^2{\beta}\) & \(-\frac{(1-\cos{\beta})\sin{\beta}}{2}\) \\
$\!d_{0,n}^{(2)}\!\) & \(\sqrt{\frac{3}{8}}\sin^2{\beta}\) & \(\sqrt{\frac{3}{8}}\sin{2\beta}\) & \(\frac{3\cos^2{\beta}-1}{2}\) & \(-\sqrt{\frac{3}{8}}\sin{2\beta}\) & \(\sqrt{\frac{3}{8}}\sin^2{\beta}\) \\
$\!d_{-1,n}^{(2)}\!\) & \(\frac{(1-\cos{\beta})\sin{\beta}}{2}\) & \(\frac{\cos{\beta}+1}{2}-\cos^2{\beta}\) & \(\sqrt{\frac{3}{8}}\sin{2\beta}\) & \(\frac{\cos{\beta}-1}{2}+\cos^2{\beta}\) & \(-\frac{(1+\cos{\beta})\sin{\beta}}{2}\) \\
$\!d_{-2,n}^{(2)}\!\) & \(\frac{(1-\cos{\beta})^2}{4}\) & \(\frac{(1-\cos{\beta})\sin{\beta}}{2}\) & \(\sqrt{\frac{3}{8}}\sin^2{\beta}\) & \(\frac{(1+\cos{\beta})\sin{\beta}}{2}\) & \(\frac{(1+\cos{\beta})^2}{4}\) \\\hline\hline
\end{tabular} 
}\caption[Second-rank Wigner functions defined in terms of reduced functions]{Second-rank Wigner functions are defined in terms of reduced functions as \(\mathfrak{D}_{m,n}^{(2)}(\alpha,\beta,\gamma)=\e^{-im\alpha}d_{m,n}^{(2)}(\beta)\e^{-in\gamma}\)}\label{TAB:Wigner}
\egroup
\end{center}
\end{table}

\begin{table}[ht]
\begin{center}
\bgroup
\def\arraystretch{2.5}
\small{
\begin{tabular}{ c c|c c c }
\hline\hline
\multicolumn{1}{c}{$\ell$} & \multicolumn{1}{c|}{$m$} & \multicolumn{1}{c}{\(Y_{\!\ell}^{m_{}^{\vphantom{\prime}}}\!(\theta,\varphi)\)} & \multicolumn{1}{c}{\(Y_{\!\ell}^{m_{}^{\vphantom{\prime}}}\!(x,y,z)\)} & \multicolumn{1}{c}{\(\hat{T}_{\ell m}(\hat{L})\)} \\\hline
0 & 0 & \(\sqrt{\frac{1}{4\pi}}\)                                         & \(\sqrt{\frac{1}{4\pi}}\)                          & \(\Unit\)\\
1 & 1 & \(-\sqrt{\frac{3}{8\pi}}\e^{i\varphi}\sin{\theta}\)               & \(-\sqrt{\frac{3}{8\pi}}\frac{x+iy}{r}\)           & \(-\sqrt{\frac{1}{2}}\hat{L}_+^{}\) \\
1 & 0 & \(\sqrt{\frac{3}{4\pi}}\cos{\theta}\)                             & \(\sqrt{\frac{3}{4\pi}}\frac{z}{r}\)               & \(\hat{L}_z^{}\) \\
1 & -1 & \(\sqrt{\frac{3}{8\pi}}\e^{-i\varphi}\sin{\theta}\)              & \(\sqrt{\frac{3}{8\pi}}\frac{x-iy}{r}\)            & \(\sqrt{\frac{1}{2}}\hat{L}_-^{}\) \\
2 & 2 & \(\sqrt{\frac{15}{32\pi}}\e^{2i\varphi}\sin^2{\theta}\)           & \(\sqrt{\frac{15}{32\pi}}\frac{(x+iy)^2}{r^2}\)    & \(\frac{1}{2}\hat{L}_+^2\) \\
2 & 1 & \(-\sqrt{\frac{15}{8\pi}}\e^{i\varphi}\sin{\theta}\cos{\theta}\)  & \(-\sqrt{\frac{15}{8\pi}}\frac{(x+iy)z}{r^2}\)     & \(-\frac{1}{2}\big(\hat{L}_z^{}\hat{L}_+^{}+\hat{L}_+^{}\hat{L}_z^{}\big)\) \\
2 & 0 & \(\sqrt{\frac{5}{16\pi}}(3\cos^2{\theta}-1)\)                     & \(\sqrt{\frac{5}{16\pi}}\frac{2z^2-x^2-y^2}{r^2}\) & \(\sqrt{\frac{2}{3}}\!\Big(\hat{L}_z^2-\frac{1}{4}\big(\hat{L}_+^{}\hat{L}_-^{}+\hat{L}_-^{}\hat{L}_+^{} \big)\!\Big)\) \\
2 & -1 & \(\sqrt{\frac{15}{8\pi}}\e^{-i\varphi}\sin{\theta}\cos{\theta}\) & \(\sqrt{\frac{15}{8\pi}}\frac{(x-iy)z}{r^2}\)      & \(\frac{1}{2}\big(\hat{L}_z^{}\hat{L}_-^{}+\hat{L}_-^{}\hat{L}_z^{}\big)\) \\
2 & -2 & \(\sqrt{\frac{15}{32\pi}}\e^{-2i\varphi}\sin^2{\theta}\)         & \(\sqrt{\frac{15}{32\pi}}\frac{(x-iy)^2}{r^2}\)    & \(\frac{1}{2}\hat{L}_-^2\) \\\hline\hline
\end{tabular} 
}\caption[Rank-1 irreducible spherical tensor operators]{Spherical harmonics may be written in terms of Cartesian coordinates, then to be replaced by corresponding Cartesian operators, to give rank-1 irreducible spherical tensor operators, for a single spin.}\label{TAB:ISTO}
\egroup
\end{center}
\end{table}

\begin{table}[ht]
\begin{center}
\bgroup
\def\arraystretch{2.5}
\small{
\begin{tabular}{c c|c c}
\hline\hline
\multicolumn{1}{c}{\(\ell\)} &\multicolumn{1}{c|}{\(m\)} & \multicolumn{1}{c}{\(a_m^{(\ell)}\)} & \multicolumn{1}{c}{\(\hat{T}_{\ell m}(\vhat{L},\vhat{S})\)} \\\hline & & \\[-6ex]
0 & 0 & \(-\frac{1}{\sqrt{3}}\big( a_{xx}^{}+a_{yy}^{}+a_{zz}^{} \big)\) & \(-\frac{1}{\sqrt{3}}\vhat{L}\ccdot\bgroup
\def\arraystretch{1.0}\begin{pmatrix}1&0&0\\0&1&0\\0&0&1\end{pmatrix}\egroup\ccdot\vhat{S}\) \\[3ex]
1 & 1 &  \(-\frac{1}{2}\big( a_{zx}^{}-a_{xz}^{}-i( a_{zy}^{}-a_{yz}^{}\, ) \!\big)\) & \(-\frac{1}{2}\vhat{L}\ccdot\bgroup
\def\arraystretch{1.0}\begin{pmatrix}0&0&-1\\0&0&-i\\1&i&0\end{pmatrix}\egroup\ccdot\vhat{S}\) \\[3ex]
1 & 0 &  \(-\frac{i}{\sqrt{2}}\big( a_{xy}^{}-a_{yx}^{} \big)\) & \(-\frac{1}{\sqrt{2}}\vhat{L}\ccdot\bgroup
\def\arraystretch{1.0}\begin{pmatrix}0&-i&0\\i&0&0\\0&0&0\end{pmatrix}\egroup\ccdot\vhat{S}\) \\[3ex]
1 & -1 &  \(-\frac{1}{2}\big( a_{zx}^{}-a_{xz}^{}+i( a_{zy}^{}-a_{yz}^{}\, ) \!\big)\) & \(-\frac{1}{2}\vhat{L}\ccdot\bgroup
\def\arraystretch{1.0}\begin{pmatrix}0&0&-1\\0&0&i\\1&-i&0\end{pmatrix}\egroup\ccdot\vhat{S}\) \\[3ex]
2 & 2 &  \(+\frac{1}{2}\big( a_{xx}^{}-a_{yy}^{}-i( a_{xy}^{}+a_{yx}^{}\, ) \!\big)\) & \(+\frac{1}{2}\vhat{L}\ccdot\bgroup
\def\arraystretch{1.0}\begin{pmatrix}1&i&0\\i&-1&0\\0&0&0\end{pmatrix}\egroup\ccdot\vhat{S}\) \\[3ex]
2 & 1 & \(-\frac{1}{2}\big( a_{xz}^{}+a_{zx}^{}-i( a_{yz}^{}+a_{zy}^{}\, ) \!\big)\) & \(-\frac{1}{2}\vhat{L}\ccdot\bgroup
\def\arraystretch{1.0}\begin{pmatrix}0&0&1\\0&0&i\\1&i&0\end{pmatrix}\egroup\ccdot\vhat{S}\) \\[3ex]
2 & 0 &  \(+\frac{i}{\sqrt{6}}\big( 2a_{zz}^{}-( a_{xx}^{}+a_{yy}^{}\, ) \!\big)\) & \(+\frac{1}{\sqrt{6}}\vhat{L}\ccdot\bgroup
\def\arraystretch{1.0}\begin{pmatrix}-1&0&0\\0&-1&0\\0&0&2\end{pmatrix}\egroup\ccdot\vhat{S}\) \\[3ex]
2 & -1 &  \(+\frac{1}{2}\big( a_{xz}^{}+a_{zx}^{}+i( a_{yz}^{}+a_{zy}^{}\, ) \!\big)\) & \(+\frac{1}{2}\vhat{L}\ccdot\bgroup
\def\arraystretch{1.0}\begin{pmatrix}0&0&1\\0&0&-i\\1&-i&0\end{pmatrix}\egroup\ccdot\vhat{S}\) \\[3ex]
2 & -2 & \(+\frac{1}{2}\big( a_{xx}^{}-a_{yy}^{}+i( a_{xy}^{}+a_{yx}^{}\, ) \!\big)\) & \(+\frac{1}{2}\vhat{L}\ccdot\bgroup
\def\arraystretch{1.0}\begin{pmatrix}1&-i&0\\-i&-1&0\\0&0&0\end{pmatrix}\egroup\ccdot\vhat{S}\) \\[1.5ex]\hline\hline
\end{tabular} 
}\caption[Rank-2 irreducible spherical tensor operators]{The coefficients required for the transformation between the standard \(3\times 3\) matrix representation and rank-2 irreducible spherical tensors operators, for 2 spins.}\label{TAB:SPericalTensor}
\egroup
\end{center}
\end{table}

\chapter[Arbitrary Waveform Generator]{The Arbitrary Waveform\linebreak Generator in Optimisation} \label{Chapter:AppendixXepr}
\renewcommand*{\CurrentPath}{./Appendix_Xepr}
\renewcommand{\figwidth}{0.775\columnwidth}

\section{AWG Schematics}\label{Section:AWGScheme}

\begin{figure}
\centering{\includegraphics{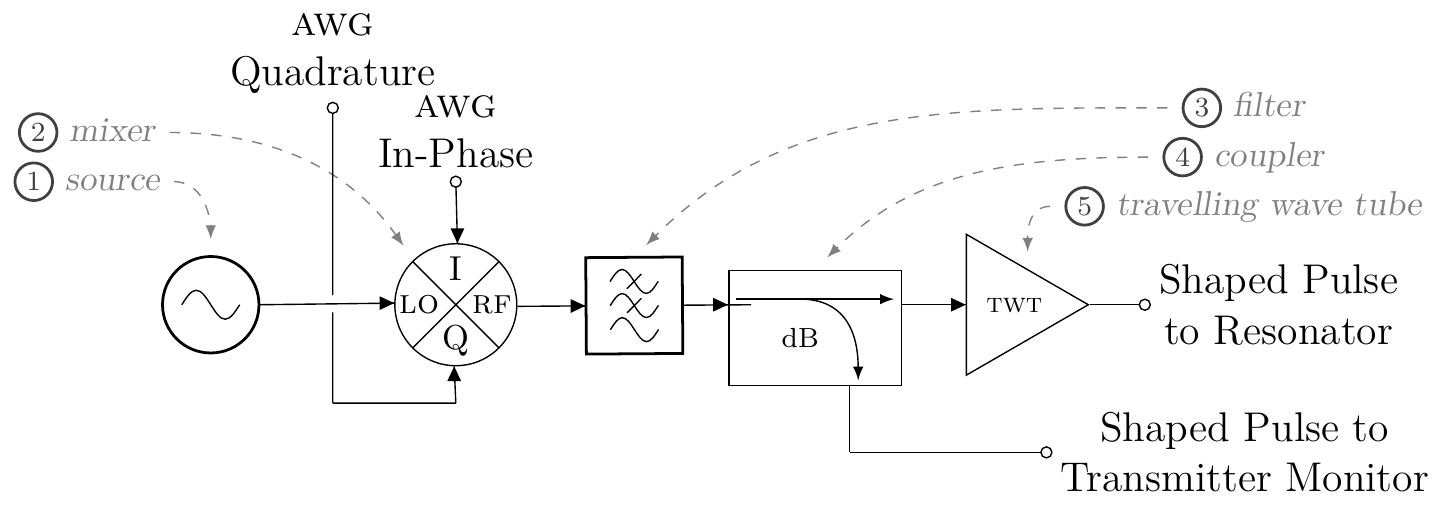}}
\caption{The Bruker SpinJet Schematics. \label{BlockDiagSchematics}}
\end{figure}

A schematic block diagram of an \textsc{awg} is shown in \cref{BlockDiagSchematics} and has the following function:
\begin{enumerate}
 \item A set of pulses, in the form of a specified code, is sent to the \textsc{awg} through an Ethernet port, processed by on-board computation hardware.
 \item Pulses are formed on an in-phase channel (I channel) and an $90^{\circ}$ out-of-phase, quadrature channel (Q channel).
 \begin{gather}
  \textsc{awg}_I=\cos{\omega_{a}^{} t}\nonumber\\
  \textsc{awg}_Q=\sin{\omega_{a}^{} t}\nonumber
 \end{gather}
 \item Two \textsc{awg} channels are connected to the I and Q ports of an IQ-mixer.
 \item The IQ-mixer takes the amplified signal from a microwave source, then splitting this into an in-phase channel and its quadrature channel
 \begin{gather}
  LO_I=\cos{\omega_{o}^{} t}\nonumber\\
  LO_Q=\sin{\omega_{o}^{} t}\nonumber
 \end{gather}
 where $\omega_{o}^{}$ is the angular frequency of the carrier, microwave signal (local oscillator). 
 \item Separately for the I and Q channels, the \textsc{awg} channel and corresponding microwave channel are mixed.
 \begin{gather}
  I=\cos{(\omega_{o}^{} -\omega_{a}^{})t}+\cos{(\omega_{o}^{} +\omega_{a}^{})t},\nonumber\\ Q=\sin{(\omega_{o}^{} -\omega_{a}^{})t}-\sin{(\omega_{o}^{} +\omega_{a}^{})t}\nonumber
 \end{gather}
 \item These two mixed signals are then recombined at the final stage of the IQ-mixer.
\end{enumerate}

\section{Xepr Python code}\label{Section:xeprScripts}

\lstset{
  basicstyle=\footnotesize\ttfamily,          
  xleftmargin=.1\columnwidth, xrightmargin=.1\columnwidth,
  language=Python,                          
  numbers=left,	                            
  frame=TB,
  numberstyle=\tiny\color{gray},
  rulecolor=\color{gray},
  commentstyle=\color[rgb]{0.0,0.6,0.0},    
  keywordstyle=\color[rgb]{0.0,0.0,1.0},    
  stringstyle=\color[rgb]{0.58,0.0,0.82},   
}

To programme a feedback loop on to the combination of \textsc{awg} and spectrometer, Python interpreted \textsc{XeprAPI} scripting can be used for operation of the \textsc{Xepr} application by Bruker BioSpin. \textsc{XeprAPI} can be easily combined with the language ProDeL\footnote{PROcedure DEscription Language embedded within the \textsc{Xepr} software} used in PulseSPEL files as a back-end to the \textsc{Xepr} console software.

\subsubsection{Load current experiment}

A header script must be invoked at the start of every \textsc{Xepr} Python script, to import libraries needed and to locate and load the \textsc{Xepr} API module. After the header, \texttt{XeprAPI.Xepr()} creates the \textsc{Xepr} object and connects to the \textsc{Xepr} software, the API has to be enabled within the \textsc{Xepr} GUI menu\footnote{Processing \(\to\) XeprAPI \(\to\) Enable Xepr} API. Also common to \textsc{Xepr} scripts is the command to create/access an experiment object of the \textsc{Xepr} application, invoked with \texttt{cur\_exp = Xepr.XeprExperiment()}, to access the current \textsc{Xepr} experiment or \texttt{config = Xepr.XeprExperiment(``AcqHidden'')} to access \textsc{Xepr}'s hidden experiment object.

The current experiment object must be imported to the current workspace, assuming there is a current experiment on the spectrometer. The hidden experiment object will become useful in modifying some parts of the experiment.

\begin{lstlisting}[language=Python]
# load the Xepr API module into Xepr structure
Xepr = XeprAPI.Xepr()

# define the current and hidden experiments
currentExp = Xepr.XeprExperiment()
hiddenExp = Xepr.XeprExperiment("AcqHidden")
\end{lstlisting}

\subsubsection{Acquire signal data}

Measure the 1D or 2D signal as an array of real numbers, corresponding to relative intensity of the measured signal. The current experiment object should already be imported to the workspace. The signal data should be stored in a plain text file to be used later, within Matlab.

Having established a connection to \textsc{Xepr} through the \textsc{XeprAPI} module, and accessed a current experiment within the \textsc{Xepr} GUI\footnote{An initial experiment should be set up within the \textsc{Xepr} GUI before these scripts are invoked; the intention is to modify values of variables already set within the \textsc{Xepr} GUI.}, the next task is to modify values of variables that will become the variables of the objective function.

\begin{lstlisting}[language=Python]
# run compiled experiment, wait until finished
currentExp.aqExpRunAndWait()

# acquire current signal dataset
dset = Xepr.XeprDataset()

# separate data into x, real(y), and imag(y)
ordinate = dset.O
x_signal = dset.X
rY_signal = ordinate.real
iy_signal = ordinate.imag
\end{lstlisting}

\subsubsection{Modify PulseSPEL variables}

PulseSPEL definitions from the definition file can be modified by initially parsing all PulseSPEL parameters contained in the experiment and then matching them with inputs to the Python script, the matched parameters can then be modified as instructed. The first input to the Python script is the variable name and the second is the new value.

\begin{lstlisting}[language=Python]
# get current PulseSPEL variables
Defs = currentExp.getParam("PlsSPELGlbTxt").value
Defs = Defs.split("\n")

# replace matched PulseSPEL variables
for value in Defs:
  if str(sys.argv[1]) in value:
    new = str(sys.argv[1])+" = "+str(sys.argv[2])
    currentExp["ftEPR.PlsSPELSetVar"].value = new
\end{lstlisting}

The script will read comma-separated variable names stored in the first line of the text file \texttt{data\_varsfile.txt}, which can be created from within Matlab, then match them to similarly ordered comma-separated variable values in the second line of the text file. Although it is not ideal to pass variables from one script to another in this way, it is a simple way which works given a waiting time between commands. The script also contains an error handling for variables defined in \texttt{data\_varsfile.txt} which have not been previously defined in the \textsc{Xepr} experiment. The command \texttt{cur\_exp[``ft\textsc{esr}.PlsSPELSetVar''].value = cmdStr} sets the current experiment's PulseSPEL variables to those scraped from the externally created \texttt{data\_varsfile.txt} file to the string \texttt{cmdStr}.

\subsubsection{Load shape file}

Assuming a plain text shape file has already been created, the code to send this shape to the experiment involves loading this shape file to the workspace, then compiling it. The argument of the Python script is the name of the shape file. When a maximum number of uploads is reached, 115, the current experiment should be reset (not shown here).

\begin{lstlisting}[language=Python]
# Load shapefile into PulseSPEL
Xepr.XeprCmds.aqPgShpLoad(str(sys.argv[1]))

# Show shapefile in PulseSPEL window
Xepr.XeprCmds.aqPgShowShp()

# Compile shapefile into experiment
Xepr.XeprCmds.aqPgCompile()
\end{lstlisting}

The final step is to run the current experiment with the modified variables with the command \texttt{cur\_exp.aqExpRunAndWait()} and wait for the experiment to finish.

Within an objective function, a Matlab optimisation should modify the values of the variables in the file \texttt{data\_varsfile.txt} at each iteration, then to call the Python script to send these modified values to the spectrometer. The result of the modified experiment should pass a metric to be maximised/minimised back to Matlab's optimisation; the objective of the objective function. A simple metric to use would be that which gives the best echo signal i.e. maximising the echo.

The echo signal data must be read from the \textsc{Xepr} software, again, this is achieved with the \textsc{XeprAPI}. Script that acquires the \textsc{Xepr} data with the command \texttt{dset = Xepr.XeprDataset()}, then to sort and store relevant data in three text files to be read by Matlab's objective function an interpreted as an echo. A simple echo signal maximisation would maximise the integral of the echo shape.

The optimal control strategy set out so far is no more than an automatic calibration, which a moderately experienced experimenter can achieve in very little time through the \textsc{Xepr} GUI. In fact, with error handling and waiting times included in the Python scripts outlined so far, each experimental modification will take a number of seconds: a simplex method would take a long time to search through this variable space to its optimal experimental settings. However, this is just a simple example to text the two-way communication between an optimisation algorithm and its ``hardware objective function''.

\section{Feedback Control Results Tables}\label{AWGResults}

\begin{table}
\begin{center}
\begin{small}
{\begin{tabular}{r c c c c c c }
\hline\hline Experiment: & \multicolumn{6}{c}{\textsc{oop-eseem}}\\\hline\\[-10pt]

Pulse sequence: & \multicolumn{6}{c}{$h\nu\to\frac{\pi}{2}\to\pi\to s(t)$}\\

Optimised Pulse: & \multicolumn{6}{c}{$\frac{\pi}{2}$-pulse}\\

$\frac{\pi}{2}$ pulse length: & \multicolumn{6}{c}{2ns$\leq$variable$\leq$160ns}\\

$\pi$ pulse length: & \multicolumn{6}{c}{32ns}\\

Phase cycling: & \multicolumn{6}{c}{16-step}\\

Transient averages: & \multicolumn{6}{c}{22}\\\hline\\[-10pt]

Algorithm: & \multicolumn{3}{c}{NM} & \multicolumn{3}{c}{MDS} \\

Pulse shape: & \multicolumn{3}{c}{wurst} & \multicolumn{3}{c}{wurst} \\
Phase offset: & \multicolumn{3}{c}{nome} & \multicolumn{3}{c}{$-\frac{\pi}{2}<$variable$\leq+\frac{\pi}{2}$} \\\hline\\[-10pt]

Iteration & \(\mX\) & \#\(f(\objv{})\) & \(\Obj{\objv{}}\) & \(\mX\) & \#\(f(\objv{})\) & \(\Obj{\objv{}}\) \\\hline\\[-10pt]

1:  & \(\mI\) & 4 & 0.20 & \(\mI\) & 5  & 1.25 \\
2:  & \(\mE\) & 2 & 0.39 & \(\mR\) & 8  & 1.29 \\
3:  & \(\mC\) & 2 & 0.56 & \(\mC\) & 16 & 1.46 \\
4:  & \(\mR\) & 2 & 0.79 & \(\mC\) & 8  & 1.63 \\
5:  & \(\mC\) & 2 & 1.03 &   &    &      \\
6:  & \(\mS\) & 7 & 1.20 &   &    &      \\
7:  & \(\mS\) & 7 & 1.20 &   &    &      \\
8:  & \(\mS\) & 7 & 1.20 &   &    &      \\
9:  & \(\mC\) & 2 & 1.20 &   &    &      \\\hline\\[-10pt]

Total experiments: & \multicolumn{3}{c}{$2\times$42} & \multicolumn{3}{c}{$2\times$37}\\

Total time: &\multicolumn{3}{c}{218s} & \multicolumn{3}{c}{99s} \\

Average f(x) time: & \multicolumn{3}{c}{5.2s} & \multicolumn{3}{c}{2.7s}\\\hline\hline 
\end{tabular}}
\end{small}
\end{center}
\caption[Feedback control results for OOP-ESEEM experiments]{Results of the feedback control optimisation of \textsc{oop-eseem} experiments using Nelder-Mead (NM) and multidirectional search (MDS) algorithms. \(\Obj{\objv{}}\) is the quality of the echo signal. \#\(f(\objv{})\) is the number of functional evaluations (2 experiments for \textsc{oop-eseem}). The step is the type of simplex \(\mX=\) at that iteration -- \(\mI=\) initial, \(\mR=\) reflect, \(\mE=\) expand, \(\mC=\) contract, \(\mS=\) shrink. The termination condition of the algorithms was a simplex size $<0.05$ or no progress was made after $150$ experiment evaluations were made during the contraction steps. In each case the objective vale was normalised to the same measure from the same experiments using conventional hard pulses.}\label{TAB-ExperimentalData2}
\end{table}

\cref{TAB-ExperimentalData1,TAB-ExperimentalData2} outline the results obtained in \cref{Section:FBresults} for the feedback control of 2-pulse Hahn echo experiments and \textsc{oop-eseem} experiments. The variables of optimisation are the piecewise triangular shape increments of a \(\pi\)-pulse in the echo experiments, and the \textsc{wurst} pulse shape parameters (pulse length, amplitude, phase offset, and pulse excitation bandwidth from \cref{EQN:wurst}) in the \textsc{oop-eseem} experiments.

\begin{table}
\begin{center}
\begin{small}
{\begin{tabular}{r c c c c c c c c c}
\hline\hline Experiment: & \multicolumn{9}{c}{2-pulse echo} \\\hline\\[-10pt]

Pulse sequence: & \multicolumn{9}{c}{$\frac{\pi}{2}\to\pi\to s(t)$} \\

Optimised Pulse: & \multicolumn{9}{c}{$\pi$-pulse} \\

$\frac{\pi}{2}$ pulse length: & \multicolumn{9}{c}{16ns} \\

$\pi$ pulse length: & \multicolumn{9}{c}{32ns} \\

Phase cycling: & \multicolumn{9}{c}{none}\\

Transient averages: & \multicolumn{9}{c}{200}\\\hline\\[-10pt]

Algorithm: & \multicolumn{3}{c}{NM} & \multicolumn{3}{c}{NM} & \multicolumn{3}{c}{NM} \\

Pulse shape: & \multicolumn{3}{c}{33 point} & \multicolumn{3}{c}{21 point} & \multicolumn{3}{c}{11 point}  \\
Phase offset: & \multicolumn{3}{c}{none} & \multicolumn{3}{c}{none} & \multicolumn{3}{c}{none}  \\\hline\\[-10pt]

Iteration & \(\mX\) & \#\(f(\objv{})\) & \(\Obj{\objv{}}\) & \(\mX\) & \#\(f(\objv{})\) & \(\Obj{\objv{}}\) & \(\mX\) & \#\(f(\objv{})\) & \(\Obj{\objv{}}\) \\\hline\\[-10pt]

1:  & \(\mI\) & 67  & 0.43 & \(\mI\) & 43 & 0.75 & \(\mI\) & 23 & 0.71  \\
2:  & \(\mR\) & 2   & 0.44 & \(\mE\) & 24 & 0.83 & \(\mE\) & 4  & 0.71  \\
3:  & \(\mE\) & 20  & 0.47 & \(\mE\) & 16 & 0.89 & \(\mE\) & 2  & 0.80  \\
4:  & \(\mE\) & 20  & 0.52 & \(\mE\) & 18 & 1.09 & \(\mE\) & 6  & 0.84  \\
5:  & \(\mE\) & 16  & 0.59 & \(\mE\) & 40 & 1.09 & \(\mE\) & 2  & 1.02  \\
6:  & \(\mE\) & 26  & 0.72 & \(\mE\) & 58 & 1.16 & \(\mE\) & 10 & 1.03  \\
7:  & \(\mE\) & 60  & 0.93 & \(\mE\) & 58 & 1.21 & \(\mE\) & 2  & 1.23  \\
8:  & \(\mE\) & 104 & 0.98 & \(\mE\) & 20 & 1.21 & \(\mE\) & 28 & 1.27  \\
9:  & \(\mR\) & 24  & 0.98 & \(\mR\) & 2  & 1.22 & \(\mR\) & 14 & 1.28  \\
10: & \(\mR\) & 2   & 0.98 & \(\mE\) & 36 & 1.24 & \(\mC\) & 26 & 1.28  \\
11: &   &     &       & \(\mR\) & 14 & 1.24 &   &    &      \\
12: &   &     &       & \(\mE\) & 16 & 1.24 &   &    &      \\
13: &   &     &       & \(\mR\) & 18 & 1.25 &   &    &      \\
14: &   &     &       & \(\mR\) & 16 & 1.27 &   &    &      \\\hline\\[-10pt]

Total experiments: & \multicolumn{3}{c}{341} & \multicolumn{3}{c}{389} & \multicolumn{3}{c}{119} \\

Total time: & \multicolumn{3}{c}{1030s} & \multicolumn{3}{c}{1273s} & \multicolumn{3}{c}{380s} \\

Average f(x) time: & \multicolumn{3}{c}{3.0s} & \multicolumn{3}{c}{3.3s} & \multicolumn{3}{c}{3.2s} \\\hline\hline
\end{tabular}}
\end{small}
\end{center}
\caption[Feedback control results for 2-pulse echo experiments]{Results of the feedback control optimisation of 2-pulse echo experiments using the Nelder-Mead (NM) algorithm. A number of contraction steps are omitted from the 2-pulse echo experiments, where no objective improvement was made. \(\Obj{\objv{}}\) is the quality of the echo signal. \#\(f(\objv{})\) is the number of functional evaluations (1 experiment for the 2-pulse echoes). The step is the type of simplex \(\mX=\) at that iteration -- \(\mI=\) initial, \(\mR=\) reflect, \(\mE=\) expand, \(\mC=\) contract, \(\mS=\) shrink. The termination condition of the algorithms was a simplex size $<0.05$ or no progress was made after $150$ experiment evaluations were made during the contraction steps. In each case the objective vale was normalised to the same measure from the same experiments using conventional hard pulses.}\label{TAB-ExperimentalData1}
\end{table}

\begin{figure}
\centering{\includegraphics{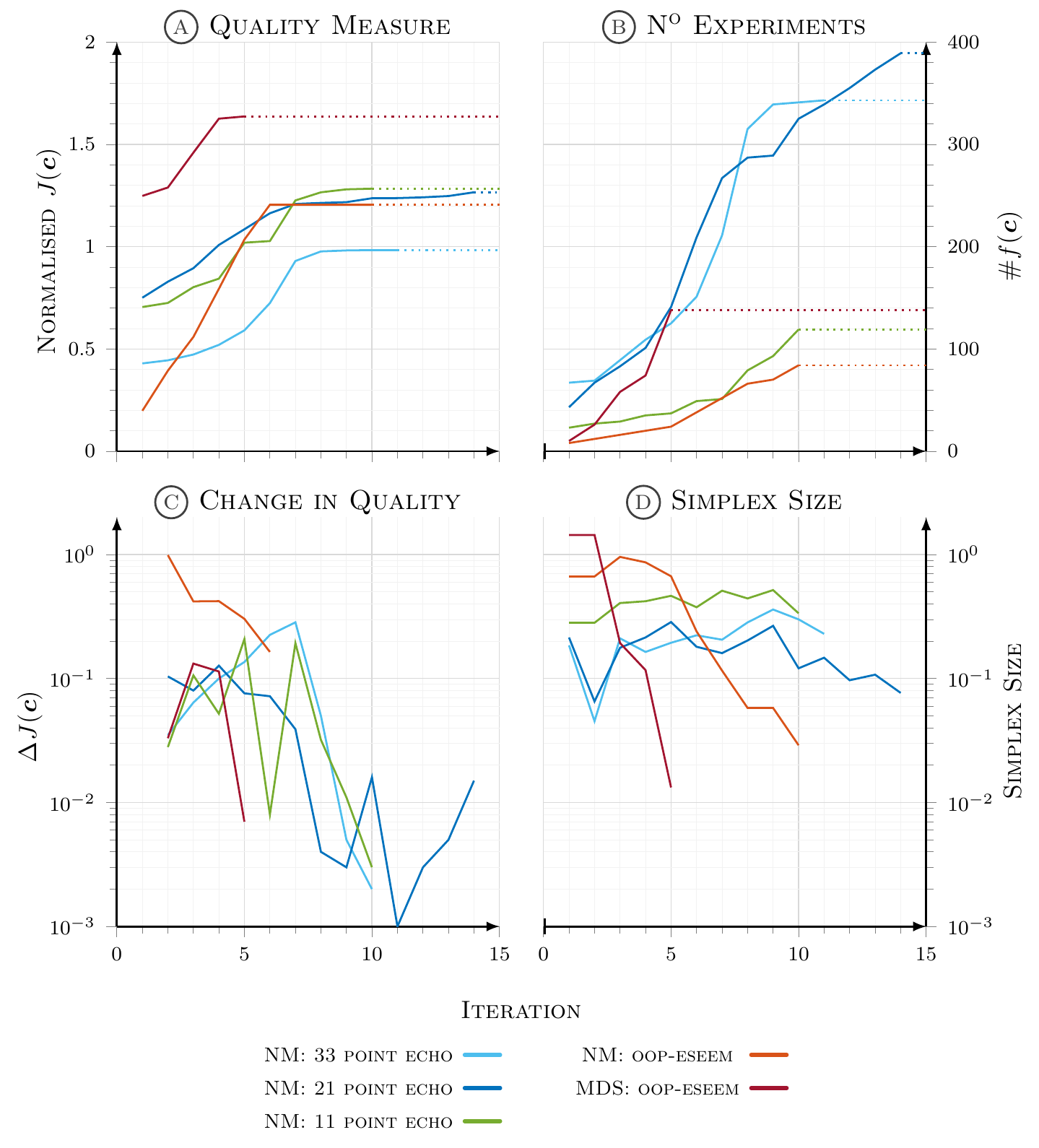}}
\caption[Convergence of Nelder-Mead and multidirectional search algorithms]{Convergence of Nelder-Mead (NM) and multidirectional search (MDS) algorithms. \plotlabel{a} shows the quality of the echo signal and \plotlabel{c} is its change. \plotlabel{b} shows the number of functional evaluations and \plotlabel{d} is a measure of simplex size. \label{FIG-ConvergenceData}}
\end{figure}

Echo experiments all start from a random \(\pi\)-pulse shape and \textsc{oop-eseem} experiments start with random parameters of the \textsc{wurst} pulse at 32~ns, seeded from the same random number. All \textsc{wurst} pulse shapes have a amplitude envelope of 25\% rise/fall time. Convergence characteristics of these feedback control experiments are shown in \cref{FIG-ConvergenceData}.

\end{appendices}
\backmatter
\newcommand{\doi}[1]{\url{https://doi.org/#1}}
\bibliographystyle{DLG_PhDThesis2017}
\bibliography{./literature/magres_general,./literature/optimcon,./literature/physics_general,./literature/numoptim,./literature/num_general,./literature/mri,./literature/esr,./literature/grape_citations,./literature/krotov_citations}

\AtEndDocument{\cleardoublepage}

\end{document}